\documentclass[opre, nonblindrev]{informs4}

\DoubleSpacedXI

\usepackage{natbib}
 \bibpunct[, ]{(}{)}{,}{a}{}{,}%
 \def\bibfont{\small}%
\usepackage[normalem]{ulem}
\useunder{\uline}{\ul}{}
\usepackage{algorithm, algorithmic, multirow, blkarray, xcolor,subcaption, subfiles, bm, diagbox, enumitem,comment,xcolor, mathtools}
\usepackage{tikz-cd}
\usepackage{pgfplots}
\usepackage{pgfplotstable}
\usetikzlibrary{calc}

\newcommand{\ldr}{\mathrm{LDR}}

\newcommand{\lt}{\left}
\newcommand{\rt}{\right}
\newcommand{\Real}{\mathbb{R}}
\newcommand{\expec}{\mathrm{E}}
\newcommand{\prob}{\mathrm{P}}

\newcommand{\bprime}{{b^{\prime}}}
\newcommand{\Oas}{{O_{\text{a.s.}}}}
\newcommand{\Tas}{{\Theta_{\text{a.s.}}}}
\newcommand{\tot}{\text{tot}}

\newcommand{\ACC}{\textup{ACC}}
\newcommand{\FN}{\textup{FN}}

\newcommand{\tk}[1]{{\color{magenta}{#1}}}

\TheoremsNumberedThrough     
\ECRepeatTheorems

\EquationsNumberedThrough    

\MANUSCRIPTNO{}

\begin{document}


\RUNAUTHOR{Kim, Kim, and Song}

\RUNTITLE{Selection of the Most Probable Best}

\TITLE{Selection of the Most Probable Best}

\ARTICLEAUTHORS{%
\AUTHOR{Taeho Kim}
\AFF{Department of Industrial and Systems Engineering, Texas A\&M University, TX 77843,\\  \EMAIL{thk5594@tamu.edu}}
\AUTHOR{Kyoung-Kuk Kim}
\AFF{College of Business, Korea Advanced Institute of Science and Technology, Seoul, Republic of Korea,\\ \EMAIL{kkim@kaist.ac.kr}} 
\AUTHOR{Eunhye Song}
\AFF{School of Industrial and Systems Engineering, Georgia Institute of Technology, GA 30332, \\ \EMAIL{eunhye.song@isye.gatech.edu}}
} 

\ABSTRACT{
We consider an expected-value ranking and selection (R\&S) problem where all $k$ solutions' simulation outputs depend on a common parameter whose uncertainty can be modeled by a distribution. We define the most probable best (MPB) to be the solution that has the largest probability of being optimal with respect to the distribution and design an efficient sequential sampling algorithm to learn the MPB when the parameter has a finite support.
We derive the large deviations rate of the probability of falsely selecting the MPB and formulate an optimal computing budget allocation problem to find the rate-maximizing static sampling ratios. The problem is then relaxed to obtain a set of optimality conditions that are interpretable and computationally efficient to verify. We devise a series of algorithms that replace the unknown means in the optimality conditions with their estimates 
and prove the algorithms' sampling ratios achieve the conditions as the simulation budget increases. 
Furthermore, we show that the empirical performances of the algorithms can be significantly improved by adopting the kernel ridge regression for mean estimation while achieving the same asymptotic convergence results. 
The algorithms are benchmarked against a state-of-the-art contextual R\&S algorithm and demonstrated to have superior empirical performances. }

%


\maketitle

%


\section{Introduction}
\label{sec:intro_RS}

When randomness in a simulation model is driven by input models estimated from finite data, the simulation output is subject to uncertainty caused by estimation errors in the input models. This additional uncertainty, distinguished from inherent stochastic simulation errors, is often referred to as \emph{input uncertainty}. If the simulation model is applied to find an optimal design or policy for the target real-world system, then input uncertainty must be accounted for to make a correct statistical inference on the performance of a selected solution. 
In this work, we consider a ranking and selection (R\&S) problem under input uncertainty focusing on the case when all $k$ solutions in comparison share a common input model with uncertain parameters. 
Then, the simulation output mean is a function of the  parameters, and consequently the optimum of the R\&S problem also depends on the input parameters. We assume that the uncertainty about the true parameter values is represented by a probability distribution function. Under this distribution, we define \emph{the most probable best} (MPB) as the solution that has the highest probability of being the best. 

Although the MPB can be defined with a continuous distribution on the parameters, we focus on the case where the distribution has a finite support with $B$ candidates in this paper. Then, finding the MPB leads to solving a nested R\&S problem; for each candidate input model, the \textit{inner-level} R\&S problem compares $k$ solutions to find the conditional best given the input model. Once all $B$ inner problems are solved, the probability of each solution being optimal can be computed. We refer to this probability as the \emph{preference probability} of each solution. The \textit{outer-level} R\&S problem selects the solution with the largest preference probability as the MPB.

The concept of the MPB has been applied in the medical decision-making literature~\citep{healthcare1,healthcare2}, where the goal is to assess the probability that each treatment is the most cost-effective given a patient's willingness-to-pay and uncertainties about a disease. These problems can be considered as an R\&S problem under input uncertainty.
If a Bayesian approach is taken so that the probability simplex on the input model is indeed a posterior distribution given data, then the MPB can be considered the optimal Bayes or maximum a posteriori decision rule~\citep{berger1985statistical, robert2007bayesian}. This decision rule returns a solution that maximizes the expected utility under the posterior; our utility is an indicator function, 
which returns $1$, if each solution is the best decision given a realized input model and $0$, otherwise. This type of decision rules have been applied to problems in finance and econometrics \citep{harvey2010portfolio, hirano2010decision}, marketing \citep{rossi2012bayesian}, and classification in machine learning \citep{bishop2006pattern}.

Several frameworks have been proposed to account for input uncertainty in R\&S and optimization via simulation in general. We categorize them into three groups according to the treatment of input uncertainty in their problem formulation. 
The first approach is to apply a risk measure with respect to input model estimation error to the performance measure of each solution, and then to compare the risk measure values;  \cite{corlu2015, wu2019, pearce2017}, and \cite{ungredda2020} use a risk-neutral measure (mean) and 
\cite{xie2015,wu2018bayesian} and \cite{helin2020} explore (conditional) value at risk. 
The second category takes the distributionally robust optimization  approach~\citep{gao2017,fan2020distributionally}; it first assumes an ambiguity set on the input model, finds the worst-case input model for each solution within the ambiguity set, and selects the solution with the best worst-case performance.  
The last category focuses on providing a probability guarantee that the best solution chosen under the best estimate of the real-world input distribution is in fact optimal; \cite{corlu2013,songnelsonhong2015} and \cite{song2019} take this view, however, point out that when estimation error of the input model is large, the probability guarantee may be lower than the desired level. The MPB formulation does not fit in any of the three categories above, which warrants its exploration. 


 The MPB formulation also lets us assess the risk of selecting a suboptimal solution in the R\&S problem due to the discrepancy between the simulator and the target system caused by input uncertainty. 
If the MPB's preference probability clearly dominates all other solutions', then the risk is low. On the other hand, if there are several solutions whose preference probabilities are close to the MPB's, then collecting additional data to reduce the uncertainty about the input model may reduce the risk of making a suboptimal decision. \cite{KimSong2022} build upon this idea to formulate a budget allocation problem for input data collection when there are several data sources to make the MPB converge to the true optimum at the fastest rate.

Our objective is to devise a sequential R\&S procedure to find the MPB efficiently. To estimate the mean performance of each solution-input model pair at any point in the algorithm, we adopt the Bayesian machinery to learn the mean performance as in~\cite{ryzhov2016convergence}; we impose the normal-normal conjugate models and independent priors for all pairs. The probability of false selection (PFS) is defined as the probability that the estimated MPB determined from the posterior means is not the true MPB. We take the fixed-budget approach that aims to minimize the PFS given the budget. However, solving the optimal computing budget allocation (OCBA) problem exactly for a finite budget is challenging even for a classical R\&S problem~\citep{chen2000simulation}. Instead, \cite{glynn2004large} derive the large deviations rate (LDR) of the PFS and find the optimality conditions for the \textit{static} sampling ratios that maximize the LDR for the classical R\&S problem. We take the same approach to derive the optimality conditions for the asymptotically optimal static sampling ratios for the MPB selection problem. These optimality conditions depend on the unknown means of simulation outputs. To reconcile, we replace the unknown means with the posterior means of the Bayesian model and utilize the sample-version of the optimality conditions to propose several sequential sampling algorithms.

Although such plug-in approaches have been widely adopted in the R\&S literature~\citep{glynn2004large, pasupathy2014stochastically,Feldman-BORS:18,Appleaget-MORS:20,gao2019selecting,chen2022BOLD},
the static allocation is only asymptotically optimal and the plug-in algorithm's performance may be far from optimal~\citep{peng2016dynamic, peng2018ranking}. Indeed, for our problem, a plug-in algorithm (Algorithm~\ref{alg:rate_opt}) may not converge to the optimal static sampling ratio asymptotically due to the complex nested problem structure that the MPB problem has.
Instead, the version (Algorithm~\ref{alg:rate_opt_posterior}) that combines the plug-in estimates and posterior sampling is shown to satisfy the optimality conditions for the static allocation ratios asymptotically.

Alternatively, a dynamic program (DP) can be formulated to maximize the expected reward (e.g., $0$-$1$ reward or negative regret) after a finite simulation budget is spent. However, solving the DP exactly is computationally challenging even for a moderate problem size. Instead, myopic or one-step optimal policies have been proposed and studied; see~\cite{chick2001new, frazier2008knowledge, ryzhov2016convergence, peng2017myopic} for instance. 

Notably, there are some recent work that bridge the gap between the two abovementioned approaches by proving that a sequential sampling algorithm based on a myopic policy indeed achieves the asymptotically optimal static allocations as the simulation budget grows. 
\cite{chen2019complete} propose a sequential sampling algorithm based on the complete expected improvement (CEI), which is originally created as a myopic sampling policy~\citep{salemi2019}, and prove that its sampling ratios for the solutions asymptotically satisfy the optimality conditions in \cite{glynn2004large}. \cite{Avic23:gECI} derive the same result for the policy that applies the gradient of the CEI with respect to the sample size as a sampling rule.

While these results are desirable, in this work, we focus on studying the optimal static allocation ratios and creating sequential policies from them as the first pass to tackle the MPB selection problem. We believe that these results will provide a foundation to create and analyze a DP-based sampling rule for the MPB formulation in a future study.



If one views a candidate input model as a context, our work can be connected to the contextual R\&S studied by \cite{gao2019selecting,shen2021ranking}, and \cite{Li:22DSCO}. 
The contextual R\&S aims to maximize the mean (or worst-case) probability of correct selection over all possible covariates. Therefore, there is no outer-level R\&S problem for the contextual R\&S as in the MPB formulation.
The difference between these two formulations manifests in the optimality conditions for the respective OCBA problems as we discuss in Section~\ref{sec:probable_best}. 

Because of the unique nested structure, deriving the LDR of the PFS for the MPB formulation has a technical challenge absent in classical or contextual R\&S. The outer-level R\&S problem's correct selection depends on how well the inner-level problems are solved under each candidate input model. However, the importance of each inner-level problem for achieving the outer-level correct selection differs; even if false selection occurs at some inner-level problems, the MPB  may still be found correctly.  
Hence, the MPB formulation requires \textit{quantifying the importance of each inner-level problem for outer-level correct selection};  there is no straightforward extension of an existing R\&S algorithm that provides this feature. We introduce the \emph{balance weight} to measure how important it is to sample each solution-input model pair in achieving the outer-level correct selection. The balance weights appear in the optimality conditions for our OCBA problem.

We also prove asymptotic convergence of the sequential sampling algorithm for the contextual R\&S problem stated in~\cite{gao2019selecting} with only a sketch of proof provided at the time of the submission of this paper. This result is an intermediate step to show our Theorem~\ref{thm:proof_BWalg} and we could not find a rigorous proof in the literature at the time of the submission of this paper.

Moreover, we further improve the finite-sample performance of our sequential algorithm by adopting kernel ridge regression (KRR) to replace the posterior mean estimates. This approach exploits spatial inference by pooling simulation outputs run at different input model parameters and improves the estimation error of the plug-in means at the early stage of the algorithm. Furthermore, we prove that the large deviations analysis leads to the same optimality condition for static allocation ratios as when posterior means are used.

A preliminary version of this work published in a conference proceeding discusses the optimality conditions for the static sampling allocations without a proof and the sequential sampling algorithms are analyzed empirically~\citep{kimkimsong}. In this work, we provide a complete set of theoretical analyses for both static and sequential sampling algorithms we propose.

The remainder of the paper is organized as follows. In Section~\ref{sec:prob_formulation},  we formally define the MPB selection problem. Section~\ref{sec:learning.model} discusses the Bayesian learning model for the MPB estimation.
Section~\ref{sec:probable_best} presents the OCBA formulation for the MPB problem and its relaxation schemes. The optimality conditions for the static sampling ratios are derived for a relaxed OCBA problem. Section~\ref{sec:sequential_learning} proposes sequential sampling algorithms and analyzes their asymptotic properties. Section~\ref{sec:GP_extension} incorporates the KRR to obtain better finite-sample performances. Section~\ref{sec:experiments} provides numerical performance analyses for all proposed algorithms. In particular, a market simulation example is presented, where the MPB formulation is applied to find the market-share maximizing product under economic uncertainties. Conclusions are given in Section~\ref{sec:conclusion}. Proofs of all theoretical results in the paper are in the Electronic Companion of this paper.


\section{Problem Definition}\label{sec:prob_formulation}



Suppose the input model of the simulator is parameterized by $\theta \in \mathcal{X}$ whose true value is unknown, but its uncertainty can be modeled by some distribution $\pi$ defined on $\mathcal{X}$. 
Let   $y_i(\theta) : = \expec\lt[Y_i(\theta) | \theta\rt]$ be the conditional mean of the $i$th solution's simulation output given $\theta$ for  $1\leq i\leq k$.  The \textit{most probable best} (MPB) is defined as
\begin{equation}\label{eq:target}
    i^* := \argmax\nolimits_{1 \leq i\leq k} \prob_{\pi}\lt\{y_i(\theta) \leq  y_j(\theta), \forall j\neq i   \rt\}.
\end{equation}
We refer to the probability in~\eqref{eq:target} as the \emph{preference probability} of the $i$th solution, the probability that Solution $i$ performs no worse than any other solutions under $\pi$. Therefore, the MPB is the  (potentially nonunique) solution with the largest preference probability under $\pi$. Definition~\eqref{eq:target} is agnostic to whether $\pi$ is constructed under a Bayesian or frequentist framework. Throughout the paper, we assume that fixed $\pi$ is given.

Although the MPB can be defined for general distribution $\pi$, in this paper, we focus on finding the MPB when $\theta$ has finite support $\Theta = \{\theta_1,\theta_2,\ldots, \theta_B \}\subseteq \mathcal{X}$.  Thus, $\pi$ can be represented as a probability simplex $\pi=\{p_1,p_2,\ldots,p_B|\sum_{b} p_b = 1\}$ on $\Theta$. Under this assumption, \eqref{eq:target} can be rewritten as
\begin{equation}\label{eq:target_approx}
    i^* = \argmax\nolimits_{1\leq i\leq k} \sum\nolimits_{b=1}^{B} p_b\mathbf{1}\lt\{y_i(\theta_b) = y_{i^b}(\theta_b)\rt\},
\end{equation}
where $i^{b}: = \argmin_{i} y_i(\theta_b)$ is the conditional optimum at $\theta_b$. Finding $i^*$ can be formulated as a nested R\&S problem; selecting $i^b$ given $\theta_b$ corresponds to an inner-level R\&S problem. The outer-level R\&S problem then compares all $k$ solutions' preference probabilities to find $i^*$.

To simplify the later analysis, we assume (i) for each $\theta_b$, $y_{i}(\theta_b) \neq y_j(\theta_b)$ holds for all $i \neq j$, which implies $i^b$ is unique for each $\theta_b$; and (ii) $i^*$ is unique. Assuming uniqueness of the optimum is common in the OCBA  literature for the rate analysis purposes.  We discuss how to modify our framework when $i^*$ is non-unique in Section~\ref{ec:nonunique.mpb}. With these assumptions, \eqref{eq:target_approx} further simplifies to $i^* = \argmax_{1\leq i\leq k} \sum_{b=1}^{B} p_b\mathbf{1}\lt\{i = i^b\rt\}$. 
Additionally, we define the \textit{favorable set} of $i$ to be 
\begin{equation}\label{eq:favorable.set}
\Theta_i=\{\theta_b: i=i^b, 1\leq b \leq B\},
\end{equation}
the set of $\theta$s  at which $i$ is optimal. The preference probability of $i$ can  be written as $\prob_\pi(\Theta_i)$.

In general, $y_i(\theta_b)$ has no known analytical expression and must be estimated via simulation.
Our objective is to create an algorithm that sequentially simulates some $(i,\theta_b)$ pair at each iteration to find the MPB in~\eqref{eq:target_approx} efficiently given a simulation budget.
Such an algorithm returns $i^*_n$ as an estimate for $i^*$ after running $n$ replications. In case of ties in the estimated preference probabilities, $i^*_n$ may be a set of solutions. 

One measure of efficiency of the algorithm is the \emph{probability of correct selection} (PCS), where the correct selection event  
is defined as CS$:= \{i_n^* = \{i^*\}\}$. Given $n$, an efficient algorithm would maximize the PCS. Equivalently, one can define the \emph{false selection} event as FS$: = \{i_n^* \neq \{i^*\}\}$ and minimize the \emph{probability of false selection} (PFS); we take this approach. Any event of a tie is included in FS as $i^*$ is assumed to be unique.

The MPB problem is closely related to the contextual R\&S problem, which regards $\theta$ as a context that parameterizes each R\&S problem and aims to maximize  
the average PCS or worst-case PCS~\citep{gao2019selecting,shen2021ranking, Li:22DSCO, Cakmak-GPCOBCA:24} with respect to possible values of $\theta$. 
Contextual R\&S puts an emphasis on solving the R\&S problem at each $\theta$ equally well, which clearly differs from solving~\eqref{eq:target_approx} to find $i^*$. Nevertheless, one can still apply a contextual R\&S algorithm to solve~\eqref{eq:target_approx}; if each R\&S problem at fixed $\theta$ is solved correctly, then $i^*$ is found correctly. However, this would not be as efficient as algorithms tailored to solve~\eqref{eq:target_approx} as our empirical results in  Section~\ref{sec:experiments} indicate.


\section{Learning Models for the MPB Estimation} \label{sec:learning.model}

As the simulation budget is allocated sequentially, we need a mechanism to estimate $y_i(\theta_b)$ from simulation results. In this section, we introduce a learning model for $y_i(\theta)$ and the MPB estimator, $i^*_n$, derived from the model. 

Let us consider  Bayesian estimator $\eta_i(\theta)$ for $y_i(\theta)$ for any $(i,\theta)$ pair by assuming 
\begin{equation}\label{eq:normaL_model}
    \begin{aligned}       
        \eta_i(\theta)  \sim  N({\mu}_{i, 0}(\theta), \sigma^2_{i, 0}(\theta)),\hspace{20pt}
        Y_i(\theta)  \sim  N(\eta_i(\theta), \lambda^2_i(\theta)),
    \end{aligned}
\end{equation}
where $\lambda_i(\theta)$ is the simulation error variance of $Y_i(\theta)$ given $\theta$, and $\mu_{i,0}(\theta)$ and $\sigma_{i,0}^2(\theta)$ are the mean and variance of the prior distribution of $\eta_i(\theta)$, respectively. Notice that~\eqref{eq:normaL_model} replaces the mean of  $Y_i(\theta)$, $y_i(\theta)$, with its Bayesian estimator $\eta_i(\theta).$ We assume all solution-parameter pairs are simulated independently, i.e., no correlation between $Y_i(\theta_b)$ and $Y_\ell(\theta_c)$ for $(i,b)\neq(\ell,c)$.

After $n$ simulation replications are made, let  $N^n_i(\theta)\geq 1$ denote
the number of replications allocated 
to $(i,\theta)$. 
Conditional on the simulation outputs, $Y_{i1}(\theta),Y_{i2}(\theta),\ldots,Y_{i N^n_i(\theta)}(\theta)$,  the posterior mean and variance, $\mu_{i,n}(\theta)$ and $\sigma_{i,n}^2(\theta)$, of $\eta_i(\theta)$ are updated as 
\begin{equation}\label{eq:update_normal}
\begin{aligned}
    \mu_{i, n}(\theta) = \sigma_{i,n}^2(\theta) \lt(\frac{\mu_{i, 0}(\theta)}{\sigma_{i,0}^2(\theta)} + \frac{N^n_i(\theta)}{\lambda_i^2(\theta)}\frac{1}{N^n_i(\theta)} \sum\nolimits_{r=1}^{N^n_i(\theta)}Y_{ir}(\theta) \rt), \;\;
        \sigma_{i,n}^2(\theta) = \lt(\frac{1}{\sigma_{i,0}^2(\theta)} + \frac{N^n_i(\theta)}{\lambda_i^2(\theta)}\rt)^{-1}.
\end{aligned}   
\end{equation}
Given the noninformative prior, $\sigma_{i, 0}^2(\theta) = \infty$, \eqref{eq:update_normal} simplifies to
$\mu_{i, n}(\theta) =  \sum_{r=1}^{N^n_i(\theta)}Y_{ir}(\theta)/N^n_i(\theta)$ and $\sigma_{i, n}^2(\theta) = \lambda_i^2(\theta)/N^n_i(\theta)$ regardless of the choice for $\mu_{i,0}(\theta)$. Then, $\mu_{i, n}(\theta)$ and $\sigma_{i, n}^2(\theta)$ are identical to the sample mean and its variance, respectively, in the frequentist's setting.
In the algorithms proposed in Section~\ref{sec:sequential_learning}, we run a warm-up experiment to obtain the initial posterior estimates.

There are different ways to utilize Model~\eqref{eq:normaL_model} to solve the MPB problem. One way is to take a purely Bayesian approach and regard $y_i(\theta)$ to be random rather than a fixed value. Under this viewpoint, the definition of the MPB in~\eqref{eq:target} needs to be recast as 
\begin{equation}\label{eq:Bayesian_MPB}
    \argmax\nolimits_{1\leq i\leq k}\prob\{\eta_i(\theta) \leq \eta_j(\theta), \forall j\neq i|\mathcal{X}_n\},
\end{equation}
where $\mathcal{X}_n$ is the set of simulation outputs accumulated up to the $n$th replication, and the probability is taken with respect to the joint distribution of $\{\eta_i(\theta)\}_{1\leq i\leq k}$ conditional on $\mathcal{X}_n$ and $\boldsymbol{\pi}$. If $\pi$ is also a posterior distribution defined on $\Theta$ given some data, then~\eqref{eq:Bayesian_MPB} can be interpreted as the solution that maximizes the posterior preference probability.

Alternatively, one may utilize Model~\eqref{eq:normaL_model} as a vehicle to obtain the point estimator, $\mu_{i,n}(\theta)$,  of $y_i(\theta)$ whose true value is \textit{fixed but unknown}; \citet{chick2001new, frazier2008knowledge} take this approach in solving the classical R\&S problem. We also take this view and plug the point estimators,  $\mu_{i,n}(\theta_b), 1\leq i \leq k, 1\leq b\leq B$, into~\eqref{eq:target_approx} to define the following estimators of $i^*$ and $i^b$:
\begin{equation}\label{def:condopt-mpb}
    i^b_n = \argmin\nolimits_{1\leq i\leq k} \mu_{i,n}(\theta_b)\;\; \text{and}\;\; i^*_n = \argmax\nolimits_{1\leq i\leq k} \sum\nolimits_{b=1}^B p_b\mathbf{1}\left\{i=i^b_n\right\}.
\end{equation}
Note that $i^b_n$ is unique almost surely under Model~\eqref{eq:normaL_model}; in fact, it suffices to have a continuous simulation output distribution.  Meanwhile, there may be ties at the outer-level problem for finite $n$, particularly when $p_b = 1/B$ for all $1\leq b\leq B$ even if $i^*$ is unique.

In Section~\ref{sec:relax_OCBA}, we briefly discuss how to extend our framework to non-normal cases. 
Note that $\lambda_i(\theta)$ is assumed to be known throughout the paper until Section~\ref{subsec:vco} for simplicity. To account for unknown variances,~\eqref{eq:normaL_model} can be replaced with a normal-gamma model; See~\ref{apdx:unknown_variance} for details.

In addition to the plug-in estimates discussed above, one of our sequential algorithms proposed in Section~\ref{sec:sequential_learning} (Algorithm~\ref{alg:rate_opt_posterior}) utilizes the full posterior distribution of $\eta_i(\theta_b)$ for some $(i,\theta_b)$ pairs to better account for the estimation error in $y_i(\theta_b)$ when making the sampling decision.

Although Model~\eqref{eq:normaL_model} assumes independent priors and simulations for all $(i,\theta_b)$ pairs, potentially significant computational efficiency can be gained by 
 (i) modeling spatial correlation between $\eta_i(\theta)$ and $\eta_i(\theta^\prime)$ for $\theta \neq \theta^\prime$; and (ii) 
adopting common random numbers (CRNs) to simulate all $k$ solutions at each $\theta_b$. We address (i) in Section~\ref{sec:GP_extension}, however, analyzing the correlations in simulation outputs induced by CRNs in the OCBA framework could be very demanding. \citet{fu2007simulation} show that even for the classical R\&S problem, the exact asymptotically optimal sampling ratios can only be derived for when $k=2$. For $k\geq 3$, they provide an approximate scheme by maximizing a surrogate of PCS. The challenge is exacerbated for the MPB problem due to the nested R\&S structure. Thus, we only consider independent simulations in this paper leaving the CRN extension to future research.

\section{Asymptotic Analysis of Sampling Allocation}
\label{sec:probable_best}

Characterizing the PFS for finite $n$ is challenging in general for an R\&S problem, which is further complicated by the nested R\&S structure for the MPB problem. To gain intuitions to design sequential sampling algorithms, this section first establishes the asymptotic convergence rate of PFS assuming all $\{(y_i(\theta_b), \lambda^2_i(\theta_b))\}$s to be fully known. Then, we formulate and solve the OCBA problem to maximize the convergence rate when $n\to\infty$ to find the asymptotically optimal static sampling ratios. These ratios are later utilized in Section~\ref{sec:sequential_learning} to design sequential sampling algorithms.

In general, the large-deviations rate (LDR) of sequence of rare events $\{E_n\}$ is defined as
$\lim_{n \rightarrow \infty} -\frac{1}{n}\log \prob(E_n) = I,$
provided that $I>0$ exists. Namely, $I$ represents the exponential decay rate of $\lt\{P(E_n)\rt\}$ as $n$ increases. When $I$ does not exist, one can consider $\lim\inf_{n \rightarrow \infty} -\frac{1}{n}\log \prob(E_n)$. 

If simulation budget $n$ is spent judiciously, the FS event becomes a rare event as $n$ increases.
We first focus on characterizing the static allocation rule that maximizes the LDR of FS. Let $\bm{\alpha}$ be the vector of $\{\alpha_i(\theta_b)\}_{1\leq i\leq k, 1\leq b\leq B}$, where each $\alpha_i(\theta_b)$ is the static sampling ratio of $n$ allocated to $(i, \theta_b)$.
In the remainder of the section, all theoretical results are shown for a deterministic sampling rule that satisfies  $\lim_{n \rightarrow \infty}{N^n_i(\theta_b)}/{n} = \alpha_i(\theta_b)$ and $N_i^n(\theta_b) \rightarrow \infty$ for all $(i, \theta_b)$ given any $\boldsymbol{\alpha}$. Note that we allow $\alpha_i(\theta_b) = 0$ for some $(i,\theta_b)$, i.e., $N^n_i(\theta_b)$ increases sublinearly in $n$. 
A propose sequential sampling algorithm proposed in Section~\ref{sec:sequential_learning} achives thse conditions in the limit.

Let $\ldr_\text{FS} := \liminf_{n\rightarrow \infty} - \frac{1}{n}\log\prob\lt(i_n^* \neq \{i^*\}\rt)$.
Then, the OCBA problem of our interest is  
\begin{equation}\label{opt:OCBA_LDR}
    \begin{aligned}
        \max\nolimits_{\bm{\alpha}}\ldr_\text{FS}
         \;\;\;\; \textrm{subject to} \;\;  \sum\nolimits_{i=1}^{k}\sum\nolimits_{b=1}^{B} \alpha_i(\theta_b) = 1, \alpha_i(\theta_b) \geq 0, 1\leq i \leq k, 1\leq b \leq B.
    \end{aligned}
\end{equation}
Problem \eqref{opt:OCBA_LDR} finds  $\bm\alpha$ that maximizes the limit infimum on the exponential decay rate of the PFS.
In Section~\ref{subsec:OCBA}, we  derive $\ldr_\text{FS}$ as a function of $\bm\alpha$.



\subsection{Exact Optimal Computing Budget Allocation Formulation}
\label{subsec:OCBA}

The FS event occurs only if at least one inner problem falsely selects its conditional optimum, but not all inner false selections are equally critical for CS. For instance, even if $i^*$ is falsely deemed suboptimal for some $\theta_b\in\Theta_{i^*}$, as long as its estimated preference probability still dominates all other solutions', then CS still occurs.
Hence, the key to characterizing $\text{LDR}_{\text{FS}}$ is to understand how critical each inner problem is to FS.

Let  $d_j$ be the  difference in preference probabilities between $i^*$ and $j$: 
\begin{equation}\label{def:d_j}
    \begin{aligned}
        d_j &:= 
        \sum\nolimits_{b=1}^{B}p_b\mathbf{1}\lt\{i^* = i^{b}\rt\} - \sum\nolimits_{b=1}^{B} p_b\mathbf{1}\lt\{j = i^{b}\rt\}. 
    \end{aligned}
\end{equation}
Namely, $d_j$ measures how much inferior $j$ is to $i^*$ at the outer-level problem; the smaller $d_j$, the more competitive $j$ is against $i^*$. Because both sums in~\eqref{def:d_j} are in $[0, 1]$,  $d_j \in [-1, 1]$. Since $i^*$ is unique by assumption, $d_j > 0 $ for all $j \neq i ^*$ and $d_{i^*} = 0$. 

To jointly consider all inner problems, we define
\begin{equation}\label{eq:set.of.M}
    \mathcal{M} := \lt\{\BFM| \BFM \in \{0, 1\}^{k \times B}, \BFM^\top\BFone_k = \BFone_B \rt\},
\end{equation} 
where $\BFone_k$ denotes the $k$-dimensional column vector of ones. For each $k\times B$ matrix $\BFM\in\mathcal M$, let $m_{i,b}$ denote its $(i,b)$th element. The equality constraint in~\eqref{eq:set.of.M} implies that only one $m_{i,b}$ in each column of $\BFM$ equals $1$.  If we regard $m_{i,b}$ to be an indicator variable to specify whether $i$ is the conditional optimum at $\theta_b$, then $\mathcal M$ is the collection of all possible conditional optima mappings for a generic $k\times B$ problem while only one element of $\mathcal M$ corresponds to the true configuration of Problem~\eqref{eq:target_approx}. Let $\BFM^*$ denote the true mapping. Then, $\mathcal M \backslash \left\{\BFM^*\right\}$ enumerates all possible ways one or more inner-level false selections jointly occur. The question is, which of these mappings lead to FS in the outer level? To answer this, we define $d_j(\BFM)$ and $I(\BFM)$ 
for each $\BFM \in \mathcal{M}$ as
\begin{equation}\label{def:d_i}
    d_j(\BFM) := \sum\nolimits_{b=1}^{B}p_b m_{i^*,b} - \sum\nolimits_{b=1}^{B}p_b m_{j,b} \;\;\; \mbox{and} \;\;\; I(\BFM) := \lt\{(i, \theta_b)| m_{i,b} = 1, i \neq i^b\rt\}.
\end{equation}
In words, $d_j(\BFM)$ measures the difference in preference probabilities of $i^*$ and $j$ under $\BFM$, and $I(\BFM)$ is the set of misspecified conditional optimum-parameter pairs under $\BFM$. By definition, $d_j(\BFM^*) = d_j$ for all $1\leq j\leq k,$ and $I(\BFM^*) = \emptyset.$ 

Let us consider realized conditional optima $\lt\{i_n^b\rt\}_{b =1}^B$ 
after $n$ simulation replications. Define $k\times B$ matrix $\BFM_n$ whose $(i,b)$th element is $m_{i,  b}^n = \mathbf{1}\{i = i_n^b\}.$ Then, there exists a unique element of $\mathcal M$ that coincides with $\BFM_n$. 
Recall that we define FS as the event such that there exists $j\neq i^*$ that ties with or dominates $i^*$ at the outer-level problem. This implies that $\BFM_n$ leads to FS, if and only if, $d_j(\BFM_n)\leq 0$ 
for some $j \neq i^*$. Namely, FS$=\bigcup_{j \neq i^*} \lt\{\BFM_n \in \mathcal{A}_j\rt\}$, where for each $j \neq i^*$
\begin{equation}\label{def:A_i}
    \mathcal{A}_j := \lt\{\BFM \in \mathcal{M}| d_j(\BFM)\leq 0\rt\} 
\end{equation}
is a collection of mappings that make $j$ as good as or better than $i^*$ at the outer-level problem. 
Observe that 
$\max_{j \neq i^*} \prob(\BFM_n \in \mathcal{A}_j) \leq$ PFS $ \leq \sum_{j \neq i^*}\prob(\BFM_n \in \mathcal{A}_j)$, which  implies
\begin{eqnarray}
    \ldr_{\mathrm{FS}} = \liminf\nolimits_{n \rightarrow \infty} -\tfrac{1}{n}\log \text{PFS} = \min\nolimits_{j \neq i^*}  \liminf\nolimits_{n \rightarrow \infty} -\tfrac{1}{n} \log \prob(\BFM_n \in \mathcal{A}_j). \label{eq:LDR_min}
\end{eqnarray}
Thus,~\eqref{eq:LDR_min} stipulates that characterizing $\ldr_{\mathrm{FS}}$ boils down to finding the LDR of $\prob\lt(\BFM_n \in \mathcal{A}_j\rt)$ for all $j\neq i^*$.
The following theorem formally states this result. 

\begin{theorem}\label{thm:Gtilde}
    Let $x^+ = \max(x, 0)$. Given fixed $\boldsymbol{\alpha}$, define 
    \small
    \begin{align}
        \widetilde{G}_i(\theta_b) := \min_{x\in [y_{i^b}(\theta_b), y_i(\theta_b)]}\bigg\{\frac{\alpha_{i^b}(\theta_b)}{2\lambda^2_{i^b}(\theta_b)}(x-y_{i^b}(\theta_b))^2 &+ \frac{\alpha_i(\theta_b)}{2\lambda_i^2(\theta_b)}(x - y_i(\theta_b))^2 \nonumber \\
        &+ \sum_{\ell\neq i^b: y_\ell(\theta_b) < y_i(\theta_b)}\frac{\alpha_\ell(\theta_b)}{2\lambda_\ell^2(\theta_b)} \lt[(x - y_\ell(\theta_b))^+\rt]^2\bigg\} \label{def:G_tilde}
    \end{align}  \normalsize
    and $\widetilde{\ldr}_{j, i^*} := \min\nolimits_{\BFM \in \mathcal{A}_j} \sum\nolimits_{(i, \theta_b) \in I(\BFM)} \widetilde{G}_{i}(\theta_b)$. 
    Then, we have (i) $\liminf_{n\rightarrow \infty}-\frac{1}{n}\log\prob(m_{i, b}^n = 1) = \widetilde{G}_i(\theta_b)$; (ii)  $\liminf\nolimits_{n \rightarrow \infty} -\tfrac{1}{n} \log \prob(\BFM_n \in \mathcal{A}_j) = \widetilde{\ldr}_{j, i^*}$ for $j\neq i^*$; and (iii) $\ldr_\textup{FS} = \min_{j \neq i^*}\widetilde{\ldr}_{j, i^*}$.
\end{theorem}

Part (i) states that $\widetilde G_i(\theta_b)$ is the LDR of the event such that $i\neq i^b$ is falsely selected as $i_n^b$. 
Observe that  $\widetilde{G}_i(\theta_b)$ depends  not only on $\alpha_i(\theta_b)$, but also on all $\alpha_\ell(\theta_b)$ such that $\ell\in\{\ell| y_\ell(\theta_b) < y_i(\theta_b)\}$. Intuitively, this makes sense because for $i\neq i^b$ to be $i^n_b$, the posterior mean of $i$ should beat all other solutions' and beating the solutions whose true means are smaller than $i$'s stipulates the LDR. Part (ii) derives the LDR of $j$ beating $i^*$ in the outer-level problem. Notice that the expression for $\widetilde{\ldr}_{j, i^*}$ depends on $\BFM \in \mathcal{A}_j$ that has the smallest sum of LDRs of all inner-level false selection events that constitute $\BFM$. Lastly, Part (iii) follows from~\eqref{eq:LDR_min}.

\noindent \textbf{Remark~1}: The expression for $\widetilde{G}_i(\theta_b)$ in~\eqref{def:G_tilde} and Part (i) of Theorem~\ref{thm:Gtilde} offer a tool to analyze the LDR of the outer-level FS event defined for a more general nested R\&S problem than the MPB framework. 
To the best of our knowledge, this is the first result in the R\&S literature that derives the LDR of the event such that \textit{a suboptimal solution is the sample-best}. The existing results (cf.~\cite{glynn2004large}) utilize the LDR of a suboptimal solution outperforming the true best, but not necessarily beating all other solutions.

From Theorem~\ref{thm:Gtilde}, Problem~\eqref{opt:OCBA_LDR} can be rewritten as follows:
\begin{equation} \label{eq:exact.ocba} 
    \max_{\boldsymbol{\alpha}} \min_{j\neq i^*} \underbrace{\min_{\mathbf{M} \in \mathcal{A}_j} \sum_{(i,\theta_b) \in I(\mathbf{M}) } \widetilde{G}_{i}(\theta_b)}_{=\widetilde{\ldr}_{j, i^*}} \ \ \mbox{subject to} \ \ \sum_{i=1}^k\sum_{b=1}^B \alpha_i(\theta_b) = 1, \alpha_i(\theta_b) \geq 0, 1\leq i \leq k, 1\leq b\leq B.
\end{equation}
In Lemma~\ref{lem:grad_Gtilde} in Appendix~\ref{ec:gap_analysis}, we show that \eqref{eq:exact.ocba} is a convex optimization problem in $\boldsymbol{\alpha}$ and therefore can be solved by a first-order method. Specifically, we adopt the entropic mirror descent algorithm~\citep{beck2003mirror} to numerically solve~\eqref{eq:exact.ocba} to evaluate the optimality gap in LDR for our proposed algorithms in Section~\ref{sec:experiments}. 
However, applying the same approach in the sequential learning setting is challenging for several reasons. First,  characterizing all   $\BFM \in \mathcal{A}_j$ is problem-specific and its computational complexity increases in $k$ and $B$. Even if the true means at all solution-parameter pairs are known, there are $k^B$ possible $\BFM$s to consider. When the posterior means are updated based on the sample, $\mathcal{A}_j$ may need to be updated, which further increases the computational cost.
Second, even if $\mathcal{A}_j$ is fully characterized, simple optimality conditions for  $\boldsymbol{\alpha}$ cannot be derived from the Karush-Kuhn-Tucker (KKT) conditions unlike in the classical problem. This makes it difficult to devise a sequential sampling algorithm whose allocation criterion is cheap to evaluate. 
While we can numerically solve the plug-in version of~\eqref{eq:exact.ocba}  applying the entropic mirror descent algorithm at each iteration (assuming $\mathcal{A}_j$s are characterized), it is undesirable to spend a large computational budget to solve an inexact problem in each iteration. In the following section, we tackle these issues by approximating~\eqref{eq:exact.ocba} with a problem that has simple optimality conditions for $\bm\alpha$ that are amenable for devising a sequential sampling rule.

\subsection{Relaxed OCBA and its Optimality Conditions}\label{sec:relax_OCBA}

To tackle the complexity caused by characterizing $\mathcal{A}_j$, we first define two solution-parameter sets: $\Xi^\text{adv} : = \lt\{(i^*, \theta_b) : \theta_b \in \Theta^c_{i^*}\rt\}$ includes all parameters in the adversarial set of $i^*$ paired with $i^*$; and $\Xi := \lt\{(i, \theta_b)| i \neq i^*, i \neq i^b\rt\}$. Thus, $\Xi \cup \Xi^{\text{adv}}$ includes all possible parameter-suboptimal solution pairs.
The following lemma states that the inner-most minimization problem of~\eqref{eq:exact.ocba}  can be formulated as a (minimization version of a) knapsack problem \citep{martello1990knapsack} with one extra constraint. 
\begin{lemma} \label{lem:knapsack}
    Let  $\BFm$ be the vectorized  $\BFM$ that consists of the elements corresponding to $\Xi\cup \Xi^{\text{adv}}$; ${\tilde{\BFg}}$ be the vector of $\{\widetilde G_i(\theta_b): (i,\theta_b)\in\Xi\cup \Xi^{\text{adv}}\}$; and $\mathbf v_j$ be the vector of $\left\{\{v_j[(i, \theta_b)]\}^+: (i,\theta_b)\in\Xi\cup \Xi^{\text{adv}}\right\}$, where
$
v_j[(i, \theta_b)] = p_b\lt(\mathbf{1}\lt\{j = i\rt\} - \mathbf{1}\lt\{j = i^b\rt\} - \mathbf{1}\lt\{i^* = i\rt\} + \mathbf{1}\lt\{i^* = i^b\rt\}\rt).
$
Suppose the elements of  $\widetilde\BFg$ and $\mathbf{v}_j$ are sorted in the same order as the corresponding elements in $\mathbf m$. Then, for each $j\neq i^*$
\end{lemma}
\begin{equation}\label{opt:knapsack}
    \widetilde\ldr_{j, i^*} \ = \ \min\; \widetilde\BFg^\top \BFm \hspace{15pt} \text{ subject to } \;   \BFv_j^\top\BFm \geq d_j, \hspace{10pt} \BFM \in \mathcal{M}.
\end{equation}

Recall from~\eqref{eq:set.of.M} that $\BFM \in \mathcal{M}$ implies the elements of $\BFM$ are binary and each column sum is one; except for the latter condition,~\eqref{opt:knapsack} has the exact structure of a knapsack problem. 
Here, $v_j[(i, \theta_b)]$ quantifies the importance of misspecifying $(i, \theta_b)$ has on the comparison between $i^*$ and $j$ at the outer-level problem. 
Thus, the knapsack formulation~\eqref{opt:knapsack} finds the set of misspecified $(i,\theta_b)$ pairs with the smallest sum of LDRs (cost) that is enough to cause $j$ to outperform or tie with $i^*$ in the outer-level problem (minimum weight constraint).


There is no closed-form expression for the optimal solution of a knapsack problem in general. Since our goal is to obtain easy-to-compute optimality conditions for $\bm\alpha$  to guide sequential sampling, we modify the constraint on $\mathbf m$ to $m_{i,b}\geq 0$
so that we can analytically characterize the optimal $\bm\alpha$ of the relaxed problem. The relaxed problem provides a lower bound for $\widetilde\ldr_{j, i^*}$ for any  $\bm\alpha$ as stated in Theorem~\ref{thm:lower_LDR}.

\begin{theorem}\label{thm:lower_LDR} Let $W_i(\theta_b) := \infty$ for all $(i, \theta_b) \in \Xi^c$. For each $(i,\theta_b)\in\Xi,$  let 
\begin{equation} \label{eq:weights}
   W_i(\theta_b) := \begin{cases} \max\lt\{\min\lt(\min_{j \neq i^*}d_j, \frac{d_i}{2}\rt)/p_b, 1\rt\}, & \mbox{if } \theta_b  \in\Theta_{i^*} \\
        \max\lt\{d_i/p_b, 1\rt\}, & \mbox{if } \theta_b  \notin\Theta_{i^*}\end{cases}
    \end{equation}
Given fixed $\boldsymbol{\alpha},$  $\min_{j \neq i^*}\widetilde{\ldr}_{j, i^*} \geq \underline{\widetilde\ldr}:=\min_{(i, \theta_b) \in \Xi}W_i(\theta_b)\widetilde{G}_i(\theta_b)$.
\end{theorem}

We refer to $W_i(\theta_b)$ as the \emph{balance weight} at $(i,\theta_b)$. 
Intuitively, $W_i(\theta_b)$ reflects the \emph{importance of misspecifying $i\neq i^b$ to be $i^b_n$ to the FS event}. If  $\widetilde G_i(\theta_b)$ is the same for all $(i,\theta_b)\in \Xi$, then the smallest $W_i(\theta_b)$ determines $\widetilde{\underline{\ldr}}$. Given $\theta_b$, the smaller $d_i$ is, the smaller $W_{i}(\theta_b)$ is; this makes sense as smaller $d_i$ implies that Solution $i$ is more competitive against $i^*$ and therefore, misspecifying $i$ as $i^b$ is more critical to FS. Given $i$, $W_{i}(\theta_b)$ is smaller when $\theta_b \in \Theta_{i^*}$; again, this is intuitive as we would like to make fewer mistakes at specifying $i^*$ to be the optima at the parameters in $\Theta_{i^*}$ to avoid underestimating its preference probability.

Using $\widetilde{\underline{\ldr}}$ defined in Theorem~\ref{thm:lower_LDR}, an approximate version of~\eqref{eq:exact.ocba} is formulated as  
\begin{equation}\label{opt:OCBA} 
    \begin{aligned}
       \max_{\bm{\alpha}} \underbrace{\min_{(i, \theta_b) \in \Xi}W_i(\theta_b)\widetilde{G}_i(\theta_b)}_{= \widetilde{\underline{\ldr}}}
       \;\; \;\; \textrm{subject to} \;\; \sum\nolimits_{i=1}^{k}\sum\nolimits_{b=1}^{B} \alpha_i(\theta_b) = 1, \alpha_i(\theta_b) \geq 0, 1\leq i \leq k, 1 \leq b \leq B.
    \end{aligned}
\end{equation} 
Theorem~\ref{thm:lower_LDR} implies  that the optimal objective function value of~\eqref{opt:OCBA} bounds that of~\eqref{eq:exact.ocba} from below. 
Since $\widetilde{G}_i(\theta_b)$ is concave in $\bm\alpha$ for each $(i,\theta_b)$, it is easy to see that $\widetilde{\underline{\ldr}}$ is  concave in $\bm{\alpha}$. Because the constraints of~\eqref{opt:OCBA} are linear, the KKT conditions provide its optimality conditions. 

However, although~\eqref{opt:OCBA} lets us avoid characterizing $\mathcal{A}_j, j\neq i^*$, the KKT conditions cannot be solved for $\boldsymbol\alpha$ analytically due to the complexity of $\widetilde{G}(i,\theta_b)$. 
Thus, we propose to replace $\widetilde{G}_i(\theta_b)$ with its lower bound to further relax~\eqref{opt:OCBA} so that the resulting problem has easy-to-compute optimality conditions for $\boldsymbol{\alpha}$. 
To this end, we adopt the following lower bound for $\widetilde G_i(\theta_b)$:
\begin{equation*}
G_{i}(\theta_b) := \left\{
\begin{array}{ll}
  \dfrac{(y_i(\theta_b) - y_{i^b}(\theta_b))^2}{2\lt(\lambda_i^2(\theta_b)/\alpha_i(\theta_b) + \lambda_{i^b}^2(\theta)/\alpha_{i^b}(\theta_b)\rt)},   & \mbox{ if } \alpha_i(\theta_b)>0 \mbox{ and } \alpha_{i^b}(\theta_b)>0, \\
    0, &  \mbox{ otherwise, } 
\end{array}
\right.
\end{equation*}
Indeed, $ G_{i}(\theta_b)$ is the LDR of $i\neq i^b$ falsely beating $i^b$ at $\theta_b$~\citep{glynn2004large}. Such an event is a subset of the event, $i_n^b=i$. Hence, from the definition of $\widetilde{G}_i(\theta_b)$, it is easy to see that $\widetilde{G}_i(\theta_b) \geq G_i(\theta_b)$. 
Proposition~\ref{prop:cond_opt_LDR} formalizes this relationship. 

\begin{proposition}\label{prop:cond_opt_LDR} For fixed $\bm\alpha$,  
$ \widetilde{G}_i(\theta_b) \geq G_i(\theta_b)$ for each $i\neq i^b$ and $\widetilde{G}_i(\theta_b) = G_i(\theta_b)$, if and only if, $\alpha_i(\theta_b) = \alpha_{i^b}(\theta_b) = 0$ or when  either $\alpha_i(\theta_b)$ or $\alpha_{i^b}(\theta_b)$ is positive and
\begin{equation}\label{eq:minimizer}
    x_i(\theta_b) : = \frac{(\alpha_i(\theta_b)/\lambda_i^2(\theta_b))y_i(\theta_b) + (\alpha_{i^b}(\theta_b)/\lambda_{i^b}^2(\theta_b))y_{i^b}(\theta_b)}{\alpha_i(\theta_b)/\lambda_i^2(\theta_b) + \alpha_{i^b}(\theta_b)/\lambda_{i^b}^2(\theta_b)} \in [y_{i^b}(\theta_b), \min\nolimits_{j \neq i^b} y_j(\theta_b)].
\end{equation}
\end{proposition}


When $i$ is the second best at $\theta_b$, \eqref{eq:minimizer} always holds.
We point out that~\eqref{eq:minimizer} bears its own theoretical interest as it characterizes when the two rate functions, $ \widetilde{G}_i(\theta_b)$  and $G_i(\theta_b)$, are equal.
From the G{\"a}rtner-Ellis theorem \citep{dembo2009large}, the LDR of a rare-event probability is determined by the most likely scenario causing the rare event. 
Since $\{\mu_{i, n}(\theta_b) \leq \mu_{j, n}(\theta_b), \forall j \neq i\} \subseteq \{\mu_{i, n}(\theta_b) \leq \mu_{i^b, n}(\theta_b)\}$, if the two events share a common most likely scenario, then $\widetilde{G}_i(\theta_b) = G_i(\theta_b)$. \cite{glynn2004large} show that the most likely scenario of $\{\mu_{i, n}(\theta_b) \leq \mu_{i^b, n}(\theta_b)\}$ is $A := \{\mu_{i, n}(\theta_b) \approx \mu_{i^b, n}(\theta_b) \approx x_i(\theta_b)\}$. Let $\widetilde{A} := A \cap \{\mu_{j, n}(\theta_b) \approx y_j(\theta_b), \forall j \neq i, i^b\}$. Then, $\widetilde{A}$  is asymptotically equivalent to  $A$ since $\prob(\mu_{j, n}(\theta_b) \approx y_j(\theta_b), \forall j \neq i, i^b) \to 1$  as $n$ increases. If~\eqref{eq:minimizer} holds, then 
\[
\widetilde{A}  = \{\mu_{i, n}(\theta_b) \approx \mu_{i^b, n}(\theta_b) \approx x_i(\theta_b) \stackrel{\text{\eqref{eq:minimizer}}}{\leq} y_j(\theta_b) \approx \mu_{j, n}(\theta_b), \forall j \neq i, i^b\}\subseteq\{\mu_{i, n}(\theta_b) \leq \mu_{j, n}(\theta_b), \forall j \neq i\}.\] 
Hence, $\widetilde{A}$ is the most likely scenario for both $\{\mu_{i, n}(\theta_b) \leq \mu_{j, n}(\theta_b), \forall j \neq i\}$ and $\{\mu_{i, n}(\theta_b) \leq \mu_{i^b, n}(\theta_b)\}$. Consequently, $G_i(\theta_b)=\widetilde{G}_i(\theta_b)$.

By replacing $\widetilde{G}_i(\theta_b)$ in~\eqref{opt:OCBA} with ${G}_i(\theta_b)$, we obtain the following OCBA problem:
\begin{equation}\label{opt:aOCBA}
    \begin{aligned} 
       \max_{\bm{\alpha}} \underbrace{\min_{(i, \theta_b) \in \Xi}W_i(\theta_b){G}_i(\theta_b)}_{:= \underline{\ldr}} \;\;         \textrm{subject to} \;\; \sum\nolimits_{i=1}^{k}\sum\nolimits_{b=1}^{B} \alpha_i(\theta_b) = 1, \alpha_i(\theta_b) \geq 0, 1\leq i \leq k, 1\leq b \leq B.
    \end{aligned}
\end{equation}
The KKT conditions for~\eqref{opt:aOCBA} can be simplified to a system of equations stated in Theorem~\ref{thm:balance}.

\begin{theorem}[Optimality conditions for \eqref{opt:aOCBA}]\label{thm:balance}
  Any ${\bm{\alpha}}$ is optimal for \eqref{opt:aOCBA}, if and only if, ${\bm{\alpha}}$ satisfies the following system of equations:
  \begin{itemize}
      \item (Global balance condition) For all $1 \leq b \leq B$,
        $ \;   {{\alpha}^2_{i^b}(\theta_b)}/{\lambda^2_{i^b}(\theta_b)} = \sum_{i \neq i^b, i^*}{{\alpha}^2_i(\theta_b)}/{\lambda^2_i(\theta_b)}.$
    \item (Pairwise balance condition) For all $(i, \theta_b), (j, \theta_{b^\prime}) \in \Xi$, $\{W_i(\theta_b)\}$ given in~\eqref{eq:weights} satisfies
      \begin{equation} \label{eq:balance.weights}
          W_i(\theta_b)G_{i}(\theta_b) = W_j(\theta_{b^\prime})G_{j}(\theta_{{b^\prime}}).
      \end{equation}
    \item (Zero asymptotic sampling ratio) ${\alpha}_{i^*}(\theta_b) = 0$ for all  $\theta_b \in \Theta^c_{i^*}$. 
  \end{itemize}
\end{theorem}

The global balance condition ensures that $i^b$ at each $\theta_b$ is correctly identified by allocating enough replications to each $i^b$ relative to other solutions. One can observe that this condition is identical to the global balance condition for the classical R\&S problem \citep{glynn2004large} as the optimal ${\alpha}_{i^*}(\theta_b) = 0$ for all  $\theta_b \in \Theta^c_{i^*}$ in our setting according to the last condition in Theorem~\ref{thm:balance}. 


The pairwise balance condition matches the product of $W_i(\theta_b)$ and $G_i(\theta_b)$ for all suboptimal (non-MPB) solution-parameter pairs.
Let $\BFW$ and $\BFG$ be $k \times B$ matrices whose $(i,b)$th entries are $W_i(\theta_b) $ and $G_{i}(\theta_b)$, respectively. For two matrices $A$ and $B$ of the same size, the Hadamard product, $A \circ B$, is defined as $(A \circ B)_{ij} = A_{ij}B_{ij}$. For ease of exposition, we define $0\cdot \infty = \infty$. Thus,~\eqref{eq:balance.weights} is equivalent to that the elements of $\mathbf{W} \circ \mathbf{G}$ corresponding to all $(i, \theta_b) \in \Xi$ are identical, and the rest of elements are all $\infty$.

The last condition states that the optimal asymptotic sampling ratio for $i^*$ at any adversarial $\theta_b$ is $0$. 
This may be surprising as the optimal sampling ratios for the classical R\&S problem are strictly positive~\citep{glynn2004large}. This stark difference has implications in our setting. Suppose $i^*$ is correctly identified as the conditional optimum at all $\theta_b\in \Theta_{i^*}$. Then, even if $i^*$ is incorrectly identified as the conditional optimum at some $\theta_b\in \Theta^c_{i^*}$, CS still occurs as long as the best among $i\neq i^*$ is correctly identified not to overestimate the preference probability of any $i\neq i^*$.

We also caution that the last condition must be interpreted carefully. For any sequential sampling procedure to achieve the optimality conditions in Theorem~\ref{thm:balance}, we need means and variances of all pairs to be consistently estimated in the limit. Thus, the procedure must sample $(i^*,\theta_b), \forall \theta_b\in\Theta_{i^*}^c$, at sublinear rates in $n$, i.e., still simulated infinitely often, but not as frequently as other pairs. 

The optimality conditions for C-OCBA in \cite{gao2019selecting} have the same global balance conditions as ours. However, our pairwise balance conditions are the same as theirs only if $W_i(\theta_b)$s equal one for all $(i, \theta_b) \in \Xi\cup\Xi^{\text{adv}}$.  This makes sense; because the objective of contextual R\&S is to minimize the PFS averaged over the contexts, all suboptimal pairs are equally important for C-OCBA. The third condition in Theorem~\ref{thm:balance} does not apply to the C-OCBA problem.

\noindent \textbf{Remark~2}: 
Although we focus on normal simulation outputs in this paper,~\eqref{opt:aOCBA} can be extended to non-normal cases. For general sampling distributions such as Bernoulli, Exponential, and Noncentral Chi-squared, we can replace $G_i(\theta_b)$ in~\eqref{opt:aOCBA} with their respective pairwise LDR expressions. The optimality conditions for such OCBA formulations can be obtained by substituting the global and pairwise balance conditions in Theorem~\ref{thm:balance} with those of the balancing optimal large deviations algorithm proposed by~\cite{chen2022BOLD}.

\noindent \textbf{Remark~3}: Theorems~\ref{thm:lower_LDR} and \ref{thm:balance} can be extended to the case when $i^*$ is non-unique. See Section~\ref{ec:nonunique.mpb}.


Due to relaxations, an optimal $\bm\alpha$ for~\eqref{opt:aOCBA} may be suboptimal for~\eqref{eq:exact.ocba}. Thus, a natural question is: how much loss is there in the LDR of PFS caused by the relaxations? To shed light on this question, we numerically assess the gap for some examples in Section~\ref{subsec:synthetic}. 
Another question of interest is: does the zero asymptotic sampling ratio condition still hold for the optimal solutions for~\eqref{eq:exact.ocba}? We devote Section~\ref{subsec:effect_relax} to discussing sufficient conditions under which the zero sampling ratios are preserved for~\eqref{eq:exact.ocba}.

\subsection{Sufficient Conditions to Preserve Zero Sampling Ratios in the Exact OCBA Formulation}\label{subsec:effect_relax}

Let $\widetilde{\boldsymbol{\alpha}}^\textsf{opt}=\{\widetilde{\alpha}^\textsf{opt}_i(\theta_b)\}$ and $\underline{\boldsymbol{\alpha}}^\textsf{opt}=\{\underline{\alpha}^\textsf{opt}_i(\theta_b)\}$ respectively denote optimal solutions of~\eqref{eq:exact.ocba} and~\eqref{opt:aOCBA}. From Theorem~\ref{thm:balance}, we have $\underline{\alpha}^\textsf{opt}_{i^*}(\theta_b) = 0$  for all $\theta_b \in \Theta_{i^*}^c$. Recall that the relaxations discussed in Section~\ref{sec:relax_OCBA} are carried out in two steps:  i) knapsack relaxation: \eqref{eq:exact.ocba}$\to$\eqref{opt:OCBA}; and ii)  bounding $\widetilde{G}_i(\theta_b)$ from below by $G_i(\theta_b)$: \eqref{opt:OCBA}$\to$\eqref{opt:aOCBA}.
To better understand how zero sampling ratios arise, we reverse the order of relaxations and consider the following intermediate problem: 
\small
\begin{eqnarray}\label{eq:opt_G_knapsack}
    \max_{\boldsymbol{\alpha}} \min_{j\neq i^*} \min_{\mathbf{M} \in \mathcal{A}_j} \sum_{(i,\theta_b) \in I(\mathbf{M}) } {G}_{i}(\theta_b)\ \ \mbox{subject to} \ \ \sum_{i=1}^k\sum_{b=1}^B \alpha_i(\theta_b) = 1, \alpha_i(\theta_b) \geq 0, 1\leq i \leq k, 1\leq b\leq B.
\end{eqnarray} \normalsize
Observe that~\eqref{eq:opt_G_knapsack} is obtained by replacing $\widetilde{G}_i(\theta_b)$ in~\eqref{eq:exact.ocba} with $G_i(\theta_b).$ Since $\widetilde{G}_i(\theta_b) \geq G_i(\theta_b)$ for each $(i,\theta_b)$, the optimal objective function value of~\eqref{eq:opt_G_knapsack} is a lower bound for that of~\eqref{eq:exact.ocba}. Moreover, we can obtain~\eqref{opt:aOCBA} from~\eqref{eq:opt_G_knapsack} via the same knapsack relaxation made in~\eqref{opt:OCBA}. 
These relationships are illustrated in the following diagram;  the constraints for $\bm\alpha$ are omitted for readability.
{\small
\begin{equation*}
    \begin{tikzcd}
\eqref{eq:exact.ocba}: \;\; \max_{\boldsymbol{\alpha}} \min_{j\neq i^*} \min_{\mathbf{M} \in \mathcal{A}_j} \sum_{(i,\theta_b) \in I(\mathbf{M}) } \widetilde{G}_{i}(\theta_b) \arrow{r}{}\;\;\; \arrow[swap]{d}{\text{Theorem~\ref{thm:lower_LDR}}} & \;\;\;\max_{\boldsymbol{\alpha}} \min_{j\neq i^*} \min_{\mathbf{M} \in \mathcal{A}_j} \sum_{(i,\theta_b) \in I(\mathbf{M}) } {G}_{i}(\theta_b) \arrow{d}{} \;\; :\eqref{eq:opt_G_knapsack}\\
\eqref{opt:OCBA}: \;\; \max_{\bm{\alpha}}\min_{(i, \theta_b) \in \Xi}W_i(\theta_b)\widetilde{G}_i(\theta_b)  \;\;\;\arrow{r}{\text{Proposition~\ref{prop:cond_opt_LDR}}} & \;\;\;\max_{\bm{\alpha}}\min_{(i, \theta_b) \in \Xi}W_i(\theta_b){G}_i(\theta_b) \;\; : \eqref{opt:aOCBA}
\end{tikzcd}
\end{equation*}}

Let $\boldsymbol{\alpha}^\textsf{opt} = \{{\alpha}^\textsf{opt}_i(\theta_b)\}$ be an optimal solution of~\eqref{eq:opt_G_knapsack}. Proposition~\ref{prop:WG_to_Gknapsack} below states that the zero sampling ratio condition we observed for $\underline{\boldsymbol{\alpha}}^\textsf{opt}$ is indeed preserved for $\boldsymbol{\alpha}^\textsf{opt}$.
\begin{proposition}\label{prop:WG_to_Gknapsack}
      For all $\theta_b \in \Theta^c_{i^*}$, $\alpha^\textsf{opt}_{i^*}(\theta_b) = 0$. 
\end{proposition}
Therefore, if the zero sampling ratio condition is to break down at all, it happens at~\eqref{eq:exact.ocba}$\to$\eqref{eq:opt_G_knapsack}. 

We proceed to establish sufficient conditions under which the zero asymptotic sampling ratio condition is preserved for~\eqref{eq:exact.ocba}. The following corollary 
states the first condition.


\begin{corollary}\label{cor:suff_0}
    If there exists $\widetilde{\boldsymbol{\alpha}}^\textsf{opt}$ that satisfies~\eqref{eq:minimizer} for all $(i, \theta_b)$, then any $\boldsymbol{\alpha}^\textsf{opt}$ is optimal for~\eqref{eq:exact.ocba} and $\widetilde{\alpha}^\textsf{opt}_{i^*}(\theta_b) = 0$ for all $\theta_b \in \Theta_{i^*}^c$. 
\end{corollary}

In fact, Corollary~\ref{cor:suff_0} implies a much stronger result as $\boldsymbol{\alpha}^\textsf{opt}$ becomes optimal for~\eqref{eq:exact.ocba}, i.e., no gap between the optimal LDRs from~\eqref{eq:exact.ocba} and~\eqref{opt:aOCBA}. Corollary~\ref{cor:suff_0} follows immediately from Proposition~\ref{prop:cond_opt_LDR} as~\eqref{eq:minimizer} implies $G_i(\theta_b)=\widetilde{G}_i(\theta_b)$. 
However,  requiring~\eqref{eq:minimizer} to hold for all $(i, \theta_b)$ may be quite demanding. 
Proposition~\ref{prop:suff_1} below provides a milder condition under which we can identify a subset of  $\Theta_{i^*}^c$ where the zero sampling ratio condition is preserved.

\begin{proposition}\label{prop:suff_1}
    For each $\theta_b \in \Theta_{i^*}^c$, if $\frac{\partial\widetilde{G}_i(\theta_b)}{\partial \alpha_{i^*}(\theta_b)}\big|_{\boldsymbol{\alpha} = \widetilde{\boldsymbol{\alpha}}^\textsf{opt}} = 0$ for all $i \notin \{i^*, i^b\}$,
    then we have $\widetilde{\alpha}^\textsf{opt}_{i^*}(\theta_b) =~0$.
\end{proposition}
The assumption of Proposition~\ref{prop:suff_1} implies that 
$\widetilde{\alpha}^\textsf{opt}_{i^*}(\theta_b)$ does not contribute to the objective function  of~\eqref{eq:exact.ocba} at all when we allocate the simulation budget according to $\widetilde{\boldsymbol{\alpha}}^\textsf{opt}$.  From this observation, one may expect $\widetilde{\alpha}^\textsf{opt}_{i^*}(\theta_b) = 0$. The proof of Proposition~\ref{prop:suff_1} establishes this intuition with rigor by showing that if $\widetilde{\alpha}^\textsf{opt}_{i^*}(\theta_b) > 0$, then one can increase the objective function value of~\eqref{eq:exact.ocba} by reallocating $\widetilde{\alpha}^\textsf{opt}_{i^*}(\theta_b)$ to some other $(i, \theta_b)$ pairs, which contradicts  the optimality of ${\widetilde{\boldsymbol{\alpha}}^\textsf{opt}}$. 

Any statement implying the condition of Proposition~\ref{prop:suff_1} can be a sufficient condition. For instance, since $G_i(\theta_b)$ does not involve $\alpha_{i^*}(\theta_b)$ if $i \neq i^*$, the condition of Corollary~\ref{cor:suff_0} implies that of Proposition~\ref{prop:suff_1}. In Proposition~\ref{prop:suff_1-1}, we provide a weaker condition on the mean-variance configuration of the problem that implies Proposition~\ref{prop:suff_1}.

We can also gain intuitions on when the zero sampling ratio condition breaks down from the contrapositive of Proposition~\ref{prop:suff_1}: if $\widetilde{\alpha}_{i^*}^\textsf{opt}(\theta_b) \neq 0$ for some $\theta_b \in \Theta^c_{i^*}$, then at least one of $\{\widetilde{G}_i(\theta_b)\}_{i \neq i^*, i^b}$ depends on $\widetilde{\alpha}_{i^*}^\textsf{opt}(\theta_b)$  at $\widetilde{\boldsymbol{\alpha}}^\textsf{opt}$. Sampling $(i^*, \theta_b)$ may increase some $\widetilde{G}_i(\theta_b)$ and possibly decrease the others. Thus, the zero sampling ratio condition is violated only when there exists $\boldsymbol{\alpha}$ that makes the former effect greater than the latter. 

Although insightful, Proposition~\ref{prop:suff_1} is difficult to verify as $\widetilde{\boldsymbol{\alpha}}^\textsf{opt}$ is unknown. Proposition~\ref{prop:suff_2} below presents a sufficient condition free of $\widetilde{\boldsymbol{\alpha}}^\textsf{opt}$. 

\begin{proposition}\label{prop:suff_2}
        Let $k^b = \argmax_{1\leq i \leq k} y_i(\theta_b)$ and fix $\theta_b \in \Theta_{i^*}^c$. If there exists $1\leq j^b \leq k$ such that 
        \begin{equation}\label{eq:suff_2}
            y_{j^b}(\theta_b) < y_{i^*}(\theta_b) \;\; \text{and} \;\; \frac{y_{k^b}(\theta_b) - y_{j^b}(\theta_b)}{\lambda_{j^b}(\theta_b)} \geq \frac{y_{k^b}(\theta_b) - y_{i^*}(\theta_b)}{\lambda_{i^*}(\theta_b)},
        \end{equation}
        then we have $\widetilde{\alpha}^\textsf{opt}_{i^*}(\theta_b) = 0$.
\end{proposition}


To gain insight, consider an assumption stricter than that of Proposition~\ref{prop:suff_2}:  at $\theta_b\in\Theta_{i^*}^c$, there exists $j^b$ such that $y_{j^b}(\theta_b) < y_{i^*}(\theta_b)$ and $\lambda_{j^b}(\theta_b) < \lambda_{i^*}(\theta_b)$ ($j^b$ is the better solution with the smaller variance). In this case, it is more beneficial to simulate $j^b$ over $i^*$ to reduce the rate at which any $i\notin \{i^b, i^*\}$ being falsely selected as the best at $\theta_b$, which makes $\widetilde{\alpha}^\textsf{opt}_{i^*}(\theta_b) = 0$ asymptotically optimal. This assumption is relaxed in~\eqref{eq:suff_2} by accounting for the relative dominance of $y_{j^b}(\theta_b)$ over $y_{i^*}(\theta_b)$ compared to the worst mean, $y_{k^b}(\theta_b)$; the more clearly dominant $j^b$ is over $i^*$ relative to $k^b$, $\widetilde{\alpha}^\textsf{opt}_{i^*}(\theta_b) = 0$  is optimal under a wider range of $\lambda_{j^b}(\theta_b)$ relative to $\lambda_{i^*}(\theta_b)$.  When  $y_{j^b}(\theta_b)$ is sufficiently dominant,  $\lambda_{j^b}(\theta_b)$ needs not be smaller than $\lambda_{i^*}(\theta_b)$.

Condition~\eqref{eq:suff_2} can be established for each $\theta_b$, independently from the outer-level R\&S problem. Thus,  if we construct a subset of $\Theta_{i^*}^c$ that consists of $\theta_b$ satisfying~\eqref{eq:suff_2}, then the zero asymptotic sampling ratio condition holds for all $\theta_b$ in the subset.
Moreover, verifying the sample-moment version of~\eqref{eq:suff_2} is fairly easy after a sequential sampling algorithm terminates.

\section{Sequential Learning Procedures}\label{sec:sequential_learning}

The optimality conditions in Theorem~\ref{thm:balance} depend on $\lt\{y_i(\theta_b)\rt\}_{1\leq i\leq k, 1\leq b\leq B}$, which must be estimated from simulations in reality. In this section, we present four sequential sampling algorithms that simultaneously learn $\lt\{y_i(\theta_b)\rt\}_{1\leq i\leq k, 1\leq b\leq B}$ and the asymptotically optimal allocations.  In all these algorithms, each $y_i(\theta_b)$ in the balance weight expressions is replaced with its plug-in estimate, $\mu_{i,n}(\theta_b)$. Thus, the resulting (plug-in) optimality conditions are inexact. Nevertheless, the hope is that as $n$ increases, the algorithms behave similarly as the static sampling rule that allocates the simulation budget according to the asymptotic optimal sampling ratios. The ideal result, which we do not show here, would be that the plug-in algorithm achieves the same LDR as the optimal static sampling rule.  However,  even for the classical R\&S problems, analyzing the plug-in algorithms' LDR is challenging and remains to be shown. Instead, several plug-in algorithms are proven to achieve what we refer to as \textit{strong consistency of sampling ratios}, i.e., the sampling ratios allocated by the algorithms asymptotically satisfy the conditions for optimal static sampling ratios almost surely~\citep{glynn2004large, pasupathy2014stochastically,Feldman-BORS:18,Appleaget-MORS:20,gao2019selecting,chen2022BOLD}. While the strong consistency of sampling ratios does not imply that plug-in algorithms achieve the optimal LDR of the static allocation rule, \cite{garivier2016optimal} show that it is a necessary condition for a sampling rule to achieve optimal efficiency. With this motivation, we analyze the asymptotic behaviors of the sampling ratios of the four plug-in algorithms we propose in this section.



To facilitate the discussion, we define some notation for sample statistics. Let $\Xi_n$, $d_{i, n}$, $\mathbf{W}_n$, and $\mathbf{G}_n$ be the plug-in versions of $\Xi$, $d_i=d_i(\BFM^*)$, $\mathbf{W}$, and $\BFG$ defined by replacing $\BFM^*$ with $\BFM_n$, $i^*$ with $i_n^*$, and $i^b$ with $i_n^b$ for all $b$, respectively. We further denote the estimated favorable set of $i_n^*$ by $\Theta_n^*:=\{\theta_b: i_n^* = i_n^b\}$ and the   fraction of $n$ allocated to $(i,\theta_b)$ by $\alpha_{i, n}(\theta_b)$.

Algorithm~\ref{alg:MPB_basic} presents the basic structure that all four algorithms follow. In  Step~\ref{step:outer}, two or more solutions may be tied when determining $i^*_n$.  Should a tie occur, we may select $i^*_n$ by applying a tie-breaking rule such as the first index rule or random sampling rule.

\setcounter{algorithm}{-1}
\begin{algorithm}[tb]
\caption{{Algorithmic framework for  selection of the most probable best}}\label{alg:MPB_basic}
\begin{algorithmic}[1]
    \STATE Given total simulation budget $N$, allocate $n_0$ initial replications at all $(i, \theta_b)$. Let $n = n_0kB$.
    \WHILE{ $n < N$}
        \STATE Calculate $\mu_{i, n}(\theta_b)$ in~\eqref{eq:update_normal} for all $(i, \theta_b)$. \label{step:alg0_start}
        \STATE Inner-level problem : For each $\theta_b$, find $i_n^{b} : = \argmin_{1\leq i\leq k} \mu_{i,n}(\theta_b)$. \label{step:inner}
        \STATE Outer-level problem : Find $i^*_n = \argmax_{1\leq i\leq k} \sum_{b=1}^B p_b\mathbf{1}\left\{i=i^b_n\right\}. 
        $\label{step:outer}
        \STATE Make the next sampling decision, $(i, \theta_b)$, and run a replication at $(i, \theta_b)$; $n \leftarrow n+1$.
    \ENDWHILE
    \RETURN $i_N^*$ as the estimated MPB.
\end{algorithmic}
\end{algorithm}




Our first sequential sampling algorithm, Algorithm~\ref{alg:rate_opt}, applies the plug-in versions of the balance conditions in Theorem \ref{thm:balance} to make sampling decisions in Steps~\ref{step:pairwise}--\ref{alg:alg1_end2}. In the event of a tie for $i^*_n$,
we apply the same tie-breaking rule in Algorithm~\ref{alg:rate_opt} and all other variants: we choose $\argmax_{j \in i_n^*}\min_{b:i_n^b \neq j} G_{j, n}(\theta_b)$ as $i^*_n$. Note that the inner minimization term represents the empirical LDR of Solution $j$ identified as one of the tied $i^*_n$ incorrectly beating a solution in its (estimated) adversarial set. Then, the outer maximization problem picks $j \in i_n^*$ for which such an event is least likely to occur (i.e, its adversarial set is most likely to be correct).
If the algorithm samples all $(i, \theta_b)$ infinitely, the event of a tie happens only finitely many times with probability $1$ under the uniqueness assumption of $i^*$ by the strong law of large numbers. 

\begin{algorithm}[tbp]
\caption{Sequential Sampling Algorithm for Selection of the MPB}
\begin{algorithmic}[1]
\STATE Initialization with Step~1 of Algorithm~\ref{alg:MPB_basic}.  
\WHILE{$n < N $}
    \STATE Run Steps~\ref{step:alg0_start}--\ref{step:outer} of Algorithm~\ref{alg:MPB_basic} and update $G_{i, n}(\theta_{b})$ for all $i \neq i_n^b$.\label{alg:alg1_start1}
    \STATE Construct $\BFW_n$ and $\BFG_n$. Find $(i, \theta_b) = \argmin \BFW_n \circ \BFG_n$. \label{step:pairwise}
    \IF {$\Large({N^n_{i^b_n}(\theta_b)}/{\lambda_{i_n^b}(\theta_b)}\Large)^2 < \sum_{j \neq i_n^b, i_n^*}\lt({N^n_j(\theta_b)}/{\lambda_j(\theta_b)}\rt)^2$} \label{step:global} 
        \STATE Let $j = i_n^b$.
    \ELSE
        \STATE Let $j = i$.
    \ENDIF \label{alg:alg1_end2}
    \STATE Run a replication at $(j, \theta_b)$; $n \leftarrow n+1$.  \label{alg:alg1_end3}
\ENDWHILE
\RETURN $i_N^*$ as the estimated MPB.
\end{algorithmic}
\label{alg:rate_opt}
\end{algorithm}

Algorithm \ref{alg:rate_opt} is easy to implement as we only need to compute $\mathbf{G}_n$, $\mathbf{W}_n$, and $\lt(N^n_i(\theta_b)/\lambda_i(\theta_b)\rt)^2$ at each iteration. The following theorem states strong consistency of Algorithm~1.
\begin{theorem}\label{cor:Alg1_convergence}
    Algorithm~1 finds the MPB almost surely as $n$ increases.
\end{theorem}
Unfortunately, Algorithm~\ref{alg:rate_opt} does not guarantee that $N^n_{i}(\theta_b) \rightarrow \infty$ for all $(i,\theta_b)$ almost surely because it tends to stop simulating $\lt\{(i_n^*, \theta_b) : i_n^b \neq i_n^*\rt\}$ once it correctly specifies the minimum number of $\Theta_{i^*}$ needed to distinguish $i_n^*$ from the second best. For instance, suppose that $B = 50$ with equiprobable $\theta_b$s, $|\Theta_{i^*}|=15$, and the second best solution is the conditional optima at $10$ $\theta_b$s. Then, it is enough to correctly specify $i^*$ as the conditional optima at $13$ out of $15$ $\theta_b$s in $\Theta_{i^*}$ to achieve CS provided that the conditional optima are correctly specified at all other parameters in $\Theta^c_{i^*}$. Even if at two remaining parameters in $\Theta_{i^*}$ the second best is incorrectly specified as the conditional optima, CS occurs. Hence, Algorithm~\ref{alg:rate_opt} may not fully characterize $\Theta_{i^*}$ in the limit even though $i_n^* \stackrel{a.s.}{\to} i^*$. This behavior may also impede empirical convergence of the algorithm; underestimating $|\Theta_{i^*}|$ means that the algorithm perceives the problem to be harder than it is. 

The next variant, Algorithm~\ref{alg:rate_opt_posterior}, guarantees that $N^n_{i^*}(\theta_b)$ increases sublinearly in $n$ for all $\theta_b \in \Theta_{i^*}^c$ so that  $N^n_{i^*}(\theta_b) \rightarrow \infty$ and $\alpha_{i^*,n}(\theta_b) \rightarrow 0$ almost surely.  
The only difference from Algorithm~\ref{alg:rate_opt} is Step~4: we estimate $\eta_{i^*_n}(\theta_b)$  for any $\theta_b$ in $(\Theta^*_n)^c$ with a posterior sample from~\eqref{eq:update_normal}, not the posterior  mean. Define event $A = \{\bar{\mu}_{i_n^*, n}(\theta_b) < \mu_{i_n^b, n}(\theta_b)\}$ for such $\theta_b$. When $A$ occurs, the updated $i_n^b$ in Step~\ref{step:posterior} must be equal to $i_n^*$ since all other plug-in estimates at $\theta_b$ are unchanged. 
This increases the estimated preference probability of $i^*_n$, and thus $i_n^*$ remains unchanged from Step~3. The posterior samples are discarded after each iteration.
Algorithm~\ref{alg:rate_opt_posterior} simulates $(i_n^*, \theta_b)$ such that $\theta_b \in (\Theta^*_n)^c$, if event $A$ occurs and Step~5 selects $(i_n^*, \theta_b)$ as the next sampling pair. Hence, the probability of sampling $(i_n^*, \theta_b)$ is strictly positive and bounded from above by $\prob(A) = \Phi\lt(\sqrt{N_{i_n^*}^n(\theta_b)}\frac{\mu_{i_n^b, n}(\theta_b) - \mu_{i_n^*, n}(\theta_b)}{\lambda_{i_n^*}(\theta_b)}\rt)$, where $\Phi(\cdot)$ is the standard normal CDF. This contrasts with Algorithm~\ref{alg:rate_opt}, which does not simulate  such  $(i_n^*, \theta_b)$.  Note that  $\prob(A)$ vanishes as $N_{i_n^*}^n(\theta_b) \rightarrow \infty$. 
Thus,  Algorithm~\ref{alg:rate_opt_posterior} allows sampling $(i_n^*, \theta_b)$ for $\theta_b \in (\Theta^*_n)^c$ recognizing that $\Theta^*_n$ may not equal $\Theta_{i^*}$ for finite $n$, whereas as $n$ increases, it behaves increasingly similarly as Algorithm~\ref{alg:rate_opt} reckoning that $\Theta^*_n$ is estimated more precisely. Formalizing this idea, Theorem~\ref{thm:balance} shows that  Algorithm~\ref{alg:rate_opt_posterior} achieves the strong consistency of sampling ratios.

\begin{algorithm}[tbp]
\caption{Hybrid of Posterior Sample for Selection of the MPB}
\begin{algorithmic}[1]
\STATE  Initialization with Step~1 of Algorithm~\ref{alg:MPB_basic}.
\WHILE{$n < N $}
    \STATE Run Step~\ref{alg:alg1_start1} of Algorithm~\ref{alg:rate_opt}.
    \STATE For all $\theta_b\in \lt(\Theta_n^*\rt)^c$, replace $\mu_{i^*_n, n}(\theta_b)$ with $\bar{\mu}_{i_n^*, n}(\theta_b) \sim N(\mu_{i^*_n, n}(\theta_b), \sigma_{i^*_n, n}^2(\theta_b))$ in~\eqref{eq:update_normal}; update $i_n^b$.  \label{step:posterior}
    \STATE Run Steps~\ref{step:pairwise}--\ref{alg:alg1_end3} of Algorithm~\ref{alg:rate_opt}.  
\ENDWHILE
\RETURN $i_N^*$ as the estimated MPB.
\end{algorithmic}
\label{alg:rate_opt_posterior}
\end{algorithm}

\begin{theorem}\label{thm:proof_EIalg}
  The sampling allocations made by Algorithm \ref{alg:rate_opt_posterior} satisfy the following almost surely:
  \begin{equation*}
      \lim_{n \rightarrow \infty} \Big\{\lt(\tfrac{\alpha_{i^b, n}(\theta_b)}{\lambda_{i^b}(\theta_b)}\rt)^2  - \sum\nolimits_{j \neq i^b, i^*}\lt(\tfrac{\alpha_{j, n}(\theta_b)}{\lambda_j(\theta_b)}\rt)^2\Big\} = 0, \forall\theta_b \mbox{ and }
\lim_{n \rightarrow \infty}\tfrac{W_{i}(\theta_b)G_{i, n}(\theta_b)}{W_{j}(\theta_c)G_{j, n}(\theta_c)} = 1, \forall (i, \theta_b) \neq (j, \theta_c) \in \Xi.
\end{equation*}
Moreover, $\limsup_{n \rightarrow \infty} N^n_i(\theta_b)=\infty$ for all $(i, \theta_b)$ and $\lim_{n \rightarrow \infty} N^n_{i^*}(\theta_b)/n = 0$ for all $(i^*, \theta_b) \in \Xi^\textup{adv}$. 
\end{theorem}
\tk{

\vspace{3mm}


}


\subsection{Algorithms for Learning Both MPB and its Favorable Set}

So far, we have focused on selecting the MPB correctly, however, one may be interested in precisely estimating the MPB's preference probability. 
This motivates us to devise algorithms that learn both $i^*$ and  $\Theta_{i^*}$ correctly. Assuming $i^*$ is learned correctly, identifying $\Theta_{i^*}$ can be viewed as a binary classification task for each $\theta_b$ to determine  whether $\theta_b \in \Theta_{i^*}$ or not. 
Thus, we can construct a confusion matrix for identifying  $\Theta_{i^*}$ as follows:
\small\begin{equation}\label{eq:confusion_matrix}
\begin{blockarray}{ccc}
\text{Actual $\backslash$ Predicted} & \theta_b \in \Theta_n^* & \theta_b \in (\Theta_n^*)^c \\
\begin{block}{c(cc)}
  \theta_b \in \Theta_{i^*} & \text{True Positive}  & \text{False Negative} \\
  \theta_b \in \Theta_{i^*}^c & \text{False Positive} & \text{True Negative}\\
\end{block}
\end{blockarray}
\end{equation}\normalsize


\noindent From~\eqref{eq:confusion_matrix}, we define the \emph{false negative rate} (FNR) and \emph{accuracy} (ACC)  as 
\begin{equation}\label{def:error}
    \text{FNR} := \frac{\prob_{\pi}\lt(\Theta_{i^*} \cap \lt(\Theta^*_{n}\rt)^c\rt)}{\prob_{\pi}\lt(\Theta_{i^*}\rt)}; \hspace{15pt}
    \text{ACC} := \prob_{\pi}\lt(\Theta_{i^*} \cap \Theta^*_n\rt) + \prob_{\pi}\lt(\Theta^c_{i^*} \cap \lt(\Theta^*_{n}\rt)^c\rt).
\end{equation}
ACC is maximized when $\Theta_n^* = \Theta_{i^*}$, i.e., the estimated preference probability of $i^*$ is  exact. This motivates us to design \textbf{Algorithm~3} that maximizes the probability of event $\text{ACC}_* := \{i_n^* = i^*\}\cap \{\Theta_n^* = \Theta_{i^*}\}$. 
ACC penalizes both false negative and false positive. Meanwhile, recall that  underestimation of $\Theta_{i^*}$ (and preference probability of $i^*$) causes performance degradation of Algorithm~\ref{alg:rate_opt}. Thus, we also consider \textbf{Algorithm 4} that penalizes only false negatives by minimizing FNR. Since FNR$=0$ when $\{\Theta^*_{n} \supseteq \Theta_{i^*}\}$, Algorithm~4 maximizes  the probability of event  
$\text{FN}_*:= \{i_n^* = i^*\}\cap \{\Theta^*_{n} \supseteq \Theta_{i^*}\}.$
In the following, we discuss the optimality conditions for the OCBA formulation when each of $\prob(\text{ACC}_*)$ and $\prob(\text{FN}_*)$ is adopted in lieu of PFS.

First, the LDR of $\text{ACC}_*^c$ can be bounded from below as
\begin{eqnarray}
   \liminf\nolimits_{n \rightarrow \infty} -\tfrac{1}{n}\log \prob(\text{ACC}_*^c)& = &\liminf\nolimits_{n \rightarrow \infty} -\tfrac{1}{n}\log \prob(\lt\{i_n^* \neq i^*\rt\} \cup \lt\{\Theta^*_n \neq \Theta_{i^*}\rt\})\nonumber\\
   & \geq & \min\lt\{\underline{\ldr}, \liminf\nolimits_{n \rightarrow \infty} -\tfrac{1}{n}\log\prob\lt(\Theta^*_n \neq \Theta_{i^*}\rt)\rt\}.\label{eq:lower_EF}
\end{eqnarray}
Thus, it only remains to find the LDR of $\prob(\Theta_n^* \neq \Theta_{i^*})$. To guarantee $\lt\{\Theta^*_n = \Theta_{i^*}\rt\}$, we need to ensure (a) for each $\theta_b$ in $\Theta_{i^*}$, the conditional optimum is correctly identified to be $i^*$; and (b) for each $\theta_b$ in $\Theta_{i^*}^c$, $i^*$ is not chosen as the conditional optimum. Consequently, we have
\begin{equation}\label{eq:LDR_set}
    \liminf_{n \rightarrow \infty} -\tfrac{1}{n}\log\prob\lt(\Theta^*_n \neq \Theta_{i^*}\rt) \geq \min\big\{\underbrace{\min\nolimits_{\lt\{i \neq i^*, \theta_b \in \Theta_{i^*}\rt\}} G_{i}(\theta_b)}_{\substack{\text{LDR of not (a)}}}, \underbrace{\min\nolimits_{\theta_b \in \Theta_{i^*}^c} G_{i^*}(\theta_b)}_{\substack{\text{LDR of not (b)}}}\big\}.
\end{equation}
\noindent
Combining~\eqref{eq:lower_EF} with~\eqref{eq:LDR_set},  $\liminf_{n\rightarrow \infty} -\frac{1}{n}\log\prob(\text{ACC}_*^c)$ is bounded from below by
\begin{equation*}\label{eq:LDR_setPCS}
    \underline{\ldr}_\ACC : = \min\lt\{\underline{\ldr}, \min\nolimits_{\lt\{i \neq i^*, \theta_b \in \Theta_{i^*}\rt\}} G_{i}(\theta_b), \min\nolimits_{\theta_b \in \Theta_{i^*}^c} G_{i^*}(\theta_b)\rt\}.
\end{equation*}
Then, the OCBA for Algorithm~3 can be formulated as
\begin{equation}\label{opt:aOCBA_EF}
    \begin{aligned}
       \underline{\ldr}_\ACC^* := \max_{\bm{\alpha}} \underline{\ldr}_\ACC \;\;
        \textrm{subject to} \;\; \sum\nolimits_{i=1}^{k}\sum\nolimits_{b=1}^{B} \alpha_i(\theta_b) = 1, \alpha_i(\theta_b) \geq 0, 1\leq i \leq k, 1 \leq b \leq B.
    \end{aligned}
\end{equation}

To maximize $\prob(\FN_*)$, note that $\lt\{\Theta_{i^*} \subseteq \Theta^*_{n}\rt\} = \lt\{\mu_{i^b_n, n}(\theta_b) \leq \mu_{i^*, n}(\theta_b), \forall \theta_b \in (\Theta^*_{n})^c \rt\}$. Thus, the LDR of $\FN_*^c$ is bounded from below as
\begin{equation*}
\begin{aligned}
   \liminf_{n \rightarrow \infty} -\tfrac{1}{n}\log\prob\lt(\FN_*^c\rt) &= \liminf_{n \rightarrow \infty} -\tfrac{1}{n}\log\prob\lt(\lt\{i_n^* \neq i^*\rt\}\cup \lt\{\mu_{i^*, n}(\theta_b) < \mu_{i^b_n, n}(\theta_b), \exists \theta_b \in (\Theta^*_n)^c \rt\}\rt) \\
   & 
   \geq \min \lt\{\underline{\ldr}, \min\nolimits_{\theta_b \in \Theta^c_{i^*}}G_{i^*}(\theta_b)\rt\}. 
\end{aligned}
\end{equation*}
Defining $\underline{\ldr}_\FN : = \min \lt\{\underline{\ldr}, \min_{\theta_b \in \Theta^c_{i^*}}G_{i^*}(\theta_b)\rt\}$, the OCBA for Algorithm~4 can be written as
\begin{equation}\label{opt:aOCBA_IF}
    \begin{aligned}
       \underline{\ldr}_\FN^* := \max_{\bm{\alpha}} \underline{\ldr}_\FN        \;\; \textrm{subject to} \;\; \sum\nolimits_{i=1}^{k}\sum\nolimits_{b=1}^{B}\alpha_i(\theta_b) = 1, \alpha_i(\theta_b) \geq 0, 1\leq i \leq k, 1 \leq b\leq B.
    \end{aligned}
\end{equation}

Since $\underline{\ldr}_\ACC$ and $\underline{\ldr}_\FN$ are concave in $\bm{\alpha}$, \eqref{opt:aOCBA_EF} and~\eqref{opt:aOCBA_IF} are convex programs. Therefore, we can derive their optimality conditions from the KKT conditions as summarized in  Theorem~\ref{thm:balance_modified}.

 \begin{theorem}[Optimality conditions for \eqref{opt:aOCBA_EF} and~\eqref{opt:aOCBA_IF}]\label{thm:balance_modified}
  Any ${\bm{\alpha}}$ is optimal for~\eqref{opt:aOCBA_EF} (respectively \eqref{opt:aOCBA_IF}), if and only if, ${\bm{\alpha}}$ satisfies the following system of equations:
  \begin{itemize}[leftmargin=*]
      \item (Global balance condition) For all $1 \leq b \leq B$,
        $ \;   {\alpha}^2_{i^b}(\theta_b)/{\lambda^2_{i^b}(\theta_b)} = \sum_{i \neq i^b}{\alpha}^2_i(\theta_b)/{\lambda^2_i(\theta_b)}.$
    \item (Pairwise balance condition) For all $(i, \theta_b), (j, \theta_{b^\prime}) \in \Xi \cup \Xi^{\textup{adv}}$, 
     $W^{\ACC}_i(\theta_b)G_{i}(\theta_b) = W^{\ACC}_j(\theta_{b^\prime})G_{j}(\theta_{{b^\prime}})$ (resp. $W^{\FN}_i(\theta_b)G_{i}(\theta_b) = W^{\FN}_j(\theta_{b^\prime})G_{j}(\theta_{{b^\prime}})$),  where \small 
    \begin{equation} \label{eq:weights.modified} 
    W^{\ACC}_i(\theta_b) := 
    \begin{cases} 1, & \mbox{if } \theta_b  \in\Theta_{i^*} \mbox{ or } i = i^*~\&~\theta_b \in \Theta^c_{i^*}\\
    W_i(\theta_b), & \mbox{if } \theta_b  \notin\Theta_{i^*}~\&~i \neq i^*
        \end{cases}, \hspace{15pt}
        W^{\FN}_i(\theta_b) = \begin{cases} W_i(\theta_b), & \mbox{if } (i, \theta_b)  \in \Xi\\
        1, & \mbox{if } (i, \theta_b) \in \Xi^{\textup{adv}}
        \end{cases},
    \end{equation}\normalsize
    and $W^\ACC_i(\theta_b)=W^\FN_i(\theta_b)=\infty$ for all $(i, \theta_b) \in \lt(\Xi\cup\Xi^\textup{adv}\rt)^c$.
  \end{itemize}
\end{theorem}

Let $\BFW_n^\ACC$ and $\BFW_n^\FN$ be the sample version of $\BFW^\ACC$ and $\BFW^\FN$ after $n$ replications. Algorithm~3 (Algorithm~4) is identical to Algorithm~\ref{alg:rate_opt} except that $\mathbf W_n$ in Step 3 and $\sum_{j \neq i_n^b, i_n^*}\lt({N^n_j(\theta_b)}/{\lambda_j(\theta_b)}\rt)^2$ in Step~\ref{step:global} are replaced with $\mathbf W^{\ACC}_n$ ($\mathbf W^{\FN}_n$) and $\sum_{j \neq i_n^b}\lt({N^n_j(\theta_b)}/{\lambda_j(\theta_b)}\rt)^2$, respectively. Theorem~\ref{thm:proof_BWalg} states that Algorithms 3 and 4 achieve the strong consistency of sampling ratios. 

\begin{theorem}\label{thm:proof_BWalg} The sampling allocations made by Algorithm~3 satisfy the following almost surely:
\[
\lim_{n \rightarrow \infty} \Big\{\lt(\tfrac{\alpha_{i^b, n}(\theta_b)}{\lambda_{i^b}(\theta_b)}\rt)^2  - \sum_{j \neq i^b}\lt(\tfrac{\alpha_{j, n}(\theta_b)}{\lambda_j(\theta_b)}\rt)^2\Big\} = 0, \forall\theta_b \mbox{ and }
\lim_{n \rightarrow \infty}\tfrac{W^{\ACC}_{i}(\theta_b)G_{i ,n}(\theta_b)}{W^{\ACC}_{j}(\theta_c)G_{j, n}(\theta_c)} = 1, \forall (i, \theta_b), (j, \theta_c) \in \Xi\cup\Xi^{\textup{adv}}
\]
and
$\limsup_{n \rightarrow \infty} {N^n_i(\theta_b)} \rightarrow \infty$ for all $(i, \theta_b)$. For Algorithm~4, the same statements hold when $W^{\ACC}_{i}(\theta_b)$ is replaced with $W^{\FN}_{i}(\theta_b)$.
\end{theorem}

\section{Improving Finite-Sample Performance with Kernel Ridge Regression}\label{sec:GP_extension}

In this section, we adopt the kernel ridge regression (KRR) to improve the performance of the proposed plug-in algorithms by reducing initial estimation errors via pooling simulation outputs obtained at different parameter values. 
Insofar, by imposing an independent prior for each $\eta_i(\theta_b)$,  $y_i(\theta_b)$ can only be learned from simulations made at $(i, \theta_b)$, i.e., we do not exploit simulation outputs from parameters other than $\theta_b$ to infer $y_i(\theta_b)$. If $y_i(\theta)$ is reasonably smooth in $\theta$, then  $y_i(\theta_b)$ may be learned more accurately by utilizing simulation outputs obtained at parameters close to $\theta_b$.

We first establish some definitions to introduce KRR. The kernel $K:\mathcal{X} \times \mathcal{X} \rightarrow \Real$ defined on the domain $\mathcal{X}$ of $\theta$ is said to be a \emph{positive-definite} (PD) kernel if it  satisfies $\sum_{\ell=1}^L\sum_{\ell^\prime = 1}^L K(\theta_\ell, \theta_{\ell^\prime})c_\ell c_{\ell^\prime} \geq 0$ for any integer $L$, $\theta_\ell \in \mathcal{X}$ and $c_\ell\in \Real$. The reproducing kernel Hilbert space (RKHS), $\mathcal{K}$, generated by  $K$ is the  closure of $\{f(\cdot)| f(\cdot) = \sum_{\ell=1}^L c_\ell K(\theta_\ell, \cdot), L \in \mathbb{Z}^+, \{\theta_\ell\}_{1\leq \ell\leq L} \subseteq \mathcal{X}\}$, and is equipped with norm $||f||_{\mathcal{K}} = \sum_{\ell=1}^L\sum_{\ell^\prime = 1}^L K(\theta_\ell, \theta_{\ell^\prime})c_\ell c_{\ell^\prime}$ for $f\in \mathcal{K}$. 

For each Solution $i$, we aim to find function $g\in \mathcal{K}$ that best fits $\{y_i(\theta_b)\}_{1\leq b\leq B}$, i.e., $y_i(\theta) \approx \beta_i + g(\theta)$, where $\beta_i$ models the constant feature  of $i$ independent of $\theta_b$. To measure the goodness-of-fit, we adopt the Gaussian negative log-likelihood loss (GNLLL) since $\mu_{i,n}(\theta_b)$ is normally distributed under Model~\eqref{eq:normaL_model}. 
After spending  $n$ replications, we seek for $g^*_{i, n}$ such that 
\begin{equation}\label{opt:kernel_regression}
    g^*_{i, n} = \argmin_{g \in \mathcal{K}} \left\{\sum_{b=1}^B \frac{N_i^n(\theta_b)}{2\lambda_i^2(\theta_b)} (\mu_{i, n}(\theta_b) - g(\theta_b) -\beta_i)^2 + \kappa ||g||^2_{\mathcal{K}}\right\}, \;\; \text{for some $\kappa > 0$,}
\end{equation}
where the first term in the objective function is the sum of the GNLLLs at each $\theta_b$ with mean $\beta_i + g(\theta_b)$ from the model and known variance $\frac{\lambda^2_i(\theta_b)}{N_i^n(\theta_b)}$; and $\kappa||g||^2_{\mathcal{K}}$ is a penalty term to prevent overfitting. Theorem~\ref{thm:prediction} below provides a closed-form expression of $g_{i, n}^*$ evaluated at $\{\theta_b\}_{1\leq b\leq B}$.

\begin{theorem}\label{thm:prediction}
    Let $\widehat{\BFmu}_{i, n} = \{g^*_{i, n}(\theta_b)\}_{1\leq b\leq B}$ be the $B$-dimensional vector of the predictive means at $\{\theta_b\}_{1\leq b\leq B}$ obtained by solving~\eqref{opt:kernel_regression} and $\Sigma_{i, n} = \text{diag}\left(\left\{\frac{\lambda_i^2(\theta_b)}{N_i^n(\theta_b)}\right\}\right)$. Then, we have
\begin{equation}\label{eq:post_update}
    \begin{aligned}
    \widehat{\boldsymbol{\mu}}_{i, n} &= \beta_{i} \BFone_B + \BFK_i^\top \left(\BFK_i + \kappa\Sigma_{i, n}\right)^{-1}\boldsymbol({\BFmu}_{i, n} - \beta_{i}\BFone_B),
    \end{aligned}
\end{equation}
where $\BFmu_{i, n} = \{\mu_{i, n}(\theta_b)\}_{1\leq b \leq B}$ and $\BFK_i$ is a $B\times B$ matrix whose $(b, \bprime)$-th element is $K(\theta_b, \theta_\bprime)$.
\end{theorem}

We propose {\bf Algorithm~5}, which replaces the posterior mean $\BFmu_{i, n}$ with~\eqref{eq:post_update} in Algorithm~\ref{alg:rate_opt_posterior}. {Once $\BFmu_{i_n^*, n}$ is updated from the posterior sampling in Line~4 of Algorithm~\ref{alg:rate_opt_posterior}, we also update $\widehat{\BFmu}_{i_n^*, n}$ via~\eqref{eq:post_update} accordingly}.
A natural question is how does this modification affect the algorithm's finite-sample and asymptotic behaviors?  As it is well known in the literature, a plug-in algorithm derived from the optimal static sampling ratios tends to suffer from poor initial estimates of the unknown means due to small sample sizes~\citep{WuZhou:18FixedRS}. Such poor estimates subsequently affect sampling ratios in the later iterations. 
Therefore, the KRR estimator works to reduce the error by linearly interpolating the posterior means at all $\{(i, \theta_b)\}_{1\leq b\leq B}$. 

On the other hand, as $n \rightarrow \infty$, $\widehat{\BFmu}_{i, n}$ falls back to $\BFmu_{i, n}$ as $\Sigma_{i,n}$ converges to a zero matrix. To investigate its convergence rate, we define the cumulant generating function (CGF) of $\widehat{\BFmu}_{i, n}$, $\widehat{\Lambda}_{i, n}(\BFt) \triangleq \log \expec[\exp(\BFt^\top \widehat{\boldsymbol{\mu}}_{i, n})]$.
Theorem~\ref{thm:LDR_GP} below provides the limiting behavior of 
$\widehat{\Lambda}_{i, n}(\BFt)$. We notice that a similar result has been reported by~\cite{Cakmak-GPCOBCA:24} with a different proof.

\begin{theorem}\label{thm:LDR_GP}
    If $\lim_{n \rightarrow \infty}\frac{N_i^n(\theta_b)}{n} = \alpha_i(\theta_b)$ and $\lim_{n \rightarrow \infty} N_i^n(\theta_b) = \infty$, then $\widehat{\Lambda}_{i, n}(\BFt)$ satisfies
    \begin{equation}\label{eq:LDR_GP}
        \lim_{n \rightarrow \infty}\frac{1}{n} \widehat{\Lambda}_{i, n}(n\BFt) 
        = \sum_{b=1}^B\left(y_i(\theta_b) t_b + \frac{\lambda_i^2(\theta_b)t_b^2}{2\alpha_i(\theta_b)}\right).
    \end{equation}
\end{theorem}
The right-hand side of~\eqref{eq:LDR_GP} is identical to the limiting CGF of $\BFmu_{i, n}$ obtained from the independent normal prior. The G\"{a}rtner-Ellis theorem states that the LDR of a rare event involving $\widehat\BFmu_{i, n}$ is given as the Fenchel-Legendre transformation of the limiting CGF. Hence, Theorem~\ref{thm:LDR_GP} implies that the asymptotically optimal static sampling ratios derived for $\{\widehat{\BFmu}_{i, n}\}_{1\leq i\leq k}$ are identical to those attained by adopting $\{\BFmu_{i, n}\}_{1\leq i\leq k}$.


Moreover, adopting $\{\widehat{\BFmu}_{i, n}\}_{1\leq i\leq k}$ can be expected to improve the finite-sample performance since the KRR may have a better mean estimator via pooling. 
We empirically evaluate the performance of Algorithm~5 in Section~\ref{subsec:vco} and observe that it significantly outperforms Algorithm 2. The improvement is more noticeable at the early stage of simulation. 

\noindent{\bf Remark~4}: 
In the contextual R\&S literature, some recent papers apply the Gaussian process (GP) regression to learn the mean, variance, and the LDR at $(i, \theta_b)$ pairs \citep{Cakmak-GPCOBCA:24,Du-COCBA:24}. 
Note that choosing $\kappa=1$ in~\eqref{eq:post_update} makes $\widehat{\boldsymbol{\mu}}_{i, n}$ identical to the predictive mean of the GP when the covariance matrix at $\{\theta_b\}_{1\leq b\leq B}$ is given as $\BFK_i$~\citep{ankenman2010:SK}. This has motivated us to adopt  $\kappa = 1$ in our empirical study. 

\section{Empirical Analysis}
\label{sec:experiments}

In this section, we present numerical performance analyses using a synthetic example and a more realistic market simulation model. The former is designed to examine the effects of the solutions' mean configuration, simulation variance, and non-normality of simulation output to the algorithms' performances using the solutions with known means and variances. The latter is to demonstrate application of MPB in a decision-making problem.
For both examples, we compare Algorithms \ref{alg:rate_opt}--4 with  Equal Allocation (EA) that samples $\lt\{(i, \theta_b)\rt\}$ uniformly given a budget and C-OCBA proposed by \cite{gao2019selecting} for contextual R\&S. Algorithm~5 is tested for the second example. 
For all algorithms, initial sample size $n_0=5$ is adopted for the synthetic example and $n_0 = 10$ for the market simulation example. We evaluate three performance metrics: (i) PFS: $\prob(i_n^* \neq i^*)$; (ii) FNR; and (iii) $1-$ACC. All metrics are reported in the base-10 logarithmic scale.

\subsection{Synthetic example}
\label{subsec:synthetic}

We consider a set of synthetic examples to test robustness of the six algorithms in comparison under different problem configurations. We adopt $k =10$ and $B = 50$, and assume $p_b = 1/B$ for all $1\leq b \leq 50$; a case with unequal probabilities can be found in Appendix~\ref{ec:unequal}. Both cases exhibit the same conclusion. All simulation variances are assumed to be known in this section.

We test four scenarios with different problem configurations. For each scenario, $10,000$ macro runs are made to compute the performance metrics. In all scenarios, we randomize the simulation output standard deviations, $\{\lambda_i(\theta_b)\}$, by sampling them from uniform distribution $U[4, 6]$ in each macro run.
In Baseline scenario, the conditional optimum at each $\theta_b$ is set to be
\begin{equation}\label{eq:synthetic_opt}
    i^b = \ell \;\; (\mbox{if } 5\ell-4\leq b \leq 5\ell, \mbox{ for some } 1\leq \ell \leq 7),\;\; i^b= 8 \;\; (\mbox{if } 36 \leq b \leq 41), \;\; i^b= 10\;\;  (\mbox{if } 42 \leq b \leq 50). 
\end{equation}
This implies $i^*=10$ with preference probability $0.18$. The second best, $i = 8$, has preference probability $0.12$. All solutions except for $i = 9$ are conditional optima at some $\theta_b$. For each macro run, we set $y_{i^b}(\theta_b)=1$ and fill in  $\lt\{y_i(\theta_b)\rt\}_{1\leq i \leq k,i \neq i^b}$ randomly without replacement from  $\lt\{2,3,\cdots,k\rt\}$ for each $1\leq b\leq 50$. 
Additionally, three modifications to Baseline instance are included in the experiment, each of which is designed to examine the algorithms' sensitivity to a problem feature: 
\begin{itemize} [leftmargin=*]
    \item Scenario 1 (more dominant $i^*$): $i^* = 10$ is made more dominant by modifying~\eqref{eq:synthetic_opt} so that $i^b = 10$ for all $36 \leq b\leq 50$. This makes any $1 \leq i \leq 7$ be the second best and the difference in preference probability compared to $i^*$ is $0.2$, which is increased from $0.06$ in Baseline.
    \item Scenario 2 (larger simulation error variance): We draw $\lambda_i(\theta_b)$ from $U[8, 12]$ for each macro run, i.e., the simulation standard deviation is twice of Baseline's on average. 
    \item Scenario 3 (non-normal simulation output): To test robustness when the normality assumption fails to hold, the simulation output is set to follow a shifted exponential distribution, i.e., $Y_i(\theta_b) \overset{D}{=} \mu_i(\theta_b) - \lambda_i(\theta_b) + \text{Exp}(\lambda_i(\theta_b))$, where $\overset{D}{=}$ means equivalence in distribution and Exp($\lambda_i(\theta_b)$) is the exponential random variable with mean $\lambda_i(\theta_b)$. 
\end{itemize}


\begin{figure}[!tb]
	\centering\resizebox{\textwidth}{!}{
	\begin{tikzpicture}[font=\footnotesize]
	\begin{semilogyaxis}[name=plot1,height=1.5in,width=1.8in,
	title={Baseline},
	ylabel={PFS},
	legend style={at={(0.97, 0.97)},
		anchor=north east, font=\scriptsize, nodes=right, style={row sep=0.0cm}},ytickten={0, -1, -2, -3},
	ymin=0.41188643151E-02,
	ymax=1E0,
    xmin=2.5, xmax=20,
	every tick label/.append style={font=\tiny},
	axis on top,
	scaled x ticks = false,
	xticklabel style={/pgf/number format/fixed},
	/pgf/number format/1000 sep={}]
	
    \addplot+[black, densely dotted, mark = none]
	table[x index=0,y index=1, col sep=comma]{plotdata/PFS_Example1.csv};
    \addplot+[black, solid, mark = none]
	table[x index=0,y index=2, col sep=comma, ]{plotdata/PFS_Example1.csv};
    \addplot+[black, mark=*, mark size=1.5pt]
	table[x index=0,y index=3, col sep=comma]{plotdata/PFS_Example1.csv};
    \addplot+[violet, mark = triangle*, mark size=1.5pt]
	table[x index=0,y index=4, col sep=comma]{plotdata/PFS_Example1.csv};
    \addplot+[green, mark = square*, mark size=1.5pt]
	table[x index=0,y index=5, col sep=comma]{plotdata/PFS_Example1.csv};
    \addplot+[red, solid, mark = diamond*, mark size=1.5pt]
	table[x index=0,y index=6, col sep=comma]{plotdata/PFS_Example1.csv};
	
	\end{semilogyaxis}
	\begin{semilogyaxis}[name=plot2,height=1.5in,width=1.8in,at={($(plot1.east)+(0.2in,0in)$)},anchor=west,
	title={Scenario 1},
	legend style={at={(0.97, 0.97)},
		anchor=north east, font=\scriptsize, nodes=right, style={row sep=0.0cm}},ytickten={0, -1, -2},
	ymin=0.41188643151E-02,
	ymax=1E0,
    yticklabels={},
    xmin=2.5, xmax=5,
	every tick label/.append style={font=\tiny},
	axis on top,
	scaled x ticks = false,
	xticklabel style={/pgf/number format/fixed},
	/pgf/number format/1000 sep={}]

    \addplot+[black, densely dotted, mark = none]
	table[x index=0,y index=1, col sep=comma]{plotdata/PFS_Example2.csv};
    \addplot+[black, solid, mark = none]
	table[x index=0,y index=2, col sep=comma, ]{plotdata/PFS_Example2.csv};
    \addplot+[black, mark=*, mark size=1.5pt]
	table[x index=0,y index=3, col sep=comma]{plotdata/PFS_Example2.csv};
    \addplot+[violet, mark = triangle*, mark size=1.5pt]
	table[x index=0,y index=4, col sep=comma]{plotdata/PFS_Example2.csv};
    \addplot+[green, mark = square*, mark size=1.5pt]
	table[x index=0,y index=5, col sep=comma]{plotdata/PFS_Example2.csv};
    \addplot+[red, solid, mark = diamond*, mark size=1.5pt]
	table[x index=0,y index=6, col sep=comma]{plotdata/PFS_Example2.csv};

    \end{semilogyaxis}

    \begin{semilogyaxis}[name=plot3,height=1.5in,width=1.8in,at={($(plot2.east)+(0.2in,0in)$)},anchor=west,
	title={Scenario 2},
	legend style={at={(0.97, 0.97)},
		anchor=north east, font=\scriptsize, nodes=right, style={row sep=0.0cm}},ytickten={0, -1, -2},
	ymin=0.41188643151E-02,
	ymax=1E0,
    yticklabels={},
    xmin=2.5, xmax=100,
	every tick label/.append style={font=\tiny},
	axis on top,
	scaled x ticks = false,
	xticklabel style={/pgf/number format/fixed},
	/pgf/number format/1000 sep={}]

    \addplot+[black, densely dotted, mark = none]
	table[x index=0,y index=1, col sep=comma]{plotdata/PFS_Example3.csv};
    \addplot+[black, solid, mark = none]
	table[x index=0,y index=2, col sep=comma, ]{plotdata/PFS_Example3.csv};
    \addplot+[black, mark=*, mark size=1.5pt]
	table[x index=0,y index=3, col sep=comma]{plotdata/PFS_Example3.csv};
    \addplot+[violet, mark = triangle*, mark size=1.5pt]
	table[x index=0,y index=4, col sep=comma]{plotdata/PFS_Example3.csv};
    \addplot+[green, mark = square*, mark size=1.5pt]
	table[x index=0,y index=5, col sep=comma]{plotdata/PFS_Example3.csv};
    \addplot+[red, solid, mark = diamond*, mark size=1.5pt]
	table[x index=0,y index=6, col sep=comma]{plotdata/PFS_Example3.csv};

    \end{semilogyaxis}

    \begin{semilogyaxis}[name=plot4,height=1.5in,width=1.8in,at={($(plot3.east)+(0.2in,0in)$)},anchor=west,
	title={Scenario 3},
	legend style={at={(0.97, 0.97)},
		anchor=north east, font=\tiny, nodes=right, style={row sep=0.0cm}},ytickten={0, -1, -2},
	ymin=0.41188643151E-02,
	ymax=1E0,
    yticklabels={},
    xmin=2.5, xmax=20,
	every tick label/.append style={font=\tiny},
	axis on top,
	scaled x ticks = false,
	xticklabel style={/pgf/number format/fixed},
	/pgf/number format/1000 sep={},
 legend entries={EA, C-OCBA, Alg 1, Alg 2, Alg 3, Alg 4},
	legend pos=outer north east]

    \addplot+[black, densely dotted, mark = none]
	table[x index=0,y index=1, col sep=comma]{plotdata/PFS_Example4.csv};
    \addplot+[black, solid, mark = none]
	table[x index=0,y index=2, col sep=comma, ]{plotdata/PFS_Example4.csv};
    \addplot+[black, mark=*, mark size=1.5pt]
	table[x index=0,y index=3, col sep=comma]{plotdata/PFS_Example4.csv};
    \addplot+[violet, mark = triangle*, mark size=1.5pt]
	table[x index=0,y index=4, col sep=comma]{plotdata/PFS_Example4.csv};
    \addplot+[green, mark = square*, mark size=1.5pt]
	table[x index=0,y index=5, col sep=comma]{plotdata/PFS_Example4.csv};
    \addplot+[red, solid, mark = diamond*, mark size=1.5pt]
	table[x index=0,y index=6, col sep=comma]{plotdata/PFS_Example4.csv};

    \end{semilogyaxis}


    \begin{semilogyaxis}[name=plot5,height=1.5in,width=1.8in, at={($(plot1.south)+(0in,-1.2in)$)}, anchor = south,
	ylabel={FNR},
	legend style={at={(0.97, 0.97)},
		anchor=north east, font=\scriptsize, nodes=right, style={row sep=0.0cm}},ytickten={0, -1, -2, -3},
	ymin=0.00063095734,
	ymax=1E0,
    xmin=2.5, xmax=40,
	every tick label/.append style={font=\tiny},
	axis on top,
	scaled x ticks = false,
	xticklabel style={/pgf/number format/fixed},
	/pgf/number format/1000 sep={}]
	
    \addplot+[black, densely dotted, mark = none]
	table[x index=0,y index=1, col sep=comma]{plotdata/FNR_Example1.csv};
    \addplot+[black, solid, mark = none]
	table[x index=0,y index=2, col sep=comma, ]{plotdata/FNR_Example1.csv};
    \addplot+[black, mark=*, mark size=1.5pt]
	table[x index=0,y index=3, col sep=comma]{plotdata/FNR_Example1.csv};
    \addplot+[violet, mark = triangle*, mark size=1.5pt]
	table[x index=0,y index=4, col sep=comma]{plotdata/FNR_Example1.csv};
    \addplot+[green, mark = square*, mark size=1.5pt]
	table[x index=0,y index=5, col sep=comma]{plotdata/FNR_Example1.csv};
    \addplot+[red, solid, mark = diamond*, mark size=1.5pt]
	table[x index=0,y index=6, col sep=comma]{plotdata/FNR_Example1.csv};
	
	\end{semilogyaxis}
	\begin{semilogyaxis}[name=plot6,height=1.5in,width=1.8in,at={($(plot5.east)+(0.2in,0in)$)},anchor=west,
	legend style={at={(0.97, 0.97)},
		anchor=north east, font=\scriptsize, nodes=right, style={row sep=0.0cm}},ytickten={0, -1, -2, -3},
	ymin=0.00063095734,
	ymax=1E0,
    yticklabels={},
    xmin=2.5, xmax=15,
	every tick label/.append style={font=\tiny},
	axis on top,
	scaled x ticks = false,
	xticklabel style={/pgf/number format/fixed},
	/pgf/number format/1000 sep={}]

    \addplot+[black, densely dotted, mark = none]
	table[x index=0,y index=1, col sep=comma]{plotdata/FNR_Example2.csv};
    \addplot+[black, solid, mark = none]
	table[x index=0,y index=2, col sep=comma, ]{plotdata/FNR_Example2.csv};
    \addplot+[black, mark=*, mark size=1.5pt]
	table[x index=0,y index=3, col sep=comma]{plotdata/FNR_Example2.csv};
    \addplot+[violet, mark = triangle*, mark size=1.5pt]
	table[x index=0,y index=4, col sep=comma]{plotdata/FNR_Example2.csv};
    \addplot+[green, mark = square*, mark size=1.5pt]
	table[x index=0,y index=5, col sep=comma]{plotdata/FNR_Example2.csv};
    \addplot+[red, solid, mark = diamond*, mark size=1.5pt]
	table[x index=0,y index=6, col sep=comma]{plotdata/FNR_Example2.csv};

    \end{semilogyaxis}

    \begin{semilogyaxis}[name=plot7,height=1.5in,width=1.8in,at={($(plot6.east)+(0.2in,0in)$)},anchor=west,
	legend style={at={(0.97, 0.97)},
		anchor=north east, font=\scriptsize, nodes=right, style={row sep=0.0cm}},ytickten={0, -1, -2, -3},
	ymin=0.00063095734,
	ymax=1E0,
    yticklabels={},
    xmin=2.5, xmax=100,
	every tick label/.append style={font=\tiny},
	axis on top,
	scaled x ticks = false,
	xticklabel style={/pgf/number format/fixed},
	/pgf/number format/1000 sep={}]

    \addplot+[black, densely dotted, mark = none]
	table[x index=0,y index=1, col sep=comma]{plotdata/FNR_Example3.csv};
    \addplot+[black, solid, mark = none]
	table[x index=0,y index=2, col sep=comma, ]{plotdata/FNR_Example3.csv};
    \addplot+[black, mark=*, mark size=1.5pt]
	table[x index=0,y index=3, col sep=comma]{plotdata/FNR_Example3.csv};
    \addplot+[violet, mark = triangle*, mark size=1.5pt]
	table[x index=0,y index=4, col sep=comma]{plotdata/FNR_Example3.csv};
    \addplot+[green, mark = square*, mark size=1.5pt]
	table[x index=0,y index=5, col sep=comma]{plotdata/FNR_Example3.csv};
    \addplot+[red, solid, mark = diamond*, mark size=1.5pt]
	table[x index=0,y index=6, col sep=comma]{plotdata/FNR_Example3.csv};

    \end{semilogyaxis}

    \begin{semilogyaxis}[name=plot8,height=1.5in,width=1.8in,at={($(plot7.east)+(0.2in,0in)$)},anchor=west,
	legend style={at={(0.97, 0.97)},
		anchor=north east, font=\tiny, nodes=right, style={row sep=0.0cm}},ytickten={0, -1, -2, -3},
	ymin=0.00063095734,
	ymax=1E0,
    yticklabels={},
    xmin=2.5, xmax=60,
	every tick label/.append style={font=\tiny},
	axis on top,
	scaled x ticks = false,
	xticklabel style={/pgf/number format/fixed},
	/pgf/number format/1000 sep={}]

    \addplot+[black, densely dotted, mark = none]
	table[x index=0,y index=1, col sep=comma]{plotdata/FNR_Example4.csv};
    \addplot+[black, solid, mark = none]
	table[x index=0,y index=2, col sep=comma, ]{plotdata/FNR_Example4.csv};
    \addplot+[black, mark=*, mark size=1.5pt]
	table[x index=0,y index=3, col sep=comma]{plotdata/FNR_Example4.csv};
    \addplot+[violet, mark = triangle*, mark size=1.5pt]
	table[x index=0,y index=4, col sep=comma]{plotdata/FNR_Example4.csv};
    \addplot+[green, mark = square*, mark size=1.5pt]
	table[x index=0,y index=5, col sep=comma]{plotdata/FNR_Example4.csv};
    \addplot+[red, solid, mark = diamond*, mark size=1.5pt]
	table[x index=0,y index=6, col sep=comma]{plotdata/FNR_Example4.csv};

    \end{semilogyaxis}



    \begin{semilogyaxis}[name=plot9,height=1.5in,width=1.8in, at={($(plot5.south)+(0in,-1.2in)$)}, anchor = south,
	ylabel={$1-\text{ACC}$},
	legend style={at={(0.97, 0.97)},
		anchor=north east, font=\scriptsize, nodes=right, style={row sep=0.0cm}},ytickten={0, -1, -2, -3},
	ymin=0.00063095734,
	ymax=1E0,
    xmin=2.5, xmax=40,
	every tick label/.append style={font=\tiny},
	axis on top,
	scaled x ticks = false,
	xticklabel style={/pgf/number format/fixed},
	/pgf/number format/1000 sep={}]
	
    \addplot+[black, densely dotted, mark = none]
	table[x index=0,y index=1, col sep=comma]{plotdata/ACC_Example1.csv};
    \addplot+[black, solid, mark = none]
	table[x index=0,y index=2, col sep=comma, ]{plotdata/ACC_Example1.csv};
    \addplot+[black, mark=*, mark size=1.5pt]
	table[x index=0,y index=3, col sep=comma]{plotdata/ACC_Example1.csv};
    \addplot+[violet, mark = triangle*, mark size=1.5pt]
	table[x index=0,y index=4, col sep=comma]{plotdata/ACC_Example1.csv};
    \addplot+[green, mark = square*, mark size=1.5pt]
	table[x index=0,y index=5, col sep=comma]{plotdata/ACC_Example1.csv};
    \addplot+[red, solid, mark = diamond*, mark size=1.5pt]
	table[x index=0,y index=6, col sep=comma]{plotdata/ACC_Example1.csv};
	
	\end{semilogyaxis}
	\begin{semilogyaxis}[name=plot10,height=1.5in,width=1.8in,at={($(plot9.east)+(0.2in,0in)$)},anchor=west,
	legend style={at={(0.97, 0.97)},
		anchor=north east, font=\scriptsize, nodes=right, style={row sep=0.0cm}},ytickten={0, -1, -2, -3},
	ymin=0.00063095734,
	ymax=1E0,
    yticklabels={},
    xmin=2.5, xmax=15,
	every tick label/.append style={font=\tiny},
	axis on top,
	scaled x ticks = false,
	xticklabel style={/pgf/number format/fixed},
	/pgf/number format/1000 sep={}]

    \addplot+[black, densely dotted, mark = none]
	table[x index=0,y index=1, col sep=comma]{plotdata/ACC_Example2.csv};
    \addplot+[black, solid, mark = none]
	table[x index=0,y index=2, col sep=comma, ]{plotdata/ACC_Example2.csv};
    \addplot+[black, mark=*, mark size=1.5pt]
	table[x index=0,y index=3, col sep=comma]{plotdata/ACC_Example2.csv};
    \addplot+[violet, mark = triangle*, mark size=1.5pt]
	table[x index=0,y index=4, col sep=comma]{plotdata/ACC_Example2.csv};
    \addplot+[green, mark = square*, mark size=1.5pt]
	table[x index=0,y index=5, col sep=comma]{plotdata/ACC_Example2.csv};
    \addplot+[red, solid, mark = diamond*, mark size=1.5pt]
	table[x index=0,y index=6, col sep=comma]{plotdata/ACC_Example2.csv};

    \end{semilogyaxis}

    \begin{semilogyaxis}[name=plot11,height=1.5in,width=1.8in,at={($(plot10.east)+(0.2in,0in)$)},anchor=west,
	legend style={at={(0.97, 0.97)},
		anchor=north east, font=\scriptsize, nodes=right, style={row sep=0.0cm}},ytickten={0, -1, -2, -3},
	ymin=0.00063095734,
	ymax=1E0,
    yticklabels={},
    xmin=2.5, xmax=100,
	every tick label/.append style={font=\tiny},
	axis on top,
	scaled x ticks = false,
	xticklabel style={/pgf/number format/fixed},
	/pgf/number format/1000 sep={}]

    \addplot+[black, densely dotted, mark = none]
	table[x index=0,y index=1, col sep=comma]{plotdata/ACC_Example3.csv};
    \addplot+[black, solid, mark = none]
	table[x index=0,y index=2, col sep=comma, ]{plotdata/ACC_Example3.csv};
    \addplot+[black, mark=*, mark size=1.5pt]
	table[x index=0,y index=3, col sep=comma]{plotdata/ACC_Example3.csv};
    \addplot+[violet, mark = triangle*, mark size=1.5pt]
	table[x index=0,y index=4, col sep=comma]{plotdata/ACC_Example3.csv};
    \addplot+[green, mark = square*, mark size=1.5pt]
	table[x index=0,y index=5, col sep=comma]{plotdata/ACC_Example3.csv};
    \addplot+[red, solid, mark = diamond*, mark size=1.5pt]
	table[x index=0,y index=6, col sep=comma]{plotdata/ACC_Example3.csv};

    \end{semilogyaxis}

    \begin{semilogyaxis}[name=plot12,height=1.5in,width=1.8in,at={($(plot11.east)+(0.2in,0in)$)},anchor=west,
	every axis x label/.append style = {at={(1.2, 0.15)}, },
	xlabel={\tiny Simulation budget ($\times 10^3$)},
	legend style={at={(0.97, 0.97)},
		anchor=north east, font=\tiny, nodes=right, style={row sep=0.0cm}},ytickten={0, -1, -2, -3},
	ymin=0.00063095734,
	ymax=1E0,
    yticklabels={},
    xmin=2.5, xmax=60,
	every tick label/.append style={font=\tiny},
	axis on top,
	scaled x ticks = false,
 xticklabel style={/pgf/number format/fixed},
	/pgf/number format/1000 sep={}]

    \addplot+[black, densely dotted, mark = none]
	table[x index=0,y index=1, col sep=comma]{plotdata/ACC_Example4.csv};
    \addplot+[black, solid, mark = none]
	table[x index=0,y index=2, col sep=comma, ]{plotdata/ACC_Example4.csv};
    \addplot+[black, mark=*, mark size=1.5pt]
	table[x index=0,y index=3, col sep=comma]{plotdata/ACC_Example4.csv};
    \addplot+[violet, mark = triangle*, mark size=1.5pt]
	table[x index=0,y index=4, col sep=comma]{plotdata/ACC_Example4.csv};
    \addplot+[green, mark = square*, mark size=1.5pt]
	table[x index=0,y index=5, col sep=comma]{plotdata/ACC_Example4.csv};
    \addplot+[red, solid, mark = diamond*, mark size=1.5pt]
	table[x index=0,y index=6, col sep=comma]{plotdata/ACC_Example4.csv};

    \end{semilogyaxis}
	
	\end{tikzpicture}}

 \caption{Performance measure estimates from 10{,}000 macro runs for each scenario. 
 }
    \label{fig:tot}
 \end{figure}
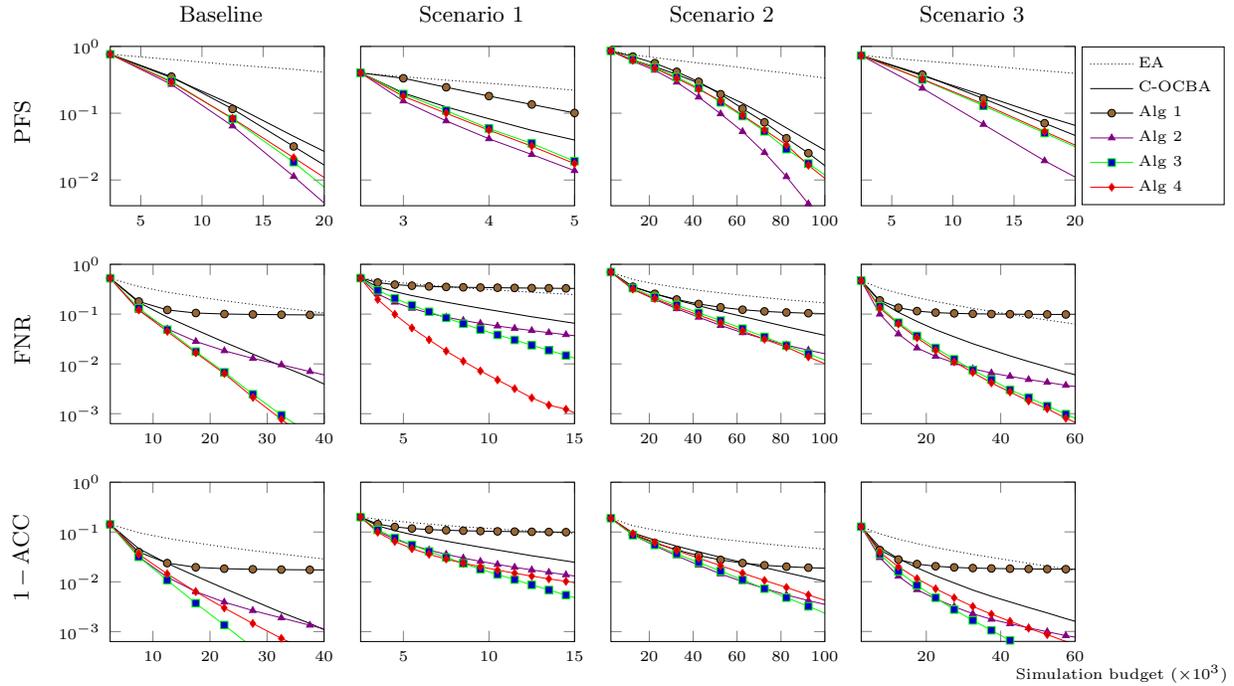

Figure~\ref{fig:tot} summarizes numerical results for all scenarios. Baseline is presented in the first column. Notice that EA and C-OCBA show slower convergence in PFS than our algorithms, where the gaps widen as the simulation budget increases.  Algorithm~\ref{alg:rate_opt} falls behind Algorithms~3 and~4 in PFS. As mentioned in Section~\ref{sec:sequential_learning}, Algorithm~1's shortfall is caused by its underestimation of the MPB's preference probability as  can be confirmed from the FNR plot. Algorithm~\ref{alg:rate_opt_posterior} corrects this behavior and achieves the fastest PFS convergence rate while showing good performances in FNR and $1-$ACC. On the other hand, Algorithms~3 and~4 exhibit the best FNR and $1-$ACC, respectively, as they are designed to maximize the respective measures. 

For Scenario 1, finding the MPB is expected to be easier than Baseline. While all algorithms show improved PFS, notice that C-COBA  outperforms Algorithm~\ref{alg:rate_opt}. This is because Algorithm~\ref{alg:rate_opt}'s underestimation of the MPB's preference probability is more severe when $i^*$ is dominant. Algorithms~\ref{alg:rate_opt_posterior},~4, and~3 continue to show the best performances in PFS, FNR, and $1-$ACC, respectively. 

\begin{figure}
    \centering
    \begin{tikzpicture}
    \begin{axis}[name=plot1,height=2in,width=2.5in, 
        title = {\footnotesize Baseline},
        ylabel = {\footnotesize Fraction allocated to $\Xi^\text{adv}$},
    	legend style={at={(0.97, 0.97)},
    		anchor=north east, font=\scriptsize, nodes=right, style={row sep=0.0cm}},ytickten={-1, -2, -3},
    	ymin=0.025,
    	ymax = 0.2,
        xmin=2.5, xmax=20,
    	every tick label/.append style={font=\tiny},
    	axis on top,
    	scaled x ticks = false,
    	xticklabel style={/pgf/number format/fixed},
            y tick label style={/pgf/number format/precision=10},
    	/pgf/number format/1000 sep={}]
    	
        \addplot+[black, densely dotted, mark = none]
    	table[x index=0,y index=1, col sep=comma]{plotdata/zero_Example1.csv};
        \addplot+[black, solid, mark = none]
    	table[x index=0,y index=2, col sep=comma]{plotdata/zero_Example1.csv};
        \addplot+[black, mark=*, mark size=1.5pt]
    	table[x index=0,y index=3, col sep=comma]{plotdata/zero_Example1.csv};
        \addplot+[violet, mark = triangle*, mark size=1.5pt]
    	table[x index=0,y index=4, col sep=comma]{plotdata/zero_Example1.csv};
        \addplot+[green, mark = square*, mark size=1.5pt]
    	table[x index=0,y index=5, col sep=comma]{plotdata/zero_Example1.csv};
        \addplot+[red, solid, mark = diamond*, mark size=1.5pt]
    	table[x index=0,y index=6, col sep=comma]{plotdata/zero_Example1.csv};
    	
    	\end{axis}
    
        \begin{axis}[name=plot2,height=2in,width=2.5in,at={($(plot1.east)+(0.6in,0in)$)},anchor=west,
        title = {\footnotesize Scenario 2},
        every axis x label/.append style = {at={(1.35, 0.1)}, },
    	xlabel={\tiny Simulation budget ($\times 10^3$)},
    	legend style={at={(0.97, 0.97)},
    		anchor=north east, font=\tiny, nodes=right, style={row sep=0.0cm}},ytickten={0, -1, -2, -3},
    	ymin=0.025,
    	ymax = 0.2,
        xmin=2.5, xmax=80,
    	every tick label/.append style={font=\tiny},
    	axis on top,
    	scaled x ticks = false,
    	xticklabel style={/pgf/number format/fixed},
            y tick label style={/pgf/number format/precision=10},
    	/pgf/number format/1000 sep={},
    legend entries={EA, C-OCBA, Alg 1, Alg 2, Alg 3, Alg 4},
    	legend pos=outer north east]
    
        \addplot+[black, densely dotted, mark = none]
    	table[x index=0,y index=1, col sep=comma]{plotdata/zero_Example3.csv};
        \addplot+[black, solid, mark = none]
    	table[x index=0,y index=2, col sep=comma, ]{plotdata/zero_Example3.csv};
        \addplot+[black, mark=*, mark size=1.5pt]
    	table[x index=0,y index=3, col sep=comma]{plotdata/zero_Example3.csv};
        \addplot+[violet, mark = triangle*, mark size=1.5pt]
    	table[x index=0,y index=4, col sep=comma]{plotdata/zero_Example3.csv};
        \addplot+[green, mark = square*, mark size=1.5pt]
    	table[x index=0,y index=5, col sep=comma]{plotdata/zero_Example3.csv};
        \addplot+[red, solid, mark = diamond*, mark size=1.5pt]
    	table[x index=0,y index=6, col sep=comma]{plotdata/zero_Example3.csv};
    
        \end{axis}
    	
    	\end{tikzpicture}
     \caption{The sum of sampling ratio allocated at the adversarial set paired with $i^*$ for Baseline and Scenario~2.}
        \label{fig:sampling.fraction}
        
    \end{figure}
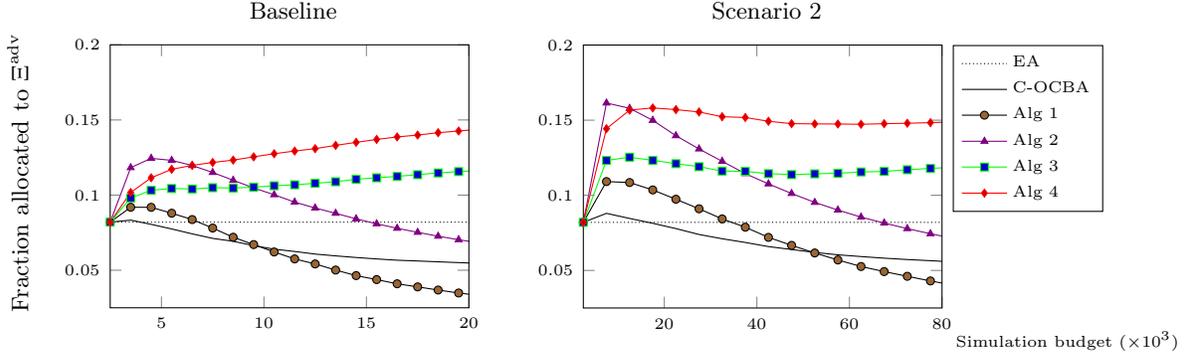

Scenario 2 has higher simulation variances, which explains why all algorithms perform worse than in Baseline given the same budget. Compared to other scenarios, Algorithm~2 exhibits relatively good performance under all three measures in Scenairo~2. To see why, Figure~\ref{fig:sampling.fraction} shows the fractional sampling efforts all six algorithms allocate to all $(i^*,\theta_b)\in\Xi^{\text{adv}}$ as $n$ increases in Baseline and Scenario~2, respectively. Recall that Algorithm~2 guarantees zero asymptotic sampling ratios to $\Xi^{\text{adv}}$, whereas Algorithms~3 and~4 allocate strictly positive sampling ratios to them. Figure~\ref{fig:sampling.fraction} illustrates these behaviors, however, notice that Algorithm~2 allocates a much higher fraction of the budget to $\Xi^{\text{adv}}$ initially compared to Algorithm~1. This can be attributed to that the simulation variances are larger in Scenario~2, so the posterior variances of $\eta_{i^*}(\theta_b), \theta_b\in\Theta_{i^*}^c$, are also larger. Because Algorithm~2 replaces $\mu_{i^*,n}(\theta_b)$ for these pairs with the respective posterior samples from $\eta_{i^*}(\theta_b)$ there is a bigger chance that $i^*$  becomes the sample-best at each $\theta_b\in \Theta_{i^*}(\theta_b)$ granting a higher chance for $(i^*,\theta_b)$ to be sampled. Although this behavior subsides as the simulation budget increases, it makes Algorithm~2 allocate more effort to the pairs that asymptotically would receive zero sampling ratios in the beginning, which manifests as better performances in ACC and FNR. 

Finally, the last column (Scenario 3) of Figure~\ref{fig:tot} shows that even if the simulation error is non-Gaussian, similar trends can be observed as in Baseline, however, all algorithms perform worse. This is not surprising as they are tailored for Gaussian simulation outputs.

Next, to demonstrate the loss in LDR due to relaxations, we compare the optimal objective function values of~\eqref{eq:exact.ocba} and~\eqref{opt:aOCBA} for Baseline and Scenario~1. As discussed in Section~\ref{sec:relax_OCBA},~\eqref{eq:exact.ocba} can be numerically solved via the entropic mirror descent algorithm in Section~\ref{ec:gap_analysis}. We derive an upper bound on the remaining optimality gap when the algorithm is stopped after a finite number of iterations in Proposition~\ref{prop:modified_upper}. 
Utilizing the upper bound, for each macro run, we assess the \emph{relative optimality gap}, $1- \frac{(\ldr_{\textup{FS}} \text{ of } \underline{\boldsymbol{\alpha}}^\textsf{opt})}{\text{(upper bound)}}$. Recall that $\underline{\boldsymbol{\alpha}}^\textsf{opt}$ is the optimal sampling ratios for~\eqref{opt:aOCBA} and its $\ldr_{\textup{FS}}$ is evaluated from the objective function of~\eqref{eq:exact.ocba}. Note that the relative optimality gap is a conservative measure since it is greater than the ground-truth relative loss. 

\begin{table}[tbp!]
\centering 
\caption{Relative optimality gaps between~\eqref{eq:exact.ocba} and~\eqref{opt:aOCBA} for Baseline and Scenario 1 obtained from $1{,}000$ macro runs; $(\gamma_t, T)$ in Algorithm~\ref{EC:alg_EMD} are set as $(1, 5\times10^6)$ and $(1, 10^6)$, respectively. 
}
\label{tab:loss_summary}
\begin{tabular}{|c|c|c|ccc|} 
\hline
\multirow{2}{*}{} & \multirow{2}{*}{$\ldr_{\textup{FS}}$ of $\underline{\boldsymbol{\alpha}}^\textsf{opt}$ } & \multirow{2}{*}{Upper bound computed from~\ref{prop:modified_upper}} & \multicolumn{3}{c|}{Relative optimality gap}                              \\ \cline{4-6} 
                  &                                                                    &                                                                                          & \multicolumn{1}{c|}{~~mean~~}  & \multicolumn{1}{c|}{~~max~~}   & \multicolumn{1}{c|}{~~min~~}   \\ \hline
Baseline          & 3.216E-4                                                           & 3.290E-4                                                                                 & \multicolumn{1}{c|}{0.023} & \multicolumn{1}{c|}{0.060} & \multicolumn{1}{c|}{0.008} \\ \hline
Scenario 1        & 7.616E-4                                                           & 8.978E-4                                                                                 & \multicolumn{1}{c|}{0.151} & \multicolumn{1}{c|}{0.208} & \multicolumn{1}{c|}{0.060} \\ \hline
\end{tabular}
\end{table}

Table~\ref{tab:loss_summary} summarizes the results obtained from $1{,}000$ macro runs. In Baseline, the estimated relative optimality gap is 2\% on average and at most 6\%, demonstrating that the optimal LDR after relaxations closely matches the true optimal LDR. On the contrary, the average relative optimality gap for Scenario~1 is about 15\%. Recall that it is more difficult to identify the MPB in Baseline than in Scenario~1. This numerical example hints at that the relative optimality gap gets smaller as  MPB identification becomes more challenging. 

\subsection{Market simulation for new product release}\label{subsec:vco}

We introduce a market simulation example to demonstrate feasible applications of the MPB framework in a realistic setting. In this example,  a decision-maker (Firm~0 hereafter) selects a new product to release among several candidates aiming to maximize the firm's market share. The relationship between profitability and market share has been a long-standing subject of marketing studies; a firm with a higher share enjoys greater market power by influencing the price of its products by exercising control over demand and supply, and higher perceived product quality by consumers~\citep{marketshare}. 
We examine the revenue market share, defined by the ratio of the firm's revenue to the total market revenue. Moreover, we consider a scenario where economic uncertainties influence the purchase behavior of consumers. For instance, General Motors makes the content decisions for the vehicle models four years before they are launched~\citep{VCO2023}. Due to such delays, it is natural to expect that the price sensitivity of the consumers when the product is released to the market differs from when the launch decision is made.
Modeling such uncertainties with parameter $\theta$, our MPB formulation returns the product that has the highest likelihood of maximizing the market share subject to economic uncertainties. 

\begin{table}[tbp!]
\renewcommand{\arraystretch}{0.8}
\centering
\caption{Product portfolios of Firm~0 and competitors.  The brand value of each firm is presented. For the competitors, the release likelihood of each candidate product is presented in parentheses in the product column. }
\label{tab:tot_portfolio}
\resizebox{0.9\textwidth}{!}{\begin{tabular}{|ccc|l|c|c|c|c|}
\cline{1-3} \cline{5-8}
\multicolumn{3}{|c|}{Firm 0 (Brand value = 450)}                                                                            & \; & Competitors (Brand value)                                                      & Product (Release probability) & Feature-levels & Price \\ \cline{1-3} \cline{5-8} 
\multicolumn{1}{|c|}{Product}                 & \multicolumn{1}{c|}{Feature-levels}                 & Price                &  & \multirow{3}{*}{\begin{tabular}[c]{@{}c@{}}Firm 1\\ (500)\end{tabular}} & Prod 1.1 (0.5)        & (2, 3, 2, 3, 3)  & 430   \\ \cline{1-3} \cline{6-8} 
\multicolumn{1}{|c|}{\multirow{2}{*}{Prod 1}} & \multicolumn{1}{c|}{\multirow{2}{*}{(3, 3, 3, 3, 3)}} & \multirow{2}{*}{490} &  &                                                                         & Prod 1.2 (0.3)        & (2, 2, 3, 3, 2)  & 420   \\ \cline{6-8} 
\multicolumn{1}{|c|}{}                        & \multicolumn{1}{c|}{}                                 &                      &  &                                                                         & Prod 1.3 (0.2)        & (3, 3, 3, 3, 3)  & 520   \\ \cline{1-3} \cline{5-8} 
\multicolumn{1}{|c|}{\multirow{2}{*}{Prod 2}} & \multicolumn{1}{c|}{\multirow{2}{*}{(2, 2, 2, 2, 2)}} & \multirow{2}{*}{350} &  & \multirow{3}{*}{\begin{tabular}[c]{@{}c@{}}Firm 2\\ (350)\end{tabular}} & Prod 2.1 (0.4)        & (2, 2, 3, 3, 3)  & 390   \\ \cline{6-8} 
\multicolumn{1}{|c|}{}                        & \multicolumn{1}{c|}{}                                 &                      &  &                                                                         & Prod 2.2 (0.4)        & (3, 2, 2, 2, 3)  & 390   \\ \cline{1-3} \cline{6-8} 
\multicolumn{1}{|c|}{\multirow{2}{*}{Prod 3}} & \multicolumn{1}{c|}{\multirow{2}{*}{(1, 1, 1, 1, 1)}} & \multirow{2}{*}{210} &  &                                                                         & Prod 2.3 (0.2)        & (3, 1, 3, 3, 3)  & 400   \\ \cline{5-8} 
\multicolumn{1}{|c|}{}                        & \multicolumn{1}{c|}{}                                 &                      &  & \multirow{4}{*}{\begin{tabular}[c]{@{}c@{}}Firm 3\\ (250)\end{tabular}} & Prod 3.1 (0.3)        & (2, 2, 2, 2, 2)  & 400   \\ \cline{1-3} \cline{6-8} 
\multicolumn{1}{|c|}{\multirow{2}{*}{Prod 4}} & \multicolumn{1}{c|}{\multirow{2}{*}{(3, 3, 2, 2, 2)}} & \multirow{2}{*}{420} &  &                                                                         & Prod 3.2 (0.3)        & (1, 2, 1, 2, 2)  & 390   \\ \cline{6-8} 
\multicolumn{1}{|c|}{}                        & \multicolumn{1}{c|}{}                                 &                      &  &                                                                         & Prod 3.3 (0.2)        & (1, 3, 1, 1, 2)  & 350   \\ \cline{1-3} \cline{6-8} 
\multicolumn{1}{|c|}{\multirow{2}{*}{Prod 5}} & \multicolumn{1}{c|}{\multirow{2}{*}{(3, 3, 1, 1, 1)}} & \multirow{2}{*}{350} &  &                                                                         & Prod 3.4 (0.2)        & (1, 2, 2, 1, 1)  & 310   \\ \cline{5-8} 
\multicolumn{1}{|c|}{}                        & \multicolumn{1}{c|}{}                                 &                      &  & \multirow{5}{*}{\begin{tabular}[c]{@{}c@{}}Firm 4\\ (150)\end{tabular}} & Prod 4.1 (0.2)        & (1, 1, 1, 1, 1)  & 230   \\ \cline{1-3} \cline{6-8} 
\multicolumn{1}{|c|}{\multirow{2}{*}{Prod 6}} & \multicolumn{1}{c|}{\multirow{2}{*}{(2, 2, 3, 3, 1)}} & \multirow{2}{*}{400} &  &                                                                         & Prod 4.2 (0.2)        & (1, 2, 1, 2, 2)  & 320   \\ \cline{6-8} 
\multicolumn{1}{|c|}{}                        & \multicolumn{1}{c|}{}                                 &                      &  &                                                                         & Prod 4.3 (0.2)        & (1, 1, 1, 2, 2)  & 270   \\ \cline{1-3} \cline{6-8} 
\multicolumn{1}{|c|}{\multirow{2}{*}{Prod 7}} & \multicolumn{1}{c|}{\multirow{2}{*}{(2, 2, 1, 1, 3)}} & \multirow{2}{*}{300} &  &                                                                         & Prod 4.4 (0.2)        & (2, 1, 2, 1, 1)  & 350   \\ \cline{6-8} 
\multicolumn{1}{|c|}{}                        & \multicolumn{1}{c|}{}                                 &                      &  &                                                                         & Prod 4.5 (0.2)        & (2, 1, 1, 1, 1)  & 300   \\ \cline{1-3} \cline{5-8} 
\end{tabular}}
\end{table}

The first three columns of Table~\ref{tab:tot_portfolio} show the seven candidate products of Firm~0; among them, one product is to be released in the market. 
Each product can be characterized by its features (e.g., color). There are five main features of the products, each offered in three levels (e.g., red, blue, black). 
We assume the firm learns the consumer utilities for product feature-levels from a conjoint analysis,  a survey designed to estimate how much a consumer values the features of a product relative to the price~\citep{train2009discrete}. In practice, respondents of a conjoint analysis are carefully recruited to reflect the consumer base in the market. We assume that the survey was conducted with a group of 100 respondents, which is a perfect stratified sample of the consumer base. Let $\boldsymbol{\beta}_n \in \mathbb{R}^{17}$ be the  vector of  estimated utility parameters of the $n$th respondent, where the first fifteen elements are the estimated utilities for the feature-levels, and the last two elements represent utilities for each firm's brand value and price, respectively. 
Tables~\ref{tab:Utility_part1} and~\ref{tab:Utility_part2} in Section~\ref{apdx:market_simulation} show the numerical values of $\boldsymbol{\beta}_n$ for $1\leq n \leq 100$ we adopted in this example. 

Five competitors with varying brand values sell the same type of products as shown in Table~\ref{tab:tot_portfolio}.
Firm~0 cannot perfectly predict the features of the competitors' products that will be available in the market at the time of the new product release, however, has probabilistic information as presented in Table~\ref{tab:tot_portfolio}; e.g., Firm~1 is likely to release Product~1.1  with probability $0.5$. 

Let $\BFx_{ j r} = (x_{jr}(\ell))_{1\leq \ell\leq 17}$ be the vector of feature-levels of Product~$r$ of Firm~$j$. The first 15 elements of $\BFx_{jr}$ are indicators denoting whether the product offers the corresponding feature-levels. The last two elements correspond to the brand value and the (negative) price, respectively.  The following linear model represents the $n$th customer's utility for Product~$r$ of Firm~$j$:
\begin{equation}\label{eq:general_choice_model}
        U_{njr} = \boldsymbol{\beta}_n^{\top}\BFx_{jr} + \epsilon_{njr}, \mbox{ where }
        \{\epsilon_{njr}\} \;\; \text{is i.i.d and} \;\; \prob(\epsilon_{njr} \leq \epsilon) = \exp(-\exp(-\epsilon)) \;\; \forall \epsilon\geq 0
\end{equation}
Note that $\epsilon_{njr}$  is the random noise that cannot be captured by the linear model. The no-purchase option can be modeled by setting $\BFx_{jr}$ to be the vector of zeroes.

We introduce parameter $\theta$ to model the change in the customer's marginal utility for the price. The adjusted utility vector of the $n$th customer, $\boldsymbol{\beta}_n(\theta)$, has $(\boldsymbol{\beta}_n(\theta))_\ell = (\boldsymbol{\beta}_{n})_\ell$ for all $1\leq \ell\leq 16$, and $(\boldsymbol{\beta}_n(\theta))_{17} = \theta\times(\boldsymbol{\beta}_{n})_{17}$; $U_{njr}(\theta)$ can be defined by replacing $ \boldsymbol{\beta}_n$ in~\eqref{eq:general_choice_model} with $\boldsymbol{\beta}_n(\theta).$
Since the marginal utility of the price is negative, $\theta>1$ implies that the value of a dollar has gone up compared to when the conjoint analysis was conducted, reducing the product utility. 
We assume that the distribution of $\theta$ is approximated by the probability simplex in  Figure~\ref{fig:probability_weight}.

A MPB problem is then formulated to select the product maximizing Firm~0's market share. 
Solution $i$, represents the $i$th candidate product and $\theta_b, 1\leq b\leq 20,$ is each value on the support of $\theta$ in Figure~\ref{fig:probability_weight}. Given $i$ and $\theta_b$, the simulated market share is denoted by $Y_i(\theta_b)$. Since Firm~0 does not know exactly which products it will compete against, we construct the consumer's choice set (the set of options to compare before making a purchase decision) by sampling each competitor's product according their release probabilities in each replication. Specifically, let $c_j, 1\leq j \leq 4,$ be Firm $j$'s product available in the market. Then, $c_1$ is sampled among Products~1.1--1.3 according to their release probabilities in Table~\ref{tab:tot_portfolio}. Once all $c_j$s  are sampled, each simulated consumer has the choice set, $\{c_0, c_1, c_2, c_3, c_4, \mbox{no purchase}\}$, where $c_0$ is Product $i$. Under Model~\eqref{eq:general_choice_model}, the probability of  the $n$th customer selecting $c_j$ can be derived as $\frac{\exp(\beta_n(\theta_b)^{\top}\BFx_{j c_j} )}{1 + \sum_{\ell=0}^{4}\exp(\beta_n(\theta_b)^{\top}\BFx_{\ell c_\ell})}$~\citep{train2009discrete}. From these probabilities, 
 the number customers who purchase $c_j$ given $\BFc = \{c_0,c_1,\ldots,c_4\}$, 
 $N_j(\BFc)$, can be generated by sampling from the multinomial distribution. Then, the  market share of Firm 0 is 
$Y_i(\theta_b) = p_{0 c_0}N_0(\BFc)/\{\sum_{j=0}^4 p_{j c_j}N_j(\BFc)\}$,
where $p_{jr}$ is the price of Firm $j$'s Product $r$. Hence,  the expected market share of Product $i$ given $\theta_b$ is $y_i(\theta_b) = \expec_{\BFc, \mathbf{N}}[Y_i(\theta_b)|\theta_b]$,   where the expectation is taken  with respect to $\mathbf{c}$ as well as $\mathbf{N} = \{N_j(\mathbf{c}), 0\leq j \leq  4\}.$ 

To evaluate the performances of the proposed algorithms,  $y_i(\theta_b)$ for all $1\leq i\leq 7$ and $1\leq b\leq 20$ are estimated via Monte Carlo simulation and presented in Table~\ref{tab:mean_ft}.   
Table~\ref{tab:criterion_results} shows the preference probabilities of each product computed from the Monte Carlo estimates. Clearly, Product~3 is the MPB as it has the largest preference probability. 


\begin{table}[tbp!]
\renewcommand{\arraystretch}{0.8}
\centering
\caption{Preference probability, expected revenue market share, and the expected sales revenue of each product. The optimum under each criterion is in bold.}
\label{tab:criterion_results}
\resizebox{0.9\textwidth}{!}{\begin{tabular}{|c|l|l|l|l|l|l|l|}
\hline
                                    & Prod 1    & Prod 2    & Prod 3          & Prod 4                                    & Prod 5    & Prod 6    & Prod 7          \\ \hline
Preference probability  & 0.189    & 0         & \textbf{0.440} & 0.140                                    & 0         & 0         & 0.231          \\ \hline
Average market share  ($\mathrm{E}_{\theta}[y_i(\theta)]$)                & 0.214    & 0.237    & 0.243          & 0.232                                    & 0.223    & 0.201    & \textbf{0.247} \\ \hline
Average sales revenue              & 5508.27 & 5665.04 & 5095.23       & {\textbf{5801.53}}                      & 5311.81 & 4905.47 & 5716.77       \\ \hline
\end{tabular}}
\end{table}


We also report two alternative decision criteria in Table~\ref{tab:criterion_results}. One is to average the estimated market shares over $\theta$; Product 7 turns out to be the best in this case. However, this criterion is difficult to interpret because the overall market size (purchasing population) changes in $\theta$.  The variability of market size over $\theta$ is detailed in Table EC.4. Interestingly, Product 7's preference probability is only a half of Product 3's.
The other decision criterion is the expected sales revenue, which is maximized when  Product 4 is adopted, however, its preference probability is only 0.14 compared to 0.44 of Product 3 implying that it has a high probability of failing to be the market leader. When $\theta$ is large, Product 3's pricing is advantageous to the competitors' products. These observations show that the MPB formulation approaches the problem from a different perspective and offers valuable business insights. 

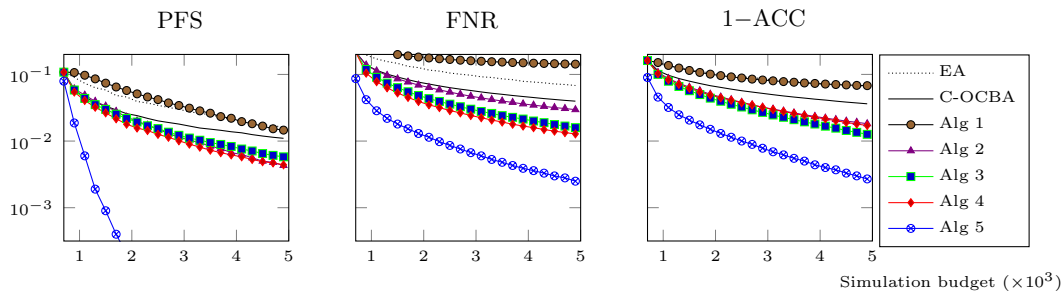
\begin{figure}
\centering
\begin{tikzpicture}
\begin{semilogyaxis}[name=plot1,height=1.6in,width=1.8in, 
    title = {\footnotesize PFS},
	legend style={at={(0.97, 0.97)},
		anchor=north east, font=\scriptsize, nodes=right, style={row sep=0.0cm}},ytickten={-1, -2, -3, -4},
	ymin=0.0003162277,
	ymax = 0.19952623149,
        xmin=0.7, xmax=5,
	every tick label/.append style={font=\tiny},
	axis on top,
	scaled x ticks = false,
	xticklabel style={/pgf/number format/fixed},
	/pgf/number format/1000 sep={}]
	
    \addplot+[black, densely dotted, mark = none]
	table[x index=0,y index=1, col sep=comma]{plotdata/KRR/MarketShare_KRR_PFS.csv};
    \addplot+[black, solid, mark = none]
	table[x index=0,y index=2, col sep=comma]{plotdata/KRR/MarketShare_KRR_PFS.csv};
    \addplot+[black, mark=*, mark size=1.5pt]
	table[x index=0,y index=3, col sep=comma]{plotdata/KRR/MarketShare_KRR_PFS.csv};
    \addplot+[violet, mark = triangle*, mark size=1.5pt]
	table[x index=0,y index=4, col sep=comma]{plotdata/KRR/MarketShare_KRR_PFS.csv};
    \addplot+[green, mark = square*, mark size=1.5pt]
	table[x index=0,y index=5, col sep=comma]{plotdata/KRR/MarketShare_KRR_PFS.csv};
    \addplot+[red, solid, mark = diamond*, mark size=1.5pt]
	table[x index=0,y index=6, col sep=comma]{plotdata/KRR/MarketShare_KRR_PFS.csv};	
     \addplot+[blue, solid, mark = otimes, mark size=1.5pt]
	table[x index=0,y index=7, col sep=comma]{plotdata/KRR/MarketShare_KRR_PFS.csv};	
	\end{semilogyaxis}
 
	\begin{semilogyaxis}[name=plot2,height=1.6in,width=1.8in,at={($(plot1.east)+(0.35in,0in)$)},anchor=west,
    title = {\footnotesize FNR},
	legend style={at={(0.97, 0.97)},
		anchor=north east, font=\scriptsize, nodes=right, style={row sep=0.0cm}},ytickten={0, -1, -2, -3},
	ymin=0.0003162277,
	ymax = 0.19952623149,
    yticklabels={},
    xmin=0.7, xmax=5,
	every tick label/.append style={font=\tiny},
	axis on top,
	scaled x ticks = false,
	xticklabel style={/pgf/number format/fixed},
	/pgf/number format/1000 sep={}]

    \addplot+[black, densely dotted, mark = none]
	table[x index=0,y index=1, col sep=comma]{plotdata/KRR/MarketShare_KRR_FNR.csv};
    \addplot+[black, solid, mark = none]
	table[x index=0,y index=2, col sep=comma, ]{plotdata/KRR/MarketShare_KRR_FNR.csv};
    \addplot+[black, mark=*, mark size=1.5pt]
	table[x index=0,y index=3, col sep=comma]{plotdata/KRR/MarketShare_KRR_FNR.csv};
    \addplot+[violet, mark = triangle*, mark size=1.5pt]
	table[x index=0,y index=4, col sep=comma]{plotdata/KRR/MarketShare_KRR_FNR.csv};
    \addplot+[green, mark = square*, mark size=1.5pt]
	table[x index=0,y index=5, col sep=comma]{plotdata/KRR/MarketShare_KRR_FNR.csv};
    \addplot+[red, solid, mark = diamond*, mark size=1.5pt]
	table[x index=0,y index=6, col sep=comma]{plotdata/KRR/MarketShare_KRR_FNR.csv};
    \addplot+[blue, solid, mark = otimes, mark size=1.5pt]
	table[x index=0,y index=7, col sep=comma]{plotdata/KRR/MarketShare_KRR_FNR.csv};	

    \end{semilogyaxis}

    \begin{semilogyaxis}[name=plot3,height=1.6in,width=1.8in,at={($(plot2.east)+(0.35in,0in)$)},anchor=west,
    title = {\footnotesize $1-$ACC},
    every axis x label/.append style = {at={(1.35, 0.1)}, },
	xlabel={\tiny Simulation budget ($\times 10^3$)},
	legend style={at={(0.97, 0.97)},
		anchor=north east, font=\tiny, nodes=right, style={row sep=0.0cm}},ytickten={0, -1, -2, -3},
	ymin= 0.0003162277,
	ymax = 0.19952623149,
    yticklabels={},
    xmin=0.7, xmax=5,
	every tick label/.append style={font=\tiny},
	axis on top,
	scaled x ticks = false,
	xticklabel style={/pgf/number format/fixed},
	/pgf/number format/1000 sep={},
        legend entries={EA, C-OCBA, Alg 1, Alg 2, Alg 3, Alg 4, Alg 5},
	legend pos=outer north east]

    \addplot+[black, densely dotted, mark = none]
	table[x index=0,y index=1, col sep=comma]{plotdata/KRR/MarketShare_KRR_ACC.csv};
    \addplot+[black, solid, mark = none]
	table[x index=0,y index=2, col sep=comma, ]{plotdata/KRR/MarketShare_KRR_ACC.csv};
    \addplot+[black, mark=*, mark size=1.5pt]
	table[x index=0,y index=3, col sep=comma]{plotdata/KRR/MarketShare_KRR_ACC.csv};
    \addplot+[violet, mark = triangle*, mark size=1.5pt]
	table[x index=0,y index=4, col sep=comma]{plotdata/KRR/MarketShare_KRR_ACC.csv};
    \addplot+[green, mark = square*, mark size=1.5pt]
	table[x index=0,y index=5, col sep=comma]{plotdata/KRR/MarketShare_KRR_ACC.csv};
    \addplot+[red, solid, mark = diamond*, mark size=1.5pt]
	table[x index=0,y index=6, col sep=comma]{plotdata/KRR/MarketShare_KRR_ACC.csv};
    \addplot+[blue, solid, mark = otimes, mark size=1.5pt]
	table[x index=0,y index=7, col sep=comma]{plotdata/KRR/MarketShare_KRR_ACC.csv};	
    \end{semilogyaxis}
	
	\end{tikzpicture}
 \caption{Performance measure estimates from 100{,}000 macro runs.} 
    \label{fig:market}   
\end{figure}

Figure~\ref{fig:market} shows the performances of the seven algorithms applied to this example. Unlike the synthetic examples in Section~\ref{subsec:synthetic}, the simulation error variances are unknown in this case, so Normal-gamma conjugate model~\eqref{eq:normal_gamma_model} in Section~\ref{apdx:unknown_variance} is adopted. Moreover, a batch of $5$ replications is run every time a sampling decision is made. In addition to Algorithms 1--4, we implement Algorithm~5 introduced in Section~\ref{sec:GP_extension} to examine the KRR's finite-sample performance improvement. We adopt the squared exponential kernel, $K(\theta, \theta^\prime) = s_i^2 \exp(-\frac{||\theta - \theta^\prime||_2^2}{2\ell_i})$, 
where the two hyperparameters, $(s_i, \ell_i)$, and the constant trend $\beta_i$ are calibrated from the initial sample means $\{\mu_{i,n_0}(\theta_b)\}_{1\leq b\leq B}$ via maximum likelihood estimation for each $1\leq i\leq k$. 

From Figure~\ref{fig:market}, observe that Algorithm~5 clearly outperforms the other algorithms in all three performance measures. This demonstrates that pooling the replications made at different input parameters via the KRR substantially improves the empirical convergence rates of all three performance measures. In particular, the PFS of Algorithm~5 decays dramatically faster compared to other algorithms. 


\section{Conclusions}\label{sec:conclusion}

We introduced the MPB as a solution approach for a R\&S problem when there is uncertainty about the true input distribution. We showed that the MPB selection problem can be formulated as nested R\&S and derived the LDR for the PFS. To find the asymptotically optimal static sampling ratios for all solution-parameter pairs, we formulated an OCBA problem, which is shown to be convex, and discussed a numerical solution approach. To obtain a computationally efficient sequential sampling algorithm, we relaxed the OCBA problem by first recognizing that it can reformulated as a knapsack problem, then by deriving a lower bound for the exact LDR. The resulting OCBA problem has easy-to-verify set of optimality conditions. We proposed four sequential sampling algorithms then showed their strong consistency. While all of them find the MPB almost surely as the simulation budget increases, Algorithms~2,~3, and~4 are respectively designed to minimize PFS, 1$-$ACC, and FNR most effectively, which can be confirmed by our numerical analyses. In particular, Algorithm~2 shows strong performance under all three measures when simulation error variances of solutions are large, which can be attributed to that it combines posterior mean prediction and posterior sampling in sampling decisions. 

To demonstrate possible applications of MPB, we introduced the market share maximization problem under economic uncertainties. In this example,  the solution with the best expected market share, where the average is taken with respect to the economic uncertainties, has a low probability of becoming the market leader, whereas the MPB can directly maximize the probability of selecting a share-maximizing product. This highlights that the MPB provides a different perspective than the risk-neutral formulation when there is uncertainty about the simulation input model. 

Another important application is optimization of the input data collection scheme for a R\&S problem when there are several unknown input distributions that must be estimated from data. This problem has two layers of sampling decisions to make; 1) the simulation sampling scheme to find the optimal \textit{simulated} system, which may not be the true optimum if the input models have large estimation error, and 2) the input data collection scheme to improve the input models. The MPB formulation is directly related to 1); if the uncertainty about $\theta$ is modeled with a posterior distribution given input data, the MPB is a maximum a posteriori estimator of the true optimum given the posterior. Expanding on this idea, \cite{KimSong2022} propose an OCBA framework to simultaneously optimize the input data and simulation sampling schemes.


There are several research questions we aim to explore in future studies. The first is to create a sequential sampling algorithm beyond the plug-in versions. A DP-based algorithm may significantly improve the finite-sample performance of the plug-in algorithms, if a suitable acquisition function can be defined. The optimal sampling ratios analyzed in this paper can serve as a benchmark to assess the large-sample performance of such an algorithm. 
Moreover, adopting CRNs may significantly improve the computational efficiency of the selection procedure. 
Incorporating a continuous parameter support and non-normal simulation output distributions is another possible avenue.  Lastly, a purely Bayesian formulation of the MPB problem discussed in Section~\ref{sec:learning.model} requires further investigations. 

\bibliographystyle{informs2014} 
\bibliography{references.bib}

\begin{thebibliography}{60}
\providecommand{\natexlab}[1]{#1}
\providecommand{\url}[1]{\texttt{#1}}
\providecommand{\urlprefix}{URL }

\bibitem[{Ankenman et~al.(2010)Ankenman, Nelson, \protect\BIBand{}
  Staum}]{ankenman2010:SK}
Ankenman B, Nelson BL, Staum J (2010) Stochastic kriging for simulation
  metamodeling. \emph{Operations Research} 58(2):371--382.

\bibitem[{Applegate et~al.(2020)Applegate, Feldman, Hunter, \protect\BIBand{}
  Pasupathy}]{Appleaget-MORS:20}
Applegate EA, Feldman G, Hunter SR, Pasupathy R (2020) Multi-objective ranking
  and selection: Optimal sampling laws and tractable approximations via
  {SCORE}. \emph{Journal of Simulation} 14(1):21--40.

\bibitem[{Avci et~al.(2023)Avci, Nelson, Song, \protect\BIBand{}
  W{\"a}chter}]{Avic23:gECI}
Avci H, Nelson BL, Song E, W{\"a}chter A (2023) Using cache or credit for
  parallel ranking and selection. \emph{ACM Transactions on Modeling and
  Computer Simulation} 33(4):1--28.

\bibitem[{Barton et~al.(2008)Barton, Briggs, \protect\BIBand{}
  Fenwick}]{healthcare2}
Barton GR, Briggs AH, Fenwick EAL (2008) Optimal cost-effectiveness
  decisions:the role of the cost-effectiveness acceptability curve ({CEAC}),
  the cost-effectiveness acceptability frontier ({CEAF}), and the expected
  value of perfection information ({EVPI}). \emph{Value in Health}
  11(5):886--897.

\bibitem[{Bechhofer et~al.(1959)Bechhofer, Elmaghraby, \protect\BIBand{}
  Morse}]{bechhofer1959single}
Bechhofer RE, Elmaghraby S, Morse N (1959) A single-sample multiple-decision
  procedure for selecting the multinomial event which has the highest
  probability. \emph{The Annals of Mathematical Statistics} 102--119.

\bibitem[{Beck \protect\BIBand{} Teboulle(2003)}]{beck2003mirror}
Beck A, Teboulle M (2003) Mirror descent and nonlinear projected subgradient
  methods for convex optimization. \emph{Operations Research Letters}
  31(3):167--175.

\bibitem[{Berger(1985)}]{berger1985statistical}
Berger JO (1985) \emph{Statistical Decision Theory and Bayesian Analysis}
  (Springer Science \& Business Media).

\bibitem[{Bhattacharya et~al.(2022)Bhattacharya, Morgan, \protect\BIBand{}
  Rego}]{marketshare}
Bhattacharya A, Morgan NA, Rego LL (2022) Examining why and when market share
  drives firm profit. \emph{Journal of Marketing} 86(4):73--94.

\bibitem[{Bishop(2006)}]{bishop2006pattern}
Bishop CM (2006) Pattern recognition. \emph{Machine learning} 128(9).

\bibitem[{Boyd \protect\BIBand{} Vandenberghe(2004)}]{boyd2004convex}
Boyd S, Vandenberghe L (2004) \emph{Convex optimization} (Cambridge University
  Press).

\bibitem[{Cakmak et~al.(2023)Cakmak, Wang, Gao, \protect\BIBand{}
  Zhou}]{Cakmak-GPCOBCA:24}
Cakmak S, Wang Y, Gao S, Zhou E (2023) Contextual ranking and selection with
  {G}aussian processes and {OCBA}. \emph{ACM Transactions on Modeling and
  Computer Simulation} Forthcoming.

\bibitem[{Chen et~al.(2000)Chen, Lin, Y{\"u}cesan, \protect\BIBand{}
  Chick}]{chen2000simulation}
Chen CH, Lin J, Y{\"u}cesan E, Chick SE (2000) Simulation budget allocation for
  further enhancing the efficiency of ordinal optimization. \emph{Discrete
  Event Dynamic Systems} 10(3):251--270.

\bibitem[{Chen(1992)}]{chen1992truncated}
Chen P (1992) Truncated selection procedures for the most probable event and
  the least probable event. \emph{Annals of the Institute of Statistical
  Mathematics} 44(4):613--622.

\bibitem[{Chen \protect\BIBand{} Ryzhov(2019)}]{chen2019complete}
Chen Y, Ryzhov IO (2019) Complete expected improvement converges to an optimal
  budget allocation. \emph{Advances in Applied Probability} 51(1):209--235.

\bibitem[{Chen \protect\BIBand{} Ryzhov(2023)}]{chen2022BOLD}
Chen Y, Ryzhov IO (2023) Balancing optimal large deviations in sequential
  selection. \emph{Management Science} 69(6):3457--3473.

\bibitem[{Chick \protect\BIBand{} Inoue(2001)}]{chick2001new}
Chick SE, Inoue K (2001) New two-stage and sequential procedures for selecting
  the best simulated system. \emph{Operations Research} 49(5):732--743.

\bibitem[{Corlu \protect\BIBand{} Biller(2013)}]{corlu2013}
Corlu C, Biller B (2013) A subset selection procedure under input parameter
  uncertainty. Pasupathy R, Kim SH, Tolk A, Hill R, Kuhl ME, eds.,
  \emph{Proceedings of the 2013 Winter Simulation Conference}, 463--473
  (Piscataway, New Jersey: IEEE).

\bibitem[{Corlu \protect\BIBand{} Biller(2015)}]{corlu2015}
Corlu CG, Biller B (2015) Subset selection for simulations accounting for input
  uncertainty. Yilmaz L, Chan WKV, I~Moon TMKR, Macal C, Rossetti MD, eds.,
  \emph{Proceedings of the 2015 Winter Simulation Conference}, 437--446
  (Piscataway, New Jersey: IEEE).

\bibitem[{Daskin et~al.(1997)Daskin, Hesse, \protect\BIBand{}
  Revelle}]{daskin1997alpha}
Daskin MS, Hesse SM, Revelle CS (1997) $\alpha$-reliable p-minimax regret: A
  new model for strategic facility location modeling. \emph{Location Science}
  5(4):227--246.

\bibitem[{Dembo \protect\BIBand{} Zeitouni(2009)}]{dembo2009large}
Dembo A, Zeitouni O (2009) \emph{Large Deviations Techniques and Applications},
  volume~38 (New York, NY, USA: Springer Science \& Business Media).

\bibitem[{Du et~al.(2024)Du, Gao, \protect\BIBand{} Chen}]{Du-COCBA:24}
Du J, Gao S, Chen CH (2024) A contextual ranking and selection method for
  personalized medicine. \emph{Manufacturing \& Service Operations Management}
  26(1):167--181.

\bibitem[{Durrett(2019)}]{durrett2019probability}
Durrett R (2019) \emph{Probability: Theory and Examples}, volume~49 (Cambridge
  University Press).

\bibitem[{Fan et~al.(2020)Fan, Hong, \protect\BIBand{}
  Zhang}]{fan2020distributionally}
Fan W, Hong LJ, Zhang X (2020) Distributionally robust selection of the best.
  \emph{Management Science} 66(1):190--208.

\bibitem[{Feldman \protect\BIBand{} Hunter(2018)}]{Feldman-BORS:18}
Feldman G, Hunter SR (2018) {SCORE} allocations for bi-objective ranking and
  selection. \emph{ACM Transactions on Modeling and Computer Simulation
  (TOMACS)} 28(1):1--28.

\bibitem[{Fenwick et~al.(2001)Fenwick, Claxton, \protect\BIBand{}
  Sculpher}]{healthcare1}
Fenwick E, Claxton K, Sculpher M (2001) Representing uncertainty: the role of
  cost-effectiveness acceptability curves. \emph{Health Economics}
  10(8):779--787.

\bibitem[{Frazier et~al.(2008)Frazier, Powell, \protect\BIBand{}
  Dayanik}]{frazier2008knowledge}
Frazier PI, Powell WB, Dayanik S (2008) A knowledge-gradient policy for
  sequential information collection. \emph{SIAM Journal on Control and
  Optimization} 47(5):2410--2439.

\bibitem[{Fu et~al.(2007)Fu, Hu, Chen, \protect\BIBand{}
  Xiong}]{fu2007simulation}
Fu MC, Hu JQ, Chen CH, Xiong X (2007) Simulation allocation for determining the
  best design in the presence of correlated sampling. \emph{INFORMS Journal on
  Computing} 19(1):101--111.

\bibitem[{Gao et~al.(2019)Gao, Du, \protect\BIBand{} Chen}]{gao2019selecting}
Gao S, Du J, Chen CH (2019) Selecting the optimal system design under
  covariates. \emph{2019 IEEE 15th International Conference on Automation
  Science and Engineering (CASE)}, 547--552 (Piscataway, New Jersey: IEEE).

\bibitem[{Gao et~al.(2017)Gao, Xiao, Zhou, \protect\BIBand{} Chen}]{gao2017}
Gao S, Xiao H, Zhou E, Chen W (2017) Robust ranking and selection with optimal
  computing budget allocation. \emph{Automatica} 81:30--36.

\bibitem[{Garivier \protect\BIBand{} Kaufmann(2016)}]{garivier2016optimal}
Garivier A, Kaufmann E (2016) Optimal best arm identification with fixed
  confidence. \emph{Conference on Learning Theory}, 998--1027 (PMLR).

\bibitem[{Glynn \protect\BIBand{} Juneja(2004)}]{glynn2004large}
Glynn P, Juneja S (2004) A large deviations perspective on ordinal
  optimization. Ingalls RG, Rossetti MD, Smith JS, Peters BA, eds.,
  \emph{Proceedings of the 2004 Winter Simulation Conference}, 577--585
  (Piscataway, New Jersey: IEEE).

\bibitem[{Harvey et~al.(2010)Harvey, Liechty, Liechty, \protect\BIBand{}
  M{\"u}ller}]{harvey2010portfolio}
Harvey CR, Liechty JC, Liechty MW, M{\"u}ller P (2010) Portfolio selection with
  higher moments. \emph{Quantitative Finance} 10(5):469--485.

\bibitem[{Hirano(2010)}]{hirano2010decision}
Hirano K (2010) Decision theory in econometrics. \emph{Microeconometrics},
  29--35 (Springer).

\bibitem[{Kim et~al.(2021)Kim, Kim, \protect\BIBand{} Song}]{kimkimsong}
Kim KK, Kim T, Song E (2021) Selection of the most probable best under input
  uncertainty. Kim S, Feng B, Smith K, Masoud S, Zheng Z, eds.,
  \emph{Proceedings of the 2021 Winter Simulation Conference} (Piscataway, New
  Jersey: IEEE).

\bibitem[{Kim \protect\BIBand{} Song(2022)}]{KimSong2022}
Kim T, Song E (2022) Optimizing input data acquisition for ranking and
  selection: A view through the most probable best. \emph{2022 Winter
  Simulation Conference}, 2258--2269.

\bibitem[{Li et~al.(2022)Li, Lam, \protect\BIBand{} Peng}]{Li:22DSCO}
Li H, Lam H, Peng Y (2022) Efficient learning for clustering and optimizing
  context-dependent designs. \emph{Operations Research} 72(2):617--638.

\bibitem[{Martello \protect\BIBand{} Toth(1990)}]{martello1990knapsack}
Martello S, Toth P (1990) \emph{Knapsack Problems: Algorithms and Computer
  Implementations} (John Wiley \& Sons, Inc.).

\bibitem[{Pasupathy et~al.(2014)Pasupathy, Hunter, Pujowidianto, Lee,
  \protect\BIBand{} Chen}]{pasupathy2014stochastically}
Pasupathy R, Hunter SR, Pujowidianto NA, Lee LH, Chen CH (2014) Stochastically
  constrained ranking and selection via {SCORE}. \emph{ACM Transactions on
  Modeling and Computer Simulation} 25(1):1--26.

\bibitem[{Pearce \protect\BIBand{} Branke(2017)}]{pearce2017}
Pearce M, Branke J (2017) Efficient expected improvement estimation for
  continuous multiple ranking and selection. Chan WKV, D’Ambrogio A,
  Zacharewicz G, Mustafee N, Wainer G, Page E, eds., \emph{Proceedings of the
  2017 Winter Simulation Conference}, 2161--2172 (Piscataway, New Jersey:
  IEEE).

\bibitem[{Peng et~al.(2016)Peng, Chen, Fu, \protect\BIBand{}
  Hu}]{peng2016dynamic}
Peng Y, Chen CH, Fu MC, Hu JQ (2016) Dynamic sampling allocation and design
  selection. \emph{INFORMS Journal on Computing} 28(2):195--208.

\bibitem[{Peng et~al.(2018)Peng, Chong, Chen, \protect\BIBand{}
  Fu}]{peng2018ranking}
Peng Y, Chong EK, Chen CH, Fu MC (2018) Ranking and selection as stochastic
  control. \emph{IEEE Transactions on Automatic Control} 63(8):2359--2373.

\bibitem[{Peng \protect\BIBand{} Fu(2017)}]{peng2017myopic}
Peng Y, Fu M (2017) Myopic allocation policy with asymptotically optimal
  sampling rate. \emph{IEEE Transactions on Automatic Control}
  62(4):2041--2047.

\bibitem[{Powell \protect\BIBand{} Ryzhov(2012)}]{powell2012optimal}
Powell WB, Ryzhov IO (2012) \emph{Optimal Learning}, volume 841 (John Wiley \&
  Sons).

\bibitem[{Robert(2007)}]{robert2007bayesian}
Robert C (2007) \emph{The Bayesian Choice: From Decision-theoretic Foundations
  to Computational Implementation} (Springer Science \& Business Media).

\bibitem[{Rossi et~al.(2012)Rossi, Allenby, \protect\BIBand{}
  McCulloch}]{rossi2012bayesian}
Rossi PE, Allenby GM, McCulloch R (2012) \emph{Bayesian Statistics and
  Marketing} (John Wiley \& Sons).

\bibitem[{Ryzhov(2016)}]{ryzhov2016convergence}
Ryzhov IO (2016) On the convergence rates of expected improvement methods.
  \emph{Operations Research} 64(6):1515--1528.

\bibitem[{Salemi et~al.(2019)Salemi, Song, Nelson, \protect\BIBand{}
  Staum}]{salemi2019}
Salemi P, Song E, Nelson BL, Staum J (2019) Gaussian markov random fields for
  discrete optimization via simulation: Framework and algorithms.
  \emph{Operations Research} 67(1):250--266.

\bibitem[{Sch{\"o}lkopf \protect\BIBand{} Smola(2002)}]{scholkopf:02Kernel}
Sch{\"o}lkopf B, Smola AJ (2002) \emph{Learning with kernels: support vector
  machines, regularization, optimization, and beyond} (MIT press).

\bibitem[{Shen et~al.(2021)Shen, Hong, \protect\BIBand{}
  Zhang}]{shen2021ranking}
Shen H, Hong LJ, Zhang X (2021) Ranking and selection with covariates for
  personalized decision making. \emph{INFORMS Journal on Computing}
  33(4):1500--1519.

\bibitem[{Song \protect\BIBand{} Nelson(2019)}]{song2019}
Song E, Nelson BL (2019) Input--output uncertainty comparisons for discrete
  optimization via simulation. \emph{Operations Research} 67(2):562--576.

\bibitem[{Song et~al.(2015)Song, Nelson, \protect\BIBand{}
  Hong}]{songnelsonhong2015}
Song E, Nelson BL, Hong LJ (2015) Input uncertainty and indifference-zone
  ranking \& selection. Yilmaz L, Chan WKV, I~Moon TMKR, Macal C, Rossetti MD,
  eds., \emph{Proceedings of the 2015 Winter Simulation Conference}, 414--424
  (Piscataway, New Jersey: IEEE).

\bibitem[{Tollefson et~al.(2014)Tollefson, Goldsman, Kleywegt,
  \protect\BIBand{} Tovey}]{tollefson2014optimal}
Tollefson E, Goldsman D, Kleywegt A, Tovey C (2014) Optimal selection of the
  most probable multinomial alternative. \emph{Sequential Analysis}
  33(4):491--508.

\bibitem[{Train(2009)}]{train2009discrete}
Train KE (2009) \emph{Discrete Choice Methods with Simulation} (Cambridge
  University Press).

\bibitem[{Ungredda et~al.(2022)Ungredda, Pearce, \protect\BIBand{}
  Branke}]{ungredda2020}
Ungredda J, Pearce M, Branke J (2022) Bayesian optimisation vs. input
  uncertainty reduction. \emph{ACM Transactions on Modeling and Computer
  Simulation} In press.

\bibitem[{Wu \protect\BIBand{} Zhou(2018)}]{WuZhou:18FixedRS}
Wu D, Zhou E (2018) Analyzing and provably improving fixed budget ranking and
  selection algorithms, \url{ https://doi.org/10.48550/arXiv.1811.12183}.

\bibitem[{Wu \protect\BIBand{} Zhou(2019)}]{wu2019}
Wu D, Zhou E (2019) Ranking and selection under input uncertainty: Fixed
  confidence and fixed budget, https://arxiv.org/abs/1708.08526.

\bibitem[{Wu et~al.(2018)Wu, Zhu, \protect\BIBand{} Zhou}]{wu2018bayesian}
Wu D, Zhu H, Zhou E (2018) A {B}ayesian risk approach to data-driven stochastic
  optimization: formulations and asymptotics. \emph{SIAM Journal on
  Optimization} 28(2):1588--1612.

\bibitem[{Wu-Smith et~al.(2023)Wu-Smith, Keenan, Owen, Norton, Kamm,
  Schumacher, Fenyes, Kiggins, Konkel, Rosen, Schmitter, Sheremet,
  \protect\BIBand{} Yochim}]{VCO2023}
Wu-Smith P, Keenan PT, Owen JH, Norton A, Kamm K, Schumacher KM, Fenyes P,
  Kiggins D, Konkel PW, Rosen W, Schmitter K, Sheremet S, Yochim L (2023)
  General motors optimizes vehicle content for customer value and
  profitability. \emph{INFORMS Journal on Applied Analytics} 53(1):59--69.

\bibitem[{Xie \protect\BIBand{} Zhou(2015)}]{xie2015}
Xie W, Zhou E (2015) Simulation optimization when facing input uncertainty.
  Yilmaz L, Chan WKV, Moon I, Roeder TMK, Macal C, Rossetti MD, eds.,
  \emph{Proceedings of the 2015 Winter Simulation Conference}, 3714--3724
  (Piscataway, New Jersey: IEEE).

\bibitem[{Zhu et~al.(2020)Zhu, Liu, \protect\BIBand{} Zhou}]{helin2020}
Zhu H, Liu T, Zhou E (2020) Risk quantification in stochastic simulation under
  input uncertainty. \emph{ACM Transactions on Modeling and Computer
  Simulation} 30(1):1--24.

\end{thebibliography}


\ECSwitch


\ECHead{Electronic Companion of ``Selection of the Most Probable Best"}
\noindent This electronic companion provides technical proofs and some experiment details on market simulation not presented in the main body of the paper. 

\section{Relations to Other Formulation}\label{apdx:connection_to_other_relation}

\cite{daskin1997alpha} consider a value-at-risk (VaR) regret minimization problem. Under our notation, this approach can be formulated as
\begin{equation}\label{eq:var_regret}
    \min\nolimits_{1 \leq i \leq k}\text{VaR}_\pi^{1-\alpha}\{y_i(\theta) - \min\nolimits_{1\leq j\leq k}y_{j}(\theta)\},
\end{equation}
where $\text{VaR}_\pi^{1-\alpha}$ is the $1-\alpha$ level VaR with respect to $\pi$. Observe that for some $i$, if $\prob_\pi(\Theta_i) \geq 1-\alpha$, then  $\text{VaR}_\pi^{1-\alpha} \{y_i(\theta) - \min_{1\leq j\leq k}y_{j}(\theta)\} = 0$. Hence, the MPB belongs to the solution set of~\eqref{eq:var_regret} when its preference probability is no less than the risk level, $1-\alpha$. 

Further, the MPB formulation is closely related to the multinomial R\&S~\citep{bechhofer1959single,chen1992truncated, tollefson2014optimal}. Their objective is to select a solution among $k$ with the largest probability such that its simulation output is the largest (or smallest).  Our problem formulation has a nested structure that the classical multinomial R\&S does not; we solve a mean-value inner R\&S problem given each parameter value, then solve a multinomial R\&S at the outer level, where the optimum (MPB) is defined as the preference-probability-maximizing solution.

\section{Bayesian Learning Model for Unknown Variance}\label{apdx:unknown_variance}

In this section, we consider the extension of Bayesian model in~\eqref{eq:normaL_model} to the unknown variance case. Suppose both mean and variance of simulation model are unknown in advance. 
The normal-gamma conjugate pair is frequently adopted in the Bayesian learning literature (e.g., see Section 5 in \cite{ryzhov2016convergence} and Section 2.3.5 in \cite{powell2012optimal}) to model the case of unknown simulation variance. The prior of this model can be written as
\begin{equation}\label{eq:normal_gamma_model}
    \begin{aligned}
        Y_i(\theta) & \sim  N(\eta_i(\theta), \lambda^2_i(\theta)), \\
        (\eta_i(\theta), \lambda^{-2}_i(\theta)) & \sim  NG({\mu}_{i, 0}(\theta), \tau_{i, 0}(\theta), \gamma_{i,0}(\theta), \omega_{i, 0}(\theta)),
    \end{aligned}
\end{equation}
where $NG$ represents the normal-gamma distribution. 
From~\eqref{eq:normal_gamma_model}, one can derive $\lambda^{-2}_{i}(\theta) \sim \text{Gamma}(\gamma_{i, 0}(\theta),\omega_{i, 0}(\theta))$ where $\omega_{i, 0}(\theta)$ is a rate parameter and $\eta_{i}(\theta)|\lambda_i^{-2}(\theta) \sim N(\mu_{i, 0}(\theta), 1/(\tau_{i, 0}(\theta)\lambda_i^{-2}(\theta)))$. From the conjugacy of Model~(\ref{eq:normal_gamma_model}), the joint posterior distribution of $(\eta_i(\theta), \lambda_i^{-2}(\theta))$ given the simulation outputs, $Y_{i1}(\theta),Y_{i2}(\theta),\ldots,Y_{i N^n_i(\theta)}(\theta)$, at $(i,\theta)$ is 
$NG(\mu_{i, n}(\theta), \tau_{i, n}(\theta), \gamma_{i, n}(\theta), \omega_{i, n}(\theta)),$
where 
\begin{equation*}
    \begin{aligned}
       \mu_{i, n}(\theta) &= \frac{\mu_{i, 0}(\theta)\tau_{i, 0}(\theta) + \sum_{r=1}^{N_i^n(\theta)}Y_{ir}(\theta)}{\tau_{i, 0}(\theta) + N_i^n(\theta)}, 
       \tau_{i, n}(\theta)  = \tau_{i, 0}(\theta) + N_i^n(\theta),
       \gamma_{i, n}(\theta)  = \gamma_{i, 0}(\theta) + \frac{1}{2}N_i^n(\theta), \\
       \omega_{i, n}(\theta) & = \omega_{i, 0}(\theta) + \frac{1}{2}\sum_{r=1}^{N_i^n(\theta)} \lt(Y_{ir}(\theta) - m_i(\theta)\rt)^2 + \frac{\tau_{i, 0}(\theta)N^n_i(\theta)}{2\lt(\tau_{i, 0}(\theta) + N^n_i(\theta)\rt)}(m_i(\theta) - \mu_{i, 0}(\theta))^2,
    \end{aligned}
\end{equation*}
and $m_i(\theta)$ is the sample mean of simulation outputs made at $(i,\theta)$. We introduce  $S^2_i(\theta): = \expec\lt[\lambda_i^{-2}(\theta) | \mathcal{E}_n\rt]^{-1}$ as an estimator for the simulation error variance at $(i,\theta)$. Then, $S^2_i(\theta)= {\omega_{i,n}(\theta)}/{\gamma_{i, n}(\theta)}$ from that the marginal distribution of $\lambda^{-2}_i(\theta)$ is Gamma$(\gamma_{i,n}(\theta), \omega_{i,n}(\theta))$. 
Since $S^{-2}_i(\theta)$ is a nonnegative martingale, it converges to $\expec[\lambda^{-2}_i(\theta)|\mathcal{E}_{\infty}] = \lambda^{-2}_i(\theta)$ almost surely (a.s.) from the martingale convergence theorem.  Adopting $S_i^2(\theta_b)$ as a plug-in estimator of $\lambda^2_i(\theta_b)$, the plug-in sample LDR, $G_{i, n}(\theta_b)$, is given as
\begin{equation*}
    G_{i, n}(\theta_b) = \dfrac{(\mu_{i, n}(\theta_b) - \mu_{i^b_n, n}(\theta_b))^2}{2\lt({S^2_i(\theta_b)}/{\alpha_{i, n}(\theta_b)} + {S^2_{i_n^b}(\theta_b)}/{\alpha_{i_n^b, n}(\theta_b)}\rt)}.
\end{equation*}
If noninformative prior $(\tau_{i, 0}(\theta), \gamma_{i, 0}(\theta), \omega_{i, 0}(\theta)) = (0, 0, 0)$ is adopted, then $\mu_{i, n}(\theta)$ and $S_i^2(\theta_b)$ coincide with the sample mean and sample variance of the simulation outputs, respectively. 

\section{Proofs of Results in Section~\ref{sec:probable_best}}\label{ec:alg_proof}

\proof{Proof of Theorem~\ref{thm:Gtilde}}.
(i) Let us fix $\theta_b$. We apply the G\"{a}rtner-Ellis theorem~\citep{dembo2009large} to obtain the  expression for $\widetilde{G}_i(\theta_b)$. Given $\BFt = (t_1,t_2,\ldots, t_k)$ and a finite sample of size $n$, the joint cumulant generating function (cgf)  of $\{\mu_{\ell, n}(\theta_b)\}_{1 \leq \ell \leq k}$ can be derived as $\Lambda_n(\BFt) = \sum_{\ell = 1}^k \left(y_\ell(\theta_b)t_\ell + \frac{\lambda^2_\ell(\theta_b)}{2N_\ell^n(\theta_b)}t_\ell^2\right)$ under the simulation output normality assumption. Then, the limiting joint cgf, $\Lambda(\BFt)$, is defined as $\Lambda(\BFt) = \lim_{n \rightarrow \infty}\frac{1}{n}\Lambda_n(n\BFt) = \sum_{\ell = 1}^k \left(y_\ell(\theta_b)t_\ell + \frac{\lambda_\ell^2(\theta_b)}{2\alpha_\ell(\theta_b)}t_\ell^2\mathbf{1}\{\alpha_\ell(\theta_b) > 0\} + \infty \cdot \mathbf{1}\{\alpha_\ell(\theta_b) = 0, t_\ell \neq 0\}\right)$ since we assume the static allocation $\lim_{n \rightarrow \infty} \frac{N_\ell^n(\theta_b)}{n} = \alpha_\ell(\theta_b)$ for all $(\ell, \theta_b)$. The last term in the summation comes from the fact that for $\alpha_\ell(\theta_b)= 0$, $\lim_{n \rightarrow \infty} \frac{1}{n} \{y_\ell(\theta_b)(nt_\ell) + \frac{\lambda_\ell^2(\theta_b)}{N_\ell^n(\theta_b)}(nt_\ell)^2\} = \infty$ if $t_\ell \neq 0$ and $ = 0$ if $t_\ell = 0$. Given input vector $\BFx = (x_1,\ldots, x_k)$, the Fenchel-Legendre transformation of $\Lambda(\BFt)$ is then derived as 
\begin{equation*}
    \begin{aligned}
        \bar{\Lambda}(\BFx) = \sup_{\BFt}\left\{\BFx^\top \BFt - \Lambda(\BFt)\right\} &= \sup_{\BFt}\left\{\sum_{\ell=1}^k\left(x_\ell t_\ell - y_\ell(\theta_b)t_\ell -\frac{\lambda_\ell^2(\theta_b)}{2\alpha_\ell(\theta_b)}t_\ell^2 - \infty \cdot \mathbf{1}\{\alpha_\ell(\theta_b) = 0, t_\ell \neq 0\}\right)\right\}\\
        & = \sum_{\ell = 1}^k \sup_{t_\ell} \left((x_\ell-y_\ell(\theta_b)) t_\ell -\frac{\lambda_\ell^2(\theta_b)}{2\alpha_\ell(\theta_b)}t_\ell^2 - \infty \cdot \mathbf{1}\{\alpha_\ell(\theta_b) = 0, t_\ell \neq 0\}\right)\\
        & = \sum_{\ell = 1}^k \frac{\alpha_\ell(\theta_b)}{2\lambda^2_\ell(\theta_b)}(x_\ell - y_\ell(\theta_b))^2,
    \end{aligned}
\end{equation*}
where the second line follows from the separability of the objective function, and the last holds since the supremum is attained at $t_\ell = \frac{\alpha_\ell(\theta_b)(x_\ell - y_\ell(\theta_b))}{\lambda_\ell^2(\theta_b)}$ if $\alpha_\ell(\theta_b)> 0$ and $t_\ell = 0$ if $\alpha_\ell(\theta_b) = 0$ for each~$\ell$. 
 Let $E$ be an arbitrary subset of $\Real^k$ and $\textup{int}(E)$ be its interior. If $\inf_{\BFx \in E}\bar{\Lambda}(\BFx) = \inf_{\BFx \in \textup{int}(E)}\bar{\Lambda}(\BFx)$, then the G\"{a}rtner-Ellis theorem stipulates the following result:
\begin{equation}\label{eq:LDP}
    \lim_{n \rightarrow \infty}-\frac{1}{n}\log\prob\left(\{\mu_{\ell,n}(\theta_b)\}_{1\leq \ell\leq k} \in E\right) = \inf_{x \in E}\bar{\Lambda}(x) = \inf_{\BFx \in E}\sum_{\ell = 1}^k \frac{\alpha_\ell(\theta_b)}{2\lambda^2_\ell(\theta_b)}(x_\ell - y_\ell(\theta_b))^2.
\end{equation}
To derive the LDR of $\prob(m_{i, b}^n = 1)$, we set $E= \{\BFx: x_i \leq x_\ell, \forall \ell \neq i\}$ since $\{\mu_{\ell,n}(\theta_b)\}_{1\leq \ell\leq k} \in E$ implies that $i$ is the sample best. Also, it is easy to show that $\inf_{\BFx \in E}\bar{\Lambda}(\BFx) = \inf_{\BFx \in \textup{int}(E)}\bar{\Lambda}(\BFx)$ in this case.
Plugging in the choice of $E$ to~\eqref{eq:LDP}, we have
\begin{equation}\label{def:tilde_G}
    \begin{aligned}
    \widetilde{G}_i(\theta_b) &= \lim_{n \rightarrow \infty} -\frac{1}{n}\log\prob(m_{i, b}^n = 1) \\
    & = \lim_{n \rightarrow \infty}-\frac{1}{n}\prob\left((\mu_{1, n}(\theta_b), \ldots, \mu_{k, n}(\theta_b)) \in \left\{\BFx: x_i \leq x_\ell, \forall \ell \neq i\right\}\right)\\
    &= \inf_{x_i \leq x_\ell, \forall \ell \neq i}\bigg\{\sum_{\ell=1}^k \frac{\alpha_\ell(\theta_b)}{2\lambda_\ell^2(\theta_b)}(x_\ell - y_\ell(\theta_b))^2\bigg\}.
    \end{aligned}
\end{equation}
The infimum of~\eqref{def:tilde_G} involves $k$  variables: $x_1,x_2,\ldots,x_k$. We first reformulate \eqref{def:tilde_G} so that it involves a single variable:
    \begin{align}
        \widetilde{G}_i(\theta_b) &= \min_{x = x_i} \min_{x \leq x_\ell, \ell \neq i} \bigg\{\frac{\alpha_i(\theta_b)}{2\lambda_i^2(\theta_b)}(x_i - y_i(\theta_b))^2 + \sum_{\ell \neq i}\frac{\alpha_\ell(\theta_b)}{2\lambda_\ell^2(\theta_b)} (x_\ell - y_\ell(\theta_b))^2 \bigg\}.\nonumber \\
        &= \min_{x} \bigg\{\frac{\alpha_i(\theta_b)}{2\lambda_i^2(\theta_b)}(x - y_i(\theta_b))^2 + \sum_{\ell \neq i}\min_{x \leq x_\ell}\left\{\frac{\alpha_\ell(\theta_b)}{2\lambda_\ell^2(\theta_b)} (x_\ell - y_\ell(\theta_b))^2\right\} \bigg\}.\nonumber \\
        &= \min_{x} \bigg\{\frac{\alpha_i(\theta_b)}{2\lambda_i^2(\theta_b)}(x - y_i(\theta_b))^2 + \sum_{\ell \neq i}\frac{\alpha_\ell(\theta_b)}{2\lambda_\ell^2(\theta_b)} \lt[(x - y_\ell(\theta_b))^+\rt]^2 \bigg\}. \label{eq:temp}
    \end{align}
The second equality holds since the inner objective function is separable. The third equality can be justified from that $\inf_{x \leq x_\ell}\frac{\alpha_\ell(\theta_b)}{2\lambda_\ell^2(\theta_b)}(x_\ell - y_\ell(\theta_b))^2$ is attained at $x_\ell = \max(x, y_\ell(\theta_b))$ for each $\ell \neq i$. Let $F(x)$ denote the  function we minimize in~\eqref{eq:temp}. For $x< y_{i^b}(\theta_b)$,  $F(x) = \frac{\alpha_i(\theta_b)}{2\lambda_i^2(\theta_b)}(x - y_i(\theta_b))^2$, which is decreasing in $x$. Hence, we can see that the solution of $\min_x F(x)$  belongs to $\{x: y_{i^b}(\theta_b) \leq x\}$. Therefore,~\eqref{eq:temp} can be rewritten as 
\small
\begin{equation}\label{eq:temp2}
    \widetilde{G}_i(\theta_b) = \min_{x:y_{i^b}(\theta_b) \leq x} \bigg\{\frac{\alpha_{i^b}(\theta_b)}{2\lambda_{i^b}^2(\theta_b)}(x - y_{i^b}(\theta_b))^2 + \frac{\alpha_i(\theta_b)}{2\lambda_i^2(\theta_b)}(x - y_i(\theta_b))^2 + \sum_{\ell \neq i^b, i}\frac{\alpha_\ell(\theta_b)}{2\lambda_\ell^2(\theta_b)} \lt[(x - y_\ell(\theta_b))^+\rt]^2 \bigg\}.
\end{equation}
\normalsize
Since all quadratic terms in~\eqref{eq:temp2} are increasing in $x$ when $x> y_i(\theta_b)$, we can replace the range of $x$ in the minimization problem of~\eqref{eq:temp2} with $x\in[y_{i^b}(\theta_b), y_i(\theta_b)]$. For all $\ell$ such that $y_\ell(\theta_b) > y_i(\theta_b)$, if for $x \in [y_{i^b}(\theta_b),y_i(\theta_b)]$, we have $(x-y_\ell(\theta_b))^+ = 0$ and therefore,~\eqref{def:G_tilde} follows. 

(ii) From the definition of $\mathcal{A}_j$, for each $j \neq i^*$,
\begin{eqnarray*}
\max_{\BFM \in \mathcal{A}_j}\prob\lt(\BFM_n = \BFM\rt) \leq \prob\lt(\BFM_n \in\mathcal{A}_j\rt) \leq \sum_{\BFM \in \mathcal{A}_j} \prob\lt(\BFM_n = \BFM\rt).
\end{eqnarray*}
Therefore, we have
\begin{eqnarray*}
    \liminf_{n \rightarrow \infty} -\frac{1}{n}\log\prob\lt(\BFM_n \in\mathcal{A}_j\rt) & = & \min_{\BFM \in \mathcal{A}_j} \liminf_{n \rightarrow \infty} -\frac{1}{n}\log \prob(\BFM_n = \BFM)\\
    &=& \min_{\BFM \in \mathcal{A}_j} \liminf_{n \rightarrow \infty} -\frac{1}{n}\log \prob\lt(m_{i,  b}^n = m_{i,b}, (i, \theta_b) \in I(\BFM)\rt)\\
    &=& \min_{\BFM \in \mathcal{A}_j} \liminf_{n \rightarrow \infty} -\frac{1}{n}\log \prod_{(i, \theta_b) \in I(\BFM)}\prob\lt(m_{i,  b}^n = 1\rt)\\
    & = & \min_{\BFM \in \mathcal{A}_j} \sum_{(i, \theta_b) \in I(\BFM)} \liminf_{n \rightarrow \infty}-\frac{1}{n}\log \prob\lt(m_{i,  b}^n = 1\rt) \\ 
    & = & \min_{\BFM \in \mathcal{A}_j} \sum_{(i, \theta_b) \in I(\BFM)} \widetilde{G}_i(\theta_b)  = \widetilde{\ldr}_{j, i^*}.
\end{eqnarray*}
The second equality holds from that only $(i, \theta_b) \in I(\BFM)$ affects the LDR of $ \prob(\BFM_n = \BFM)$. The third equality holds because simulation outputs are independent across $\theta_b$. 

Lastly, (iii) follows from combining (ii) with~\eqref{eq:LDR_min}. \halmos
\endproof

\proof{Proof of Lemma~\ref{lem:knapsack}.}
For any $\BFM \in \mathcal{M}$, we have
\begin{eqnarray}
    d_j - d_j(\BFM) &=& \sum_{b=1}^{B}p_b\lt(m_{j, b} - \mathbf{1}\lt\{j = i^b\rt\} - m_{i^*,b} + \mathbf{1}\{i^* = i^b\}\rt) \nonumber\\
    & = & \sum_{b=1}^{B}\sum_{i=1}^{k} p_b\lt[\lt(m_{j, b} - \mathbf{1}\lt\{j = i^b\rt\} - m_{i^*,b} + \mathbf{1}\{i^* = i^b\}\rt)\rt]m_{i,b} \nonumber\\
    &=& \sum_{(i, \theta_b) \in \Xi\cup\Xi^\text{adv}} v_j[(i, \theta_b)]m_{i, b}, \label{eq:breaking}
\end{eqnarray}
where the first equality holds from the definition of $d_j(\BFM)$ in~\eqref{def:d_i} and the second follows as only one of $m_{1,b}, m_{2,b},\ldots,m_{k, b}$ equals one while all else are zeroes;~\eqref{eq:breaking} holds from the definition of $v_j[(i, \theta_b)]$. From~\eqref{eq:breaking},~\eqref{def:A_i} can be rewritten as $\mathcal{A}_j = \lt\{\BFM : d_j - \sum\nolimits_{(i, \theta_b) \in \Xi\cup\Xi^\text{adv}} v_j[(i, \theta_b)]m_{i,b} \leq 0\rt\}$ for each $j \neq i^*$.  As $ v_j[(i, \theta_b)] \leq v_j[(i, \theta_b)]^+$, we have for any $j\neq i^*$
\begin{eqnarray}
&&\mathcal{A}_j \subseteq \widetilde{\mathcal{A}}_j : =  \bigg\{\BFM \in \mathcal{M}: d_j - \sum_{(i, \theta_b) \in \Xi\cup\Xi^\text{adv}} \left\{v_j[(i, \theta_b)]\right\}^+m_{i, b}\leq 0 \bigg\}, \nonumber\\
&&\widetilde{\ldr}_{j, i^*} \overset{\text{Theorem~\ref{thm:Gtilde}}}{=} \;\min_{\BFM \in \mathcal{A}_j}\sum_{(i, \theta_b) \in \Xi\cup\Xi^\text{adv}}\widetilde{G}_i(\theta_b)m_{i,b} \geq \min_{\BFM \in \widetilde{\mathcal{A}}_j}\sum_{(i, \theta_b) \in \Xi\cup\Xi^\text{adv}}\widetilde{G}_i(\theta_b)m_{i, b}.\label{eq:lower}
\end{eqnarray}
Lemma~\ref{lem:relax_A_i} below shows that   the inequality in~\eqref{eq:lower} indeed holds  as equality, i.e.,  $\widetilde{\ldr}_{j, i^*}= \min_{\BFM \in \widetilde{\mathcal{A}}_j}\sum_{(i, \theta_b) \in \Xi\cup\Xi^\text{adv}}\widetilde{G}_i(\theta_b)m_{i, b}$. From the definitions of $\BFm$ and $\BFv_j$, $\BFm \in \widetilde{\mathcal{A}}_j$ is equivalent to $d_j - \BFv_j^\top\BFm \leq 0$. Furthermore, we have $\sum_{(i, \theta_b) \in \Xi\cup\Xi^\text{adv}}\widetilde{G}_i(\theta_b)m_{i, b} = \widetilde{\BFg}^\top \BFm$, which completes the proof. 
\halmos
\endproof 
The subsequent discussions require the following definition. 
\begin{definition}\label{def:order_relation}
Partial order relation $\preceq$ between two matrices in $\mathcal{M}$ is defined as: 
\begin{equation}\label{eq:def_order}
     \BFM_1 \preceq \BFM_2, \text{ if $I(\BFM_1) \subseteq I(\BFM_2)$}.
 \end{equation}
\end{definition}
In words, $\preceq$ orders $\BFM\in\mathcal{M}$ by the number of conditional optima that  $\BFM$ misspecifies.  All required properties of a partial order relation (reflexivity, antisymmetry, and transitivity) can be shown to hold from~\eqref{eq:def_order}. 

The following lemma stipulates that the inequality in~\eqref{eq:lower} indeed holds as equality. 

\begin{lemma}\label{lem:relax_A_i}
For each $j\neq i^*,$ we have 
$\min_{\BFM \in \mathcal{A}_j}\sum_{(i, \theta_b) \in I(\BFM)}\widetilde{G}_i(\theta_b) = \min_{\BFM \in \widetilde{\mathcal{A}}_j}\sum_{(i, \theta_b) \in I(\BFM)}\widetilde{G}_i(\theta_b)$.
\end{lemma}
\proof{Proof.}
Notice that if $\BFM_1 \preceq \BFM_2$ for some $\BFM_1, \BFM_2 \in \widetilde{\mathcal{A}}_j$, then we have $\sum\nolimits_{(i, \theta_b) \in I(\BFM_1)} \widetilde{G}_i(\theta_b) \leq \sum\nolimits_{(i, \theta_b) \in I(\BFM_2)} \widetilde{G}_i(\theta_b)$. This implies that $\widetilde{\ldr}_{j, i^*}$ remains unchanged when $\widetilde{\mathcal{A}}_j$ is replaced with $\widetilde{\mathcal{A}}_j - \lt\{\BFM_2\rt\}$; that is, we can remove any non-minimal elements with respect to $\preceq$ from $\mathcal{A}_j$ without affecting $\widetilde{\ldr}_{j, i^*}$. 
Therefore, it suffices to show that any $\BFM\in \widetilde{\mathcal{A}}_j\setminus\mathcal{A}_j$ is non-minimal of $\widetilde{\mathcal{A}}_j$. 

For each $\BFM\in \widetilde{\mathcal{A}}_j\setminus\mathcal{A}_j$, we can always construct $\BFM_1 \in \mathcal{M}$ such that $I(\BFM_1) = \lt\{(i, \theta_b) \in I(\BFM)| v_j[(i, \theta_b)] > 0\rt\}$ by taking the misspecified pairs from $\BFM$ such that $v_j[(i,\theta_b)]>0$ and setting all other misspecified pairs in $\BFM$ with $v_j[(i,\theta_b)]=0$ to be correctly mapped. 
Then,  we have $\BFM_1 \preceq \BFM$ and  $\BFM_1 \in \mathcal{A}_j$ as $d_j(\BFM_1)= d_j - \sum_{(i, \theta_b) \in \Xi\cup\Xi^{\text{adv}}}\left\{v_j[(i, \theta_b)]\right\}^+m_{i, b} \leq 0$. 
This implies $\BFM$ is non-minimal, as desired.  \halmos
\endproof

Table~\ref{tab:partition} summarizes classification of each $(i, \theta_b)$ in $\Xi\cup\Xi^\text{adv}$ according to $v_j[(i, \theta_b)]^+$. Since $v_j[(i, \theta_b)]/p_b = \mathbf{1}\lt\{j = i\rt\} - \mathbf{1}\lt\{j = i^b\rt\} - \mathbf{1}\lt\{i^* = i\rt\} + \mathbf{1}\lt\{i^* = i^b\rt\}$, each $v_j/p_b$ ranges from -2 to 2. Thus, the value in Table~\ref{tab:partition}, $v_j[(i, \theta_b)]^+/p_b$, must be 0, 1, or 2. From Table~\ref{tab:partition}, one can confirm that $v_j[(i^*, \theta_b)]^+ = 0$ holds for all $j$ and $(i^*, \theta_b) \in \Xi^\text{adv}$. Similar to Lemma~\ref{lem:relax_A_i}, for all $j \neq i^*$, we can rule out $\BFM \in \widetilde{\mathcal{A}}_j$ such that $m_{i^*, b} = 1$ holds for some $(i^*, \theta_b) \in \Xi^\text{adv}$. In this regard, we can restrict our attention to $\Xi$ only.

\begin{table}[h]
    \centering
    \caption{Given $j\neq i^*$, $\Xi$ is partitioned into five subsets according to the row/column conditions. All $(i,\theta_b)$ in each subset have an identical value of  $v_j[(i, \theta_b)]^+/p_b$ given in each cell. The first cell is marked with N/A as no pair in $\Xi$ satisfies the two conditions (i.e., null set). }    \label{tab:partition}
    \begin{tabular}{|c |c| c| c|} \hline
    \multicolumn{1}{|c|}{\backslashbox{$i$}{$\theta_b$}} & \multicolumn{1}{c|}{{ } $i^b=i$ { }}       & \multicolumn{1}{c|}{{ } $i^b\neq i $ \& $i^b\neq i^*$ { }} &  \multicolumn{1}{c|}{{ }  $i^b=i^*$ { }} \\ \cline{1-4}
    $i = j$  & N/A & $1$ & \multicolumn{1}{c|}{$2$} \\\cline{1-4}
    $i\neq j$  & $0$ & $ 0$ & \multicolumn{1}{c|}{$1$} \\\cline{1-4}
    \end{tabular}
\end{table}

\proof{Proof of Theorem~\ref{thm:lower_LDR}.}

For $j\neq i^*$, define $\widetilde{\Xi}_j := \{(i, \theta_b) \in \Xi| d_j(\BFM) \leq 0, I(\BFM) = \lt\{(i, \theta_b)\rt\}\}.$
Namely, misspecifying each solution-parameter pair in $\widetilde{\Xi}_j$ makes  $j$ outperform or tie with $i^*$. 
We further define two disjoint subsets of $\widetilde{\mathcal{A}}_j$:
$\mathcal{B}_1 :=\{\BFM \in \widetilde{\mathcal{A}}_j| I(\BFM) \cap \widetilde{\Xi}_j = \emptyset\}$ and $\mathcal{B}_2 :=\{\BFM \in \widetilde{\mathcal{A}}_j| I(\BFM) \subseteq \widetilde{\Xi}_j\}.$
Suppose there exists $\BFM \in \widetilde{\mathcal{A}}_j$ such that $\BFM \notin \mathcal{B}_1\cup\mathcal{B}_2$. Then, we can construct  $\BFM_1$ such that  
$I(\BFM_1) = I(\BFM) \cap \widetilde{\Xi}_j$ and we have $\BFM_1 \preceq \BFM$ from Definition~\ref{def:order_relation}. Hence, any element in $\widetilde{\mathcal{A}}_j\setminus\mathcal{B}_1 \cup \mathcal{B}_2$ can be excluded from deriving $\widetilde\ldr_{j, i^*}$ from which the second equality below follows: 
\[
\widetilde\ldr_{j, i^*} = \min_{\BFM \in \widetilde{\mathcal{A}}_j}\sum_{(i, \theta_b) \in I(\BFM)}\widetilde{G}_i(\theta_b)= \min_{\BFM \in \mathcal{B}_1 \cup\mathcal{B}_2}\sum_{(i, \theta_b) \in I(\BFM)}\widetilde{G}_i(\theta_b).
\]
Since $\mathcal{B}_1$ and $\mathcal{B}_2$ are disjoint, we further obtain
\begin{eqnarray}
    \widetilde\ldr_{j, i^*} &=& \min\lt\{\min_{\BFM \in \mathcal{B}_1}\sum_{(i, \theta_b) \in I(\BFM)}\widetilde{G}_i(\theta_b), \min_{\BFM \in \mathcal{B}_2}\sum_{(i, \theta_b) \in I(\BFM)}\widetilde{G}_i(\theta_b)\rt\}\nonumber \\
    &=& \min\lt\{\min_{\BFM \in \mathcal{B}_1}\sum_{(i, \theta_b) \in I(\BFM)}\widetilde{G}_i(\theta_b), \min_{(i,\theta_b) \in \widetilde{\Xi}_j}\widetilde{G}_i(\theta_b)\rt\}, \label{eq:splitted}
\end{eqnarray}
where~\eqref{eq:splitted}  holds because a single element of $\widetilde\Xi_j$ is sufficient to make $j$ outperform or tie with $i^*$.

Next, we consider the first inner minimization problem in~\eqref{eq:splitted}. Since any $\BFM \in \mathcal{B}_1$ satisfies $I(\BFM) \cap \widetilde{\Xi}_j = \emptyset$, we can construct $\widetilde{\mathcal{B}}_1$, a superset of $\mathcal{B}_1$ such that
\begin{equation*}
    \mathcal{B}_1 
    \subseteq \widetilde{\mathcal{B}}_1 := \left\{\BFM \in \mathcal{M}: d_j - \sum_{(i, \theta_b) \in (\Xi \cup \Xi^\text{adv})\setminus \widetilde{\Xi}_j} \{v_j[(i, \theta_b)]\}^+m_{i, b} \leq 0 \right\}.
\end{equation*}
Then, we have $\min_{\BFM \in \mathcal{B}_1}\sum_{(i, \theta_b) \in I(\BFM)}\widetilde{G}_i(\theta_b) \geq \min_{\BFM \in \widetilde{\mathcal{B}}_1}\sum_{(i, \theta_b) \in I(\BFM)}\widetilde{G}_i(\theta_b)$. Let $\widetilde{\BFv}_j$ be a vector defined by replacing all elements of $\BFv_j$ corresponding to $(i, \theta_b) \in \widetilde{\Xi}_j$ with zero. Then, we have
\begin{equation}\label{opt:knapsack_new}
\min_{\BFM \in \widetilde{\mathcal{B}}_1}\sum_{(i, \theta_b) \in I(\BFM)}\widetilde{G}_i(\theta_b) \;\; = \;\; 
    \min\; \widetilde{\BFg}^\top \BFm \hspace{15pt} \text{ subject to } \; \widetilde{\BFv}_j^\top\BFm - d_j \geq 0, \; \BFM\in \mathcal{M}.
\end{equation}
Since there is no analytical expression for the optimal solution for~\eqref{opt:knapsack_new} 
in general, we further drop the integrality and equality constraints in $\mathcal{M}$ to obtain: 
\begin{equation}\label{opt:IP_relaxed_OCBA2}
    \min\; \widetilde{\BFg}^\top \BFm \hspace{15pt} \text{ subject to } \; \widetilde{\BFv}_j^\top\BFm - d_j \geq 0, \; \BFm\geq 0,
\end{equation}
where $\BFm\geq 0$ means element-wise nonnegativity. Note that~\eqref{opt:IP_relaxed_OCBA2} is a linear program and thus can be easily solved by deriving its KKT conditions. The optimal objective function value of~\eqref{opt:IP_relaxed_OCBA2} turns out to be  $d_j\lt\{{\min_{(i, \theta_b) \in \Xi\setminus \widetilde{\Xi}_j} \widetilde{G}_i(\theta_b)}/{v_j[(i, \theta_b)]^+}\rt\}$. Therefore, we have  
\begin{equation}
     \min_{\BFM \in \mathcal{B}_1}\sum_{(i, \theta_b) \in I(\BFM)}\widetilde{G}_i(\theta_b) \geq \lt[d_j\lt\{{\min_{(i, \theta_b) \in \Xi\setminus \widetilde{\Xi}_j} \widetilde{G}_i(\theta_b)}/{v_j[(i, \theta_b)]^+}\rt\}\rt], \label{eq:xi>1}
 \end{equation}


Note that $(i, \theta_b) \in \widetilde{\Xi}_j$, if and only if, $d_j/v_j[(i, \theta_b)]^+\leq 1$. 
Combining this observation with~\eqref{eq:xi>1},~\eqref{eq:splitted} can be bounded from below by
\begin{equation*}
    \widetilde{\ldr}_{j, i^*} \geq \min_{(i, \theta_b) \in \Xi}\lt\{\max\lt[1, \frac{d_j}{v_j[(i, \theta_b)]^+}\rt]\widetilde{G}_i(\theta_b)\rt\}. 
\end{equation*}

Putting all $\widetilde\ldr_{j, i^*}$ terms together, we  obtain {\small 
\begin{equation*}
    \begin{aligned}
    \min_{j \neq i^*} \widetilde{\ldr}_{j, i^*} \geq \min_{j \neq i^*}\min_{(i, \theta_b) \in \Xi}\lt\{\max\lt[1, \frac{d_j}{v_j[(i, \theta_b)]^+}\rt]\widetilde{G}_i(\theta_b)\rt\} = \min_{(i, \theta_b) \in \Xi}\lt\{\max\lt[1, \min_{j \neq i^*}\frac{d_j}{v_j[(i, \theta_b)]^+}\rt]\widetilde{G}_i(\theta_b)\rt\}.
     \end{aligned} \normalsize
\end{equation*}}
The equality holds since $\min_{1\leq n \leq N}\max(1, a_n) = \max(1, \min_{1\leq n\leq N}a_n)$ for all finite $N$.



Finally, we show that $\max\lt[1, \min_{j \neq i^*}{d_j}/{v_j[(i, \theta_b)]^+}\rt] = W_i(\theta_b)$ holds for all $(i, \theta_b) \in \Xi$. For $\theta_b \in \Theta_{i^*}$ (or $i^b=i^*)$, we have  $v_i[(i, \theta_b)]^+/p_b =~2$ and $v_j[(i, \theta_b)]^+/p_b = 1$ for all $j \neq i, i^*$ from Table~\ref{tab:partition}. Then, from~\eqref{eq:weights}, we have 
$ W_i(\theta_b) 
 = \max\lt[1, \min \lt\{{d_i}/{2}, \min_{j \neq i^*} d_j\rt\}/p_b\rt] =\max\lt[1, \min_{j \neq i^*}{d_j}/{v_j[(i, \theta_b)]^+}\rt]$. 
For $\theta_b \notin \Theta_{i^*}$ (or $i^b \neq i^*$), we have $v_i[(i, \theta_b)]^+/p_b = 1$ and $v_j[(i, \theta_b)]^+/p_b = 0$ for all $j \neq i, i^*$ from Table~\ref{tab:partition}.  Therefore, again from~\eqref{eq:weights}, $  W_i(\theta_b) = \max[1, d_i/p_b] = \max\lt[1, \min_{j \neq i^*}{d_j}/{v_j[(i, \theta_b)]^+}\rt]$, as desired. \halmos

\endproof

\proof{Proof of Proposition~\ref{prop:cond_opt_LDR}.} Let us fix $\theta_b$ and $i\neq i^b$. Suppose both $\alpha_{i}(\theta_b)$ and $\alpha_{i^b}(\theta_b)$ are positive. Then,  we have 
\[
\widetilde{G}_i(\theta_b) = \liminf_{n \rightarrow \infty} -\frac{1}{n}\log \prob\lt(m_{i,  b}^n = 1\rt) \geq \lim_{n \rightarrow \infty} -\frac{1}{n}\log \prob(\mu_{i,n}(\theta_b) < \mu_{i^b,n}(\theta_b)) = G_{i}(\theta_b),
\]
where the first inequality follows from $\lt\{m_{i,  b}^n = 1\rt\} \subseteq \lt\{\mu_{i, n}(\theta_b) < \mu_{i^b, n}(\theta_b)\rt\}$. 
If either $\alpha_i(\theta_b)$ or $\alpha_{i^b}(\theta_b)$ is $0$, then $G_i(\theta_b)=0$ by definition. Therefore,  $\liminf_{n \rightarrow \infty} - \frac{1}{n}\log \prob\lt(m_{i,  b}^n = 1\rt)\geq G_i(\theta_b)$ as desired. 

Next, we prove the equality condition~\eqref{eq:minimizer}. Recall that~\eqref{eq:temp2} shows $\widetilde{G}_i(\theta_b) = \min_x F(x)$ where $F(x) = \frac{\alpha_{i^b}(\theta_b)}{2\lambda_{i^b}^2(\theta_b)}(x - y_{i^b}(\theta_b))^2 + \frac{\alpha_i(\theta_b)}{2\lambda_i^2(\theta_b)}(x - y_i(\theta_b))^2 + \sum_{\ell \neq i^b, i}\frac{\alpha_\ell(\theta_b)}{2\lambda_\ell^2(\theta_b)} \lt[(x - y_\ell(\theta_b))^+\rt]^2$. If $\alpha_i(\theta_b) = \alpha_{i^b}(\theta_b) = 0$, we have $F(y_{i^b}(\theta_b)) = 0$, which implies $\widetilde{G}_i(\theta_b) = 0$. Otherwise, let $\widetilde{x}_i$ denote the left-hand-side of~\eqref{eq:minimizer}; $\widetilde{x}_i$ is well-defined in this case.    If $\widetilde{x}_i \in [y_{i^b}(\theta_b), \min_{i \neq i^b} y_i(\theta_b)]$, then we have $[\widetilde{x}_i - \mu_j(\theta_b)]^+ = 0$ for all $j \neq i^b, i$. Furthermore, $\widetilde{G}_i(\theta_b) \leq F(\widetilde{x}_i) = \frac{\alpha_{i^b}(\theta_b)}{2\lambda_{i^b}^2(\theta_b)}(\widetilde{x}_i - y_{i^b}(\theta_b))^2 + \frac{\alpha_i(\theta_b)}{2\lambda_i^2(\theta_b)}(\widetilde{x}_i - y_i(\theta_b))^2 = G_i(\theta_b)$. Combining this with $\widetilde{G}_i(\theta_b) \geq G_i(\theta_b)$ yields $\widetilde{G}_i(\theta_b) = G_i(\theta_b)$, as desired. \halmos
\endproof

\proof{Proof of Theorem \ref{thm:balance}.} 

From the definition of $\underline{\ldr}$,~\eqref{opt:aOCBA} is equivalent to
\begin{equation}\label{opt:intermediate}
    \begin{aligned}
       \max \quad & C \\
       \textrm{s.t.} \quad &  C \leq W_i(\theta_b)G_{i}(\theta_b), \forall (i, \theta_b) \in \Xi,\\
       & \sum_{i=1}^{k}\sum_{b=1}^{B} \alpha_i(\theta_b) = 1, \\\
                      \quad & \alpha_i(\theta_b) \geq 0. 
    \end{aligned}
\end{equation}

Note that harmonic mean function $f(x, y) = 1/(1/x + 1/y)$ is concave in both $x$ and $y$, so each $G_{i}(\theta_b)$ is concave; 
thus,~\eqref{opt:intermediate} is a convex program. On the one hand, 
$G_i(\theta_b)$ for $(i, \theta_b) \in \Xi^\text{adv}$ does 
not appear in this program. For any such $(i,\theta_b)$, suppose $\alpha_i(\theta_b) > 0$. Then, we can always strictly improve the objective function value by setting $\alpha_i(\theta_b) = 0$ and increasing $\alpha_{i^\prime}(\theta_{b^\prime})$ at $(i^\prime, \theta_{b^\prime})$ pairs for which the first constraint of~\eqref{opt:intermediate} is binding.
Hence  ${\alpha}_{i^*}(\theta_b) = 0$ for all $\theta_b \in \Theta^c_{i^*}$ at optimality.  
This proves the last optimality condition.

To solve~\eqref{opt:intermediate}, we set up a Lagrangian function $\mathcal{L}(\bm{\rho}, \gamma, \bm{\kappa}; C, \bm{\alpha})$ defined as
\begin{equation}\label{eq:largrangian_ver0}
    \begin{split}
        \mathcal{L}(\bm{\rho}, \gamma, \bm{\kappa}; C, \bm{\alpha}) = - C + &\sum\nolimits_{(i, \theta_b) \in \Xi}\rho_i(\theta_b)(C - W_i(\theta_b)G_i(\theta_b)) \\
        &+ \gamma\lt(\sum\nolimits_{(i, \theta_b) \in (\Xi^\text{adv})^c} \alpha_i(\theta_b)- 1\rt) - \sum\nolimits_{(i, \theta_b) \in (\Xi^\text{adv})^c} \kappa_i(\theta_b)\alpha_i(\theta_b).
    \end{split}
\end{equation}
where $\boldsymbol{\rho} = \lt\{\rho_i(\theta_b)\rt\}$, $\gamma$, and $\bm{\kappa} = \lt\{\kappa_i(\theta_b)\rt\}$ are the Lagrangian multipliers of the first, second, and third constraints in~\eqref{opt:intermediate}, respectively. Note that in the third and fourth terms summations are over $(i, \theta_b) \in (\Xi^\text{adv})^c$ thanks to the last optimality condition. 
Next, we argue that $\kappa_i(\theta_b) = 0$ for all $(i, \theta_b) \in (\Xi^{\text{adv}})^c$. 
Suppose $\alpha_i(\theta_b) = 0$ for some $(i,\theta_b) \in \lt(\Xi^{\text{adv}}\rt)^c$. Then, $G_{i}(\theta_b) = 0$ or $G_{j}(\theta_b)$ = 0 for some $j$ (when $i = i^b$). This implies $C \leq 0$. However, if we set $\alpha_i(\theta_b) = 1/kB$ for all $1 \leq i \leq k, 1\leq b \leq B$, the optimal $C$ must be positive; for all $(i, \theta_b) \in \Xi$, $W_i(\theta_b) >0$ and positive $\alpha_i(\theta_b)$'s make $G_i(\theta_b)$ positive. Therefore, optimal $\alpha_i(\theta_b)$ must be positive for all $(i, \theta_b) \in \lt(\Xi^{\text{adv}}\rt)^c$. As a result, Lagrangian multipliers associated with the nonnegativity constraint should be zero as they are not binding, i.e., $\kappa_i(\theta_b) = 0$. Further, the uniform allocation, $\alpha_i(\theta_b) = 1/kB$, with $C = 0$ becomes an interior point, which is strictly feasible. Hence, Slater's condition implies that the strong duality holds; thus, Karush-Kuhn-Tucker (KKT) conditions become necessary and sufficient conditions for optimality. Consequently, we derive the KKT conditions for the following Lagrangian function instead of~\eqref{eq:largrangian_ver0}
\begin{equation}\label{eq:lagrangian}
    \mathcal{L}(\boldsymbol{\rho}, \gamma;C, \boldsymbol{\alpha}) := - C + \sum\nolimits_{(i, \theta_b) \in \Xi}\rho_i(\theta_b)\lt(C - W_i(\theta_b) G_{i}(\theta_b)\rt) + \gamma\lt(\sum\nolimits_{(i, \theta_b) \in \lt(\Xi^\text{adv}\rt)^c} \alpha_i(\theta_b) - 1\rt)
\end{equation}
as follows:
\begin{eqnarray}
    \rho_i(\theta_b) &\geq& 0,\forall (i, \theta_b) \in \Xi, \label{eq:dual_feas}\\
    \frac{\partial  \mathcal{L}}{\partial C} & = & - 1 + \sum_{(i, \theta_b) \in \Xi}\rho_i(\theta_b) = 0, \label{eq:stat_K}\\
    \frac{\partial  \mathcal{L}}{\partial \alpha_i( \theta_b)} &=& -\rho_i(\theta_b) W_i(\theta_b)\frac{\partial G_{i}(\theta_b)}{\partial \alpha_i(\theta_b)} + \gamma = 0, \forall (i, \theta_b) \in \Xi, \label{eq:stat_alpha}\\
    \frac{\partial  \mathcal{L}}{\partial \alpha_{i^b} (\theta_b)} &=& -\sum_{i \neq i^b, i^*}\rho_i(\theta_b) W_i(\theta_b)\frac{\partial G_{i}(\theta_b)}{\partial \alpha_{i^b} (\theta_b)} + \gamma = 0, \forall b, \label{eq:stat_alpha_2}\\
    \rho_i(\theta_b)\lt(C - W_i(\theta_b) G_{i}(\theta_b)\rt) &=& 0, \forall (i, \theta_b) \in \Xi. \label{eq:com_slack}
\end{eqnarray}
From \eqref{eq:dual_feas} and \eqref{eq:stat_K}, there exists $(i, \theta_b) \in \Xi$ such that $\rho_i(\theta_b) > 0$. Therefore, we further obtain {$\gamma~>~0$} from~\eqref{eq:stat_alpha}, which in turn implies that $\rho_i(\theta_b) > 0$ for all $(i, \theta_b) \in \Xi$. Finally,~\eqref{eq:com_slack} implies that $C =W_i(\theta_b) G_{i}(\theta_b)$ for all $(i, \theta_b) \in \Xi$, which corresponds to the pairwise balance condition. Combining \eqref{eq:stat_alpha} with \eqref{eq:stat_alpha_2}, we have
\begin{equation}\label{eq:der_bal}
    \begin{aligned}
        1 &= \frac{1}{\gamma}\sum_{i \neq i^b, i^*}\rho_i(\theta_b) W_i(\theta_b)\frac{\partial G_{i}(\theta_b)}{\partial \alpha_{i^b} (\theta_b)} = \sum_{i \neq i^b, i^*}\frac{\rho_i(\theta_b) W_i(\theta_b)\frac{\partial G_{i}(\theta_b)}{\partial \alpha_{i^b} (\theta_b)}}{\rho_i(\theta_b) W_i(\theta_b)\frac{\partial G_{i}(\theta_b)}{\partial \alpha_i( \theta_b)}} \\
        &= \sum_{i \neq i^b, i^*} \lt(\frac{\partial G_{i}(\theta_b)}{\partial \alpha_{i^b} (\theta_b)}\middle/\frac{\partial G_{i}(\theta_b)}{\partial \alpha_i(\theta_b)}\rt) =  \sum_{i \neq i^b, i^*}\frac{\lambda^2_{i^b}(\theta_b)/\alpha^2_{i^b}(\theta_b)}{\lambda^2_{i}(\theta_b)/\alpha^2_i( \theta_b)}.
    \end{aligned}
\end{equation}
For the third equality, we exploit that
$    \dfrac{\partial G_{i}(\theta_b)}{\partial \alpha_{i} (\theta_b)} = G_i(\theta_b)   \dfrac{\lambda^2_i(\theta_b)/\alpha_i^2(\theta_b)}{\lambda^2_i(\theta_b)/\alpha_i^2(\theta_b) + \lambda^2_{i^b}(\theta_b)/\alpha_{i^b}^2(\theta_b)}.$ 
\halmos
\endproof

\proof{Proof of Proposition~\ref{prop:WG_to_Gknapsack}.} From Lemma~\ref{lem:relax_A_i}, we have $$\min_{\BFM \in \mathcal{A}_j}\sum_{(i, \theta_b) \in I(\BFM)}G_i(\theta_b) = \min_{\BFM \in \widetilde{\mathcal{A}}_j}\sum_{(i, \theta_b) \in I(\BFM)}G_i(\theta_b).$$
From Table~\ref{tab:partition}, observe that any matrix $\BFM$ with $m_{i^*, b} = 1$ for some $\theta_b \in \Theta_{i^*}^c$ does not belong to $\widetilde{\mathcal{A}}_j$ for any $j \neq i^*$. Hence, $\min_{j \neq i^*}\min_{\BFM \in \widetilde{\mathcal{A}}_j}\sum_{(i, \theta_b) \in I(\BFM)}G_i(\theta_b)$ does not involve $\{G_{i^*}(\theta_b), \theta_b \in \Theta_{i^*}^c\}$ at all, which implies that $\{\alpha_{i^*}(\theta_b)\}_{\theta \in \Theta_{i*}^c}$ would not appear in the objective function of~\eqref{eq:opt_G_knapsack}. Therefore, $\alpha^\textsf{opt}_{i^*}(\theta_b) = 0$ holds for all $\theta_b \in \Theta_{i^*}^c$.  
\halmos
\endproof

 Before we discuss the proof of Propositions~\ref{prop:suff_1} and~\ref{prop:suff_2}, Lemma~\ref{lem:grad_Gtilde} below derives the gradient of $\widetilde{G}_i(\theta_b)$ with respect to~$\boldsymbol{\alpha}$.

\begin{lemma}\label{lem:grad_Gtilde}
    For each $(i, \theta_b) \in \Xi$, the gradient of $\widetilde{G}_i(\theta_b)$ is given as 
    \[
    \frac{\partial \widetilde{G}_i(\theta_b)}{\partial\alpha_j(\theta_b)} = \left\{
    \begin{array}{cc}
    \dfrac{[(\widetilde{x}_{i, b}(\boldsymbol{\alpha}) - y_j(\theta_b))^+]^2}{2\lambda^2_j(\theta_b)}, & \mbox{ if }  j \neq i, \\
    \dfrac{(\widetilde{x}_{i, b}(\boldsymbol{\alpha}) - y_j(\theta_b))^2}{2\lambda^2_j(\theta_b)},     & \mbox{ if } j=i,
    \end{array}\right.
    \]
    where $\widetilde{x}_{i, b}(\boldsymbol{\alpha})$ is the minimizer of~\eqref{def:G_tilde}.
\end{lemma}
\proof{Proof.} 
Define $F_{i, b}(x, \boldsymbol{\alpha}) = \frac{\alpha_{i^b}(\theta_b)}{2\lambda^2_{i^b}(\theta_b)}(x-y_{i^b}(\theta_b))^2 + \frac{\alpha_i(\theta_b)}{2\lambda_i^2(\theta_b)}(x - y_i(\theta_b))^2+ \sum_{\ell \neq i, i^b}\frac{\alpha_\ell(\theta_b)}{2\lambda_\ell^2(\theta_b)} \lt[(x - y_\ell(\theta_b))^+\rt]^2$. Then, $\widetilde{G}_i(\theta_b) = \min_x F_{i, b}(x, \boldsymbol{\alpha})$ from~\eqref{def:G_tilde}. 
Observe that $F_{i, b}(x, \boldsymbol{\alpha})$ is differentiable with respect to $x$ and its derivative is given as
\begin{equation}\label{eq:1st_derivative}
    \frac{\partial}{\partial x}F_{i, b}(x, \boldsymbol{\alpha}) = \frac{\alpha_{i^b}(\theta_b)}{\lambda_{i^b}(\theta_b)^2}(x-y_{i^b}(\theta_b)) + \frac{\alpha_i(\theta_b)}{\lambda_i^2(\theta_b)}(x-y_i(\theta_b)) + \sum_{\ell\neq i, i^b} \frac{\alpha_\ell(\theta_b)}{\lambda^2_\ell(\theta_b)}(x-y_\ell(\theta_b))^+.
\end{equation}
Recall that $\widetilde{x}_{i, b}(\bm{\alpha})$ as the minimizer of $F_{i, b}(x, \boldsymbol{\alpha})$ when $\boldsymbol{\alpha}$ is fixed. The first-order condition implies $\frac{\partial}{\partial x} F_{i, b}(x, \boldsymbol{\alpha})\big|_{x = \widetilde{x}_{i, b}(\bm{\alpha})} =~0$. By definition, we also have
\begin{equation*}
    \widetilde{G}_i(\theta_b) = \frac{\alpha_{i^b}(\theta_b)}{2\lambda_{i^b}^2(\theta_b)}(\widetilde{x}_{i, b}(\bm{\alpha}) - y_{i^b}(\theta_b))^2 + \frac{\alpha_i(\theta_b)}{2\lambda_i^2(\theta_b)}(\widetilde{x}_{i, b}(\bm{\alpha}) - y_i(\theta_b))^2 + \sum_{\ell\neq i, i^b} \frac{\alpha_\ell(\theta_b)}{2\lambda_\ell^2(\theta_b)}[(\widetilde{x}_{i, b}(\bm{\alpha}) - y_\ell(\theta_b))^+]^2.
\end{equation*}
Taking the partial derivative of both sides with respect to $\alpha_{i^b}(\theta_b)$, we obtain 
\begin{equation*}
    \frac{\partial\widetilde{G}_i(\theta_b)}{\partial \alpha_{i^b}(\theta_b)} = \frac{1}{2\lambda_{i^b}^2(\theta_b)}(\widetilde{x}_{i, b}(\bm{\alpha}) - y_{i^b}(\theta_b))^2 + \frac{\partial}{\partial x} F_{i, b}(x, \boldsymbol{\alpha})\big|_{x = \widetilde{x}_{i, b}(\bm{\alpha})} \frac{\partial \widetilde{x}_{i, b}(\bm{\alpha})}{\partial \alpha_{i^b}(\theta_b)} = \frac{(\widetilde{x}_{i, b}(\bm{\alpha}) - y_{i^b}(\theta_b))^2}{2\lambda_{i^b}^2(\theta_b)},
\end{equation*}
where the last equality follows by $\frac{\partial}{\partial x} F_{i, b}(x, \boldsymbol{\alpha})\big|_{x = \widetilde{x}_{i, b}(\bm{\alpha})} = 0$.
Proceeding similarly, we can derive
\begin{equation*}
    \begin{aligned}
        \frac{\partial\widetilde{G}_i(\theta_b)}{\partial \alpha_\ell(\theta_b)} = \frac{[(\widetilde{x}_{i, b}(\bm{\alpha}) - y_\ell(\theta_b))^+]^2}{2\lambda_\ell^2(\theta_b)} \;\; \mbox{for all } \ell \neq i \;\; \text{and} \;\; 
        \frac{\partial\widetilde{G}_i(\theta_b)}{\partial \alpha_i(\theta_b)} = \frac{(\widetilde{x}_{i, b}(\bm{\alpha}) - y_i(\theta_b))^2}{2\lambda_i^2(\theta_b)},
    \end{aligned}
\end{equation*}
which completes the proof. \halmos
\endproof

In the proof of Propositions~\ref{prop:suff_1} and~\ref{prop:suff_2}, we denote $\widetilde{G}_i(\theta_b)$ evaluated at ${\boldsymbol{\alpha}}$ by $\widetilde{G}_i(\boldsymbol{\alpha};\theta_b)$ to explicitly represent the dependence between $\widetilde{G}_i(\theta_b)$ and $\boldsymbol{\alpha}$. For both proofs, we apply the following logic to show that $\widetilde{\alpha}^{\textsf{opt}}_{i^*}(\theta_b)=0$ for all $\theta_b\in\Theta_{i^*}^c$. 
For any $\boldsymbol{\alpha}$, if we can reallocate $\alpha_{i^*}(\theta_b)$ to other $\{\alpha_i(\theta_b)\}_{i \neq i^*}$ while increasing $\{\widetilde{G}_i(\boldsymbol{\alpha};\theta_b)\}_{i \neq i^*}$, then $\min_{j\neq i^*}\widetilde{\ldr}_{j, i^*}$ increases as well. This is because any $\BFM$ such that $(i^*, \theta_b) \in I(\BFM)$ is not a minimal element with respect to $\preceq$, which in turn implies that $\widetilde{G}_{i^*}(\boldsymbol{\alpha}; \theta_b)$ do not affect $\min_{j\neq i}\widetilde{\ldr}_{j, i^*}$. Hence, if such reallocation exists, then the optimal sampling ratio made at $(i^*, \theta_b)$ for $\theta_b\in\Theta_{i^*}^c$ must be zero; otherwise, one can increase $\min_{j\neq i}\widetilde{\ldr}_{j, i^*}$ by decreasing $\alpha_{i^*}(\theta_b)$.

\proof{Proof of Proposition~\ref{prop:suff_1}.}
If $\frac{\partial\widetilde{G}_i(\theta_b)}{\partial \alpha_{i^*}(\theta_b)}\big|_{\boldsymbol{\alpha} = \widetilde{\boldsymbol{\alpha}}^\textsf{opt}} = 0$, then we must have $\widetilde{x}_{i, b}(\widetilde{\boldsymbol{\alpha}}^\textsf{opt}) \leq y_{i^*}(\theta_b)$ from Lemma~\ref{lem:grad_Gtilde}. We proceed to show  $\widetilde{\alpha}^\textsf{opt}_{i^*}(\theta_b) = 0$. By means of contradiction, assume $\widetilde{\alpha}^\textsf{opt}_{i^*}(\theta_b) > 0$. 
Define $\boldsymbol{\beta}$ as
\begin{equation*}
    \beta_{j}(\theta_\bprime) = \begin{cases} \widetilde{\alpha}^\textsf{opt}_j(\theta_\bprime) & \mbox{if } \bprime \neq b,\\
    \frac{\sum_{\ell = 1}^k \widetilde{\alpha}^\textsf{opt}_\ell(\theta_b)}{\sum_{\ell \neq i^*}\widetilde{\alpha}^\textsf{opt}_\ell(\theta_b)} \times \widetilde{\alpha}^\textsf{opt}_j(\theta_b) & \mbox{if } \bprime = b, j \neq i^*,\\
    0 & \mbox{if } \bprime = b, j = i^*.
    \end{cases}
\end{equation*}
Combining~\eqref{eq:1st_derivative} with $\widetilde{x}_{i, b}(\widetilde{\boldsymbol{\alpha}}^\textsf{opt}) \leq y_{i^*}(\theta_b)$, we have $\frac{\partial F_{i, b}}{\partial x}(\widetilde{x}_{i, b}(\widetilde{\boldsymbol{\alpha}}^\textsf{opt}), \boldsymbol{\beta}) = 0$. Hence, from the first-order condition, one can confirm that $\widetilde{x}_{i,b}(\widetilde{\boldsymbol{\alpha}}^\textsf{opt}) =\widetilde{x}_{i, b}(\boldsymbol{\beta})$ and $\widetilde{G}_i(\boldsymbol{\beta};\theta_b) = \frac{\sum_{\ell = 1}^k \widetilde{\alpha}^\textsf{opt}_\ell(\theta_b)}{\sum_{\ell \neq i^*}\widetilde{\alpha}^\textsf{opt}_\ell(\theta_b)}\widetilde{G}_i(\widetilde{\boldsymbol{\alpha}}^\textsf{opt};\theta_b) > \widetilde{G}_i(\widetilde{\boldsymbol{\alpha}}^\textsf{opt};\theta_b)$ holds for all $i \neq i^*, i^b$. Since $\boldsymbol{\beta}$ increases all $\widetilde{G}_i(\widetilde{\boldsymbol{\alpha}}^\textsf{opt};\theta_b)$ such that $i\neq i^*$, $\boldsymbol{\beta}$ increases $\min_{j \neq i^*}\widetilde{\ldr}_{j, i^*}$. This contradicts that $\widetilde{\boldsymbol{\alpha}}^\textsf{opt}$ is the optimal allocation maximizing $\min_{j \neq i^*}\widetilde{\ldr}_{j, i^*}$. Consequently, we have $\widetilde{\alpha}^\textsf{opt}_{i^*}(\theta_b) = 0$.  \halmos
\endproof

\vspace{5mm}

The following proposition is an example of the sufficient condition implying Proposition~\ref{prop:suff_1}.

\begin{proposition}\label{prop:suff_1-1}
    Let us fix $\theta_b \in \Theta_{i^*}^c$. If the following inequality
    \begin{equation}\label{eq:suff_1}
        \frac{(\widetilde{\alpha}^\textsf{opt}_{i}(\theta_b)/\lambda^2_{i}(\theta_b))y_{i}(\theta_b) + \sum_{\ell: y_{\ell}(\theta_b) \in [y_{i^b}(\theta_b), y_{i^*}(\theta_b))}(\widetilde{\alpha}^\textsf{opt}_{\ell}(\theta_b)/\lambda^2_{\ell}(\theta_b))y_{\ell}(\theta_b)}{\widetilde{\alpha}^\textsf{opt}_{i}(\theta_b)/\lambda^2_{i}(\theta_b) +  \sum_{\ell: y_{\ell}(\theta_b) \in [y_{i^b}(\theta_b), y_{i^*}(\theta_b))}\widetilde{\alpha}^\textsf{opt}_{\ell}(\theta_b)/\lambda^2_{\ell}(\theta_b)} \leq y_{i^*}(\theta_b)
    \end{equation}
    holds for all $i$ such that $y_i(\theta_b) > y_{i^*}(\theta_b)$, then $\widetilde{\alpha}^\textsf{opt}_{i^*}(\theta_b) = 0$. Furthermore,~\eqref{eq:minimizer} implies~\eqref{eq:suff_1}.
\end{proposition}

\proof{Proof of Proposition~\ref{prop:suff_1-1}.}
Consider $\theta_b\in\Theta_{i^*}^c$. 
From Proposition~\ref{prop:suff_1}, it suffices to show that $\widetilde{G}_i(\widetilde{\boldsymbol{\alpha}}^\textsf{opt}; \theta_b)$ does not depend on $\widetilde{\alpha}^\textsf{opt}_{i^*}(\theta_b)$ for all $i \neq i^*$. This is straightforward when $y_i(\theta_b) < y_{i^*}(\theta_b)$ from the expression of $\widetilde{G}_{i}(\widetilde{\boldsymbol{\alpha}}^\textsf{opt};\theta_b)$ in~\eqref{def:G_tilde}. For $i$ such that $y_i(\theta_b) > y_{i^*}(\theta_b)$, let us define $x_{i, b}(\boldsymbol{\alpha})$ as follows:
\begin{equation}\label{eq:interpolation}
    x_{i, b}(\boldsymbol{\alpha})= \frac{(\alpha_{i}(\theta_b)/\lambda^2_{i}(\theta_b))y_{i}(\theta_b) + \sum_{\ell: y_{\ell}(\theta_b) \in [y_{i^b}(\theta_b), y_{i^*}(\theta_b))}(\alpha_{\ell}(\theta_b)/\lambda^2_{\ell}(\theta_b))y_{\ell}(\theta_b)}{\alpha_{i}(\theta_b)/\lambda^2_{i}(\theta_b) +  \sum_{\ell: y_{\ell}(\theta_b) \in [y_{i^b}(\theta_b), y_{i^*}(\theta_b))}\alpha_{\ell}(\theta_b)/\lambda^2_{\ell}(\theta_b)}.
\end{equation}
From the assumption of the proposition, we have $x_{i, b}(\widetilde{\boldsymbol{\alpha}}^\textsf{opt}) \leq y_{i^*}(\theta_b)$  for all $i$ such that $y_i(\theta_b) > y_{i^*}(\theta_b)$. 
Observe that $\frac{\partial F_{i, b}}{\partial x}(x, \boldsymbol{\alpha})$ in~\eqref{eq:1st_derivative}
is an increasing function in $x$, and a straightforward computation yields
$$\frac{\partial F_{i, b}}{\partial x}(y_{i^*}(\theta_b), \widetilde{\boldsymbol{\alpha}}^\textsf{opt}) = \left(\frac{\widetilde{\alpha}^\textsf{opt}_{i}(\theta_b)}{\lambda^2_{i}(\theta_b)} +  \sum_{\ell: y_{\ell}(\theta_b) \in [y_{i^b}(\theta_b), y_{i}(\theta_b))}\frac{\widetilde{\alpha}^\textsf{opt}_{\ell}(\theta_b)}{\lambda^2_{\ell}(\theta_b)}\right)(y_{i^*}(\theta_b) - x_{i, b}(\widetilde{\boldsymbol{\alpha}}^\textsf{opt})) \geq 0,$$ 
where the inequality follows from the assumption.
Hence, $\frac{\partial F_{i, b}}{\partial x}(x, \widetilde{\boldsymbol{\alpha}}^\textsf{opt}) > 0$ for all $x > y_{i^*}(\theta_b)$. Combining this with the fact that $\min_{x} F_{i, b}(x, \widetilde{\boldsymbol{\alpha}}^\textsf{opt})$ must be achieved at $x$ such that $\frac{\partial F_{i, b}}{\partial x}(x, \widetilde{\boldsymbol{\alpha}}^\textsf{opt})=~0$, we have $\argmin_x F_{i, b}(x, \widetilde{\boldsymbol{\alpha}}^\textsf{opt}) \in [y_{i^b}(\theta_b), y_{i^*}(\theta_b)]$, which in turn implies that $\frac{\partial\widetilde{G}_{i}(\boldsymbol{\alpha};\theta_b)}{\partial \alpha_{i^*}(\theta_b)}\big|_{\boldsymbol{\alpha} = \widetilde{\boldsymbol{\alpha}}^\textsf{opt}} = 0$.

Finally, let us show that~\eqref{eq:minimizer} implies~\eqref{eq:suff_1}. If~\eqref{eq:minimizer} holds for some $i$ and $\theta_b\in\Theta_{i^*}^c$, then we have
{\small{\begin{equation}\label{eq:intermediate.inequality}
    \begin{aligned}
        \frac{\widetilde{\alpha}^\textsf{opt}_i(\theta_b)}{\lambda_i^2(\theta_b)} y_i(\theta_b) + \frac{\widetilde{\alpha}^\textsf{opt}_{i^b}(\theta_b)}{\lambda_{i^b}^2(\theta_b)} y_{i^b}(\theta_b) \leq \min_{\ell \neq i^b}y_\ell(\theta_b) \left(\frac{\widetilde{\alpha}^\textsf{opt}_i(\theta_b)}{\lambda_i^2(\theta_b)} + \frac{\widetilde{\alpha}^\textsf{opt}_{i^b}(\theta_b)}{\lambda_{i^b}^2(\theta_b)} \right) \leq y_{i^*}(\theta_b) \left( \frac{\widetilde{\alpha}^\textsf{opt}_i(\theta_b)}{\lambda_i^2(\theta_b)} + \frac{\widetilde{\alpha}^\textsf{opt}_{i^b}(\theta_b)}{\lambda_{i^b}^2(\theta_b)}\right).
    \end{aligned}
\end{equation}}}
The first inequality follows from~\eqref{eq:minimizer} and the second from  that $i^*\neq i^b$ at this $\theta_b.$ 
Consequently, 
\begin{equation*}
    \begin{aligned}
        &\frac{\widetilde{\alpha}^\textsf{opt}_{i}(\theta_b)}{\lambda^2_{i}(\theta_b)}y_{i}(\theta_b) + \sum_{\ell: y_{\ell}(\theta_b) \in [y_{i^b}(\theta_b), y_{i^*}(\theta_b))}\frac{\widetilde{\alpha}^\textsf{opt}_{\ell}(\theta_b)}{\lambda^2_{\ell}(\theta_b)}y_{\ell}(\theta_b)\\ 
        &= \frac{\widetilde{\alpha}^\textsf{opt}_i(\theta_b)}{\lambda_i^2(\theta_b)} y_i(\theta_b) + \frac{\widetilde{\alpha}^\textsf{opt}_{i^b}(\theta_b)}{\lambda_{i^b}^2(\theta_b)} y_{i^b}(\theta_b) + \sum_{\ell: y_{\ell}(\theta_b) \in (y_{i^b}(\theta_b), y_{i^*}(\theta_b))}\frac{\widetilde{\alpha}^\textsf{opt}_{\ell}(\theta_b)}{\lambda^2_{\ell}(\theta_b)}y_{\ell}(\theta_b)\\
        &\leq y_{i^*}(\theta_b) \left( \frac{\widetilde{\alpha}^\textsf{opt}_i(\theta_b)}{\lambda_i^2(\theta_b)} + \frac{\widetilde{\alpha}^\textsf{opt}_{i^b}(\theta_b)}{\lambda_{i^b}^2(\theta_b)}\right) + \sum_{\ell: y_{\ell}(\theta_b) \in (y_{i^b}(\theta_b), y_{i^*}(\theta_b))}\frac{\widetilde{\alpha}^\textsf{opt}_{\ell}(\theta_b)}{\lambda^2_{\ell}(\theta_b)}y_{\ell}(\theta_b)\\
        &\leq \left(\frac{\widetilde{\alpha}^\textsf{opt}_{i}(\theta_b)}{\lambda^2_{i}(\theta_b)} +  \sum_{\ell: y_{\ell}(\theta_b) \in [y_{i^b}(\theta_b), y_{i}(\theta_b))}\frac{\widetilde{\alpha}^\textsf{opt}_{\ell}(\theta_b)}{\lambda^2_{\ell}(\theta_b)}\right) y_{i^*}(\theta_b),
    \end{aligned}
\end{equation*}
where the first inequality follows from \eqref{eq:intermediate.inequality} and the last  follows from $y_\ell(\theta_b) < y_{i^*}(\theta_b)$. This proves
 $x_{i, b}(\widetilde{\boldsymbol{\alpha}}^\textsf{opt}) \leq y_{i^*}(\theta_b)$ as desired.  \halmos

\endproof

\proof{Proof of Proposition~\ref{prop:suff_2}.} 

Let us fix $\theta_b \in \Theta_{i^*}^c$.  In the following, we show that for any $\boldsymbol{\alpha}$ such that $\alpha_{i^*}(\theta_b)>0$, we can construct the following $\boldsymbol{\beta} = (\beta_\ell(\theta_{b_1}))$ from $\boldsymbol{\alpha}$ that improves $\ldr_\text{FS}$, if~\eqref{eq:suff_2} is valid: 
\begin{equation}\label{eq:def_beta}
    \beta_\ell(\theta_{b_1}) = \begin{cases}
         \alpha_{\ell}(\theta_{b_1}), &\mbox{if } \left\{\ell \notin\left\{j^b, i^*\right\}\;\; \text{\&}\;\; b_1 = b\right\} \;\; \text{or} \;\; b_1 \neq b,\\
         \alpha_{j^b}(\theta_b) + \alpha_{i^*}(\theta_b), &\mbox{if } \ell = j^b\;\; \text{\&}\;\;  b_1 = b, \\
         0, &\mbox{if } \ell = i^*\;\; \text{\&}\;\;  b_1 = b.\\
    \end{cases}
\end{equation}

If~\eqref{eq:suff_2} holds, one can confirm that the following inequality holds
\begin{equation}\label{eq:suff_cond}
    \frac{(x-y_{j^b}(\theta_b))^+}{\lambda_{j^b}(\theta_b)} \geq \frac{(x-y_{i^*}(\theta_b))^+}{\lambda_{i^*}(\theta_b)} \;\; \text{for all } x \in [y_{i^b}(\theta_b), y_{k^b}(\theta_b)].
\end{equation}
To ease the notational burden, we omit $\theta_b$ from all quantities below and denote $\widetilde{G}_i(\boldsymbol{\alpha};\theta_b) = \widetilde{G}_i(\boldsymbol{\alpha})$.
Let us consider $\ell$ with $y_\ell > y_{i^*}$ first. From the definition of $\widetilde{G}_\ell(\boldsymbol{\alpha})$ in~\eqref{def:G_tilde} and~\eqref{eq:def_beta}, we have 
\begin{equation*}
    \begin{aligned}
        \widetilde{G}_\ell({\boldsymbol{\alpha}}) &= \min_{x:y_{i^b}\leq x\leq y_\ell} \bigg\{\frac{\alpha_{i^b}}{2\lambda_{i^b}^2}(x - y_{i^b})^2 + \frac{\alpha_\ell}{2\lambda_\ell^2}(x - y_\ell)^2 + \sum_{\ell_1: y_{i^b}< y_{\ell_1} < y_\ell}\frac{\alpha_{\ell_1}}{2\lambda_{\ell_1}^2} \lt[(x - y_{\ell_1})^+\rt]^2\bigg\}\\
        &=\min_{x:y_{i^b}\leq x\leq y_\ell} \bigg\{\frac{\alpha_{i^b}}{2\lambda_{i^b}^2}(x - y_{i^b})^2 + \frac{\alpha_\ell}{2\lambda_\ell^2}(x - y_\ell)^2 +  \frac{\alpha_{i^*}}{2\lambda_{i^*}^2}[(x - y_{i^*})^+]^2 + \sum_{\ell_1 \neq i^*: y_{i^b}< y_{\ell_1} < y_\ell}\frac{\alpha_{\ell_1}}{2\lambda_{\ell_1}^2} \lt[(x - y_{\ell_1})^+\rt]^2\bigg\}\\
        & \leq \min_{x:y_{i^b}\leq x\leq y_\ell} \bigg\{\frac{\alpha_{i^b}}{2\lambda_{i^b}^2}(x - y_{i^b})^2 + \frac{\alpha_\ell}{2\lambda_\ell^2}(x - y_\ell)^2 +  \frac{\alpha_{i^*}}{2\lambda_{j^b}^2}[(x - y_{j^b})^+]^2 + \sum_{\ell_1 \neq i^*: y_{i^b}< y_{\ell_1} < y_\ell}\frac{\alpha_{\ell_1}}{2\lambda_{\ell_1}^2} \lt[(x - y_{\ell_1})^+\rt]^2\bigg\}\\
        & = \widetilde{G}_\ell({\boldsymbol{\beta}}).
    \end{aligned}
\end{equation*}
The inequality follows from~\eqref{eq:def_beta} and the last equality follows since $\beta_m = \alpha_m$ for all $m \neq j, i^*$, $\beta_{j^b} = \alpha_{j^b} + \alpha_{i^*}$, and $\beta_{i^*} = 0$ in~\eqref{eq:def_beta}. For $\ell$ such that $y_\ell < y_{i^*}$, one can see that
\begin{equation*}
    \begin{aligned}
        \widetilde{G}_\ell({\boldsymbol{\alpha}}) &= \min_{x:y_{i^b}\leq x\leq y_\ell} \bigg\{\frac{\alpha_{i^b}}{2\lambda_{i^b}^2}(x - y_{i^b})^2 + \frac{\alpha_\ell}{2\lambda_\ell^2}(x - y_\ell)^2 + \sum_{\ell_1: y_{i^b}< y_{\ell_1} < y_\ell}\frac{\alpha_{\ell_1}}{2\lambda_{\ell_1}^2} \lt[(x - y_{\ell_1})^+\rt]^2\bigg\}\\
        &\leq \widetilde{G}_\ell({\boldsymbol{\beta}}) = \min_{x:y_{i^b}\leq x\leq y_\ell} \bigg\{\frac{\beta_{i^b}}{2\lambda_{i^b}^2}(x - y_{i^b})^2 + \frac{\beta_\ell}{2\lambda_\ell^2}(x - y_\ell)^2 + \sum_{\ell_1: y_{i^b}< y_{\ell_1} < y_\ell}\frac{\beta_{\ell_1}}{2\lambda_{\ell_1}^2} \lt[(x - y_{\ell_1})^+\rt]^2\bigg\},
    \end{aligned}
\end{equation*}
as $\beta_{\ell} \geq \alpha_\ell$ holds for all $\ell \neq i^*$ from~\eqref{eq:def_beta}. Therefore, any $\boldsymbol{\alpha}^\textsf{opt}$ has $\alpha^\textsf{opt}_{i^*}=0$ for all $\theta_b\in\Theta^c_{i^*}$. 
\halmos
\endproof

\section{Extension to Non-unique MPBs}
\label{ec:nonunique.mpb}
In this section, we discuss the extensions of Theorems~\ref{thm:lower_LDR} and~\ref{thm:balance} in case there are multiple MPBs. Let $i^* = \argmax_j P_j$ be the set of the MPBs. 
    For each $\ell \in i^*$ and each conditional optima mapping $\BFM=\{m_{i,b}\}\in \mathcal{M}$, let $d_j^\ell(\BFM) := \sum_{b=1}^B p_b (m_{\ell,b}-m_{j,b})$. Given the true mapping, $\BFM^*,$ $d_j^\ell := d_j^\ell(\BFM^*)$.
    We extend the definition of the balance weight in Theorem~\ref{thm:lower_LDR} as follows: for each $\ell\in i^*$ let $\Xi_\ell:= \{(i, \theta_b)| i \neq i^b, i \neq \ell\}$. For each $(i, \theta_b) \in \Xi_\ell$ define 
    \begin{equation} \label{eq:weights_new}
        W_i^\ell(\theta_b) := \begin{cases} \max\lt\{\min\lt(\min_{j \neq \ell}d^\ell_j, \frac{d^\ell_i}{2}\rt)/p_b, 1\rt\}, & \mbox{if } \theta_b  \in\Theta_{\ell}, \\
        \max\lt\{d^\ell_i/p_b, 1\rt\}, & \mbox{if } \theta_b  \notin\Theta_{\ell},
        \end{cases}
    \end{equation}
    and let $W_i^\ell(\theta_b) = \infty$ for $(i,\theta_b)\in \Xi_\ell^c$.

    Theorem~\ref{thm:nonunique} is a counterpart of Theorem~\ref{thm:lower_LDR} extending it to the case when $i^*$ is non-unique.
    
    \begin{theorem}[Extension to the non-unique MPB case] \label{thm:nonunique}
        Let $i^* = \argmax_j P_j$. Then, the FS event, $\{i_n^*\neq i^*\}$, has the following lower bound for its LDR:
        \begin{equation*}
            \liminf_{n \rightarrow \infty} -\frac{1}{n}\log\prob\left(i_n^* \neq i^*\right) \geq \min_{(i, \theta_b) \in \cup_{\ell \in i^*}\Xi^\ell}\lt(\min_{\ell \in i^*}W^\ell_i(\theta_b)\rt)G_i(\theta_b),
        \end{equation*}
        where $W^\ell_i(\theta_b)$ is the balance weight defined in~\eqref{eq:weights_new}.
    \end{theorem}

    \proof{Proof.}
         Recall that $\BFM_n$ is the conditional optima mapping given the posterior means after $n$ simulations are made. Then, $\{i_n^* \neq i^*\} = \{\BFM_n \in \text{FS}_1 \cup \text{FS}_2\}$, where 
        \begin{equation*}
            \begin{aligned}
                 \text{FS}_1 &:= \{\BFM:\text{$\exists j \notin i^*$ such that $d^\ell_j(\BFM) \leq 0$ for some $\ell\in i^*$}\} \\
                 \text{FS}_2 &:= \{\BFM:\text{$\exists \ell_1, \ell_2 \in i^*$ such that $d^{\ell_2}_{\ell_1}(\BFM) < 0$}\}.
            \end{aligned}
        \end{equation*}
        In words, $\text{FS}_1$ occurs when some non-MPB solution $j$ ties with or outperforms an MPB; and $\text{FS}_2$ occurs when some MPB is excluded from $i^*_n$. 
        From this, we have $\liminf_{n\rightarrow \infty} -\frac{1}{n}\log\prob(i_n^* \neq i^*) = \min\left\{\liminf_{n\rightarrow \infty}-\frac{1}{n}\log\prob(\BFM_n \in \text{FS}_1), \liminf_{n\rightarrow \infty}-\frac{1}{n}\log\prob(\BFM_n \in \text{FS}_2)\right\}$. 

         Let us first consider the LDR of $\text{FS}_1$. Recall that $\BFM$ is defined as a $k\times B$ matrix with binary entries in the paper. For each $\ell \in i^*$, suppose we remove all other solutions in $i^*$ from the R\&S problem so that $\ell$ is the unique MPB. Equivalently, for each $\BFM\in\mathcal{M}$, we define $\BFM^\ell$ by deleting rows of $\BFM$ corresponding to other solutions in $i^*$ and let $\mathcal{A}_j^\ell := \{\BFM: d_j^\ell(\BFM^\ell) \leq 0\}$. 
         Then, the LDR of $\BFM_n \in\bigcup_{j \notin i^*} \mathcal{A}_j^\ell$ can be derived using the same argument when the MPB is unique. As a result, we have
         \begin{equation*}
             \liminf_{n\rightarrow \infty}-\frac{1}{n}\log\prob(\BFM_n \in \cup_{j \notin i^*} \mathcal{A}_j^\ell) \geq \min_{(i, \theta_b) \in \Xi^\ell, i \notin i^*} W^\ell_i(\theta_b)G_i(\theta_b).
         \end{equation*}
         From the definition of $\mathcal{A}_j^\ell$, we have $\text{FS}_1 = \bigcup_{\ell \in i^*}\bigcup_{j \notin i^*} \mathcal{A}_j^\ell.$
         Consequently,
         \begin{equation}\label{eq:ldr_fs1}
            \begin{aligned}
                \liminf_{n\rightarrow \infty}-\frac{1}{n}\log\prob(\BFM_n \in \text{FS}_1) &\geq \min_{\ell \in i^*}\min_{(i, \theta_b) \in \Xi^\ell, i \notin i^*} W^\ell_i(\theta_b)G_i(\theta_b)\\
                & = \min_{(i, \theta_b) \in \cup_{\ell \in i^*}\Xi^\ell, i \notin i^*}\lt(\min_{\ell \in i^*}W^\ell_i(\theta_b)\rt)G_i(\theta_b).
            \end{aligned}      
         \end{equation}
         
         To derive the LDR of $\text{FS}_2$, let us define $\mathcal{B} := \{(i, \theta_b) \in \cup_{\ell \in i^*}\Xi^\ell| i \notin i^*, i^b \notin i^*\}$, i.e., $\mathcal{B}$ collects all $(i, \theta_b)$ that do not impact estimating $\{P_\ell\}_{\ell \in i^*}$ even if $i$ is incorrectly specified as $i_n^b$. Hence, one can see that any $\BFM \in \text{FS}_2$ such that $I(\BFM) \cap \mathcal{B} \neq \emptyset$ can be excluded from the partial order relation $\preceq$ defined in Definition EC.1. The minimal set of $\text{FS}_2$, say $\widetilde{\text{FS}}_2$, satisfies $I(\BFM) \cap \mathcal{B} = \emptyset$ for all $\BFM \in \widetilde{\text{FS}}_2$. Equivalently, $I(\BFM) \subseteq \mathcal{B}^c = \{(i, \theta_b)\in \cup_{\ell \in i^*}\Xi^\ell| i \in i^* \;\; \text{or} \;\; i^b \in i^*\}$ holds for all $\BFM \in \widetilde{\text{FS}}_2$. 
         
         We proceed to show that each $\BFM \in \widetilde{\text{FS}}_2$ misspecifies exactly one solution-parameter pair in $\mathcal{B}^c$. To see this, let us consider $\BFM_1$ such that $I(\BFM_1) = \{(i, \theta_b)\} \subseteq \mathcal{B}^c$. If $i \in i^*$ and $i^b \notin i^*$, we obtain $d^\ell_i(\BFM) < 0$ for all $\ell \in i^*, \ell \neq i$, which implies $\BFM_1 \in \widetilde{\text{FS}}_2$. When $i^b \in i^*$, $d^\ell_{i^b}(\BFM_1) < 0$ holds for all $\ell \neq i^b, \ell \in i^*$. Again, we have $\BFM_1 \in \widetilde{\text{FS}}_2$. From this observation, the LDR of $\text{FS}_2$ can be written as
         \begin{equation}\label{eq:ldr_fs2}
             \begin{aligned}
                \liminf_{n\rightarrow \infty}-\frac{1}{n}\log\prob(\BFM_n \in \text{FS}_2) &\geq \min_{(i, \theta_b) \in \mathcal{B}^c}G_i(\theta_b) = \min_{(i, \theta_b) \in \mathcal{B}^c} \lt(\min_{\ell \in i^*}W_i^\ell(\theta_b)\rt)G_i(\theta_b)
            \end{aligned}   
         \end{equation}
         The second equality follows from the fact that $\min_{\ell \in i^*}W_i^\ell(\theta_b) = 1$ for all $(i, \theta_b) \in \mathcal{B}^c$, which can be seen from the definition of $W_i^\ell(\theta_b)$ in~\eqref{eq:weights_new}. Since we have $\{(i, \theta_b) \in \cup_{\ell \ne i^*}\Xi^\ell, i \in i^*\} \subseteq \mathcal{B}^c$, combining~\eqref{eq:ldr_fs1} with~\eqref{eq:ldr_fs2} completes the proof. \halmos
    \endproof

    Note that the LDR's lower bound in Theorem~\ref{thm:nonunique} equals that of Theorem~\ref{thm:lower_LDR} if $i^*$ is indeed a singleton. 
    Theorem~\ref{thm:nonunique} implies that the optimality conditions similar to those in Theorem~\ref{thm:balance} can be derived once we replace $W_i(\theta_b)$ with $\min_{\ell \in i^*}W_i^\ell(\theta_b)$ for each $(i,\theta_b)$. Similarly, the sequential sampling algorithms can be modified by replacing $\BFW_n$  
    with $\min_{\ell \in i_n^*}\{\BFW_n^\ell\}$, where $\BFW_n^\ell$ is the matrix of the sample versions of the balance weights defined in~\eqref{eq:weights_new} and the minimum operation is taken elementwise. 

\section{Proofs of Results in Section~\ref{sec:sequential_learning}}\label{apdx:conv_proofs}

For two positive random sequences $\lt\{a_n\rt\}_{n \in \mathbb{N}}$ and $\lt\{b_n\rt\}_{n \in \mathbb{N}}$, we adopt the notation, $a_n = \Oas(b_n)$, if $\limsup_{n\rightarrow \infty} {a_n}/{b_n} < \infty$ a.s.
We further define $a_n = \Tas(b_n)$, if $a_n = \Oas(b_n)$ and $b_n = \Oas(a_n)$. Equivalently, $a_n = \Tas(b_n)$ holds, if and only if, 
$0 < \liminf_{n \rightarrow \infty}{a_n}/{b_n} \leq \limsup_{n \rightarrow \infty}{a_n}/{b_n} < \infty$ a.s. Do not confuse $\Tas$ with the set notation defined for $\theta$. 

We reverse the order to show Theorem~\ref{thm:proof_BWalg} first. Showing Theorem~\ref{thm:proof_EIalg} is more involved as Algorithm~\ref{alg:rate_opt_posterior} combines both posterior mean plug in and posterior sampling to decide which solution-parameter pair to simulate next. Theorem~\ref{cor:Alg1_convergence} follows as a consequence of Theorem~\ref{thm:proof_EIalg}.

\proof{Proof of Theorem \ref{thm:proof_BWalg}.}

For each sample path, we define two partitions of the set of positive integers $\mathbb{N}$,  $\lt\{\mathcal{N}_i(\theta_b)\rt\}_{1\leq i\leq k, 1 \leq b \leq B}$ and $\lt\{\mathcal{N}(\theta_b)\rt\}_{1 \leq b \leq B}$, where
\[
\mathcal{N}_i(\theta_b) := \{n \in \mathbb{N} | \text{$(i, \theta_b)$ is sampled at the $(n+1)$th simulation run}\} \text{ and } \mathcal{N}(\theta_b):= \bigcup_{i=1}^{k}\mathcal{N}_i(\theta_b).
\]
Equivalently,  $\mathcal{N}(\theta_b) = \{n \in \mathbb{N} | \min\nolimits_{1\leq i\leq k,i\neq i_n^b} W_{i, n}(\theta_b) G_{i,n}(\theta_b) < \min\nolimits_{1\leq i\leq k,i\neq i_n^{b^\prime}} W_{i,n}(\theta_\bprime)G_{i,n}(\theta_{\bprime}), \forall \bprime \neq b\}$ from the sampling rule. We also have $\bigcup_{i=1}^{k}\bigcup_{b=1}^{B}\mathcal{N}_i(\theta_b) = \bigcup_{b=1}^{B}\mathcal{N}(\theta_b) = \mathbb{N}$. Further, the total allocation to $\theta_b$ by the $n$th replication is denoted by $N_{\tot}^n(\theta_b) := \sum_{i=1}^{k} N_i^n(\theta_b)$.

For each $\theta_b$, define $S_1(\theta_b) = \{i: \lim_{n \rightarrow \infty}N^n_i(\theta_b) = \infty\}$ and $S_2(\theta_b) = \{i: \lim_{n \rightarrow \infty}N^n_i(\theta_b) < \infty\}$; together, they form a partition of the solution set. Observe that both $\argmin_{i \in S_1(\theta_b)} \mu_{i, n}(\theta_b)$ and $\argmin_{i \in S_2(\theta_b)} \mu_{i, n}(\theta_b)$ converge given the sample path; the former is by the SLLN and latter is due to the finiteness of $N^n_i(\theta_b)$ for all $i \in S_2(\theta_b)$. Hence,  $i_n^b = i^b_0$ holds for some $i^b_0$ for all sufficiently large $n$. Corresponding to $\{i_0^b\}$, we  define $\Xi_0, \Xi_0^{\text{adv}}$ and $\{W_{i, 0}(\theta_b)\}$  so that for sufficiently large $n$, $\Xi_n = \Xi_0$, $\Xi_n^{\text{adv}} = \Xi_0^{\text{adv}}$, and $W_{i,n}(\theta_b) = W_{i, 0}(\theta_b)$. Note that we cannot guarantee $i^b_0 = i^b$ since $i^b_0$ depends on sample paths assigned to $(i, \theta_b)$ with $i \in S_2(\theta_b)$ as long as $S_2(\theta_b) \neq \emptyset$. 
As a first step, we show that for all $\theta_b$, $i_0^b = i^b$ holds almost surely.


\noindent\textbf{Step 1.} $N^n_i(\theta_b) \rightarrow \infty$ a.s. as $n \rightarrow \infty$ for all $1 \leq i \leq k, 1 \leq b \leq B$.

Since $\sum_{b=1}^B N^n_{\text{tot}}(\theta_b) = n$, there exists at least one $\theta_b$ such that $N^n_{\text{tot}}(\theta_b) \rightarrow \infty$ as $n \rightarrow \infty$. For such $\theta_b$, suppose there exists $1 \leq j \leq k$ such that $\lim_{n \rightarrow \infty} N^n_j(\theta_b) < \infty$. We first argue that $N_{i_0^b}^n(\theta_b) \rightarrow \infty$. By means of contradiction, suppose otherwise. 
Because $\lim_{n \rightarrow \infty} N^n_{i_0^b}(\theta_b) < \infty$, 
\begin{equation}\label{eq:global_bal}
    \lt(\dfrac{N^n_{i_0^b}(\theta_b)}{\lambda_{i_0^b}(\theta_b)}\rt)^2 < \sum\nolimits_{i \neq i_0^b}\lt(\dfrac{N^n_i(\theta_b)}{\lambda_i(\theta_b)}\rt)^2
\end{equation}
must hold at finitely many $n\in\mathcal{N}(\theta_b)$ because otherwise the algorithm would allocate infinite simulation effort to $i_0^b$. This implies that $\lim_{n \rightarrow \infty} N^n_i(\theta_b) < \infty$ for all $1 \leq i \leq k$, which contradicts that $N^n_{\text{tot}}(\theta_b) \rightarrow \infty$. Consequently, we have $j \neq i_0^b$, and $N^n_{i_0^b}(\theta_b) \rightarrow \infty$ follows.

To achieve $N^n_{i_0^b}(\theta_b) \rightarrow \infty$, \eqref{eq:global_bal} should hold infinitely often. Hence, there exists $i_1 \neq i_0$ such that $N^n_{i_1}(\theta_b) \rightarrow \infty$ that makes the right-hand side of~\eqref{eq:global_bal} diverge.
For such $i_1$, we can see 
\begin{equation}\label{eq:LDR.pairwise}
    nG_{i_1,n}(\theta_b) = \dfrac{(\mu_{i_1,n}(\theta_b) - \mu_{i_0^b, n}(\theta_b))^2}{2\lt(\dfrac{\lambda^2_{i_1}(\theta_b)}{N_{i_1}^n(\theta_b)} + \dfrac{\lambda^2_{i_0^b}(\theta_b)}{N_{i_0^b}^n(\theta_b)}\rt)} \rightarrow \infty.
\end{equation}
Observe that we have $\lim_{n\rightarrow \infty}nG_{j, n}(\theta_b) < \infty$ as $\lim_{n\rightarrow \infty} N_j^n(\theta_b) < \infty$, which in turn implies that $W_{j, 0}(\theta_b)G_{j, n}(\theta_b) < W_{i_1, 0}(\theta_b)G_{i_1, n}(\theta_b)$ holds for all sufficiently large $n$. Then, Steps~\ref{step:pairwise} and~\ref{step:global} of Algorithm~\ref{alg:rate_opt} would never select $(i_1, \theta_b)$ as the next sampling pair, which contradicts $N_{i_1}^n(\theta_b) \rightarrow \infty$. Hence, we conclude that $N^n_i(\theta_b) \rightarrow \infty$ for all $1\leq i\leq k$.

Now, it remains to show that  $\{N_i(\theta_b)\}_{1\leq i \leq k}$ go to infinity for all $\theta_b$. By Step~\ref{step:pairwise} of Algorithm~\ref{alg:rate_opt} again, we claim that $nG_{i^\prime,n}(\theta_{b^\prime}) \rightarrow \infty$ for all $(i^\prime, \theta_{\bprime}) \in \Xi_0\cup\Xi_0^{\text{adv}}$ from the existence of $\theta_b$ such that $N_{\text{tot}}^n(\theta_b)\rightarrow \infty$; otherwise, the existence of $(i^\prime, \theta_\bprime)$ satisfying $\limsup_{n \rightarrow \infty} nW_{i^\prime, 0}(\theta_\bprime)G_{i^\prime, n}(\theta_\bprime) <~\infty$ implies $W_{i^\prime, 0}(\theta_\bprime)G_{i^\prime, n}(\theta_\bprime) < \min_{i \neq i_0^b} W_{i, 0}(\theta_b)G_{i, n}(\theta_b)$ for all sufficiently large $n$, and this makes $n \in \mathcal{N}(\theta_b)$ hold for only finitely many $n$, which contradicts $N_{\text{tot}}^n(\theta_b) \rightarrow \infty$. Observe that in order to have $nG_{i^\prime, n}(\theta_\bprime) \rightarrow \infty$ a.s., $N_{i^\prime}^n(\theta_\bprime)$ must diverge as seen in~\eqref{eq:LDR.pairwise}. {\qed~(Step~1)} 


\vspace{12pt}

Since $N^n_{i}(\theta) \rightarrow \infty$ for any $(i, \theta_b)$, the SLLN implies that we can characterize all conditional optima for sufficiently large $n$. That is, we always have $i^b_n =i^b$, $i^*_n = i^*$, and $W_{i, n}(\theta_b) = W_i(\theta_b)$ for sufficiently large $n$. Thus, we drop $n$ from these quantities in the remainder of the proof. 

Note that for any sequence $\lt\{a_n\rt\}_{n \in \mathbb{N}}$ satisfying $\limsup_{n \rightarrow \infty} a_n = a$ with $a \in \Real \cup \lt\{\infty \rt\}$, there exists subsequence $\lt\{m_\ell\rt\}_{\ell \in \mathbb{N}}$ such that $\lim_{\ell \rightarrow \infty} a_{m_\ell} = a$. In the following, we repeatedly make this \emph{convergent subsequence argument} to verify the limit results.  

\noindent\textbf{Step 2.} For each $b$, $\limsup_{n \rightarrow \infty}{N^n_i(\theta_b)}/{N^n_j(\theta_b)} < \infty$ holds for all $i \neq j$. 

First, let us consider when $j = i^b$ and $i \neq i^b$. Suppose $\limsup_{n \rightarrow \infty}{N^n_i(\theta_b)}/{N^n_{i^b}(\theta_b)} = \infty$ by means of contradiction. Then, there exists subsequence $\lt\{n_\ell\rt\}_{\ell \in \mathbb{N}}$ such that
$\lim_{\ell \rightarrow \infty}{N^{n_\ell}_i(\theta_b)}/{N^{n_\ell}_{i^b}(\theta_b)} = \infty$. We further define another subsequence $\lt\{m_\ell\rt\}_{\ell \in \mathbb{N}}$ such that
$m_\ell := \sup\lt\{m \leq n_\ell : m \in \mathcal{N}_i(\theta_b)\rt\}$.
Note that $\lt\{m_\ell\rt\}_{\ell \in \mathbb{N}}$ is nondecreasing and $m_\ell \rightarrow \infty$ since $N^n_i(\theta_b) \rightarrow \infty$.
We also have $\lim_{\ell \rightarrow \infty}{N^{m_\ell}_i(\theta_b)}/{N^{m_\ell}_{i^b}(\theta_b)} \geq \lim_{\ell \rightarrow \infty}{N^{n_\ell}_i(\theta_b)}/{N^{n_\ell}_{i^b}(\theta_b)} = \infty$ since $N_{i^b}^{n_\ell}(\theta_b) \geq N_{i^b}^{m_\ell}(\theta_b)$ and $N_i^{n_\ell}(\theta_b) = N_i^{m_\ell}(\theta_b)$ from the definition of $m_\ell$.
Thus, for large $\ell$, we obtain that
\begin{equation*}
    \lt(\dfrac{N_{i^b}^{m_\ell}(\theta_b)}{\lambda_{i^b}(\theta_b)}\rt)^2 < \lt(\dfrac{N_{i}^{m_\ell}(\theta_b)}{\lambda_{i}(\theta_b)}\rt)^2 <\sum_{j \neq i^b}\lt(\dfrac{N^{m_\ell}_j(\theta_b)}{\lambda_{j}(\theta_b)}\rt)^2,
\end{equation*}
where the last inequality holds since $i \neq i^b$. 
However, Step 4 in Algorithm~\ref{alg:rate_opt} forces $(i^b,\theta_b)$ to be sampled next should $\theta_b$ is sampled at all. This contradicts that $m_\ell \in \mathcal{N}_i(\theta_b)$. Therefore, we have $\limsup_{n \rightarrow \infty}{N^n_i(\theta_b)}/{N^n_{i^b}(\theta_b)} < \infty$. 

Let us consider when $i, j \neq i^b$. Again, suppose $\limsup_{n \rightarrow \infty}{N^n_i(\theta_b)}/{N^n_{j}(\theta_b)} = \infty$ by means of contradiction. Similarly, we can take subsequence $\lt\{m_\ell\rt\}_{\ell \in \mathbb{N}} \subseteq \mathcal{N}_i(\theta_b)$ such that $\lim_{\ell \rightarrow\infty}{N^{m_\ell}_i(\theta_b)}/{N^{m_\ell}_j(\theta_b)} = \infty$. Furthermore, we have
\begin{equation}\label{eq:G_bound}
    \begin{aligned}
        nG_{i, m_\ell}(\theta_b) &= \dfrac{(\mu_{i,m_\ell}(\theta_b) - \mu_{i^b, m_\ell}(\theta_b))^2}{2\lt(\lambda^2_i(\theta_b) +\lambda^2_{i^b}(\theta_b)N^{m_\ell}_i(\theta_b)/N^{m_\ell}_{i^b}(\theta_b)\rt)} N^{m_\ell}_i(\theta_b),\\
        nG_{j,m_\ell}(\theta_b) &\leq \dfrac{(\mu_{j,m_\ell}(\theta_b) - \mu_{i^b, m_\ell}(\theta_b))^2}{2\lambda^2_j(\theta_b)} N^{m_\ell}_j(\theta_b),\\
    \end{aligned}
\end{equation}
where the inequality follows because $\lambda^2_j(\theta_b)/N^{m_\ell}_j(\theta_b) +\lambda^2_{i^b}(\theta_b)/N^{m_\ell}_{i^b}(\theta_b) \geq \lambda^2_j(\theta_b)/N^{m_\ell}_j(\theta_b)$ holds.
Combining~\eqref{eq:G_bound} with the product rule of limit superior, we can derive that \small
\begin{eqnarray}
    \limsup_{\ell \rightarrow \infty}\dfrac{G_{j,m_\ell}(\theta_b)}{G_{i,m_\ell}(\theta_b)} &\leq & \frac{\limsup_{\ell \rightarrow \infty}\dfrac{(\mu_{j,m_\ell}(\theta_b) - \mu_{i^b, m_\ell}(\theta_b))^2}{2\lambda^2_j(\theta_b)} }{\liminf_{\ell \rightarrow \infty}\dfrac{(\mu_{i,m_\ell}(\theta_b) - \mu_{i^b, m_\ell}(\theta_b))^2}{2\lt(\lambda^2_i(\theta_b) +\lambda^2_{i^b}(\theta_b)N^{m_\ell}_i(\theta_b)/N^{m_\ell}_{i^b}(\theta_b)\rt)}} \limsup_{\ell \rightarrow \infty} \dfrac{N^{m_\ell}_j(\theta_b)}{N^{m_\ell}_i(\theta_b)} \label{eq:breaking_liminf}\\
    & =& C_1\limsup_{\ell \rightarrow \infty} \dfrac{N^{m_\ell}_j(\theta_b)}{N^{m_\ell}_i(\theta_b)} = 0, \nonumber 
\end{eqnarray} \normalsize
where $C_1$ is a finite constant obtained by computing the limit. Note that $C_1$ is positive; the liminf term in~\eqref{eq:breaking_liminf} is positive since $\limsup_{\ell \rightarrow \infty} N^{m_\ell}_i(\theta_b)/N^{m_\ell}_{i^b}(\theta_b) < \infty$ by the first case and the other terms are also positive and finite by the SLLN. Hence, we get $\lim_{\ell \rightarrow \infty}{G_{j,m_\ell}(\theta_b)}/{G_{i,m_\ell}(\theta_b)} = 0$. Then, for sufficiently large $\ell$, $W_{j}(\theta_b)G_{j,m_\ell}(\theta_b) < W_i(\theta_b)G_{i, m_\ell}(\theta_b)$ holds and it implies that $m_\ell \notin \mathcal{N}_i(\theta_b)$, which contradicts the definition of $\lt\{m_\ell\rt\}_{\ell \in \mathbb{N}}$. Thus, we obtain $\limsup_{n \rightarrow \infty}{N^n_i(\theta_b)}/{N^n_{j}(\theta_b)} < \infty$ for all $i, j \neq i^b$. 

Finally, it only remains to show the case when $i = i^b$ and $j\neq i^b$. We argue again by contradiction; suppose $\limsup_{n \rightarrow \infty}N^n_{i^b}(\theta_b)/{N^n_{j}(\theta_b)} = \infty$. Proceeding similarly, we can take subsequence $\lt\{m_\ell\rt\}_{\ell \in \mathbb{N}} \subseteq \mathcal{N}_{i^b}(\theta_b)$ such that $\lim_{\ell \rightarrow \infty}N^{m_\ell}_{i^b}(\theta_b)/N^{m_\ell}_{j}(\theta_b) = \infty$. For any 
$i^\prime, j^\prime \neq i^b$,
we have shown that $\limsup_{n \rightarrow \infty}N^n_{i^\prime}(\theta_b)/N^n_{j^\prime}(\theta_b) < \infty$. From this, we obtain $\limsup_{\ell \rightarrow \infty}{N^{m_\ell}_{i^b}(\theta_b)}/{N^{m_\ell}_{j^\prime}(\theta_b)} = \infty$ for all $j^\prime \neq i^b$. Repeating the same logic, we can take subsequence $\lt\{m^\prime_\ell\rt\}_{\ell \in \mathbb{N}} \subseteq \mathcal{N}_{i^b}(\theta_b)$ satisfying $\lim_{\ell \rightarrow \infty}N^{m^\prime_\ell}_{i^b}(\theta_b)/N^{m^\prime_\ell}_{j^\prime}(\theta_b) = \infty$ for any $j^\prime \neq i^b$. Then, for sufficiently large $\ell$, the global balance condition, $\lt({N_{i^b}^{m^\prime_\ell}(\theta_b)}/{\lambda_{i^b}(\theta_b)}\rt)^2 > \sum_{j^\prime \neq i^b}\lt({N^{m^\prime_\ell}_{j^\prime}(\theta_b)}/{\lambda_{j^\prime}(\theta_b)}\rt)^2$, always holds. Then, from Step 4 of Algorithm~\ref{alg:rate_opt}, $m^\prime_\ell \notin \mathcal{N}_{i^b}(\theta_b)$, which contradicts the definition of $m^\prime_\ell$. {\qed~(Step~2)}

\noindent\textbf{Step 3.} $\limsup_{n \rightarrow \infty}{N^n_i(\theta_b)}/{N^n_j(\theta_c)} < \infty$ holds for all $(i,\theta_b) \neq (j, \theta_c)$. 

From Step 2, we have $N^n_i(\theta_b) = \Tas(N^n_{\tot}(\theta_b))$. Hence, it suffices to show 
$\limsup_{n\rightarrow \infty}{N^n_{\tot}(\theta_b)}/{N^n_{\tot}(\theta_c)} < \infty$. Suppose otherwise. Then, we have subsequence $\lt\{m_\ell\rt\}_{\ell \in \mathbb{N}} \subseteq \mathcal{N}(\theta_b)$ such that $\lim_{\ell\rightarrow \infty}N^{m_\ell}_{\tot}(\theta_b)/N^{m_\ell}_{\tot}(\theta_c) = \infty$. Step 2 implies that $m_\ell G_{i,m_\ell}(\theta_{b}) = \Tas(N^{m_\ell}_{\tot}(\theta_{b}))$
for all $1\leq i \leq k, 1\leq b \leq B$. As a byproduct, we derive 
\begin{equation}\label{eq:between.theta}
    \frac{\min_{1\leq i\leq k, i \neq i^c}G_{i,m_\ell}(\theta_c)}{\min_{1\leq i\leq k, i \neq i^b}G_{i,m_\ell}(\theta_b)} = \frac{\Tas(N^{m_\ell}_{\tot}(\theta_c))}{\Tas(N^{m_\ell}_{\tot}(\theta_b))} = \Tas\lt(\frac{N^{m_\ell}_{\tot}(\theta_c)}{N^{m_\ell}_{\tot}(\theta_b)}\rt) \rightarrow 0.
\end{equation}
Hence, for sufficiently large $\ell$,~\eqref{eq:between.theta} implies that $\min_{1\leq i\leq k, i \neq i^c}W_i(\theta_c)G_{i,m_\ell}(\theta_c) < \min_{1\leq i\leq k, i \neq i^b}W_i(\theta_b)G_{i,m_\ell}(\theta_b)$, that is, $m_\ell \notin \mathcal{N}(\theta_b)$, which contradicts the definition of $m_\ell$. {\qed~(Step~3)}

From Step 3, we have $N_i^n(\theta_b) = \Tas(n)$ for all $(i, \theta_b)$ and $N^n_{\tot}(\theta_b) = \Tas(n)$ for all $\theta_b$.

\noindent\textbf{Step 4.} The global balance condition holds in the limit.

In this step, we aim to show that for each $1 \leq b \leq B$,
\begin{equation}\label{eq:Delta_global}
    \lim_{n \rightarrow \infty} \Delta_{n}(\theta_b)  = 0,
\end{equation}
where auxiliary variable $\Delta_n(\theta_b)$ is defined as
\begin{equation*}
    \begin{aligned}
    \Delta_n(\theta_b) : = \lt(\frac{n}{N^n_{\tot}(\theta_b)}\rt)^2\lt\{\lt(\dfrac{\alpha_{i^b,n}(\theta_b)}{\lambda_{i^b}(\theta_b)}\rt)^2  - \sum_{\ell \neq i^b}\lt(\dfrac{\alpha_{\ell,n}(\theta_b)}{\lambda_\ell(\theta_b)}\rt)^2\rt\}.
    \end{aligned}
\end{equation*}
Combining with $N^n_{\tot}(\theta_b) = \Tas(n)$, \eqref{eq:Delta_global} implies that $$\lim\nolimits_{n \rightarrow \infty}\lt\{\lt({\alpha_{i^b,n}(\theta_b)}/{\lambda_{i^b}(\theta_b)}\rt)^2  - \sum_{\ell \neq i^b}\lt({\alpha_{\ell,n}(\theta_b)}/{\lambda_\ell(\theta_b)}\rt)^2\rt\} =~0.$$ To show~\eqref{eq:Delta_global}, we extend Lemma 4.1 in \cite{chen2019complete} in Lemma~\ref{lem:modfied_lemma} below. The  proof of Lemma~\ref{lem:modfied_lemma} can be found in Section~\ref{apdx:aux_proofs}.


\begin{lemma}[Extension of Lemma 4.1 in \cite{chen2019complete}]\label{lem:modfied_lemma}
We have
\begin{equation}
    \Delta_{n+1}(\theta_b) - \Delta_n(\theta_b)  \begin{cases} > 0, & \mbox{ if $n \in \mathcal{N}_{i^b}(\theta_b)$,} \\
    < 0, & \mbox{ if $n \in \mathcal{N}(\theta_b)\setminus \mathcal{N}_{i^b}(\theta_b)$,} \\
    = 0, & \mbox{ otherwise.}\end{cases}
\end{equation}
\end{lemma}
\vspace{11pt}

Let $\lambda_{\min} = \min_{1\leq i\leq k}\lambda_i(\theta_b)$. For arbitrary $\epsilon>0$, there exists $n_1$ such that $N^n_{\tot}(\theta_b) > 2/\lambda_{\min}^2\epsilon - 1$ for all $n \geq n_1$. If $n \in \mathcal{N}_{i^b}(\theta_b)$ and $n \geq n_1$, then $\Delta_n(\theta_b) < 0$ and we get
\begin{eqnarray}
        \Delta_{n+1}(\theta_b) &= &\lt(\dfrac{N^n_{i^b}(\theta_b)+1}{(N^n_{\tot}(\theta_b)+1)\lambda_{i^b}(\theta_b)}\rt)^2 -\sum_{j \neq i^b}\lt(\dfrac{N^n_{j}(\theta_b)}{(N^n_{\tot}(\theta_b)+1)\lambda_j(\theta_b)}\rt)^2 \nonumber \\
        & = &\lt(\dfrac{N^n_{i^b}(\theta_b)+1}{(N^n_{\tot}(\theta_b)+1)\lambda_{i^b}(\theta_b)}\rt)^2 - \lt(\dfrac{N^n_{i^b}(\theta_b)}{(N^n_{\tot}(\theta_b)+1)\lambda_{i^b}(\theta_b)}\rt)^2 + \underbrace{\lt(\frac{N^n_\tot(\theta_b)}{N^n_\tot(\theta_b) + 1}\rt)^2 \Delta_n(\theta_b)}_{\text{$\leq 0$}} \nonumber\\
        & \leq & \dfrac{2N^n_{i^b}(\theta_b) + 1}{(N^n_{\tot}(\theta_b)+1)^2\lambda^2_{i^b}(\theta_b)} \leq \dfrac{2}{(N^n_{\tot}(\theta_b)+1) \lambda^2_{\min}} < \epsilon.\label{eq:Delta_upper}
\end{eqnarray}
Similarly, when $n \in \mathcal{N}(\theta_b)\setminus\mathcal{N}_{i^b}(\theta_b)$ and $n \geq n_1$, we can verify that $\Delta_{n+1}(\theta_b) > - \epsilon$. 

We argue there exists $n_2$ $(\geq n_1)$ such that $-\epsilon < \Delta_{n_2}(\theta_b) < \epsilon$. To see this, suppose $\Delta_{n_1}(\theta_b) < 0$. Selecting $n_2 = \inf\lt\{n \geq n_1 : \Delta_n(\theta_b) \geq 0\rt\}$, we have $n_2 -1 \in \mathcal{N}_{i^b}(\theta_b)$ as $\Delta_{n_2-1}(\theta_b) <0$ and $\Delta_{n_2}(\theta_b) > 0$. Hence, \eqref{eq:Delta_upper} shows that $0< \Delta_{n_2}(\theta_b) < \epsilon$. A similar argument can be made when $\Delta_{n_1}(\theta_b) \geq 0$ to show $-\epsilon<\Delta_{n_2}(\theta_b)<0$ by choosing $n_2 = \inf\lt\{n \geq n_1 : \Delta_n(\theta_b) < 0\rt\}$.

If $-\epsilon<\Delta_{n_2}(\theta_b) < 0$, then $n_2 \in \mathcal{N}_{i^b}(\theta_b)$ or $n_2 \in \mathcal{N}(\theta_c)$ for some $\theta_c \neq \theta_b$ since $\Delta_{n_2}(\theta_b) < 0$ implies that $(i^b, \theta_b)$ should be sampled at $n_2$ if $n_2 \in \mathcal{N}(\theta_b)$. When $n_ 2\in \mathcal{N}_{i^b}(\theta_b)$, Lemma~\ref{lem:modfied_lemma} and~\eqref{eq:Delta_upper} imply that we obtain $-\epsilon < \Delta_{n_2}(\theta_b) < \Delta_{n_2+1}(\theta_b) < \epsilon$. When $n_2 \in \mathcal{N}(\theta_c)$ for some $\theta_c \neq \theta_b$, Lemma~\ref{lem:modfied_lemma} shows $-\epsilon < \Delta_{n_2}(\theta_b) = \Delta_{n_2+1}(\theta_b) < 0$. In other words, $\Delta_{n_2+1}(\theta_b) \in (-\epsilon, \epsilon)$ always holds. We can show an analogous result for when $0 < \Delta_{n_2}(\theta_b) <\epsilon$. Hence, we obtain $\Delta_n(\theta_b) \in (-\epsilon, \epsilon)$ for all $n \geq n_2$ by induction, which implies $\lim_{n \rightarrow \infty} \Delta_n(\theta_b) = 0$. {\qed~(Step~4)}


\noindent\textbf{Step 5.} The pairwise balance condition holds in the limit.

\noindent \textbf{Remark:} This proof is applicable for more general algorithms with positive balance weights not limited to Algorithms~3 and~4. For example, notice that our sampling rule is identical to the C-OCBA discussed in~\cite{gao2019selecting} if $W_i(\theta_b) = 1$ for all $(i, \theta_b)\in \Xi\cup\Xi^{\text{adv}}$. Theorem 2 in~\cite{gao2019selecting} states that the sampling ratios from their plug-in algorithm achieves convergence of global and pairwise balance conditions. However, this is stated without a proof (only a brief sketch of proof is provided). Our proof of Step~5 presented below also proves Theorem 2 in~\cite{gao2019selecting} by simply setting $W_i(\theta_b) = 1$ for all $(i, \theta_b)\in \Xi\cup\Xi^{\text{adv}}$, which involves expending the technical results in \cite{chen2022BOLD, chen2019complete}.

First, we aim to show that
\begin{equation}\label{eq:limsup_rn}
    \limsup_{n \rightarrow \infty} r_n(\theta_b, \theta_{b^\prime}) \leq 1, \;\; \text{ for all $\bprime \neq b$,}
\end{equation}
where $r_n(\theta_b, \theta_{\bprime}) : = {\min_{i \neq i^b}W_{i}(\theta_b)G_{i, n}(\theta_b)}/{\min_{j \neq i^{\bprime}}W_j(\theta_\bprime)G_{j,n}(\theta_{\bprime})}$, which is the ratio between the smallest LDR functions of $\theta_b$ and $\theta_{\bprime}$. For convenience, we define $\Gamma_{i, n}(\theta_b) = nW_i(\theta_b)G_{i, n}(\theta_b)$; note that $\Gamma_{i, n}(\theta_b) = \Tas(n)$ as $0 < \liminf_{n \rightarrow \infty} G_{i, n}(\theta_b) \leq \limsup_{n \rightarrow \infty} G_{i, n}(\theta_b) < \infty$. Once~\eqref{eq:limsup_rn} holds, we further have $\liminf_{n \rightarrow \infty} r_n(\theta_b, \theta_{\bprime}) = 1/\limsup_{n \rightarrow \infty} r_n(\theta_{\bprime}, \theta_b) \geq 1$ by symmetry, which in turn implies $\lim_{n \rightarrow \infty} r_n(\theta_b, \theta_\bprime) = 1$ for all $b \neq \bprime$ and shows the pairwise balance condition holds.

To show~\eqref{eq:limsup_rn}, the following two lemmas are introduced. They provide uniform upper bounds on the increment of scaled LDR, $\Gamma_{i,n}(\theta_b)$, and the number of replications allocated to each input parameter, respectively. Their proofs are presented in Section~\ref{apdx:aux_proofs}. To state these lemmas, suppose $n \in \mathcal{N}(\theta_b)$ and define $m_n(\theta_b) := \inf\lt\{\ell > 0 : n + \ell \in \mathcal{N}(\theta_b) \rt\}$, the number of replications until the next time algorithm allocates a replication to any of $\lt\{(i, \theta_b)\rt\}_{1\leq i \leq k}$.

\begin{lemma}\label{lem:LIL_LDR}
$\lt|\min_{i \neq i^b} \Gamma_{i, n+m_n(\theta_b)}(\theta_b) - \min_{i \neq i^b} \Gamma_{i, n}(\theta_b)\rt| = \Oas(\sqrt{n\log\log n})$ for $n \in \mathcal{N}(\theta_b)$.
\end{lemma}

\begin{lemma}\label{lem:between_sample}
Given $n\in\mathcal{N}(\theta_b)$, for each $\bprime \neq b$, the number of samples allocated to $\theta_\bprime$ is bounded as
$     \lt|\mathcal{N}(\theta_\bprime) \cap (n, n+m_n(\theta_b))\rt| = \Oas(\sqrt{n\log\log n}).
$
\end{lemma}

\noindent Because Lemma~\ref{lem:between_sample} holds for all $\bprime \neq b$, it implies that $m_n(\theta_b) = \Oas(\sqrt{n\log\log n})$. From this, for each $0 \leq t \leq m_n(\theta_b)$, we have $\Gamma_{i, n+t}(\theta_\bprime) = \Tas(n + t) = \Tas(n)$ and $\min_{i \neq i^\bprime}\Gamma_{i, n+t}(\theta_\bprime) = \Tas(n)$. With these observations and Lemma~\ref{lem:LIL_LDR}, we further obtain  
\begin{equation*}
    \begin{aligned}
        & r_{n+t}(\theta_b,\theta_\bprime) - r_n(\theta_b, \theta_\bprime)\\
        &= \frac{\min_{i \neq i^b}\Gamma_{i, n+t}(\theta_b)}{\min_{i \neq i^{\bprime}}\Gamma_{i,n+t}(\theta_{\bprime})} - \frac{\min_{i \neq i^b}\Gamma_{i, n}(\theta_b)}{\min_{i \neq i^{\bprime}}\Gamma_{i,n}(\theta_{\bprime})} \\
        & = \frac{\min_{i \neq i^b}\Gamma_{i, n+t}(\theta_b)\min_{i \neq i^{\bprime}}\Gamma_{i,n}(\theta_{\bprime}) - \min_{i \neq i^{\bprime}}\Gamma_{i,n+t}(\theta_{\bprime})\min_{i \neq i^b}\Gamma_{i, n}(\theta_b)}{\min_{i \neq i^{\bprime}}\Gamma_{i,n+t}(\theta_{\bprime})\min_{i \neq i^{\bprime}}\Gamma_{i,n}(\theta_{\bprime})} \\
        & \leq \frac{\lt|\min_{i \neq i^b}\Gamma_{i, n+t}(\theta_b) - \min_{i \neq i^b}\Gamma_{i, n}(\theta_b)\rt|\min_{i \neq i^{\bprime}}\Gamma_{i,n}(\theta_{\bprime})}{\min_{i \neq i^{\bprime}}\Gamma_{i,n+t}(\theta_{\bprime})\min_{i \neq i^{\bprime}}\Gamma_{i,n}(\theta_{\bprime})} \\
        & \quad \quad +\frac{ \lt| \min_{i \neq i^{\bprime}}\Gamma_{i,n+t}(\theta_{\bprime}) - \min_{i \neq i^{\bprime}}\Gamma_{i,n}(\theta_{\bprime})\rt| \min_{i \neq i^b}\Gamma_{i, n}(\theta_b)}{\min_{i \neq i^{\bprime}}\Gamma_{i,n+t}(\theta_{\bprime})\min_{i \neq i^{\bprime}}\Gamma_{i,n}(\theta_{\bprime})}\\
        &\leq \frac{\Oas(\sqrt{n\log\log n})\Tas(n) + \lt|\min_{i \neq i^{\bprime}}\Gamma_{i,n+t}(\theta_{\bprime}) - \min_{i \neq i^{\bprime}}\Gamma_{i,n}(\theta_{\bprime})\rt| \Tas(n)}{\Tas(n)\Tas(n)} \\
        & = \Oas\lt(\sqrt{\frac{\log\log n}{n}}\rt) + \frac{\lt|\min_{i \neq i^{\bprime}}\Gamma_{i,n+t}(\theta_{\bprime}) - \min_{i \neq i^{\bprime}}\Gamma_{i,n}(\theta_{\bprime})\rt|}{\Tas(n)}. 
    \end{aligned}
\end{equation*}
To bound the second term, we need the following intermediate result.
\begin{lemma}\label{lem:sub_rn}
Given  $n\in\mathcal{N}(\theta_b),$ we have
\begin{equation}\label{eq:sub_rn}
    \sup_{0 \leq t \leq m_n(\theta_b)}\lt|\min_{i \neq i^{\bprime}}\Gamma_{i,n+t}(\theta_{\bprime}) - \min_{i \neq i^{\bprime}}\Gamma_{i,n}(\theta_{\bprime})\rt| \leq \Oas(\sqrt{n\log\log n}).
\end{equation} 
\end{lemma}
Then, combining $r_n(\theta_b, \theta_\bprime)\leq 1$ (because $n \in \mathcal{N}(\theta_b)$) and~\eqref{eq:sub_rn}, we can obtain that
\begin{equation*}
    \sup_{0 \leq t\leq m_n(\theta_b)} r_{n+t}(\theta_b, \theta_\bprime) \leq r_n(\theta_b, \theta_\bprime) + \Oas\lt(\sqrt{\frac{\log\log n}{n}}\rt) \leq 1 + \Oas\lt(\sqrt{\frac{\log\log n}{n}}\rt).
\end{equation*}
With $m_n(\theta_b) = \Oas(\sqrt{n\log\log n})$, the above inequality shows $\limsup_{n \rightarrow \infty} r_n(\theta_b, \theta_\bprime) \leq 1$. Hence it suffices to show Lemma~\eqref{lem:sub_rn} holds, which is provided in Appendix~\ref{apdx:aux_proofs}.  

Finally, it only remains to show the pairwise balance condition holds for $(i, \theta_b)$ and $(i^\prime, \theta_b)$  such that $i \neq i^\prime$ and $i, i^\prime \neq i^b$. For fixed $\theta_b$, each inner-R\&S problem allocates replications to the $k$ solutions at the iterations in $\mathcal{N}(\theta_b)$ only, which is guided by the weighted LDR functions. When $W_i(\theta_b) = 1$ for all $i \neq i^b$, this result is a special case of Theorem 4 in \cite{chen2022BOLD} when the sampling distributions are normal. The following corollary states that the same conclusion holds even if $\{W_i(\theta_b)\}_{i\neq i^b}$ are unequal. 

\begin{corollary}[Extended Theorem 4 in~\cite{chen2022BOLD}]\label{cor:extended_chen}
For each $\theta_b$, assume we allocate simulation budget with the balance weight $\{W_i(\theta_b)\}$. Then, we have $\lim_{n \rightarrow \infty}\frac{W_{i}(\theta_b)G_{i, n}(\theta_b)}{W_{j}(\theta_b)G_{j, n}(\theta_b)} = 1$ for all $i, j \neq i^b$. 
\end{corollary}
\proof{Proof.} See \cite{chen2022BOLD} for the complete proof. We only highlight changes to be made to incorporate the balance weights.
As done for $r_n(\theta_b, \theta_\bprime)$ above, Theorem 4 of~\cite{chen2022BOLD} showed that $\liminf_{n \rightarrow \infty}r_n(i, j) \geq~1$ holds where $r_n(i, j) = \frac{G_{i, n}(\theta_b)}{G_{j, n}(\theta_b)}$ for $i, j \neq i^b$. In this corollary, we extend the definition of $r_n(i, j)$ as $r_n(i, j) := \frac{W_i(\theta_b)G_{i, n}(\theta_b)}{W_j(\theta_b)G_{j, n}(\theta_b)}$. The original proof is derived from Lemmas 7 and 8 in the same paper, which provide the upper bounds on the number of replications required until the algorithm selects the same alternative to sample next. Similar to Lemma~\ref{lem:between_sample}, those bounds depend on the increment of $nG_{i, n}(\theta_b)$. Since  $W_i(\theta_b)$'s are constant, the bound is preserved since the increments of $nG_{i, n}(\theta_b)$ and $nW_i(\theta_b)G_{i, n}(\theta_b)$ are equal up to a constant. Hence, we can see that the desired result also holds even when we have unequal balance weights. \halmos
\endproof
\vspace{12pt}
Combining Corollary~\ref{cor:extended_chen} with $\lim_{n\rightarrow \infty} r_n(\theta_b, \theta_\bprime) = 1$, we have $\lim_{n \rightarrow \infty}\frac{W_{i}(\theta_b)G_{i, n}(\theta_b)}{W_{j}(\theta_\bprime)G_{j, n}(\theta_\bprime)} = 1$ for $(i, \theta_b) \neq (j, \theta_\bprime) \in \Xi \cup \Xi^{\text{adv}}$. {\qed~(Step~5)}  \halmos


\endproof

\vspace{12pt}
Two variants of the Borel-Cantelli lemma are stated below, which are needed to show Theorem~\ref{thm:proof_EIalg}.




\begin{lemma}[Second Borel-Cantelli Lemma II, Theorem 4.3.4 in~\cite{durrett2019probability}]\label{lem:borel.cantelli} Let $\mathcal{F}_n, n \geq 0$ be a filtration with $\mathcal{F}_0 = \{\emptyset, \Omega\}$ and let $B_n, n\geq 1$ a sequence of events with $B_n \in \mathcal{F}_n$. Then,
\begin{equation*}
    \{B_n \;\; i.o.\} = \left\{\sum_{n=1}^{\infty} \prob(B_n|\mathcal{F}_{n-1}) = \infty\right\}.
\end{equation*}
\end{lemma}

\begin{lemma}[Second Borel-Cantelli Lemma III, Theorem 4.5.5 in~\cite{durrett2019probability}]\label{lem:borel.cantelli.3} Let $B_n \in \mathcal{F}_n, n \geq 0$ be a filtration with $\mathcal{F}_0 = \{\emptyset, \Omega\}$ and $p_n = \prob(B_n|\mathcal{F}_{n-1})$. Then,
\begin{equation*}
    \dfrac{\sum_{m=1}^{n} \mathbf{1}_{B_m}}{\sum_{m=1}^n p_m} \rightarrow 1 \;\; a.s. \;\text{conditional on} \;\; \left\{\sum_{m=1}^\infty p_m = \infty\right\}.
\end{equation*}
\end{lemma}


\proof{Proof of Theorem~\ref{thm:proof_EIalg}.} 
Recall that in Step~\ref{step:posterior} of Algorithm~\ref{alg:rate_opt_posterior}, for all $\theta_b \in (\Theta_n^*)^c$, $\mu_{i_n^*, n}(\theta_b)$ is replaced with $\bar{\mu}_{i_n^*, n}(\theta_b)$ sampled from the posterior and $i_n^b$ is updated accordingly. In this proof, we denote the updated version by $\bar{i}_n^b$ for all $\theta_b$ (including those in $\Theta_n^*$) to avoid having to specify whether $\theta \in \Theta_n^*$ or not. Recall that $i_n^*$ is unchanged even if we adopt $\{\bar{i}_n^b\}$. Let $\bar{\Theta}^*_n = \lt\{\theta_b : \bar{i}_n^b = i_n^*\rt\}$ be the estimated favorable set of $i^*_n$, $\bar{\Xi}_n = \{(i, \theta_b): i \neq \bar{i}_n^b, i \neq i_n^*\}$, and $\bar{\Xi}_n^{\text{adv}} = \{(i_n^*, \theta_b): \theta_b \in (\bar{\Theta}_n^*)^c\}$. Then, $\Xi_n \subseteq \bar{\Xi}_n$ and $\bar{\Xi}^{\text{adv}}_n \subseteq \Xi^\text{adv}_n$ hold. 


Let us fix a sample path. Recall that $\{i_0^b\},\Xi_0,$ and $\Xi_0^\text{adv}$ are defined for each sample path. As done in the proof of Theorem~\ref{thm:proof_BWalg}, we have that $i_n^b = i_0^b, \Xi_n = \Xi_0$, and $\Xi_n^{\text{adv}} = \Xi_0^{\text{adv}}$ hold for all sufficiently large $n$. Recall that we cannot guarantee $i_0^b = i^b$, $\Xi_0 = \Xi$, and $\Xi_0^\text{adv} = \Xi^{\text{adv}}$ if there exists $(i, \theta_b)$ such that $\lim_{n\rightarrow \infty}N_i^n(\theta_b) < \infty$; however, for every sample path, we can fix those quantities. Consequently, the MPB and its favorable set can be fixed; we denote them by $i_0^*$ and $\Theta_0^*$, respectively. Recall that the fixed quantities could be different with the ground-truth depending a simulation-output sample path if there exists $(i, \theta_b)$ such that $\lim_{n \rightarrow \infty} N_i^{n}(\theta_b) < \infty$. 

The $n$th iteration's posterior sampling distribution depends on the filtration generated by all historical simulation outputs collected up to the $(n-1)$th iteration,  $\mathcal{F}_{n-1}$. Let  $\{B_n\}$ represent the sequence of events that depend on $\{\mathcal{F}_{n-1}\}$ and the posterior sampling at the $n$th iteration for which we aim to show that they happen infinitely often. Lemma~\ref{lem:borel.cantelli} provides a sufficient condition: 
 $\lim_{n \rightarrow \infty} \prob(B_n|\mathcal{F}_{n-1}) > 0$. Hence, our strategy is to show this  condition holds for various $B_n$'s we define.

We proceed by proving Claims (I)--(III) made below, which together complete the proof. In most of our arguments, we fix the sample path to show almost sure convergence unless otherwise mentioned.

\noindent{\bf (I).} \textit{$N^n_i(\theta_b) \rightarrow \infty$ for all $1 \leq i\leq k$ and $1\leq b \leq B$.}

\textbf{Step 1.} $N^n_i(\theta_b) \rightarrow \infty$ holds for all $(i, \theta_b) \in \lt(\Xi_0^{\text{adv}}\rt)^c$. 

Observe that for each $\theta_b \in \lt(\Theta_0^*\rt)^c$, the probability of the posterior sample does not beat $i_0^b$ is always strictly positive since $\prob(\bar{i}_n^b = i_0^b|\mathcal{F}_{n-1}) = \prob(\bar{\mu}_{i_0^*, n}(\theta_b) > {\mu}_{i_0^b, n}(\theta_b)|\mathcal{F}_{n-1}) \geq 1/2$. 
We have $\prob(\bar{i}_n^b = i_0^b, \; \forall \theta_b \in \lt(\Theta_0^*\rt)^c|\mathcal{F}_{n-1}) \geq (1/2)^{\lt|\lt(\Theta_0^*\rt)^c\rt|} >0$. Hence, Lemma~\ref{lem:borel.cantelli} implies $|\mathcal{J}| = \infty$, where $\mathcal{J}:= \{n|\bar{i}_n^b = i_0^b, \;\forall \theta_b \in \lt(\Theta_0^*\rt)^c\}$. That is, there are infinitely many iterations at which the conditional optima are unchanged after posterior sampling. The sampling decision made at $n\in \mathcal{J}$ would choose pair $(i, \theta_b) \in \lt(\Xi_0^{\text{adv}}\rt)^c$ by the construction of Algorithm~2. Consequently, we can conclude that the total simulation budget allocated to all pairs in $\lt(\Xi_0^{\text{adv}}\rt)^c$ is infinite. Restricting our attention on $n \in \mathcal{J}$, we can see that $N_i^n(\theta_b) \rightarrow \infty$ holds for all $(i, \theta_b) \in \lt(\Xi_0^{\text{adv}}\rt)^c$ in a way similar to Step 1 of the proof of Theorem~6. {\qed~(Step~1)} 

\textbf{Step 2.} $N^n_{i_0^*}(\theta_b) \rightarrow \infty$ holds for all $(i_0^*, \theta_b) \in \Xi_0^{\text{adv}}$.

Suppose otherwise. Then, there exists $\theta_b \in \lt(\Theta^*_0\rt)^c$ such that $\lim_{n \rightarrow \infty}N_{i_0^*}^n(\theta_b) < \infty$. Observe that an event, where $\bar{i}_n^b = i_0^*$ with $\bar{\mu}_{i^*_0,n}(\theta_b) > \mu_{i_0^b, n}(\theta_b) - \epsilon$ for some $\epsilon >0$ and $\bar{i}_n^\bprime = i_0^\bprime$ for all $\theta_\bprime \in (\Theta_0^*)^c$ with $\bprime \neq b$, has positive probability in the limit:
\begin{equation*}
    \begin{aligned}
    &\prob(\bar{\mu}_{i^*_0,n}(\theta_b) > \mu_{i_0^b, 0}(\theta_b) - \epsilon, \bar{i}_n^b = i_0^*, \bar{i}_n^\bprime = i_0^\bprime, \;\forall \theta_\bprime(\neq\theta_b) \in \lt(\Theta_0^*\rt)^c|\mathcal{F}_{n-1})\\
    &=  \prob(- \epsilon < \bar{\mu}_{i_0^*, n}(\theta_b) - \mu_{i_0^b, n}(\theta_b) < 0|\mathcal{F}_{n-1})\prod\nolimits_{\bprime \neq b, \theta_\bprime \in \lt(\Theta_0^*\rt)^c}\prob(\bar{\mu}_{i_0^*, n}(\theta_\bprime) > \mu_{i_0^\bprime, n}(\theta_\bprime)|\mathcal{F}_{n-1}) \\
    &\geq \lt(\Phi\lt(-\sqrt{N_{i_0^*}^n(\theta_b)}\frac{\mu_{i_0^*, n}(\theta_b) - \mu_{i_0^b, n}(\theta_b)}{\lambda_{i_0^*}(\theta_b)}\rt) - \Phi\lt(-\sqrt{N_{i_0^*}^n(\theta_b)}\frac{\mu_{i_0^*, n}(\theta_b) - \mu_{i_0^b, n}(\theta_b) + \epsilon}{\lambda_{i_0^*}(\theta_b)}\rt)\rt)\lt(1/2\rt)^{|(\Theta^*_0)^c| - 1}.
    \end{aligned}
\end{equation*}
Since $N_{i_0^*}^n(\theta_b)$ is finite by the assumption, the lower bound above is strictly positive. Hence, there exist infinitely many $n$ such that $\bar{i}_n^b = i_0^*$ and $\bar{i}_n^\bprime = i_0^\bprime$ for all $\theta_\bprime \in \lt(\Theta^*_0\rt)^c$. 
Denote the set of such $n$ by $\mathcal{J}_1$. For each $n \in \mathcal{J}_1$, we have for all $j \neq i_0^*$,
\begin{equation*}
    nG_{j, n}(\theta_b) = \dfrac{(\bar{\mu}_{i_0^*, n}(\theta_b) - \mu_{j, n}(\theta_b))^2}{2\lt({\lambda^2_{i_0^*}(\theta_b)}/{N^n_{i_0^*}(\theta_b)} + {\lambda^2_{j}(\theta_b)}/{N^n_{j}(\theta_b)}\rt)}\leq \dfrac{(\mu_{j, n}(\theta_b) - \mu_{i_0^b, n}(\theta_b) + \epsilon)^2}{2\lambda^2_{i_0^*}(\theta_b)/N^n_{i_0^*}(\theta_b)} < \infty,
\end{equation*}
where the first inequality comes from $0 \leq \mu_{j, n}(\theta_b) - \bar{\mu}_{i_0^*, n}(\theta_b) \leq \mu_{j, n}(\theta_b) - \mu_{i_0^b, n}(\theta_b) + \epsilon$ and the second is straightforward since $N^n_{i_0^*}(\theta_b)$ is bounded. 

On the other hand, Step~1 implies $\min_{(i, \theta_\bprime) \in \Xi_0, \bprime \neq b} {W_{i, 0}(\theta_\bprime)G_{i, n}(\theta_\bprime)}$ diverges. Consequently, we have $n \in \mathcal{N}(\theta_b)$ for all sufficiently large $n \in \mathcal{J}_1$. Furthermore, from $\lim_{n\rightarrow \infty}N_{i^*_0}^n(\theta_b) < \infty$ and $N_{i}^n(\theta_b) \rightarrow \infty$ for all $i \neq i_0^*$, we can see that $\lt(N_{i^*_0}^n(\theta_b)/\lambda_{i^*_0}(\theta_b)\rt)^2 < \sum\nolimits_{j \neq i_0^*}\lt(N_{j}^n(\theta_b)/\lambda_{j}(\theta_b)\rt)^2$ holds for all sufficiently large $n \in\mathcal{J}_1$. As a result, for such $n$ values, we have $n \in \mathcal{N}_{i_0^*}(\theta_b)$, which implies $(i_0^*, \theta_b)$ is chosen infinitely many times. However, this is a contradiction. {\qed~(Step~2)}

\vspace{12pt}
From (I), the SLLN implies that we have $i_n^b = i^b, i_n^* = i^*, \Theta_n^* = \Theta_{i^*}$, and $\Xi_n^{\text{adv}}~=~\Xi^{\text{adv}}$ for all sufficiently large $n$. However, we cannot drop $n$ from $W_{i, n}(\theta_b), \bar{i}_n^b,$ and $\bar{\Xi}_n^{\text{adv}}$ since these quantities depend on the posterior sample. 
\vspace{12pt}

\noindent{\bf (II)}. \textit{We have $\lim_{n\rightarrow \infty}|\mathcal{C}_n(\theta_b)|/N^n_{\textup{tot}}(\theta_b) = 0$ for all $\theta_b \in \Theta^c_{i^*}$, where $\mathcal{C}_n(\theta_b):= \{m\in \mathcal{N}(\theta_b)| \bar{\mu}_{i^*, m}(\theta_b) < \mu_{i^b, m}(\theta_b), m\leq n\}$.}

Let $\mathcal{C}(\theta_b) := \lim_{n\rightarrow\infty} \mathcal{C}_n(\theta_b)$. Similarly, we introduce $\mathcal{D}_n(\theta_b) := \{m\in\mathcal{N}(\theta_b):m \leq n\}\setminus \mathcal{C}_n(\theta_b)$ and $\mathcal{D}(\theta_b):=\lim_{n\rightarrow \infty} \mathcal{D}_n(\theta_b)$. Then, it suffices to show that $|\mathcal{D}_n(\theta_b)|/N_{\text{tot}}^n(\theta_b) \rightarrow~1$. For each $m \in \mathcal{N}(\theta_b)$, observe that $p_m:= \prob(m \in \mathcal{D}(\theta_b)|\mathcal{F}_{m-1}) \geq 1/2$ as $p_m = \prob(\bar{\mu}_{i^*, m}(\theta_b) \geq \mu_{i^b, m}(\theta_b)|\mathcal{F}_{m-1}) = \Phi(\sqrt{N_{i^*}^m(\theta_b)}(\mu_{i^*, m}(\theta_b) - \mu_{i^b, m}(\theta_b))/\lambda_{i^*}(\theta_b))$ and $\mu_{i^*, m}(\theta_b) - \mu_{i^b, m}(\theta_b) \geq 0$; hence, we have $\sum_{m \in \mathcal{N}(\theta_b)} p_m = \infty$ a.s.. Therefore, Lemma~\ref{lem:borel.cantelli.3} implies that $|\mathcal{D}_n(\theta_b)|/\sum_{m \in \mathcal{N}(\theta_b), m \leq n} p_m \rightarrow 1$ a.s.. Next, we show $\lim_{n \rightarrow\infty}\sum_{m \in \mathcal{N}(\theta_b), m\leq n} p_m/N^n_\text{tot}(\theta_b) \rightarrow~1$ a.s..  

To show this, we exploit the convergence result of the Cesaro mean: if there exists a sequence $\{a_n\}_{n \in \mathbb{N}}$ such that $\lim_{n \rightarrow \infty} a_n = a$ and define $b_n = \frac{1}{n}\sum_{i = 1}^n a_i$, then $\lim_{n \rightarrow \infty} b_n = a$. By treating $p_m$ and $\sum_{m \in \mathcal{N}(\theta_b), m\leq n} p_m/N^n_\text{tot}(\theta_b)$ as $a_n$ and $b_n$, we obtain $\lim_{n \rightarrow \infty}\sum_{m \in \mathcal{N}(\theta_b), m\leq n} p_m/N^n_\text{tot}(\theta_b) = \lim_{m \in \mathcal{N}(\theta_b), m \rightarrow \infty} p_m = 1$. The last equality holds since we have shown that $N^n_{i^*}(\theta_b) \rightarrow \infty$ as $n \rightarrow \infty$, which in turn implies $p_m = \prob(\bar{\mu}_{i^*, m}(\theta_b) \geq \mu_{i^b, m}(\theta_b)|\mathcal{F}_{m-1}) = \Phi(\sqrt{N_{i^*}^m(\theta_b)}(\mu_{i^*, m}(\theta_b) - \mu_{i^b, m}(\theta_b))/\lambda_{i^*}(\theta_b)) \rightarrow 1$ as $m \rightarrow \infty$.  {\qed~(II)}

\vspace{12pt}
For each $(i^*, \theta_b) \in {\Xi}^{\text{adv}}$, $\{\bar{\mu}_{i^*, n}(\theta_b) < \mu_{i^b, n}(\theta_b)\}$ must be satisfied to guarantee $n \in \mathcal{N}_{i^*}(\theta_b)$. In other words, we have $\mathcal{N}_{i^*}(\theta_b) \subseteq \mathcal{C}(\theta_b)$. From  (II), we conclude that $N_{i^*}^n(\theta_b) = o(N_{\text{tot}}^n(\theta_b)) = o(n)$ for all $\theta_b \in \Theta^c_{i^*}$. Also, we can ignore iterations in $\mathcal{C}(\theta_b)$ without affecting the asymptotic allocation  ratios and focus on $n \in \lt(\bigcup_{\theta_b \in \Theta_{i^*}^c} \mathcal{C}(\theta_b)\rt)^c$, i.e., the iterations in which the posterior sampling does not change the conditional optimum. 

Now, it remains to show that the global and pairwise balance conditions hold in the limit. The following conjecture lets us reuse the arguments in  Steps 2--5 in the proof of Theorem~\ref{thm:proof_BWalg} with minimal modifications.


{\bf (III).} For each $\theta_b \in \Theta_{i^*}^c$ and all sufficiently large $n\in \mathcal{N}(\theta_b)$, if $n \in \mathcal{N}(\theta_b)\setminus\mathcal{N}_{i^*}(\theta_b)$, then $n \in \mathcal{D}(\theta_b)$. 

We prove the complement of the claim: for each $\theta_b \in \Theta_{i_0^*}^c$ and all sufficiently large $n\in \mathcal{N}(\theta_b)$, if $n\in\mathcal{C}(\theta_b)$, then $n\in\mathcal{N}_{i^*}(\theta_b)$.
To show this, define $\widetilde{\Delta}_n(\theta_b): = (N_{\text{tot}}^n(\theta_b))^{-2}\lt[\sum_{j \neq i^*}\lt({N_j^n(\theta_b)}/{\lambda_j(\theta_b)}\rt)^2 -\lt({N_{i^*}^n(\theta_b)}/{\lambda_{i^*}(\theta_b)}\rt)^2\rt]$, which corresponds to the sample-version of the global balance condition when the posterior sample, $\bar{\mu}_{i^*, n}(\theta_b)$, beats $\mu_{i^b, n}(\theta_b)$; observe that if $\widetilde{\Delta}_n(\theta_b) > 0$ and $n \in \mathcal{C}(\theta_b)$, then $n \in \mathcal{N}_{i^*}(\theta_b)$. Thus, we aim to show that $\widetilde{\Delta}_n(\theta_b) > 0$ for all sufficiently large $n\in\mathcal{C}(\theta_b)$. Because $N^n_{i^*}(\theta_b) = o_{\text{a.s.}}(N^n_{\text{tot}}(\theta_b))$ and from the definition of $N_{\text{tot}}^n(\theta_b)$, there exists $j_1 \neq i^*$ such that $\liminf\nolimits_{n \rightarrow \infty} N_{j_1}^n(\theta_b)/N_{\text{tot}}^n(\theta_b) > 0$. Then, we have
\begin{equation*}
    \begin{aligned}
        \liminf_{n \rightarrow \infty} \widetilde{\Delta}_n(\theta_b) &= \liminf_{n \rightarrow \infty}\lt[\frac{1}{(N_{\text{tot}}^n(\theta_b))^2}\sum_{j \neq i^*}\lt(\frac{N_j^n(\theta_b)}{\lambda_j(\theta_b)}\rt)^2 - \frac{1}{\lt(N_{\text{tot}}^n(\theta_b)\rt)^2}\lt(\frac{N_{i^*}^n(\theta_b)}{\lambda_{i^*}(\theta_b)}\rt)^2\rt]\\
        & \geq \liminf_{n \rightarrow\infty}\frac{1}{\lambda^2_{j_1}(\theta_b)}\lt(\frac{N_{j_1}^n(\theta_b)}{N_{\text{tot}}^n(\theta_b)}\rt)^2 - \limsup_{n\rightarrow \infty}\frac{1}{\lambda^2_{i^*}(\theta_b)}\lt(\frac{N_{i^*}^n(\theta_b)}{N_{\text{tot}}^n(\theta_b)}\rt)^2 > 0.
    \end{aligned}
\end{equation*}
The last inequality follows since $\liminf\nolimits_{n \rightarrow \infty} N_{j_1}^n(\theta_b)/N_{\text{tot}}^n(\theta_b) > 0$ and $N^n_{i^*}(\theta_b) = o_{\text{a.s.}}(N^n_{\text{tot}}(\theta_b))$. Hence, $n \in \mathcal{C}(\theta_b)$ implies $n \in \mathcal{N}_{i^*}(\theta_b)$ for all sufficiently large $n \in \mathcal{N}(\theta_b)$. 
{\qed~(III)}
\vspace{12pt}

Claim (III) lets us apply the subsequence arguments made in  Parts 2--5 of the proof of  Theorem~\ref{thm:proof_BWalg} with slight modifications to complete the proof of Theorem~\ref{thm:proof_EIalg}. For any subsequence $\{n_\ell\}\subseteq \mathcal{N}_i(\theta_b)$ for some $(i, \theta_b) \in \lt(\Xi^{\text{adv}}\rt)^c$, we can take another subsequence $\{m_\ell\} \subseteq \{n_\ell\}$ such that $m_\ell \in \mathcal{D}(\theta_b)$ thanks to (III). 
Thus, at each $m_\ell$, posterior sampling can be ignored. 
From this observation, we are able to use the argument made in the proof of Theorem~\ref{thm:proof_BWalg} to verify $\limsup_{n \rightarrow \infty}N_i^n(\theta_b)/N_j^n(\theta_c) < \infty$ for all $(i, \theta_b), (j, \theta_c) \in \lt(\Xi^\text{adv}\rt)^c$.

Since the global balance condition is modified when $\theta_b\in \Theta_{i^*}^c$, Step~4 of the proof of Theorem~\ref{thm:proof_BWalg} needs to be modified. We re-define $\Delta_n(\theta_b)$ as follows:
\begin{equation*}
    \Delta_n(\theta_b) = \begin{cases}\lt(\dfrac{n}{N^n_{\tot}(\theta_b)}\rt)^2\lt\{\lt(\dfrac{\alpha_{i^b,n}(\theta_b)}{\lambda_{i^b}(\theta_b)}\rt)^2  - \sum_{\ell \neq i^b}\lt(\dfrac{\alpha_{\ell,n}(\theta_b)}{\lambda_\ell(\theta_b)}\rt)^2\rt\} & \mbox{if $\theta_b \in \Theta_{i^*}$,}\\
    \lt(\dfrac{n}{N^n_{\tot}(\theta_b) - N^n_{i^*}(\theta_b)}\rt)^2\lt\{\lt(\dfrac{\alpha_{i^b,n}(\theta_b)}{\lambda_{i^b}(\theta_b)}\rt)^2  - \sum_{\ell \neq i^b, i^*}\lt(\dfrac{\alpha_{\ell,n}(\theta_b)}{\lambda_\ell(\theta_b)}\rt)^2\rt\} & \mbox{if $\theta_b \in \Theta_{i^*}^c$.}
    \end{cases}
\end{equation*}
For each $\theta_b$, it suffices to show that $\lim\nolimits_{n\rightarrow \infty}\Delta_n(\theta_b) = 0$. To avoid repetition, we omit its details; one can apply the logic similar to Step~4 of the proof of Theorem~\ref{thm:proof_BWalg} using the new $\Delta_n(\theta_b)$. Furthermore, similar arguments on $r_n(\theta_b, \theta_\bprime)$ can be made by restricting the solution-parameter pairs to $\Xi$, instead of $\Xi \cup \Xi^{\text{adv}}$. 
From this, we can conclude that the strong consistency of pairwise balance conditions holds for Algorithm~\ref{alg:rate_opt_posterior}.
\halmos
\endproof

\vspace{12pt}
\noindent {\bf Remark.} Theorems~\ref{thm:proof_EIalg} and~\ref{thm:proof_BWalg} state that $\lt\{\alpha_{i, n}(\theta_b)\rt\}$ asymptotically satisfies both global and pairwise balance conditions. Both theorems do not guarantee that $\lt\{\alpha_{i, n}(\theta_b)\rt\}$ converges. For instance, if Problem~\eqref{opt:aOCBA} has multiple maximizers, then $\lt\{\alpha_{i, n}(\theta_b)\rt\}$ may oscillate in the subset of the maximizers in the limit.

\proof{Proof of Theorem~\ref{cor:Alg1_convergence}.} We continue to use the  notation  in the proofs of Theorems~\ref{thm:proof_EIalg} and~\ref{thm:proof_BWalg}; for each sample path, $\{i_0^b\}_{1\leq b\leq B}$ represents the set of sample conditional best solutions at all $\theta_b$ in the limit, and $i_0^*$ and $\Theta_0^*$  denote the estimated MPB and its favorable set computed from $\{i_0^b\}$, respectively. Similarly, $\Xi_0^\text{adv}$ is defined from $\{i_0^b\}$. 
Thus, it suffices to show $i_0^* = i^*$ for each sample path.

Let us fix a sample path. Following Step~1 of (I) in the proof of Theorem~\ref{thm:proof_EIalg}, we have $N_{i}^n(\theta_b) \rightarrow \infty$ for all $(i, \theta_b) \in (\Xi_0^\text{adv})^c$. By means of contradiction, assume $i_0^* \neq i^*$. Combining $N_i^n(\theta_b) \rightarrow \infty$ for all $(i, \theta_b) \in (\Xi_0^\text{adv})^c$ with the SLLN, we obtain $i_0^b = i^b$ for all $\theta_b$ such that $i_0^b = i_0^*$. Hence, the estimated preference probability of Solution $i_0^*$ is bounded from above by the true preference probability of Solution $i_0^*$: that is, $\sum_{b:i_0^b = i_0^*} p_b \leq \sum_{b:i^b = i_0^*} p_b$. 

Let us show the following set of inequalities, where the last is shown above:
\begin{equation*}
    \sum_{b:i_0^b = i^*} p_b \geq \sum_{b: i^b = i^*} p_b > \sum_{b: i^b = i_0^*} p_b \geq \sum_{b:i_0^b = i_0^*} p_b.
\end{equation*}
The second inequality holds from the definition of the MPB as $i^*$ must have the highest preference probability among all solutions. If the first inequality is valid, then it contradicts that $i_0^* \neq i^*$ is the sample MPB based on $\{i_0^b\}_{1\leq b\leq B}$. To achieve this, it is enough to show that $i_0^b = i^*$ holds for all $\theta_b \in \Theta_{i^*}$. For each $\theta_b \in \Theta_{i^*}$, we have $(i_0^*, \theta_b) \in \Xi_0^\text{adv}$. Thus,  $i_0^*\neq i_0^b$ at $\theta_b \in \Theta_{i^*}$. 
Since $N^n_{i}(\theta_b) \rightarrow \infty$ for all $i\neq i_0^*$, all other conditional means at such $\theta_b$ are ordered correctly. Therefore,  we  conclude that $i_0^b = i^b=i^*$ holds  for all $\theta_b \in \Theta_{i^*}$, which completes the proof.
\halmos
\endproof

In the following, we present the proof of Theorem~\ref{thm:balance_modified}. It suffices to see how balance weights in~\eqref{eq:weights.modified} are derived since the remaining parts are straightforward from the proof of Theorem~\ref{thm:balance}.
\proof{Proof of Theorem~\ref{thm:balance_modified}.}
Consider~\eqref{opt:aOCBA_EF} first. The formulation~\eqref{opt:aOCBA_EF} is equivalent to
\begin{equation*}
    \begin{aligned}
       \max \quad & C \\
       \textrm{s.t.} \quad &  C \leq W_i(\theta_b)G_{i}(\theta_b), \forall (i, \theta_b) \in \Xi,  \\
       & C \leq G_{i}(\theta_b), \forall i \neq i^*, \theta_b \in \Theta_{i^*},\\ 
       & C \leq G_{i^*}(\theta_b), \forall \theta_b \in \Theta_{i^*}^c,\\ 
       & \sum_{i=1}^{k}\sum_{b=1}^{B} \alpha_i(\theta_b) = 1, \\\
                      \quad & \alpha_i(\theta_b) \geq 0.
    \end{aligned}
\end{equation*}
As in~\eqref{opt:intermediate}, we rewrite the first two constraints to obtain
\begin{equation}\label{opt:intermediate_EF}
    \begin{aligned}
       \max \quad & C \\
       \textrm{s.t.} \quad &  C \leq \min\{W_i(\theta_b), 1\} G_{i}(\theta_b), \forall (i, \theta_b) \in \Xi \;\;\textup{where}\;\; \theta_b \in \Theta_{i^*}\\ 
       &  C \leq W_i(\theta_b) G_{i}(\theta_b), \forall (i, \theta_b) \in \Xi \;\;\textup{where}\;\; \theta_b \in \Theta^c_{i^*}\\
       & C \leq G_{i^*}(\theta_b), \forall \theta_b \in \Theta_{i^*}^c,\\ 
       & \sum_{i=1}^{k}\sum_{b=1}^{B} \alpha_i(\theta_b) = 1, \\\
                      \quad & \alpha_i(\theta_b) \geq 0. 
    \end{aligned}
\end{equation}
Unlike Theorem~\ref{thm:balance}, all $G_{i^*}(\theta_b)$ with $\theta_b \in \Theta^c_{i^*}$ appear in~\eqref{opt:intermediate_EF} from the third constraint in~\eqref{opt:intermediate_EF}. Therefore, the sampling ratios $\lt\{\alpha_{i^*}(\theta_b), \theta_b \in \Theta^c_{i^*}\rt\}$ are positive in this case. From the definition of $\lt\{W^\ACC_i(\theta_b)\rt\}$ in~\eqref{eq:weights.modified},~\eqref{opt:intermediate_EF} is equivalent to
\begin{equation}\label{opt:EF_final}
    \begin{aligned}
       \max \quad & C \\
       \textrm{s.t.} \quad &  C \leq W^\ACC_i(\theta_b)G_{i}(\theta_b), \forall (i, \theta_b) \in \Xi \cup \Xi^{\text{adv}},\\
       & \sum_{i=1}^{k}\sum_{b=1}^{B} \alpha_i(\theta_b) = 1, \\
      \quad & \alpha_i(\theta_b) \geq 0. 
    \end{aligned}
\end{equation}
Again, from the KKT conditions, we can derive the pairwise and global balance conditions; note that the only difference here from the proof of Theorem~\ref{thm:balance} is the Lagrangian function for~\eqref{opt:EF_final}, which has the same from as~\eqref{eq:lagrangian} except that $\lt\{W_i(\theta_b)\rt\}$ are substituted with $\lt\{W^\ACC_i(\theta_b)\rt\}$. For the FN case in~\eqref{opt:aOCBA_IF}, the proof proceeds similarly. 
 \halmos
\endproof

\section{Proofs of Results in Section~\ref{sec:GP_extension}}

We first state the ``Representer Theorem"~\citep{scholkopf:02Kernel}, which lets us solve a function minimization defined on the RKHS associated with kernel space $\mathcal{K}$ easily. In general, optimization on a function space is an infinite-dimensional problem. However, the following theorem recasts~\eqref{opt:kernel_regression} into a finite-dimensional program. 

\begin{theorem}[Representer Theorem~\citep{scholkopf:02Kernel}]\label{thm:ec_representer}
    Given $y_b \in \mathcal{Y}$ for all $1\leq b\leq B$, consider the following program:
    \begin{equation}\label{opt:kernel_general}
    \min_{g \in \mathcal{K}} \left\{\sum_{b=1}^B R(y_b, g(\theta_b))+ \kappa ||g||^2_{\mathcal{K}}\right\}, \;\; \text{for some $\kappa > 0$,}
\end{equation}
where $R:\mathcal{Y}\times \Real \rightarrow \Real$ is an arbitrary function. Then, the solution to~\eqref{opt:kernel_general}, say $g^*$, is written as $g^*(\cdot) = \sum_{b=1}^B c_b K(\theta_b, \cdot)$ for some constants $\{c_b\}_{1\leq b\leq B}$. 
\end{theorem}

\proof{Proof of Theorem~\ref{thm:prediction}.}
Thanks to Theorem~\ref{thm:ec_representer}, the solution to~\eqref{opt:kernel_regression} evaluated at $\theta_1, \theta_2, \ldots, \theta_B$ is given as $\BFK_i\BFa^*$ for some vector of constant $\BFa^* \in \Real^B$. Then,~\eqref{opt:kernel_regression} can be rewritten as
\begin{equation}\label{eq:finite_opt}
    \argmin_{\BFa \in \Real^B} \left\{(\BFmu_{i, n}- \BFK_i\BFa - \beta_i\BFone_{B})^\top\Sigma_{i, n}^{-1}(\BFmu_{i, n}-\BFK_i\BFa - \beta_i\BFone_{B})+ \kappa \BFa^\top\BFK_i\BFa\right\},
\end{equation}
which boils down to finding the minimizer $\BFa$. Since the objective function of~\eqref{eq:finite_opt} is convex in $\BFa$, $\BFa^*$ is the root of the following first-order condition $2\BFK_i^\top\Sigma_{i, n}^{-1}(\BFK_i\BFa^* -(\BFmu_{i, n} - \beta_i\BFone_B))  + 2\kappa\BFK_i \BFa^* = 0.$
After some algebras, we obtain
\begin{equation}\label{eq:Ka}
    \BFK_i\BFa^* = (\BFK_i^\top\Sigma_{i,n}^{-1} + \kappa I)^{-1}\BFK_i^\top\Sigma_{i, n}^{-1}(\BFmu_{i, n}-\beta_i\BFone_B).
\end{equation}
Observe that 
\begin{equation}\label{eq:Ka_simple}
    \begin{aligned}
        (\BFK_i^\top\Sigma_{i,n}^{-1} + \kappa I)^{-1}\BFK_i^\top\Sigma_{i, n}^{-1} = I - (\BFK_i^\top\Sigma_{i,n}^{-1}/\kappa + I)^{-1}= I - \left(I - \frac{\BFK_i^\top(\Sigma_{i, n} + \BFK_i/\kappa)^{-1}}{\kappa}\right)
        = \BFK_i^\top(\BFK_i + \kappa\Sigma_{i, n} )^{-1}.
    \end{aligned}
\end{equation}
The first and third equalities follow from straightforward calculations. The second is derived from that $(\BFK_i^\top\Sigma_{i,n}^{-1}/\kappa + I) \left(I - \frac{\BFK_i^\top(\Sigma_{i, n} + \BFK_i/\kappa)^{-1}}{\kappa}\right) = I$. Combining~\eqref{eq:Ka} with~\eqref{eq:Ka_simple} yields $g^*_{i, n} = \beta_i\BFone_B + \BFK_i^\top(\BFK_i + \kappa\Sigma_{i, n} )^{-1}(\BFmu_{i, n} - \beta_i\BFone_B)$, which completes the proof. \halmos
\endproof

\proof{Proof of Theorem~\ref{thm:LDR_GP}.}
    Let $\widetilde{\Sigma}_n = \BFK_i + \kappa\Sigma_{i, n}$. Note that $\BFK_i\widetilde{\Sigma}_n^{-1} \rightarrow I$. One can rewrite~\eqref{eq:post_update} as
    \begin{equation}\label{eq:post_update_new}
        \begin{aligned}
        \widehat{\boldsymbol{\mu}}_{i, n} = \beta_i \mathbf{1}_k + \BFK_i^\top \widetilde{\Sigma}^{-1}(\boldsymbol{\mu}_{i, n} - \beta_i \mathbf{1}_k)
    = \beta_i \mathbf{1}_k + \BFK_i^\top \widetilde{\Sigma}_n^{-1}(\BFy - \beta_i\mathbf{1}_k) + \BFK_i^\top \widetilde{\Sigma}^{-1}_n(\boldsymbol{\mu}_{i, n} - \BFy),       
        \end{aligned}
    \end{equation}
    where  $\BFy= \{y_i(\theta_b)\}_{1\leq b\leq B}$. Since we have $\boldsymbol{\mu}_{i, n}- \BFy \sim MVN(\mathbf{0}_B, \text{diag}\left(\left\{\frac{\lambda_i^2(\theta_b)}{N_i^n(\theta_b)}\right\}\right))$, $\widehat{\Lambda}_{i, n}(\BFt)$ can be written as
    \begin{equation*}
        \widehat{\Lambda}_{i, n}(\BFt) = \BFt^\top\left(\beta_i \mathbf{1}_k + \BFK_i^\top \widetilde{\Sigma}_n^{-1}(\BFy - \beta_i\mathbf{1}_k)\right) + \frac{1}{2}\BFt^\top \BFK_i^\top \widetilde{\Sigma}_n^{-1}\text{diag}\left(\left\{\frac{\lambda_i^2(\theta_b)}{N_i^n(\theta_b)}\right\}\right) \lt(\BFK_i^\top \widetilde{\Sigma}_n^{-1}\rt)^\top\BFt.
    \end{equation*}
    Combining this with $\BFK_i^\top \widetilde{\Sigma}^{-1}_n \rightarrow I$ and $N_i^n(\theta_b)/n \rightarrow \alpha_i(\theta_b)$ yields 
    \begin{equation*}
        \lim_{n \rightarrow \infty}\frac{1}{n}\widehat{\Lambda}_{i, n}(n\BFt) = \BFt^\top\BFy + \frac{1}{2}\BFt^\top \text{diag}\left(\left\{\frac{\lambda_i^2(\theta_b)}{\alpha_i(\theta_b)}\right\}\right)\BFt = \sum_{b=1}^B\left(y_i(\theta_b) t_b + \frac{\lambda_i^2(\theta_b)t_b^2}{2\alpha_i(\theta_b)}\right). \halmos
    \end{equation*} 
\endproof

\section{Proofs of Auxiliary Results in Section~\ref{apdx:conv_proofs}}\label{apdx:aux_proofs}

\proof{Proof of Lemma~\ref{lem:modfied_lemma}.}
 
 The last case is straightforward from the observation that the quantities $\lt\{N^n_i(\theta_b)\rt\}_{1 \leq i \leq k}$ and $N^n_{\tot}(\theta_b)$ are unchanged if $n \in \mathcal{N}(\theta_\bprime)$ for some $\theta_\bprime \neq \theta_b$. To show the first case, suppose $n \in \mathcal{N}_{i^b}(\theta_b)$. Then, one can see that
 \begin{equation*}
     \begin{aligned}
        \Delta_{n+1}(\theta_b)- \Delta_{n}(\theta_b) &= \underbrace{\lt(\dfrac{N^n_{i^b}(\theta_b)+1}{(N^n_{\tot}(\theta_b)+1)\lambda_{i^b}(\theta_b)}\rt)^2 - \lt(\dfrac{N^n_{i^b}(\theta_b)}{N^n_{\tot}(\theta_b)\lambda_{i^b}(\theta_b)}\rt)^2}_{\text{$\geq 0$}}\\ &- \sum_{\ell \neq i^b}\bigg\{\underbrace{\lt(\dfrac{N^n_{\ell}(\theta_b)}{(N^n_{\tot}(\theta_b)+1)\lambda_\ell(\theta_b)}\rt)^2 - \lt(\dfrac{N^n_{\ell}(\theta_b)}{N^n_{\tot}(\theta_b)\lambda_\ell(\theta_b)}\rt)^2}_{\text{$\leq 0$}}\bigg\} > 0,
     \end{aligned}
 \end{equation*}
 as desired. If $n \in \mathcal{N}_j(\theta_b)$ holds for some $j \neq i^b$, then we obtain \small
 \begin{equation*}
     \begin{aligned}
         & \Delta_{n+1}(\theta_b)- \Delta_{n}(\theta_b)\\
         &= \lt(\dfrac{N^n_{i^b}(\theta_b)}{(N^n_{\tot}(\theta_b)+1)\lambda_{i^b}(\theta_b)}\rt)^2 - \lt(\dfrac{N^n_{i^b}(\theta_b)}{N^n_{\tot}(\theta_b)\lambda_{i^b}(\theta_b)}\rt)^2 - \sum_{\ell \neq i^b, j}\lt\{\lt(\dfrac{N^n_{\ell}(\theta_b)}{(N^n_{\tot}(\theta_b)+1)\lambda_\ell(\theta_b)}\rt)^2 - \lt(\dfrac{N^n_{\ell}(\theta_b)}{N^n_{\tot}(\theta_b)\lambda_\ell(\theta_b)}\rt)^2\rt\}\\
         & \;\;\; - \lt(\dfrac{N^n_j(\theta_b) + 1}{(N^n_{\tot}(\theta_b)+1)\lambda_j(\theta_b)}\rt)^2 + \lt(\dfrac{N^n_j(\theta_b)}{N^n_{\tot}(\theta_b)\lambda_j(\theta_b)}\rt)^2 \\
         & = \lt(\dfrac{N^n_{i^b}(\theta_b)}{(N^n_{\tot}(\theta_b)+1)\lambda_{i^b}(\theta_b)}\rt)^2 - \lt(\dfrac{N^n_{i^b}(\theta_b)}{N^n_{\tot}(\theta_b)\lambda_{i^b}(\theta_b)}\rt)^2 - \sum_{\ell \neq i^b}\lt\{\lt(\dfrac{N^n_{\ell}(\theta_b)}{(N^n_{\tot}(\theta_b)+1)\lambda_\ell(\theta_b)}\rt)^2 - \lt(\dfrac{N^n_{\ell}(\theta_b)}{N^n_{\tot}(\theta_b)\lambda_\ell(\theta_b)}\rt)^2\rt\} \\
         & \;\;\; \underbrace{- \lt(\dfrac{N^n_j(\theta_b) + 1}{(N^n_{\tot}(\theta_b)+1)\lambda_j(\theta_b)}\rt)^2 + \lt(\dfrac{N^n_j(\theta_b)}{(N^n_{\tot}(\theta_b)+1)\lambda_j(\theta_b)}\rt)^2}_{\text{$\leq 0$}} \\
         & \leq \lt(\dfrac{N^n_{i^b}(\theta_b)}{(N^n_{\tot}(\theta_b)+1)\lambda_{i^b}(\theta_b)}\rt)^2 - \lt(\dfrac{N^n_{i^b}(\theta_b)}{N^n_{\tot}(\theta_b)\lambda_{i^b}(\theta_b)}\rt)^2 - \sum_{\ell \neq i^b}\lt\{\lt(\dfrac{N^n_{\ell}(\theta_b)}{(N^n_{\tot}(\theta_b)+1)\lambda_\ell(\theta_b)}\rt)^2 - \lt(\dfrac{N^n_{\ell}(\theta_b)}{N^n_{\tot}(\theta_b)\lambda_\ell(\theta_b)}\rt)^2\rt\} \\
         & = \lt(\lt(\frac{N^n_{\tot}(\theta_b)}{N^n_{\tot}(\theta_b) + 1}\rt)^2 - 1\rt) \Delta_n(\theta_b) < 0.
     \end{aligned}
 \end{equation*} \normalsize
Note that $\Delta_n(\theta_b) > 0$ since $n \in \mathcal{N}_j(\theta_b)$. \halmos
\endproof

\noindent To verify Lemmas~\ref{lem:LIL_LDR} and~\ref{lem:between_sample}, we exploit the following property; for all sufficiently large $n$, any pair $(i, \theta_b)$ with $i \neq i^b$, and any integer $\ell \geq 0$,
\begin{eqnarray}
        && \lt|(\mu_{i, n+\ell}(\theta_b) - \mu_{i^b, n+\ell}(\theta_b))^2 - (y_{i}(\theta_b) - y_{i^b}(\theta_b))^2\rt|\nonumber\\
        && \lt|\lt(\mu_{i, n+\ell}(\theta_b) - \mu_{i^b, n+\ell}(\theta_b) - y_{i}(\theta_b) + y_{i^b}(\theta_b)\rt)\lt(\mu_{i, n+\ell}(\theta_b) - \mu_{i^b, n+\ell}(\theta_b) + y_{i}(\theta_b) - y_{i^b}(\theta_b)\rt)\rt|\nonumber\\
        && \leq  \lt(\lt|\mu_{i, n+\ell}(\theta_b) -  y_{i}(\theta_b)\rt| + \lt|\mu_{i^b, n+\ell}(\theta_b)- y_{i^b}(\theta_b)\rt|\rt) \times \Oas(1) \nonumber \\
        && \leq \Oas\lt(\sqrt{\frac{\log\log (n+\ell)}{n+\ell}}\rt) = \Oas\lt(\sqrt{\frac{\log\log n}{n}}\rt), \label{eq:mean_LIL}
\end{eqnarray}
where the latter inequality follows from $N^{n+\ell}_i(\theta_b) = \Tas(n+\ell)$ and the law of the iterated logarithm; ${|\mu_{i, n}(\theta_b) - y_i(\theta_b)|} = \Oas\lt(\sqrt{\log\log N_i^n(\theta_b) /N_i^n(\theta_b)}\rt)$. The last equality follows because $\sqrt{\log\log(n+\ell)/(n+\ell)}$ is decreasing in $\ell$.

\proof{Proof of Lemma~\ref{lem:LIL_LDR}.}
Recall that $n \in \mathcal{N}(\theta_b)$. Then, there are two possibilities: $n \in \mathcal{N}_i(\theta_b)$ for some $i \neq i^b$ or $n \in \mathcal{N}_{i^b}(\theta_b)$. Consider the former case.
By~\eqref{eq:mean_LIL} and $N_i^n(\theta_b) = \Tas(n)$, we have
\begin{equation*}
    \begin{aligned}
         &\lt|\min_{i \neq i^b} \Gamma_{i, n+m_n(\theta_b)}(\theta_b) - \min_{i \neq i^b} \Gamma_{i, n}(\theta_b)\rt| \\
         &\leq W_i(\theta_b)\lt|\dfrac{(\mu_{i, n+1}(\theta_b) - \mu_{i^b, n+1}(\theta_b))^2}{2\lt(\dfrac{\lambda^2_i(\theta_b)}{N_i^n(\theta_b) +1} + \dfrac{\lambda^2_{i^b}(\theta_b)}{N_{i^b}^{n}(\theta_b)}\rt)} - \dfrac{(\mu_{i, n}(\theta_b) - \mu_{i^b, n}(\theta_b))^2}{2\lt(\dfrac{\lambda^2_i(\theta_b)}{N_i^n(\theta_b)} + \dfrac{\lambda^2_{i^b}(\theta_b)}{N_{i^b}^n(\theta_b)}\rt)}\rt|\\
        &\leq W_i(\theta_b)\lt|\dfrac{(y_{i}(\theta_b) - y_{i^b}(\theta_b))^2}{2\lt(\dfrac{\lambda^2_i(\theta_b)}{N_i^n(\theta_b) + 1} + \dfrac{\lambda^2_{i^b}(\theta_b)}{N_{i^b}^{n}(\theta_b)}\rt)} - \dfrac{(y_i(\theta_b) - y_{i^b}(\theta_b))^2}{2\lt(\dfrac{\lambda^2_i(\theta_b)}{N_i^n(\theta_b)} + \dfrac{\lambda^2_{i^b}(\theta_b)}{N_{i^b}^n(\theta_b)}\rt)}\rt|+ O(\sqrt{n\log\log n})\\
        & = \Tas(1) + \Oas(\sqrt{n\log\log n}) = \Oas(\sqrt{n\log\log n})
    \end{aligned}
\end{equation*}
as desired. When $n \in \mathcal{N}_{i^b}(\theta_b)$, we can proceed similarly.  \halmos

\endproof

\proof{Proof of Lemma~\ref{lem:between_sample}.}

Define $m_{i,n}(\theta_{\bprime}) = \lt|\lt\{m \in \mathcal{N}_i(\theta_{\bprime}): n < m < n + m_n(\theta_b)\rt\}\rt|$. Then, it suffices to show
\begin{equation}\label{eq:m_in}
    m_{i,n}(\theta_{\bprime}) = \Oas(\sqrt{n\log\log n}) \;\; \text{for all }1 \leq i \leq k, \;\;\bprime \neq b,
\end{equation}

To verify~\eqref{eq:m_in}, consider the case when $i \neq i^{\bprime}$ first. If $m_{i,n}(\theta_\bprime) = 0$, it is trivial. Otherwise, let $\ell = \sup\lt\{t < m_n(\theta_b)| n+t \in \mathcal{N}_i(\theta_\bprime)\rt\}$. By definition, we have $N^{n+\ell}_i(\theta_\bprime) = N^{n}_i(\theta_\bprime) + m_{i,n}(\theta_\bprime) - 1$.
Further, the pairwise balance condition implies that $\Gamma_{i, n + \ell}(\theta_\bprime) < \min_{j \neq i^b}\Gamma_{j, n+\ell}(\theta_b) = \min_{j \neq i^b}\Gamma_{j, n+m_n(\theta_b)}(\theta_b)$ and $\min_{j \neq i^b}\Gamma_{j, n}(\theta_b) \leq \min_{j \neq i^\bprime}\Gamma_{j, n}(\theta_\bprime)$. Subtracting two terms, we have
\begin{equation}\label{eq:diff_between}
    \min_{j \neq i^\bprime}\Gamma_{j, n + \ell}(\theta_\bprime) - \min_{j \neq i^\bprime}\Gamma_{j, n}(\theta_\bprime) < \min_{j \neq i^b}\Gamma_{j, n+m_n(\theta_b)}(\theta_b) - \min_{j \neq i^b}\Gamma_{j, n}(\theta_b) \leq \Oas(\sqrt{n\log\log n}),
\end{equation}
where the last inequality follows from Lemma~\ref{lem:LIL_LDR}. In addition, we have
\begin{equation}\label{eq:diff_between_lower}
    \begin{aligned}
        &\min_{j \neq i^\bprime}\Gamma_{j, n + \ell}(\theta_\bprime) - \min_{j \neq i^\bprime}\Gamma_{j, n}(\theta_\bprime)\\
        &\quad = \Gamma_{i, n + \ell}(\theta_\bprime) - \min_{i \neq i^\bprime}\Gamma_{i, n}(\theta_\bprime)\\
        &\quad\geq \Gamma_{i, n + \ell}(\theta_\bprime) - \Gamma_{i, n}(\theta_\bprime) = W_i(\theta_b)\lt(\dfrac{(\mu_{i, n+\ell}(\theta_\bprime) - \mu_{i^\bprime, n+\ell}(\theta_\bprime))^2}{2\lt(\dfrac{\lambda^2_i(\theta_\bprime)}{N_i^{n + \ell}(\theta_\bprime)} + \dfrac{\lambda^2_{i^\bprime}(\theta_\bprime)}{N_{i^\bprime}^{n+\ell}(\theta_\bprime)}\rt)} - \dfrac{(\mu_{i, n}(\theta_\bprime) - \mu_{i^\bprime, n}(\theta_\bprime))^2}{2\lt(\dfrac{\lambda^2_i(\theta_\bprime)}{N_i^n(\theta_\bprime)} + \dfrac{\lambda^2_{i^\bprime}(\theta_\bprime)}{N_{i^\bprime}^n(\theta_\bprime)}\rt)}\rt)\\
        &\quad\geq W_i(\theta_b)\lt(\dfrac{(\mu_{i, n+\ell}(\theta_\bprime) - \mu_{i^\bprime, n+\ell}(\theta_\bprime))^2}{2\lt(\dfrac{\lambda^2_i(\theta_\bprime)}{N_i^{n + \ell}(\theta_\bprime)} + \dfrac{\lambda^2_{i^\bprime}(\theta_\bprime)}{N_{i^\bprime}^{n}(\theta_\bprime)}\rt)} - \dfrac{(\mu_{i, n}(\theta_\bprime) - \mu_{i^\bprime, n}(\theta_\bprime))^2}{2\lt(\dfrac{\lambda^2_i(\theta_\bprime)}{N_i^n(\theta_\bprime)} + \dfrac{\lambda^2_{i^\bprime}(\theta_\bprime)}{N_{i^\bprime}^n(\theta_\bprime)}\rt)}\rt) \\
        &\quad\geq W_i(\theta_b)\lt(\dfrac{(y_{i}(\theta_\bprime) - y_{i^\bprime}(\theta_\bprime))^2}{2\lt(\dfrac{\lambda^2_i(\theta_\bprime)}{N_i^{n + \ell}(\theta_\bprime)} + \dfrac{\lambda^2_{i^\bprime}(\theta_\bprime)}{N_{i^\bprime}^{n}(\theta_\bprime)}\rt)} - \dfrac{(y_i(\theta_\bprime) - y_{i^\bprime}(\theta_\bprime))^2}{2\lt(\dfrac{\lambda^2_i(\theta_\bprime)}{N_i^n(\theta_\bprime)} + \dfrac{\lambda^2_{i^\bprime}(\theta_\bprime)}{N_{i^\bprime}^n(\theta_\bprime)}\rt)}\rt) - \Oas(\sqrt{n\log\log n}).
    \end{aligned}
\end{equation}
The third inequality is derived from
\begin{equation*}
    \begin{aligned}
    &\dfrac{\lt|(y_{i}(\theta_\bprime) - y_{i^\bprime}(\theta_\bprime))^2 - (\mu_{i, n+\ell}(\theta_\bprime) - \mu_{i^\bprime, n+\ell}(\theta_\bprime))^2 \rt|}{2\lt(\dfrac{\lambda^2_i(\theta_\bprime)}{N_i^{n + \ell}(\theta_\bprime)} + \dfrac{\lambda^2_{i^\bprime}(\theta_\bprime)}{N_{i^\bprime}^{n}(\theta_\bprime)}\rt)}\\ &\quad \leq \dfrac{N_{i^\bprime}^n(\theta_\bprime)}{2\lambda^2_{i^\bprime}(\theta_\bprime)}\lt|(y_{i}(\theta_\bprime) - y_{i^\bprime}(\theta_\bprime))^2 - (\mu_{i, n+\ell}(\theta_\bprime) - \mu_{i^\bprime, n+\ell}(\theta_\bprime))^2\rt|\\
    &\quad = \Tas(n)\Oas(\sqrt{(\log\log n)/n}) = \Oas(\sqrt{n\log\log n}),
    \end{aligned}
\end{equation*}
where the first equality follows from Steps 1--3 of Theorem~6 and~\eqref{eq:mean_LIL}. Combining~\eqref{eq:diff_between} with~\eqref{eq:diff_between_lower}, we then have
\begin{equation}\label{eq:lower_bound_temp}
    \lt(\dfrac{\lambda^2_i(\theta_\bprime)}{N_i^{n + \ell}(\theta_\bprime)} + \dfrac{\lambda^2_{i^\bprime}(\theta_\bprime)}{N_{i^\bprime}^{n}(\theta_\bprime)}\rt)^{-1} - \lt(\dfrac{\lambda^2_i(\theta_\bprime)}{N_i^n(\theta_\bprime)} + \dfrac{\lambda^2_{i^\bprime}(\theta_\bprime)}{N_{i^\bprime}^n(\theta_\bprime)}\rt)^{-1} =  \Oas(\sqrt{n\log\log n}).
\end{equation}
Observe that the LHS of~\eqref{eq:lower_bound_temp} can be re-written as
\begin{equation}\label{eq:temp_2}
    \begin{aligned}
        \text{LHS of~\eqref{eq:lower_bound_temp}} &= \dfrac{\dfrac{m_{i, n}(\theta_\bprime) - 1}{N_i^{n+\ell}(\theta_\bprime)N_i^n(\theta_\bprime)}}{\lt(\dfrac{\lambda^2_i(\theta_\bprime)}{N_i^{n + \ell}(\theta_\bprime)} + \dfrac{\lambda^2_{i^\bprime}(\theta_\bprime)}{N_{i^\bprime}^{n}(\theta_\bprime)}\rt)\lt(\dfrac{\lambda^2_i(\theta_\bprime)}{N_i^n(\theta_\bprime)} + \dfrac{\lambda^2_{i^\bprime}(\theta_\bprime)}{N_{i^\bprime}^n(\theta_\bprime)}\rt)}\\
        &= \Tas(1) \lt(\dfrac{\lambda_i^2(\theta_\bprime)}{m_{i, n}(\theta_\bprime) - 1} + \dfrac{N^{n+\ell}_{i}(\theta_\bprime)\lambda^2_{i^\bprime}(\theta_\bprime)}{(m_{i, n}(\theta_\bprime) -1)N_{i^\bprime}^n(\theta_\bprime)}\rt)^{-1}.
    \end{aligned}
\end{equation}
The second equality comes from $N_i^n(\theta_\bprime)\lt(\dfrac{\lambda^2_i(\theta_\bprime)}{N_i^n(\theta_\bprime)} + \dfrac{\lambda^2_{i^\bprime}(\theta_\bprime)}{N_{i^\bprime}^n(\theta_\bprime)}\rt) = \Tas(1)$ as $N^n_i(\theta_\bprime) = \Tas(n)$ for all $(i, \theta_\bprime)$. From~\eqref{eq:lower_bound_temp} and~\eqref{eq:temp_2}, there exists constant $C_1 > 0$ such that
\begin{equation}\label{eq:temp_3}
    \dfrac{\lambda_i^2(\theta_\bprime)}{m_{i, n}(\theta_\bprime) - 1} + \dfrac{N^{n+\ell}_{i}(\theta_\bprime)\lambda^2_{i^\bprime}(\theta_\bprime)}{(m_{i, n}(\theta_\bprime) -1)N_{i^\bprime}^n(\theta_\bprime)} \geq \frac{1}{C_1\sqrt{n\log\log n}}
\end{equation}
for all sufficiently large $n$. We re-write the LHS of~\eqref{eq:temp_3} as
\begin{equation}\label{eq:temp_4}
    \begin{aligned}
       \dfrac{\lambda_i^2(\theta_\bprime)}{m_{i, n}(\theta_\bprime) - 1} + \dfrac{N^{n+\ell}_{i}(\theta_\bprime)\lambda^2_i(\theta_\bprime)}{(m_{i, n}(\theta_\bprime) -1)N_{i^\bprime}^n(\theta_\bprime)} &= \dfrac{\lambda_i^2(\theta_\bprime)}{m_{i, n}(\theta_\bprime) - 1} + \dfrac{N_i^n(\theta_\bprime)}{N_{i^\bprime}^n(\theta_\bprime)} \dfrac{N^{n+\ell}_{i}(\theta_\bprime)\lambda^2_{i^\bprime}(\theta_\bprime)}{(m_{i, n}(\theta_\bprime) -1)N_i^n(\theta_\bprime)} \\
       & = \dfrac{\lambda_i^2(\theta_\bprime)}{m_{i, n}(\theta_\bprime) - 1} + \Tas(1) \lt(\dfrac{1}{m_{i,n}(\theta_\bprime) - 1} + \dfrac{1}{N_i^n(\theta_\bprime)}\rt) \\
       & \leq \dfrac{C_2}{m_{i, n}(\theta_\bprime) - 1} + \dfrac{C_3}{n}
    \end{aligned}
\end{equation}
for some positive constants $C_2$ and $C_3$. For the second equality, we use $N_i^n(\theta_\bprime)/N^n_{i^\bprime}(\theta_\bprime) = \Tas(1)$ and $N_i^{n+\ell}(\theta_\bprime) = N_i^n(\theta_\bprime) + m_{i, n}(\theta_\bprime) - 1$. For the last inequality, we exploit $N_i^n(\theta_\bprime) = \Tas(n)$ again. Combining~\eqref{eq:temp_3} with~\eqref{eq:temp_4}, we obtain that
\begin{equation}\label{eq:temp_5}
    \dfrac{C_2}{m_{i, n}(\theta_\bprime) - 1} \geq \frac{1}{C_1\sqrt{n\log\log n}}-\dfrac{C_3}{n} \geq \dfrac{1}{2C_1 \sqrt{n\log\log n}}
\end{equation}
holds for all sufficiently large $n$. The second inequality follows from $\sqrt{n\log\log n}/n = o(1)$. From~\eqref{eq:temp_5}, we can conclude that there exists positive constant $C_5$ such that $m_{i, n}(\theta_\bprime) \leq C_5\sqrt{n\log\log n}$ for all sufficiently large $n$, which means $m_{i, n}(\theta_\bprime) = \Oas(\sqrt{n\log\log n})$.

Let us consider the case when $i = i^\bprime$. Then,~\eqref{eq:diff_between} still holds. Let $j_1 = \argmin_{i \neq i^\bprime} \Gamma_{i, n+\ell}(\theta_\bprime)$. Similar to~\eqref{eq:diff_between_lower}, one can derive that
\begin{equation}\label{eq:diff_between_lower2}
    \begin{aligned}
        &\min_{j \neq i^\bprime}\Gamma_{j, n + \ell}(\theta_\bprime) - \min_{j \neq i^\bprime}\Gamma_{j, n}(\theta_\bprime)\\
        &\quad = \Gamma_{j_1, n + \ell}(\theta_\bprime) - \min_{j \neq i^\bprime}\Gamma_{j, n}(\theta_\bprime)\\
        &\quad\geq \Gamma_{j_1, n + \ell}(\theta_\bprime) - \Gamma_{j_1, n}(\theta_\bprime) = W_{j_1}(\theta_b)\lt(\dfrac{(\mu_{j_1, n+\ell}(\theta_\bprime) - \mu_{i^\bprime, n+\ell}(\theta_\bprime))^2}{2\lt(\dfrac{\lambda^2_{j_1}(\theta_\bprime)}{N_{j_1}^{n + \ell}(\theta_\bprime)} + \dfrac{\lambda^2_{i^\bprime}(\theta_\bprime)}{N_{i^\bprime}^{n+\ell}(\theta_\bprime)}\rt)} - \dfrac{(\mu_{j_1, n}(\theta_\bprime) - \mu_{i^\bprime, n}(\theta_\bprime))^2}{2\lt(\dfrac{\lambda^2_{j_1}(\theta_\bprime)}{N_{j_1}^n(\theta_\bprime)} + \dfrac{\lambda^2_{i^\bprime}(\theta_\bprime)}{N_{i^\bprime}^n(\theta_\bprime)}\rt)}\rt)\\
        &\quad\geq W_{j_1}(\theta_b)\lt(\dfrac{(y_{j_1}(\theta_\bprime) - y_{i^\bprime}(\theta_\bprime))^2}{2\lt(\dfrac{\lambda^2_{j_1}(\theta_\bprime)}{N_{j_1}^{n}(\theta_\bprime)} + \dfrac{\lambda^2_{i^\bprime}(\theta_\bprime)}{N_{i^\bprime}^{n +\ell}(\theta_\bprime)}\rt)} - \dfrac{(y_{j_1}(\theta_\bprime) - y_{i^\bprime}(\theta_\bprime))^2}{2\lt(\dfrac{\lambda^2_{j_1}(\theta_\bprime)}{N_{j_1}^n(\theta_\bprime)} + \dfrac{\lambda^2_{i^\bprime}(\theta_\bprime)}{N_{i^\bprime}^n(\theta_\bprime)}\rt)}\rt) - \Oas(\sqrt{n\log\log n}).
    \end{aligned}
\end{equation}
Applying an argument similar to that made for~\eqref{eq:diff_between_lower} to~\eqref{eq:diff_between_lower2}, we have $m_{i^\bprime, n}(\theta_\bprime) = \Oas(\sqrt{n\log\log n})$ as desired. \halmos
\endproof

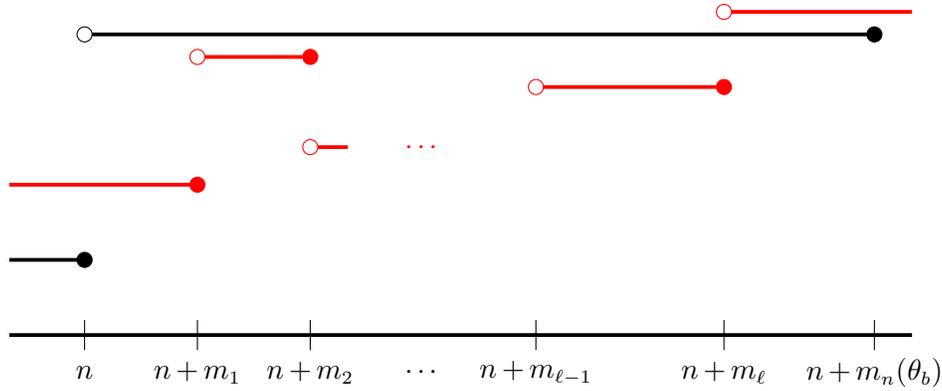
\begin{figure}[H]
\begin{center}
\begin{tikzpicture}
\coordinate (a0) at (0,0);
\coordinate (a1) at (12,0);
\coordinate (a2) at (12,4);
\coordinate (a3) at (1.1, 0);
\coordinate (a4) at (11.4, 4);

\draw[line width = 0.5mm] (a0) -- (a1);
\draw[line width = 0.5mm] (a3 |- a2) -- (a4);
\draw[fill = black] (11.5, 4) circle (0.1);

\draw (1, 4) circle (0.1);
\draw (1, 0.2) -- (1, -0.2);
\node at (1, -0.5) {$n$};
\draw (2.5, 0.2) -- (2.5, -0.2);
\node at (2.5, -0.5) {$n+m_1$};
\draw (4, 0.2) -- (4, -0.2);
\node at (4, -0.5){$n+m_2$};
\draw (7, 0.2) -- (7, -0.2);
\node at (7, -0.5){$n+m_{\ell - 1}$};
\draw (9.5, 0.2) -- (9.5, -0.2);
\node at (9.5, -0.5){$n+m_{\ell}$};
\draw (11.5, 0.2) -- (11.5, -0.2);
\node at (11.5, -0.5) {$n+m_n(\theta_b)$};

\draw[line width = 0.5mm] (0, 1) -- (0.9, 1);
\draw[fill = black] (1, 1) circle (0.1);
\draw[line width = 0.5mm, color = red] (0, 2) -- (2.4, 2);
\draw[fill = red, color = red] (2.5, 2) circle (0.1);
\draw[line width = 0.5mm, color = red] (2.6, 3.7) -- (3.9, 3.7);
\draw[color = red] (2.5, 3.7) circle (0.1);
\draw[fill = red, color = red] (4, 3.7) circle (0.1);
\draw[line width = 0.5mm, color = red] (4.1, 2.5) -- (4.5, 2.5);
\draw[color = red] (4, 2.5) circle (0.1);
\draw[line width = 0.5mm, color = red] (7.1, 3.3) -- (9.4, 3.3);
\draw[color = red] (7, 3.3) circle (0.1);
\draw[fill = red, color = red] (9.5, 3.3) circle (0.1);
\draw[line width = 0.5mm, color = red] (9.6, 4.3) -- (12, 4.3);
\draw[color = red] (9.5, 4.3) circle (0.1);

\node [color = red] at (5.5, 2.5) {$\cdots$};
\node at (5.5, -0.5) {$\cdots$};

\end{tikzpicture}
\end{center}
\caption{Illustration of $\min_{i \neq i^b}\Gamma_{i, n+t}(\theta_b)$ (black line) and $\min_{i \neq i^\bprime}\Gamma_{i, n+t}(\theta_\bprime)$ (red line) when $\mathcal{M}$ is nonempty.}
\label{fig:Gamma_description}
\end{figure}

\proof{Proof of Lemma~\ref{lem:sub_rn}}
Let $\mathcal{M}= \mathcal{N}(\theta_\bprime) \cap (n, n+m_n(\theta_b))$. If $\mathcal{M}$ is empty, it is straightforward since the left-hand side of~\eqref{eq:sub_rn} is always zero within $(n, n + m_n(\theta_b))$. Otherwise, let $\mathcal{M} = \lt\{n+m_1, n+m_2,\cdots, n + m_\ell\rt\}$ where $n + m_\ell$ is the largest element of $\mathcal{M}$. Figure~\ref{fig:Gamma_description} depicts the changes in $\min_{i \neq i^b} \Gamma_{i, n+t}(\theta_b)$ and $\min_{i \neq i^\bprime} \Gamma_{i, n+t}(\theta_\bprime)$ at each $n + t$. Since $n\in \mathcal{N}(\theta_b)$ and $n+t \notin \mathcal{N}(\theta_b)$ for all $0< t < m_n(\theta_b)$, $\min_{i \neq i^b}\Gamma_{i, n}(\theta_b)$ must be constant in this interval. On the other hand, $\min_{i \neq i^\bprime}\Gamma_{i, n+t}(\theta_\bprime)$ may change within the range, $0 < t \leq m_n(\theta_b)$, but remains constant on  $(n+m_i, n+m_{i+1}]$. In addition, to satisfy $n+m_i \in \mathcal{N}(\theta_\bprime)$ for $i = 1,2, \ldots, \ell$, $\min_{i \neq i^\bprime}\Gamma_{i, n+t}(\theta_\bprime)$ must be smaller than $\min_{i \neq i^b}\Gamma_{i, n+m_n(\theta_b)}(\theta_b)$ for $0<t \leq m_\ell$. After the last replication is made at $\theta_\bprime$, this relation should be reversed since $n+m_n(\theta_b) \in \mathcal{N}(\theta_b)$.

We show~\eqref{eq:sub_rn} by partitioning the interval into two parts: $t\in [0, m_\ell]$ or $t \in (m_\ell, m_n(\theta_b)]$. In the former case, we have $0< \min_{i \neq i^{\bprime}}\Gamma_{i,n+t}(\theta_{\bprime}) - \min_{i \neq i^{\bprime}}\Gamma_{i,n}(\theta_{\bprime})$ and 
\begin{equation}\label{eq:lower_bound1}
    \min_{i \neq i^{\bprime}}\Gamma_{i,n+t}(\theta_{\bprime}) - \min_{i \neq i^{\bprime}}\Gamma_{i,n}(\theta_{\bprime}) < \min_{i \neq i^{b}}\Gamma_{i,n+m_n(\theta_b)}(\theta_{b}) - \min_{i \neq i^{b}}\Gamma_{i,n}(\theta_{b}) \leq \Oas(\sqrt{n\log\log n}).
\end{equation}
The first inequality follows from $\min_{i \neq i^{\bprime}}\Gamma_{i,n+t}(\theta_{\bprime}) \leq \min_{i \neq i^{b}}\Gamma_{i,n+m_n(\theta_b)}(\theta_{b})$ and $\min_{i \neq i^{b}}\Gamma_{i,n}(\theta_{b}) < \min_{i \neq i^{\bprime}}\Gamma_{i,n}(\theta_{\bprime})$,
where two inequalities can be confirmed from Figure~\ref{fig:Gamma_description}. The last inequality is derived from Lemma~\ref{lem:between_sample}. For $m_\ell < t \leq m_n(\theta_b)$, 
\begin{equation}\label{eq:lower_bound2}
    \begin{aligned}
    &\min_{i \neq i^{\bprime}}\Gamma_{i,n+t}(\theta_{\bprime}) - \min_{i \neq i^{\bprime}}\Gamma_{i,n}(\theta_{\bprime})\\
    &\quad<\min_{i \neq i^{\bprime}}\Gamma_{i,n+m_\ell}(\theta_{\bprime}) + \Oas\lt(\sqrt{(n+m_\ell)\log\log(n+m_\ell)}\rt) - \min_{i \neq i^{\bprime}}\Gamma_{i,n}(\theta_{\bprime})\\
    &\quad = \lt(\min_{i \neq i^{\bprime}}\Gamma_{i,n+m_\ell}(\theta_{\bprime}) - \min_{i \neq i^{\bprime}}\Gamma_{i,n}(\theta_{\bprime})\rt) + \Oas\lt(\sqrt{(n+m_\ell)\log\log(n+m_\ell)}\rt)\\ 
    &\quad< \lt(\min_{i \neq i^{\bprime}}\Gamma_{i,n+m_\ell}(\theta_{b}) - \min_{i \neq i^{\bprime}}\Gamma_{i,n}(\theta_{b})\rt) + \Oas\lt(\sqrt{(n+m_\ell)\log\log(n+m_\ell)}\rt)\\
    &\quad\leq \Oas(\sqrt{n\log\log n}) + \Oas\lt(\sqrt{(n+m_\ell)\log\log(n+m_\ell)}\rt) = \Oas(\sqrt{n\log\log n}).
    \end{aligned}
\end{equation}
The first inequality uses Lemma~\ref{lem:LIL_LDR} by replacing $n$ and $\theta_b$ with $n+m_\ell$ and $\theta_\bprime$, respectively. The second inequality comes from $\min_{i \neq i^{\bprime}}\Gamma_{i,n+m_\ell}(\theta_{\bprime}) < \min_{i \neq i^{b}}\Gamma_{i,n+m_\ell}(\theta_b)$ and $\min_{i \neq i^{b}}\Gamma_{i,n}(\theta_{b})<\min_{i \neq i^{\bprime}}\Gamma_{i,n}(\theta_{\bprime})$. For the last equation, we use $m_\ell \leq m_n(\theta_b) = \Oas(\sqrt{n\log\log n})$. 

Proceeding similarly to~\eqref{eq:diff_between_lower}, the difference $\min_{i \neq i^{\bprime}}\Gamma_{i,n+t}(\theta_{\bprime}) - \min_{i \neq i^{\bprime}}\Gamma_{i,n}(\theta_{\bprime})$ for any $0 < t < m_n(\theta_b)$ can be bounded from below by
\begin{equation}
    \begin{aligned}
        &\min_{i \neq i^\bprime}\Gamma_{i, n + t}(\theta_\bprime) - \min_{i \neq i^\bprime}\Gamma_{i, n}(\theta_\bprime)\\
        & \quad\geq W_j(\theta_b)\lt(\dfrac{(y_{j}(\theta_\bprime) - y_{i^\bprime}(\theta_\bprime))^2}{2\lt(\dfrac{\lambda^2_j(\theta_\bprime)}{N_j^{n + t}(\theta_\bprime)} + \dfrac{\lambda^2_{i^\bprime}(\theta_\bprime)}{N_{i^\bprime}^{n}(\theta_\bprime)}\rt)} - \dfrac{(y_j(\theta_\bprime) - y_{i^\bprime}(\theta_\bprime))^2}{2\lt(\dfrac{\lambda^2_j(\theta_\bprime)}{N_j^n(\theta_\bprime)}+ \dfrac{\lambda^2_{i^\bprime}(\theta_\bprime)}{N_{i^\bprime}^n(\theta_\bprime)}\rt)}\rt)
        - \Oas(\sqrt{n\log\log n})\\
        &\quad \geq -\Oas(\sqrt{n\log\log n}),\label{eq:lower_bound3}
    \end{aligned}
\end{equation}
where $j = \argmin_{i\neq i^\bprime} \Gamma_{i, n+t}(\theta_\bprime)$. The second inequality follows since the first term is positive. Combining~\eqref{eq:lower_bound1},~\eqref{eq:lower_bound2} with~\eqref{eq:lower_bound3} completes the proof.
\endproof

\section{Numerical optimality gap analysis between the exact  and the relaxed OCBA formulations}\label{ec:gap_analysis}

This section discusses a numerical method for quantifying the gap between the optimal objective function values of Problems~\eqref{eq:exact.ocba} and~\eqref{opt:OCBA}. We obtain an upper bound on the optimality gap by solving Problem~\eqref{eq:exact.ocba} with a numerical method that has a known finite-step suboptimality bound. 

Recall that $F(\boldsymbol{\alpha}) := \min_{j \neq i^*} \widetilde{\ldr}_{j, i^*} = \min_{\BFM \in \bigcup_{j \neq i^*}  \mathcal{A}_j} \sum_{(i, \theta_b) \in I(\BFM)} \widetilde{G}_i(\theta_b)$. Then, the objective function of~\eqref{eq:exact.ocba} can be replaced with $\max_{\boldsymbol{\alpha}} F(\boldsymbol{\alpha})$. Lemma~\ref{lem:concavity} below shows that $F(\boldsymbol{\alpha})$ is a concave function.

\begin{lemma}\label{lem:concavity}
    For each $(i, \theta_b) \in \Xi$, $\widetilde{G}_i(\theta_b)$ is concave in $\bm\alpha$, which implies $F(\bm\alpha)$ is concave in  $\bm\alpha$. Thus, ~\eqref{eq:exact.ocba} is a convex optimization problem. 
\end{lemma}
\proof{Proof.} 
Let us first confirm the concavity of $\widetilde{G}_i(\theta_b)$. Define $F_{i, b}(x) = \frac{\alpha_{i^b}(\theta_b)}{2\lambda^2_{i^b}(\theta_b)}(x-y_{i^b}(\theta_b))^2 + \frac{\alpha_i(\theta_b)}{2\lambda_i^2(\theta_b)}(x - y_i(\theta_b))^2+ \sum_{\ell \neq i, i^b}\frac{\alpha_\ell(\theta_b)}{2\lambda_\ell^2(\theta_b)} \lt[(x - y_\ell(\theta_b))^+\rt]^2$. Then, $\widetilde{G}_i(\theta_b) = \min_x F_{i, b}(x)$ from~\eqref{def:G_tilde}. Since $F_{i, b}(x)$ is concave in $\boldsymbol{\alpha}$ for fixed $x$, its infimum over $x$ over any fixed $\boldsymbol{\alpha}$ is concave in $\boldsymbol{\alpha}$. Consequently, $F(\bm\alpha)$ is concave in $\bm\alpha$ since the minimum and summation operations preserve concavity. This proves that~\eqref{eq:exact.ocba} is a convex program. \halmos
\endproof

Since $F(\boldsymbol{\alpha})$ is concave, we can apply a first-order method to solve~\eqref{eq:exact.ocba} by computing a supergradient of $F(\boldsymbol{\alpha})$. In general, a supergradient of function $f$ at $x$ is defined as vector $p$ such that $f(x) + p^\top(y-x) \geq f(y)$ for all $y$ in the domain of $f$.
However, computing the supergradient requires characterizing all $\BFM \in \mathcal{A}_j$ for each $j \neq i^*$, which is computationally expensive as discussed in Section~\ref{subsec:OCBA}. 

In the special case when $p_b = 1/B$ for all $1 \leq b\leq B$ which we assume through this section, we can evaluate $F(\boldsymbol{\alpha})$ and its supergradient without having to enumerate all $\BFM \in \mathcal{M}$ as described below. For each $j \neq i^*$, we find subsets  $V_{j, 1}$ and $V_{j, 2}$ of $\Xi$, where $V_{j, \ell} := \{(i, \theta_b) \in \Xi| v_j[(i, \theta_b)]^+/p_b = \ell\}$. For $\BFM \in \mathcal{A}_j$, let $A_\ell := I(\BFM)\cap V_{j, \ell}$ and $ N_\ell := |A_\ell|$. Then, $A_{1}$ and $A_{2}$ partition $I(\BFM)$, and we have $\sum_{(i, \theta_b) \in I(\BFM)} \widetilde{G}_i(\theta_b) = \sum_{(i, \theta_b) \in A_1}\widetilde{G}_i(\theta_b) + \sum_{(i, \theta_b) \in A_2}\widetilde{G}_i(\theta_b)$. Furthermore, from the definition of $\mathcal{A}_j$, we have $N_1 + 2N_2 \geq d_j$ and $N_2$ ranges from $0$ to $\min\{\lceil\frac{d_j}{2}\rceil, |V_{j, 2}|\}$. 
Because $F(\bm\alpha)$ is determined by a minimal element of $\mathcal{A}_j$ for some $j\neq i^*$ in  partial-order relation $\preceq$ defined in the proof of Lemma~\ref{lem:knapsack}, let us focus on minimal elements of $\mathcal{A}_j$. For such an $\BFM,$
if we fix $N_2 = \nu$, then we have $N_1 = d_j - 2\nu$. Thus, we can enumerate all $I(\BFM)$ such that $(N_1, N_2) \in \{(\max(d_j - 2\nu, 0), \nu)| \nu = 0, \ldots, \max\{\lceil \frac{d_i}{2} \rceil, |V_{j, 2}|\}, \; \forall j \neq i^*\}$ to find the smallest $\sum_{(i, \theta_b) \in I(\BFM)} \widetilde{G}_i(\theta_b)$ to characterize $F(\boldsymbol{\alpha})$. Once we find $I(\BFM) = A_1 \cup A_2$ such that $F(\boldsymbol{\alpha}) = \sum_{(i, \theta_b) \in I(\BFM)} \widetilde{G}_i(\theta_b)$,  we can construct its supergradient following the procedure in Algorithm~\ref{EC:alg_function_eval}.

\setcounter{algorithm}{5}
\begin{algorithm}[tbp!]
\caption{Computation of $F(\boldsymbol{\alpha})$ and its supergradient $\partial F(\boldsymbol{\alpha})$}\label{EC:alg_function_eval}
\begin{algorithmic}[1]
    \STATE \textbf{Initialization:} Given $\boldsymbol{\alpha}$, $\{y_{i}(\theta_b)\}$, $\{\lambda_i(\theta_b)\}$, true MPB $i^*$, and favorable set $\Theta_{i^*}$.
    \STATE Set $F^\textup{min} = \infty$ and evaluate $\widetilde{G}_i(\theta_b)$ for all $1\leq i \leq k, 1\leq b\leq B$ at $\boldsymbol{\alpha}$.
    \FOR{$j = 1$ to $k$ such that $j\neq i^*$}\label{alg:for_begin}
        \STATE Construct two sets $V_{j, \ell} := \{(i, \theta_b) \in \Xi: v_j[(i, \theta_b)]^+/p_b = \ell\}$ for $\ell = 1, 2$.
        \FOR{$N_2 = 0$ to $\min\{\lceil\frac{d_j}{2}\rceil, |V_{j, 2}|\}$}
            \STATE Construct $\mathcal{V} = \{A_2: A_2\subseteq V_{j, 2}, |A_2| = N_2\}$ and let $N_1 = \max(d_j - 2N_2, 0)$.\label{line:V_construction}
            \WHILE{$\mathcal{V}$ is non-empty}\label{line:while_start}
                \STATE Pick $A_2 \in \mathcal{V}$ and let $V_{j, 1}(A_2) = \{(i, \theta_b) \in V_{j, 1}: (i_1, \theta_b) \notin A_2, \;\; \forall 1\leq i_1 \leq k\}$. \label{line:pick_delete}
                \STATE Let ${\widetilde{V}}_{j, 1}(A_2) = \{(i, \theta_b) \in V_{j, 1}(A_2): i = \argmin_{\ell: (\ell,\theta_b) \in V_{j, 1}(A_2)} \widetilde{G}_{\ell}(\theta_b)\}$. \label{line:taking_cond_opt}
                \STATE Construct $A_1$ by selecting $N_1$ $(i,\theta_b)$ pairs in $\widetilde{V}_{j, 1}(A_2)$ with the smallest $\widetilde{G}_i(\theta_b)$.\label{line:minimal}
                \IF{$F^\text{min} > \sum_{(i, \theta_b) \in A_1\cup A_2}\widetilde{G}_i(\theta_b)$}
                    \STATE Update $F^\text{min} \leftarrow \sum_{(i, \theta_b) \in A_1\cup A_2}\widetilde{G}_i(\theta_b)$ and $\BFM^{\text{min}}$ such that $I(\BFM^{\text{min}}) = A_1 \cup A_2$.
                \ENDIF
                \STATE Update $\mathcal{V} \leftarrow \mathcal{V}- A_2$.
            \ENDWHILE\label{line:while_end}
        \ENDFOR
    \ENDFOR\label{alg:for_end}
    \RETURN $F(\boldsymbol{\alpha}) = F^\text{min}$ and $k\times B$ matrix $\BFg \in \partial F(\boldsymbol{\alpha})$, where its $(i, b)$th element is $\BFg_{i, b} = \sum_{(j, \theta_\bprime) \in I(\BFM^\text{min})}\frac{\partial\widetilde{G}_j(\theta_\bprime)}{\partial \alpha_{i}(\theta_b)}$.\label{line:subgrad}
\end{algorithmic}
\end{algorithm}

Note that $\mathcal{V}$ constructed in Line~\ref{line:V_construction} of Algorithm~\ref{EC:alg_function_eval} contains all size-$N_2$ subsets of $V_{j,2}$.  Line~\ref{line:pick_delete} first picks an arbitrary $A_2 \in \mathcal{V}$. The algorithm then proceeds to find size-$N_1$ $A_1 \subseteq V_{j, 1}$  with the smallest $\sum_{(i, \theta_b) \in A_1} \widetilde{G}_i(\theta_b)$. Instead of enumerating all size-$N_1$ subsets of $V_{j,1}$, we first remove all elements of $V_{j,1}$ that should not be included in $A_1$ according to the definition of $A_1$ to improve efficiency. Recall that $A_1$ and $A_2$ partition $I(\BFM)$, which means that the same $\theta_b$ cannot be included in $A_1$ and $A_2$ simultaneously. Thus, Line~\ref{line:pick_delete} first constructs a subset of $V_{j, 1}$ denoted by $V_{j,1}(A_2)$, which contains solution-parameter pairs whose parameters are not included in $A_2$. Note that multiple solution-parameter pairs in $V_{j,1}$ may contain the same $\theta_b$. Therefore, Line~\ref{line:taking_cond_opt} further improves efficiency by 
down-selecting the solution $\ell$ that minimizes 
$\widetilde{G}_\ell(\theta_b)$ for each $\theta_b$ included in $V_{j,1}$. 
Line~\ref{line:minimal} finally selects $N_1$ $(i,\theta_b)$ pairs in $\widetilde{V}_{j, 1}(A_2)$ with the smallest $\widetilde{G}_i(\theta_b)$. 

To summarize, Lines~\ref{line:pick_delete}--\ref{line:minimal} eliminate several 
$\BFM \in \bigcup_{j \neq i^*}\mathcal{A}_j$ before enumerating them to find $\BFM$ that minimizes $ \sum_{(i, \theta_b) \in I(\BFM)} \widetilde{G}_i(\theta_b)$. 
However, this procedure needs to be rerun whenever mean-variance configuration or $\boldsymbol{\alpha}$ changes.
This makes it challenging to apply Algorithm~\ref{EC:alg_function_eval} to develop a sequential sampling scheme, where the sample means and the sampling ratios dynamically change every iteration. 

Line~\ref{line:subgrad} of Algorithm~\ref{EC:alg_function_eval} returns a gradient at $\sum_{(i, \theta_b) \in I(\BFM^\text{min})} \widetilde{G}_i(\theta_b)$. To understand why this is a valid supergradient, consider the  function,  $f(x) = \max_{1\leq m\leq M}\{f_m(x)\}$, where  $f_m (1\leq m\leq M)$ is convex for each $m$. Let us define $\partial f_m(x)$ to be the subdifferential, i.e., the set of all subgradients of $f_m$ at $x$. Then, any vector lies within the convex hull of $\bigcup_{f_m(x) = f(x)}\partial f_m(x)$ is a subgradient of $f(x)$ at $x$~\citep{boyd2004convex}. 
Thus, once we find $m$ such that $f_m(x) = f(x)$ and $f_m(x)$ is differentiable at $x$, the gradient of $f_m(x)$ can be used as a subgradient of $f$ at $x$. 
Because the negative of a concave function is a convex function, one can derive a similar result by replacing the maximum in the definition of $f(x)$ above with the minimum.

We adopt the \emph{entropic mirror descent} algorithm by~\cite{beck2003mirror} to minimize $F(\bm\alpha)$; see Algorithm~\ref{EC:alg_EMD} for the full details. The authors provide an upper bound on the optimality gap in terms of hyperparameter $\{\gamma_t\}$ for any fixed $T$. However, we find the upper bound to be too loose so that an extremely large $T$ (e.g., $T = 10^{10}$) is required to obtain a desirable numerical gap such as $3\times 10^{-5}$. To resolve this, we modify the upper bound to be lower and computable, as in the following proposition.

\begin{algorithm}[tbp]
\caption{{Entropic mirror descent for solving $\max_{\boldsymbol{\alpha}} F(\boldsymbol{\alpha})$}}\label{EC:alg_EMD}
\begin{algorithmic}[1]
    \STATE \textbf{Initialization:} $\boldsymbol{\alpha}^1  = (\alpha^1_i(\theta_b))$ with $\alpha^1_i(\theta_b) = 1/kB$ for all $(i, \theta_b)$, iteration counter $t=1$, learning rate $\gamma_t$, and the maximum iteration $T$.
    \STATE Let $F^{\textrm{best}} = F(\boldsymbol{\alpha}^1)$ via Algorithm~\ref{EC:alg_function_eval}.
    \WHILE{ $t \leq T$}
        \STATE Evaluate $F(\boldsymbol{\alpha}^t)$ and pick $\BFg^t = (\BFg^t_{i, b}) \in \partial F(\boldsymbol{\alpha}^t)$ via Algorithm~\ref{EC:alg_function_eval}.
        \STATE Update $F^{\textrm{best}} \leftarrow \max\{F^{\textrm{best}}, F(\boldsymbol{\alpha}^t)\}$.
        \STATE Update $\boldsymbol{\alpha}^{t+1} =(\alpha^{t+1}_i(\theta_b))$ as $\alpha^{t+1}_i(\theta_b) = \frac{{\alpha}_i^{t}(\theta_b)\exp(\gamma_t \BFg^t_{i, b})}{\sum^k_{j=1}\sum^B_{\bprime = 1} {\alpha}_j^{t}(\theta_\bprime)\exp(\gamma_t \BFg^t_{j, \bprime})}$ for all $(i, \theta_b)$.
        \STATE $t \leftarrow t+1$.
    \ENDWHILE
    \RETURN $F^{\textrm{best}}$ as the optimal value.
\end{algorithmic}
\end{algorithm}

\begin{proposition}\label{prop:modified_upper}
Let $\{\boldsymbol{\alpha}_t\}$ be the sequence of solutions generated by Algorithm~\ref{EC:alg_EMD}. Then, we have
\begin{equation}\label{eq:modeified_upper}
    \max_{\boldsymbol{\alpha}}F(\boldsymbol{\alpha}) \leq \max_{1\leq t \leq T} F(\boldsymbol{\alpha}^t) + \frac{\log(kB) - \sum_{t = 1}^T \{\gamma_t} \langle \boldsymbol{\alpha}^t - \boldsymbol{\alpha}^{t+1}, \BFg^t\rangle + \mathsf{D}(\boldsymbol{\alpha}^{t+1}, \boldsymbol{\alpha}^t)\}{\sum_{t=1}^T \gamma_t},
\end{equation}
where $\mathsf{D}(\boldsymbol{\alpha}, \boldsymbol{\beta}):= \sum_{i=1}^k \sum_{b=1}^B \alpha_i(\theta_b)\log\lt(\frac{\alpha_i(\theta_b)}{\beta_i(\theta_b)}\rt)$, and $\langle \boldsymbol{\alpha}, \boldsymbol{\beta} \rangle = \sum_{i=1}^k \sum_{b=1}^B \alpha_i(\theta_b)\beta_i(\theta_b)$ is the Frobenius inner product, and $\BFg^t \in \partial F(\boldsymbol{\alpha}^t)$. 
\end{proposition}

\proof{Proof.}
Let $\boldsymbol{\alpha}^\textsf{opt} \in \argmax_{\boldsymbol{\alpha}} F(\boldsymbol{\alpha})$. Section~5 in~\cite{beck2003mirror} shows that we can apply the mirror descent algorithm based on $\mathsf{D}(\boldsymbol{\alpha}, \boldsymbol{\beta})$ because $\mathsf{D}(\boldsymbol{\alpha}, \boldsymbol{\beta}) = \psi(\boldsymbol{\alpha}) - \psi(\boldsymbol{\beta}) - \langle \boldsymbol{\alpha}-\boldsymbol{\beta}, \nabla\psi(\boldsymbol{\beta})\rangle $ holds when $\psi(\boldsymbol{\alpha}) = \sum_{i=1}^k \sum_{b=1}^B \alpha_i(\theta_b)\log \alpha_i(\theta_b)$. From the proof of Theorem 4.1 in~\cite{beck2003mirror} with $f = -F$, we can see that
\begin{equation}\label{eq:temp1}
    \gamma_t(F(\boldsymbol{\alpha}^\textsf{opt}) - F(\boldsymbol{\alpha}^t)) \leq \mathsf{D}(\boldsymbol{\alpha}^\textsf{opt}, \boldsymbol{\alpha}^t) - \mathsf{D}(\boldsymbol{\alpha}^\textsf{opt}, \boldsymbol{\alpha}^{t+1}) - \mathsf{D}(\boldsymbol{\alpha}^{t+1}, \boldsymbol{\alpha}^t) +\gamma_t \langle \boldsymbol{\alpha}^t - \boldsymbol{\alpha}^{t+1}, -\BFg^t\rangle.
\end{equation}
Note that the upper bound in \cite{beck2003mirror} further bounds the last term in~\eqref{eq:temp1}, which makes the resulting optimality gap bound too loose. Instead, we sum~\eqref{eq:temp1} over all $1\leq t \leq T$ to obtain
\begin{eqnarray}
    \sum_{t=1}^T \gamma_t(F(\boldsymbol{\alpha}^\textsf{opt}) - F(\boldsymbol{\alpha}^t)) &\leq \mathsf{D}(\boldsymbol{\alpha}^\textsf{opt}, \boldsymbol{\alpha}^1) - \mathsf{D}(\boldsymbol{\alpha}^\textsf{opt}, \boldsymbol{\alpha}^{T+1})  - \sum_{t = 1}^T\{\gamma_t \langle \boldsymbol{\alpha}^t - \boldsymbol{\alpha}^{t+1}, \BFg^t\rangle + \mathsf{D}(\boldsymbol{\alpha}^{t+1}, \boldsymbol{\alpha}^t)\} \nonumber\\
    & \leq \mathsf{D}(\boldsymbol{\alpha}^\textsf{opt}, \boldsymbol{\alpha}^1) - \sum_{t = 1}^T\{\gamma_t \langle \boldsymbol{\alpha}^t - \boldsymbol{\alpha}^{t+1}, \BFg^t\rangle + \mathsf{D}(\boldsymbol{\alpha}^{t+1}, \boldsymbol{\alpha}^t)\} \nonumber\\
    & \leq \log(kB) - \sum_{t = 1}^T\{\gamma_t \langle \boldsymbol{\alpha}^t - \boldsymbol{\alpha}^{t+1}, \BFg^t\rangle + \mathsf{D}(\boldsymbol{\alpha}^{t+1}, \boldsymbol{\alpha}^t)\}.\label{eq:temp_upper}
\end{eqnarray}
The last inequality holds since $\mathsf{D}(\boldsymbol{\alpha}^\textsf{opt}, \boldsymbol{\alpha}^1) = \log(kB) + \sum_{i=1}^k \sum_{b=1}^B \alpha^\textsf{opt}_i(\theta_b)\log\lt({\alpha^\textsf{opt}_i(\theta_b)}\rt) \leq \log(kB)$. Moreover, we have
\begin{equation}\label{eq:temp_lower}
    \sum_{t=1}^T \gamma_t(F(\boldsymbol{\alpha}^\textsf{opt}) - F(\boldsymbol{\alpha}^t)) \geq \lt(\sum_{t=1}^T \gamma_t\rt) \lt(F(\boldsymbol{\alpha}^\textsf{opt})- \max_{1\leq t \leq T} F(\boldsymbol{\alpha}^t)\rt),
\end{equation}
which completes the proof. \halmos
\endproof

\section{Synthetic example with unequal probability weights}\label{ec:unequal}
This section presents additional numerical results of when $\{p_b\}$ are unequal for a synthetic example to complement the results in Section~\ref{subsec:synthetic}. We define Scenario 4 by specifying 
\begin{equation}\label{eq:synthetic_opt_unequal}
    (i^b, p_b) = \begin{cases} (\ell, 0.016), & \mbox{ if } 5\ell-4\leq b \leq 5\ell, \mbox{ for some } 2\leq \ell \leq 7, \\ (\ell, 0.032), & \mbox{ if } 5\ell-4 \leq b \leq 5\ell, \mbox{ for some } 8\leq \ell \leq 9,\\ (10, 0.02), & \mbox{ if } 41 \leq b \leq 50. \\\end{cases}
\end{equation}
Therefore, $i^* = 10$ whose preference probability is $0.2$. All solutions except for Solution 1 is a conditional optimum for some $\theta_b$. For $y_i(\theta_b)$ and $\lambda_i(\theta_b)$, the same settings as in  Section~\ref{subsec:synthetic} are applied. Additionally, we define Scenario~5 by modifying $\{(i^b, p_b), 36\leq b\leq 50\}$ of Scenario~4 to $(i^b, p_b) = (9, 0.016)$ when $36\leq b\leq 45$ and $ = (10, 0.04)$ when $46\leq b \leq 50$. Note that Scenarios~4 and~5 have same preference probabilities for all solutions, however, Scenario~5 requires identifying a larger number of $i^b$ correctly than Scenario~4 since Scenario~5 has larger $|\Theta_{i^*}|$. Hence, Scenario~4 is an easier problem than Scenario~5 even if both have the same preference probabilities for all solutions. 


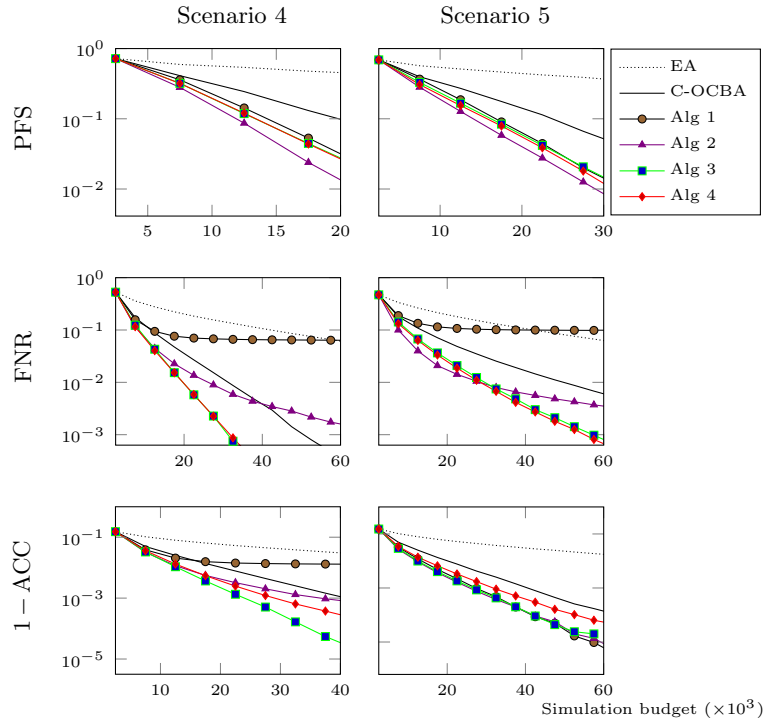
\begin{figure}[!tb]
	\centering
	\begin{tikzpicture}[font=\footnotesize]
	\begin{semilogyaxis}[name=plot1,height=1.5in,width=1.8in,
	title={Scenario 4},
	ylabel={PFS},
	legend style={at={(0.97, 0.97)},
		anchor=north east, font=\scriptsize, nodes=right, style={row sep=0.0cm}},ytickten={0, -1, -2, -3},
	ymin=0.41188643151E-02,
	ymax=1E0,
    xmin=2.5, xmax=20,
	every tick label/.append style={font=\tiny},
	axis on top,
	scaled x ticks = false,
	xticklabel style={/pgf/number format/fixed},
	/pgf/number format/1000 sep={}]
	
    \addplot+[black, densely dotted, mark = none]
	table[x index=0,y index=1, col sep=comma]{plotdata/PFS_Example5.csv};
    \addplot+[black, solid, mark = none]
	table[x index=0,y index=2, col sep=comma, ]{plotdata/PFS_Example5.csv};
    \addplot+[black, mark=*, mark size=1.5pt]
	table[x index=0,y index=3, col sep=comma]{plotdata/PFS_Example5.csv};
    \addplot+[violet, mark = triangle*, mark size=1.5pt]
	table[x index=0,y index=4, col sep=comma]{plotdata/PFS_Example5.csv};
    \addplot+[green, mark = square*, mark size=1.5pt]
	table[x index=0,y index=5, col sep=comma]{plotdata/PFS_Example5.csv};
    \addplot+[red, solid, mark = diamond*, mark size=1.5pt]
	table[x index=0,y index=6, col sep=comma]{plotdata/PFS_Example5.csv};
	
	\end{semilogyaxis}

    \begin{semilogyaxis}[name=plot2,height=1.5in,width=1.8in,at={($(plot1.east)+(0.2in,0in)$)},anchor=west,
	title={Scenario 5},
	legend style={at={(0.97, 0.97)},
		anchor=north east, font=\tiny, nodes=right, style={row sep=0.0cm}},ytickten={0, -1, -2},
	ymin=0.41188643151E-02,
	ymax=1E0,
    yticklabels={},
    xmin=2.5, xmax=30,
	every tick label/.append style={font=\tiny},
	axis on top,
	scaled x ticks = false,
	xticklabel style={/pgf/number format/fixed},
	/pgf/number format/1000 sep={},
 legend entries={EA, C-OCBA, Alg 1, Alg 2, Alg 3, Alg 4},
	legend pos=outer north east]

    \addplot+[black, densely dotted, mark = none]
	table[x index=0,y index=1, col sep=comma]{plotdata/PFS_Example6.csv};
    \addplot+[black, solid, mark = none]
	table[x index=0,y index=2, col sep=comma, ]{plotdata/PFS_Example6.csv};
    \addplot+[black, mark=*, mark size=1.5pt]
	table[x index=0,y index=3, col sep=comma]{plotdata/PFS_Example6.csv};
    \addplot+[violet, mark = triangle*, mark size=1.5pt]
	table[x index=0,y index=4, col sep=comma]{plotdata/PFS_Example6.csv};
    \addplot+[green, mark = square*, mark size=1.5pt]
	table[x index=0,y index=5, col sep=comma]{plotdata/PFS_Example6.csv};
    \addplot+[red, solid, mark = diamond*, mark size=1.5pt]
	table[x index=0,y index=6, col sep=comma]{plotdata/PFS_Example6.csv};

    \end{semilogyaxis}


    \begin{semilogyaxis}[name=plot3,height=1.5in,width=1.8in, at={($(plot1.south)+(0in,-1.2in)$)}, anchor = south,
	ylabel={FNR},
	legend style={at={(0.97, 0.97)},
		anchor=north east, font=\scriptsize, nodes=right, style={row sep=0.0cm}},ytickten={0, -1, -2, -3},
	ymin=0.00063095734,
	ymax=1E0,
    xmin=2.5, xmax=60,
	every tick label/.append style={font=\tiny},
	axis on top,
	scaled x ticks = false,
	xticklabel style={/pgf/number format/fixed},
	/pgf/number format/1000 sep={}]
	
    \addplot+[black, densely dotted, mark = none]
	table[x index=0,y index=1, col sep=comma]{plotdata/FNR_Example5.csv};
    \addplot+[black, solid, mark = none]
	table[x index=0,y index=2, col sep=comma, ]{plotdata/FNR_Example5.csv};
    \addplot+[black, mark=*, mark size=1.5pt]
	table[x index=0,y index=3, col sep=comma]{plotdata/FNR_Example5.csv};
    \addplot+[violet, mark = triangle*, mark size=1.5pt]
	table[x index=0,y index=4, col sep=comma]{plotdata/FNR_Example5.csv};
    \addplot+[green, mark = square*, mark size=1.5pt]
	table[x index=0,y index=5, col sep=comma]{plotdata/FNR_Example5.csv};
    \addplot+[red, solid, mark = diamond*, mark size=1.5pt]
	table[x index=0,y index=6, col sep=comma]{plotdata/FNR_Example5.csv};
	
	\end{semilogyaxis}

    \begin{semilogyaxis}[name=plot4,height=1.5in,width=1.8in,at={($(plot3.east)+(0.2in,0in)$)},anchor=west,
	legend style={at={(0.97, 0.97)},
		anchor=north east, font=\tiny, nodes=right, style={row sep=0.0cm}},ytickten={0, -1, -2, -3},
	ymin=0.00063095734,
	ymax=1E0,
    yticklabels={},
    xmin=2.5, xmax=60,
	every tick label/.append style={font=\tiny},
	axis on top,
	scaled x ticks = false,
	xticklabel style={/pgf/number format/fixed},
	/pgf/number format/1000 sep={}]

    \addplot+[black, densely dotted, mark = none]
	table[x index=0,y index=1, col sep=comma]{plotdata/FNR_Example4.csv};
    \addplot+[black, solid, mark = none]
	table[x index=0,y index=2, col sep=comma, ]{plotdata/FNR_Example4.csv};
    \addplot+[black, mark=*, mark size=1.5pt]
	table[x index=0,y index=3, col sep=comma]{plotdata/FNR_Example4.csv};
    \addplot+[violet, mark = triangle*, mark size=1.5pt]
	table[x index=0,y index=4, col sep=comma]{plotdata/FNR_Example4.csv};
    \addplot+[green, mark = square*, mark size=1.5pt]
	table[x index=0,y index=5, col sep=comma]{plotdata/FNR_Example4.csv};
    \addplot+[red, solid, mark = diamond*, mark size=1.5pt]
	table[x index=0,y index=6, col sep=comma]{plotdata/FNR_Example4.csv};

    \end{semilogyaxis}



    \begin{semilogyaxis}[name=plot5,height=1.5in,width=1.8in, at={($(plot3.south)+(0in,-1.2in)$)}, anchor = south,
	ylabel={$1-\text{ACC}$},
	legend style={at={(0.97, 0.97)},
		anchor=north east, font=\scriptsize, nodes=right, style={row sep=0.0cm}},ytickten={-1, -3, -5},
	ymin=0.00000316227,
	ymax=1E0,
    xmin=2.5, xmax=40,
	every tick label/.append style={font=\tiny},
	axis on top,
	scaled x ticks = false,
	xticklabel style={/pgf/number format/fixed},
	/pgf/number format/1000 sep={}]
	
    \addplot+[black, densely dotted, mark = none]
	table[x index=0,y index=1, col sep=comma]{plotdata/ACC_Example5.csv};
    \addplot+[black, solid, mark = none]
	table[x index=0,y index=2, col sep=comma, ]{plotdata/ACC_Example5.csv};
    \addplot+[black, mark=*, mark size=1.5pt]
	table[x index=0,y index=3, col sep=comma]{plotdata/ACC_Example5.csv};
    \addplot+[violet, mark = triangle*, mark size=1.5pt]
	table[x index=0,y index=4, col sep=comma]{plotdata/ACC_Example5.csv};
    \addplot+[green, mark = square*, mark size=1.5pt]
	table[x index=0,y index=5, col sep=comma]{plotdata/ACC_Example5.csv};
    \addplot+[red, solid, mark = diamond*, mark size=1.5pt]
	table[x index=0,y index=6, col sep=comma]{plotdata/ACC_Example5.csv};
	
	\end{semilogyaxis}

    \begin{semilogyaxis}[name=plot6,height=1.5in,width=1.8in,at={($(plot5.east)+(0.2in,0in)$)},anchor=west,
	every axis x label/.append style = {at={(1.2, 0.15)}, },
	xlabel={\tiny Simulation budget ($\times 10^3$)},
	legend style={at={(0.97, 0.97)},
		anchor=north east, font=\tiny, nodes=right, style={row sep=0.0cm}},ytickten={-1, -3, -5},
	ymin=6.30957344E-7,
	ymax=1E0,
    yticklabels={},
    xmin=2.5, xmax=60,
	every tick label/.append style={font=\tiny},
	axis on top,
	scaled x ticks = false,
    xticklabel style={/pgf/number format/fixed},
	/pgf/number format/1000 sep={}]

    \addplot+[black, densely dotted, mark = none]
	table[x index=0,y index=1, col sep=comma]{plotdata/ACC_Example6.csv};
    \addplot+[black, solid, mark = none]
	table[x index=0,y index=2, col sep=comma, ]{plotdata/ACC_Example6.csv};
    \addplot+[black, mark=*, mark size=1.5pt]
	table[x index=0,y index=3, col sep=comma]{plotdata/ACC_Example6.csv};
    \addplot+[violet, mark = triangle*, mark size=1.5pt]
	table[x index=0,y index=4, col sep=comma]{plotdata/ACC_Example6.csv};
    \addplot+[green, mark = square*, mark size=1.5pt]
	table[x index=0,y index=5, col sep=comma]{plotdata/ACC_Example6.csv};
    \addplot+[red, solid, mark = diamond*, mark size=1.5pt]
	table[x index=0,y index=6, col sep=comma]{plotdata/ACC_Example6.csv};

    \end{semilogyaxis}
	
	\end{tikzpicture}
 \caption{Performance measure estimates from 10{,}000 macro runs for Scenarios~4 and~5.}
    \label{fig:tot_unequal}
 \end{figure}

Figure~\ref{fig:tot_unequal} displays experiment results for Scenarios~4 and~5. Overall algorithmic behaviors are similar to those in Figure~\ref{fig:tot} showing robustness of Algorithms~\ref{alg:rate_opt}--4 for the case of unequal $\{p_b\}$. For PFS, observe that all algorithms converge faster for Scenario~4  than for Scenario~5, as expected. 
For the same simulation budget, all algorithms show slightly lower FNR and $1-$ACC for Scenario~5 than for Scenario~4.
This is because it is necessary to identify all elements of $\Theta_{i^*}$ correctly to find the correct MPB and  $|\Theta_{i^*}|$ is larger in Scenario~5. 
Thus, guarding against FS naturally forces all Algorithms 1--4 to estimate $\Theta_{i^*}$ more precisely, which also explains why Algorithm~2 is dominant in all three measures for Scenario~5 even if it is designed to minimize PFS only.
From these observations, we conclude that the algorithmic behaviors may depend on $\{p_b\}_{1\leq b\leq B}$ even if the preference probabilities for all solutions are identical.

\section{Further details on market simulation}\label{apdx:market_simulation}

\begin{figure}[hbt!]
    \centering
    \includegraphics[width = 0.4\textwidth]{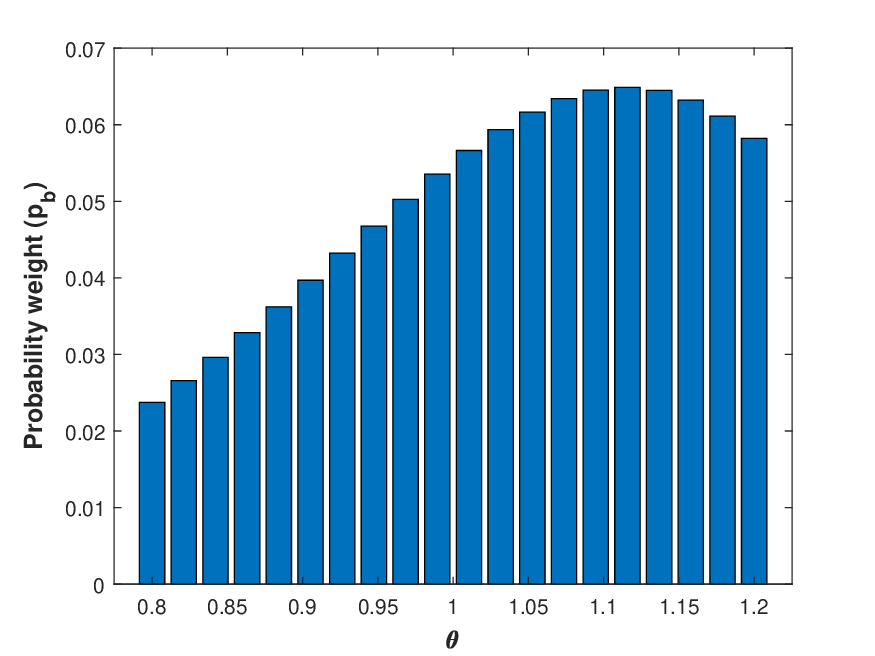}
    \caption{A probability distribution of $\theta_b$.}
    \label{fig:probability_weight}
\end{figure}

Figure~\ref{fig:probability_weight} displays the probability simplex that represents the uncertainty about $\theta_b$. The expected $\theta_b$ is approximately $1.1$, which implies that the customer dollar value is predicted to increase. As a result, the lower price would be preferred in the market.

Table~\ref{tab:mean_ft} shows the expected revenue MS at each solution-parameter pair. These quantities are estimated from $5\times 10^5$ Monte Carlo runs for each $(i, \theta_b)$. Additionally, we present the estimated no-purchase probability and sales revenue in Tables~\ref{tab:no_purchase_rate} and~\ref{tab:sales_rev}, respectively. The no-purchase probability and sales revenue can be written as
\begin{equation}
\begin{aligned}
    \text{Expected ratio of no purchase} & = \expec\lt[1 - \frac{\sum_{\ell=0}^4 N_{\ell}(\BFc)}{100}\rt], \\
    \text{Sales revenue of Prod $i$} & = \expec\lt[p_{i0}N_0(\BFc)\rt],
\end{aligned}
\end{equation}
where $N_{\ell}(\BFc)$ and $N_0(\BFc)$ are realized numbers of purchasing Firm $\ell$'s product as defined in Section~\ref{subsec:vco}. 

\begin{table}[hbt!]
\centering
\renewcommand{\arraystretch}{0.8}
\caption{Estimated mean revenue market share at each solution-parameter pair. The relative errors of all pairs are below $0.05\%$. }
\label{tab:mean_ft}
\resizebox{0.5\textwidth}{!}{\begin{tabular}{|c|lllllll|}
\hline
              & \multicolumn{1}{c}{Prod 1} & \multicolumn{1}{c}{Prod 2} & \multicolumn{1}{c}{Prod 3} & \multicolumn{1}{c}{Prod 4} & \multicolumn{1}{c}{Prod 5} & \multicolumn{1}{c}{Prod 6} & \multicolumn{1}{c|}{Prod 7} \\ \hline
$\theta_{1}$  & 0.27276                    & 0.22017                    & 0.15890                    & 0.25361                    & 0.20768                    & 0.21078                    & 0.20451                     \\
$\theta_{2}$  & 0.26710                    & 0.22202                    & 0.16567                    & 0.25189                    & 0.20933                    & 0.21016                    & 0.20853                     \\
$\theta_{3}$  & 0.26141                    & 0.22383                    & 0.17264                    & 0.25012                    & 0.21096                    & 0.20953                    & 0.21258                     \\
$\theta_{4}$  & 0.25574                    & 0.22555                    & 0.17975                    & 0.24832                    & 0.21253                    & 0.20886                    & 0.21660                     \\
$\theta_{5}$  & 0.25023                    & 0.22727                    & 0.18702                    & 0.24650                    & 0.21405                    & 0.20814                    & 0.22061                     \\
$\theta_{6}$  & 0.24471                    & 0.22891                    & 0.19444                    & 0.24467                    & 0.21556                    & 0.20737                    & 0.22460                     \\
$\theta_{7}$  & 0.23923                    & 0.23053                    & 0.20205                    & 0.24277                    & 0.21701                    & 0.20655                    & 0.22859                     \\
$\theta_{8}$  & 0.23382                    & 0.23208                    & 0.20978                    & 0.24083                    & 0.21840                    & 0.20563                    & 0.23254                     \\
$\theta_{9}$  & 0.22846                    & 0.23360                    & 0.21772                    & 0.23885                    & 0.21980                    & 0.20475                    & 0.23648                     \\
$\theta_{10}$ & 0.22320                    & 0.23509                    & 0.22572                    & 0.23683                    & 0.22111                    & 0.20380                    & 0.24042                     \\
$\theta_{11}$ & 0.21806                    & 0.23653                    & 0.23381                    & 0.23483                    & 0.22239                    & 0.20283                    & 0.24430                     \\
$\theta_{12}$ & 0.21288                    & 0.23788                    & 0.24202                    & 0.23284                    & 0.22359                    & 0.20181                    & 0.24821                     \\
$\theta_{13}$ & 0.20787                    & 0.23918                    & 0.25040                    & 0.23076                    & 0.22474                    & 0.20070                    & 0.25206                     \\
$\theta_{14}$ & 0.20287                    & 0.24039                    & 0.25881                    & 0.22861                    & 0.22580                    & 0.19961                    & 0.25592                     \\
$\theta_{15}$ & 0.19796                    & 0.24158                    & 0.26733                    & 0.22649                    & 0.22675                    & 0.19851                    & 0.25970                     \\
$\theta_{16}$ & 0.19314                    & 0.24276                    & 0.27596                    & 0.22439                    & 0.22776                    & 0.19743                    & 0.26343                     \\
$\theta_{17}$ & 0.18843                    & 0.24381                    & 0.28459                    & 0.22227                    & 0.22864                    & 0.19623                    & 0.26710                     \\
$\theta_{18}$ & 0.18381                    & 0.24484                    & 0.29332                    & 0.22009                    & 0.22953                    & 0.19502                    & 0.27069                     \\
$\theta_{19}$ & 0.17927                    & 0.24586                    & 0.30212                    & 0.21785                    & 0.23042                    & 0.19379                    & 0.27431                     \\
$\theta_{20}$ & 0.17476                    & 0.24678                    & 0.31094                    & 0.21563                    & 0.23125                    & 0.19249                    & 0.27788                     \\ \hline
\end{tabular}}
\end{table}

\begin{table}[hbt!]
\centering
\caption{Estimated mean no-purchase probabilities at each solution-parameter pair. The relative errors of all pairs are below $0.05\%$.  }
\label{tab:no_purchase_rate}
\resizebox{0.5\textwidth}{!}{\begin{tabular}{|c|lllllll|}
\hline
              & \multicolumn{1}{c}{Prod 1} & \multicolumn{1}{c}{Prod 2} & \multicolumn{1}{c}{Prod 3} & \multicolumn{1}{c}{Prod 4} & \multicolumn{1}{c}{Prod 5} & \multicolumn{1}{c}{Prod 6} & \multicolumn{1}{c|}{Prod 7} \\ \hline
$\theta_{1}$  & 0.1491                     & 0.1487                     & 0.1459                     & 0.1480                     & 0.1494                     & 0.1527                     & 0.1468                      \\
$\theta_{2}$  & 0.1640                     & 0.1623                     & 0.1579                     & 0.1622                     & 0.1631                     & 0.1670                     & 0.1598                      \\
$\theta_{3}$  & 0.1798                     & 0.1767                     & 0.1706                     & 0.1772                     & 0.1776                     & 0.1822                     & 0.1736                      \\
$\theta_{4}$  & 0.1964                     & 0.1918                     & 0.1837                     & 0.1931                     & 0.1929                     & 0.1982                     & 0.1880                      \\
$\theta_{5}$  & 0.2139                     & 0.2077                     & 0.1974                     & 0.2098                     & 0.2089                     & 0.2149                     & 0.2031                      \\
$\theta_{6}$  & 0.2322                     & 0.2243                     & 0.2116                     & 0.2273                     & 0.2256                     & 0.2324                     & 0.2188                      \\
$\theta_{7}$  & 0.2513                     & 0.2415                     & 0.2262                     & 0.2454                     & 0.2430                     & 0.2506                     & 0.2352                      \\
$\theta_{8}$  & 0.2710                     & 0.2595                     & 0.2413                     & 0.2644                     & 0.2610                     & 0.2694                     & 0.2522                      \\
$\theta_{9}$  & 0.2913                     & 0.2779                     & 0.2567                     & 0.2838                     & 0.2797                     & 0.2888                     & 0.2697                      \\
$\theta_{10}$ & 0.3121                     & 0.2968                     & 0.2725                     & 0.3038                     & 0.2987                     & 0.3087                     & 0.2876                      \\
$\theta_{11}$ & 0.3334                     & 0.3162                     & 0.2886                     & 0.3242                     & 0.3183                     & 0.3290                     & 0.3059                      \\
$\theta_{12}$ & 0.3548                     & 0.3359                     & 0.3048                     & 0.3450                     & 0.3381                     & 0.3496                     & 0.3246                      \\
$\theta_{13}$ & 0.3766                     & 0.3559                     & 0.3213                     & 0.3661                     & 0.3583                     & 0.3704                     & 0.3436                      \\
$\theta_{14}$ & 0.3984                     & 0.3761                     & 0.3379                     & 0.3872                     & 0.3786                     & 0.3914                     & 0.3628                      \\
$\theta_{15}$ & 0.4203                     & 0.3964                     & 0.3547                     & 0.4085                     & 0.3991                     & 0.4125                     & 0.3821                      \\
$\theta_{16}$ & 0.4422                     & 0.4168                     & 0.3715                     & 0.4300                     & 0.4197                     & 0.4338                     & 0.4015                      \\
$\theta_{17}$ & 0.4641                     & 0.4373                     & 0.3882                     & 0.4513                     & 0.4403                     & 0.4549                     & 0.4210                      \\
$\theta_{18}$ & 0.4858                     & 0.4577                     & 0.4050                     & 0.4726                     & 0.4608                     & 0.4760                     & 0.4405                      \\
$\theta_{19}$ & 0.5073                     & 0.4781                     & 0.4218                     & 0.4937                     & 0.4813                     & 0.4969                     & 0.4599                      \\
$\theta_{20}$ & 0.5283                     & 0.4983                     & 0.4385                     & 0.5146                     & 0.5017                     & 0.5175                     & 0.4794                      \\ \hline
\end{tabular}}
\end{table}

\begin{table}[hbt!]
\centering
\caption{Estimated mean sales revenue at each solution-parameter pair. The relative errors of all pairs are below $0.05\%$.}
\label{tab:sales_rev}
\resizebox{0.5\textwidth}{!}{\begin{tabular}{|c|lllllll|}
\hline
              & \multicolumn{1}{c}{Prod 1} & \multicolumn{1}{c}{Prod 2} & \multicolumn{1}{c}{Prod 3} & \multicolumn{1}{c}{Prod 4} & \multicolumn{1}{c}{Prod 5} & \multicolumn{1}{c}{Prod 6} & \multicolumn{1}{c|}{Prod 7} \\ \hline
$\theta_{1}$  & 9412.90                    & 6981.05                    & 4546.33                    & 8416.91                    & 6586.10                    & 6858.40                    & 6270.67                     \\
$\theta_{2}$  & 9026.70                    & 6912.49                    & 4641.37                    & 8203.47                    & 6518.00                    & 6708.66                    & 6277.88                     \\
$\theta_{3}$  & 8639.09                    & 6834.30                    & 4731.37                    & 7982.58                    & 6441.53                    & 6553.44                    & 6276.57                     \\
$\theta_{4}$  & 8253.18                    & 6745.91                    & 4814.57                    & 7755.30                    & 6355.25                    & 6391.27                    & 6265.09                     \\
$\theta_{5}$  & 7872.78                    & 6649.45                    & 4891.12                    & 7522.61                    & 6260.22                    & 6223.04                    & 6243.62                     \\
$\theta_{6}$  & 7493.86                    & 6542.87                    & 4960.21                    & 7285.40                    & 6157.19                    & 6048.36                    & 6212.63                     \\
$\theta_{7}$  & 7119.43                    & 6428.32                    & 5022.80                    & 7042.43                    & 6046.15                    & 5868.74                    & 6172.03                     \\
$\theta_{8}$  & 6752.20                    & 6304.77                    & 5076.93                    & 6795.35                    & 5926.36                    & 5683.19                    & 6121.10                     \\
$\theta_{9}$  & 6391.10                    & 6174.47                    & 5124.32                    & 6545.98                    & 5801.21                    & 5496.16                    & 6061.38                     \\
$\theta_{10}$ & 6039.25                    & 6037.01                    & 5162.39                    & 6294.49                    & 5668.70                    & 5305.18                    & 5993.27                     \\
$\theta_{11}$ & 5697.86                    & 5893.82                    & 5191.97                    & 6043.73                    & 5529.75                    & 5113.12                    & 5916.04                     \\
$\theta_{12}$ & 5363.85                    & 5743.16                    & 5213.83                    & 5793.52                    & 5385.09                    & 4919.32                    & 5830.96                     \\
$\theta_{13}$ & 5042.77                    & 5588.07                    & 5227.94                    & 5542.82                    & 5235.21                    & 4723.49                    & 5738.24                     \\
$\theta_{14}$ & 4731.73                    & 5427.40                    & 5233.15                    & 5293.89                    & 5081.28                    & 4529.84                    & 5638.33                     \\
$\theta_{15}$ & 4432.31                    & 5264.04                    & 5230.70                    & 5048.98                    & 4922.56                    & 4337.63                    & 5531.49                     \\
$\theta_{16}$ & 4145.92                    & 5098.30                    & 5220.36                    & 4807.93                    & 4763.61                    & 4147.36                    & 5418.20                     \\
$\theta_{17}$ & 3871.96                    & 4928.39                    & 5202.33                    & 4571.62                    & 4600.75                    & 3958.11                    & 5298.73                     \\
$\theta_{18}$ & 3610.59                    & 4757.94                    & 5177.06                    & 4339.36                    & 4437.96                    & 3771.38                    & 5173.65                     \\
$\theta_{19}$ & 3361.70                    & 4587.12                    & 5145.05                    & 4111.29                    & 4275.01                    & 3588.21                    & 5045.44                     \\
$\theta_{20}$ & 3124.99                    & 4415.03                    & 5105.73                    & 3890.15                    & 4111.45                    & 3408.55                    & 4912.59                     \\ \hline
\end{tabular}}
\end{table}

\begin{table}[hbt!]
\centering
\renewcommand{\arraystretch}{0.9}
\centering
\caption{Part-worth utilities for Person ID 1 to 50.}
\resizebox{0.8\textwidth}{!}{\begin{tabular}{|c|ccc|ccc|ccc|ccc|ccc|c|c|}
\hline
\multirow{2}{*}{Person ID} & \multicolumn{3}{c|}{Atrribute 1} & \multicolumn{3}{c|}{Atrribute 2} & \multicolumn{3}{c|}{Atrribute 3} & \multicolumn{3}{c|}{Atrribute 4} & \multicolumn{3}{c|}{Atrribute 5} & \multirow{2}{*}{Brand} & \multirow{2}{*}{Price} \\ \cline{2-16}
                           & Level 1   & Level 2   & Level 3  & Level 1   & Level 2   & Level 3  & Level 1   & Level 2   & Level 3  & Level 1   & Level 2   & Level 3  & Level 1   & Level 2   & Level 3  &                        &                        \\ \hline
1                          & 0.7485    & 1.4975    & 2.8942   & 0.2627    & 0.4020    & 0.5365   & 0.2805    & 0.5584    & 0.9540   & 0.1521    & 0.6204    & 1.1181   & 0.2683    & 0.6421    & 0.7566   & 0.0013                 & -0.0190                \\
2                          & 0.5555    & 1.4246    & 2.1816   & 0.3360    & 0.4514    & 0.5014   & 0.3962    & 0.8144    & 1.1762   & 0.1730    & 0.3980    & 1.0982   & 0.3784    & 0.3849    & 0.6624   & 0.0005                 & -0.0168                \\
3                          & 1.8821    & 2.0716    & 2.7521   & 0.2151    & 0.3719    & 0.4682   & 0.5014    & 0.5778    & 0.8112   & 0.1459    & 0.2688    & 0.3819   & 0.5669    & 0.8833    & 0.9761   & 0.0005                 & -0.0095                \\
4                          & 0.5299    & 1.0186    & 2.0850   & 0.2346    & 0.4562    & 0.4836   & 0.6169    & 0.7331    & 1.3286   & 0.2159    & 0.7652    & 0.8423   & 0.2346    & 0.4321    & 0.5384   & 0.0018                 & -0.0157                \\
5                          & 0.9872    & 1.7764    & 2.3515   & 0.3389    & 0.3460    & 0.5650   & 0.2501    & 0.9749    & 0.9922   & 0.4055    & 0.4645    & 1.0077   & 1.2845    & 1.4967    & 1.6817   & 0.0005                 & -0.0158                \\
6                          & 0.5297    & 1.2890    & 2.1414   & 0.3428    & 0.3572    & 0.5348   & 0.2712    & 0.5245    & 1.0707   & 0.1299    & 0.3161    & 0.3777   & 0.4890    & 0.8045    & 0.9258   & 0.0013                 & -0.0146                \\
7                          & 1.0811    & 1.1288    & 2.0340   & 0.2882    & 0.5168    & 0.5466   & 0.3598    & 0.7703    & 0.9067   & 0.1954    & 0.2897    & 0.5116   & 0.1889    & 0.2789    & 0.4442   & 0.0006                 & -0.0109                \\
8                          & 0.8671    & 1.5268    & 2.3854   & 0.2589    & 0.4066    & 0.4692   & 0.4091    & 0.8655    & 1.3052   & 0.1265    & 0.2935    & 0.5520   & 0.1606    & 0.5524    & 0.6231   & 0.0012                 & -0.0159                \\
9                          & 0.6708    & 1.3621    & 2.1285   & 0.2816    & 0.3107    & 0.4519   & 0.4105    & 1.2916    & 1.6756   & 0.1476    & 0.2533    & 0.5344   & 0.2564    & 0.6047    & 0.8057   & 0.0008                 & -0.0171                \\
10                         & 1.0620    & 1.6128    & 2.1591   & 0.1558    & 0.3729    & 0.5748   & 0.4075    & 0.7460    & 0.9405   & 0.2753    & 0.4210    & 0.7970   & 0.2859    & 0.4920    & 0.5055   & 0.0006                 & -0.0135                \\
11                         & 0.6565    & 1.5002    & 2.0440   & 0.2060    & 0.3892    & 0.5182   & 0.2842    & 0.6052    & 0.9374   & 0.3538    & 0.5074    & 0.9179   & 0.5369    & 0.6463    & 0.6599   & 0.0012                 & -0.0154                \\
12                         & 1.8005    & 1.9586    & 2.3032   & 0.2233    & 0.3250    & 0.5852   & 0.2830    & 0.5338    & 0.8742   & 0.1662    & 0.3195    & 0.3887   & 0.1384    & 0.4475    & 0.5492   & 0.0006                 & -0.0106                \\
13                         & 1.1585    & 1.3136    & 2.1946   & 0.1678    & 0.4602    & 0.5539   & 0.2955    & 0.5839    & 1.0791   & 0.2507    & 0.4981    & 0.6117   & 0.3374    & 0.4656    & 0.4780   & 0.0010                 & -0.0123                \\
14                         & 1.4831    & 1.7591    & 2.0295   & 0.1978    & 0.4885    & 0.4925   & 0.3746    & 0.5507    & 0.8072   & 0.2822    & 0.4775    & 0.5999   & 0.3914    & 0.4612    & 0.4641   & 0.0007                 & -0.0095                \\
15                         & 0.9471    & 1.5361    & 2.0568   & 0.2342    & 0.3074    & 0.5333   & 0.2838    & 0.5197    & 0.8785   & 0.2227    & 0.4289    & 0.5417   & 0.5550    & 0.6266    & 0.8384   & 0.0013                 & -0.0135                \\
16                         & 1.1430    & 1.6385    & 2.0603   & 0.3643    & 0.4560    & 0.5837   & 0.5081    & 0.8755    & 0.9402   & 0.1459    & 0.2705    & 0.6478   & 0.4378    & 0.6363    & 0.6884   & 0.0005                 & -0.0118                \\
17                         & 1.0607    & 1.6292    & 2.5980   & 0.2279    & 0.3131    & 0.4848   & 0.3494    & 0.6060    & 0.9475   & 0.5326    & 0.8872    & 1.2893   & 0.3025    & 0.5398    & 0.5637   & 0.0012                 & -0.0155                \\
18                         & 0.8693    & 1.1041    & 2.2684   & 0.1594    & 0.5082    & 0.5781   & 0.2570    & 0.7135    & 1.2572   & 0.2736    & 0.3489    & 0.4651   & 0.2538    & 0.5780    & 0.6510   & 0.0011                 & -0.0145                \\
19                         & 0.6313    & 1.5008    & 2.1846   & 0.3008    & 0.4186    & 0.5046   & 0.3099    & 0.6066    & 1.2659   & 0.2911    & 0.3089    & 0.5849   & 0.3198    & 0.3249    & 0.5103   & 0.0011                 & -0.0148                \\
20                         & 1.1260    & 1.3561    & 2.0708   & 0.1780    & 0.3532    & 0.5522   & 0.4208    & 0.5928    & 0.9391   & 0.1654    & 0.4462    & 0.5971   & 0.2753    & 0.4674    & 0.5272   & 0.0009                 & -0.0118                \\
21                         & 0.1578    & 0.3050    & 0.6365   & 1.5740    & 1.8612    & 2.2676   & 0.3792    & 0.6664    & 0.7683   & 0.3343    & 0.5102    & 0.8327   & 0.2536    & 0.3753    & 0.4623   & 0.0002                 & -0.0114                \\
22                         & 0.1564    & 0.3034    & 0.5251   & 1.0735    & 1.6164    & 2.0193   & 0.6567    & 0.8526    & 1.4890   & 0.2587    & 0.6425    & 0.9493   & 0.2705    & 0.4247    & 0.6164   & 0.0003                 & -0.0149                \\
23                         & 0.1562    & 0.3009    & 0.5489   & 0.5324    & 1.7510    & 2.3262   & 0.3736    & 1.1346    & 1.6247   & 0.2686    & 0.5213    & 1.1293   & 0.2344    & 0.4730    & 0.5643   & 0.0003                 & -0.0220                \\
24                         & 0.1870    & 0.4048    & 0.6304   & 1.2644    & 1.4546    & 2.1306   & 0.3523    & 0.7164    & 0.8550   & 0.4196    & 0.6173    & 0.9079   & 0.3444    & 0.3966    & 0.5888   & 0.0001                 & -0.0125                \\
25                         & 0.3225    & 0.3646    & 0.4546   & 1.1403    & 1.7551    & 2.1912   & 0.4458    & 0.5752    & 0.8740   & 0.3272    & 0.5545    & 0.7728   & 0.3825    & 0.5005    & 0.5130   & 0.0003                 & -0.0118                \\
26                         & 0.2216    & 0.3045    & 0.6564   & 1.4343    & 1.7851    & 2.0303   & 0.2569    & 0.7201    & 0.8749   & 0.5414    & 0.6646    & 1.0050   & 0.3634    & 0.4900    & 0.4981   & 0.0001                 & -0.0113                \\
27                         & 0.2011    & 0.3435    & 0.6536   & 1.3024    & 1.6326    & 2.0361   & 0.5535    & 0.5849    & 0.7967   & 0.3083    & 0.7453    & 0.8543   & 0.3178    & 0.4014    & 0.6200   & 0.0005                 & -0.0114                \\
28                         & 0.2497    & 0.4469    & 0.4558   & 1.4464    & 1.6915    & 2.1717   & 0.4629    & 0.9533    & 1.0862   & 0.3570    & 0.6119    & 0.7506   & 0.1792    & 0.2525    & 0.5825   & 0.0000                 & -0.0121                \\
29                         & 0.2201    & 0.4902    & 0.6427   & 0.6046    & 1.1177    & 2.1616   & 0.5380    & 0.9971    & 1.5596   & 0.3659    & 0.5618    & 0.7721   & 0.2317    & 0.4258    & 0.4759   & 0.0001                 & -0.0166                \\
30                         & 0.1856    & 0.3678    & 0.7729   & 1.9917    & 2.0928    & 2.2819   & 0.5086    & 0.6773    & 0.7900   & 0.2995    & 0.5435    & 0.8434   & 0.1631    & 0.4523    & 0.5350   & 0.0004                 & -0.0106                \\
31                         & 0.2233    & 0.4186    & 0.4626   & 0.6491    & 1.1810    & 2.2032   & 0.3000    & 0.5103    & 1.1438   & 0.2652    & 0.7555    & 0.8720   & 0.4281    & 0.5270    & 0.5662   & 0.0001                 & -0.0155                \\
32                         & 0.3343    & 0.4250    & 0.6252   & 0.8732    & 1.6407    & 2.6684   & 0.4090    & 0.5627    & 1.0338   & 0.4316    & 1.1151    & 1.3415   & 0.2837    & 0.3922    & 0.5578   & 0.0001                 & -0.0175                \\
33                         & 0.2332    & 0.4481    & 0.4992   & 0.8376    & 1.2509    & 2.1039   & 0.5743    & 0.7688    & 1.4396   & 0.2810    & 0.7458    & 0.7839   & 0.2856    & 0.3840    & 0.3858   & 0.0002                 & -0.0140                \\
34                         & 0.1978    & 0.4849    & 0.4999   & 1.8630    & 2.2289    & 2.8489   & 0.4243    & 0.8057    & 1.0833   & 0.6237    & 0.6245    & 0.8321   & 0.6378    & 0.7447    & 0.8245   & 0.0002                 & -0.0124                \\
35                         & 0.2575    & 0.3110    & 0.5861   & 0.7920    & 1.3630    & 2.1375   & 0.2877    & 0.5030    & 1.0199   & 0.2861    & 0.7635    & 0.8026   & 0.3041    & 0.3953    & 0.5128   & 0.0003                 & -0.0143                \\
36                         & 0.1754    & 0.5984    & 0.6750   & 1.0901    & 1.4195    & 2.2372   & 0.3241    & 0.7809    & 1.1493   & 0.3798    & 0.5576    & 0.7538   & 0.2685    & 0.3582    & 0.4851   & 0.0002                 & -0.0147                \\
37                         & 0.1592    & 0.4604    & 0.4693   & 1.0649    & 1.2499    & 2.0105   & 0.5076    & 0.5395    & 0.9355   & 0.3017    & 0.7435    & 0.8152   & 0.4988    & 0.6056    & 0.6295   & 0.0002                 & -0.0115                \\
38                         & 0.3643    & 0.4268    & 0.5779   & 0.8137    & 1.2121    & 2.0904   & 0.5111    & 0.6201    & 1.0742   & 0.2877    & 0.7463    & 0.8681   & 0.2754    & 0.5028    & 0.5874   & 0.0001                 & -0.0134                \\
39                         & 0.1893    & 0.4497    & 0.7091   & 1.0989    & 1.6009    & 2.2613   & 0.2732    & 0.5193    & 0.9288   & 0.3765    & 0.5454    & 0.8698   & 0.3869    & 0.6741    & 0.7225   & 0.0004                 & -0.0153                \\
40                         & 0.2852    & 0.3505    & 0.4932   & 1.1580    & 1.8046    & 2.2617   & 0.4719    & 0.6194    & 1.1009   & 0.2820    & 0.7806    & 0.7933   & 0.1609    & 0.2798    & 0.4314   & 0.0007                 & -0.0138                \\
41                         & 0.4072    & 0.5326    & 0.9902   & 0.2345    & 0.5756    & 0.8656   & 1.3652    & 1.8757    & 2.0468   & 0.2815    & 0.4164    & 0.4810   & 0.2495    & 0.3295    & 0.6281   & 0.0002                 & -0.0116                \\
42                         & 0.3370    & 0.6156    & 0.8176   & 0.2056    & 0.4570    & 1.0284   & 1.2979    & 1.8977    & 2.1128   & 0.3169    & 0.3589    & 0.5106   & 0.2788    & 0.2800    & 0.4840   & 0.0001                 & -0.0125                \\
43                         & 0.5391    & 0.7505    & 1.0833   & 0.1619    & 0.7489    & 1.3450   & 1.3528    & 2.2486    & 2.6686   & 0.2144    & 0.5677    & 0.6220   & 0.2282    & 0.3037    & 0.8597   & 0.0005                 & -0.0188                \\
44                         & 0.4870    & 0.5332    & 1.1627   & 0.2377    & 0.6206    & 0.6334   & 1.2695    & 1.8428    & 2.0796   & 0.2580    & 0.3807    & 0.5382   & 0.2622    & 0.6303    & 0.8199   & 0.0004                 & -0.0125                \\
45                         & 0.2657    & 0.6330    & 1.1618   & 0.4524    & 0.8200    & 0.8512   & 0.6347    & 1.1304    & 2.2392   & 0.4539    & 0.4926    & 0.4929   & 0.2624    & 0.4484    & 0.7227   & 0.0009                 & -0.0151                \\
46                         & 0.3297    & 0.7457    & 0.8909   & 0.4080    & 0.7516    & 1.1499   & 1.3418    & 1.5610    & 2.1898   & 0.3006    & 0.3540    & 0.6682   & 0.1822    & 0.3419    & 0.3987   & 0.0000                 & -0.0127                \\
47                         & 0.2762    & 0.6309    & 0.8075   & 0.4369    & 0.6931    & 0.9302   & 1.2392    & 1.6126    & 2.5856   & 0.3619    & 0.4053    & 0.4348   & 0.3745    & 0.7674    & 0.7784   & 0.0003                 & -0.0128                \\
48                         & 0.2898    & 0.5712    & 0.8267   & 0.2573    & 0.5877    & 1.1261   & 1.6558    & 2.0047    & 2.1915   & 0.2543    & 0.4893    & 0.7822   & 0.1334    & 0.3286    & 0.4540   & 0.0000                 & -0.0132                \\
49                         & 0.2691    & 0.7399    & 0.7544   & 0.3739    & 0.5802    & 1.4418   & 1.7158    & 2.1471    & 2.5744   & 0.1261    & 0.3877    & 0.4436   & 0.2693    & 0.3047    & 0.4690   & 0.0002                 & -0.0141                \\
50                         & 0.4179    & 0.5746    & 0.8183   & 0.2184    & 0.6647    & 0.8756   & 1.7087    & 1.8426    & 2.5133   & 0.3318    & 0.4232    & 0.6966   & 0.1908    & 0.3365    & 0.4442   & 0.0001                 & -0.0116                \\ \hline
\end{tabular}}
\label{tab:Utility_part1}
\end{table}

\begin{table}[hbt!]
\centering
\renewcommand{\arraystretch}{0.9}
\caption{Part-worth utilities for Person ID 51 to 100.}
\resizebox{0.8\textwidth}{!}{\begin{tabular}{|c|ccc|ccc|ccc|ccc|ccc|c|c|}
\hline
\multirow{2}{*}{Person ID} & \multicolumn{3}{c|}{Atrribute 1} & \multicolumn{3}{c|}{Atrribute 2} & \multicolumn{3}{c|}{Atrribute 3} & \multicolumn{3}{c|}{Atrribute 4} & \multicolumn{3}{c|}{Atrribute 5} & \multirow{2}{*}{Brand} & \multirow{2}{*}{Price} \\ \cline{2-16}
                           & Level 1   & Level 2   & Level 3  & Level 1   & Level 2   & Level 3  & Level 1   & Level 2   & Level 3  & Level 1   & Level 2   & Level 3  & Level 1   & Level 2   & Level 3  &                        &                        \\ \hline
51                         & 0.3424    & 0.5813    & 0.9729   & 0.2855    & 0.6815    & 0.7298   & 1.1016    & 1.3359    & 2.0735   & 0.1731    & 0.3380    & 0.5728   & 0.1772    & 0.5030    & 0.5459   & 0.0004                 & -0.0129                \\
52                         & 0.3457    & 0.5329    & 0.9438   & 0.1936    & 0.4174    & 0.5977   & 1.6995    & 2.2526    & 2.6267   & 0.3191    & 0.5821    & 0.6521   & 0.1886    & 0.2844    & 0.4307   & 0.0006                 & -0.0123                \\
53                         & 0.2662    & 0.5017    & 0.9308   & 0.2600    & 0.6725    & 0.7167   & 2.0299    & 2.0357    & 2.7924   & 0.1482    & 0.3140    & 0.4354   & 0.2097    & 0.3706    & 0.4052   & 0.0007                 & -0.0114                \\
54                         & 0.2802    & 0.5773    & 0.8963   & 0.5951    & 0.8589    & 1.3853   & 1.8355    & 2.3054    & 2.8002   & 0.6727    & 0.7645    & 0.8004   & 0.2698    & 0.3232    & 0.4192   & 0.0007                 & -0.0132                \\
55                         & 0.3532    & 0.8092    & 2.0664   & 0.1912    & 0.6144    & 0.7141   & 0.7887    & 1.4796    & 2.1017   & 0.3615    & 0.4386    & 0.5992   & 0.2122    & 0.4790    & 0.5487   & 0.0012                 & -0.0175                \\
56                         & 0.5753    & 0.6177    & 1.1018   & 0.5815    & 0.8800    & 0.9722   & 1.4021    & 1.9666    & 2.2141   & 0.1938    & 0.3391    & 0.4881   & 0.1320    & 0.3402    & 0.4378   & 0.0003                 & -0.0111                \\
57                         & 0.3007    & 0.5882    & 0.8167   & 0.1823    & 0.5778    & 0.8441   & 0.9147    & 1.2027    & 2.1101   & 0.1952    & 0.3596    & 0.4353   & 0.1541    & 0.4591    & 0.5107   & 0.0005                 & -0.0136                \\
58                         & 0.3705    & 0.5586    & 1.0873   & 0.2684    & 0.6041    & 0.8136   & 1.1905    & 1.9836    & 2.2176   & 0.2661    & 0.4351    & 0.6397   & 0.3502    & 0.4691    & 0.7446   & 0.0004                 & -0.0143                \\
59                         & 0.2590    & 0.6255    & 0.7595   & 0.2776    & 0.5059    & 0.8499   & 1.3031    & 1.3430    & 2.2655   & 0.1948    & 0.3825    & 0.6001   & 0.2016    & 0.4571    & 0.5361   & 0.0002                 & -0.0126                \\
60                         & 0.2596    & 0.6077    & 0.8511   & 0.4602    & 0.8319    & 1.2705   & 1.0329    & 1.5966    & 2.1749   & 0.2112    & 0.3516    & 0.4631   & 0.1755    & 0.4653    & 0.6174   & 0.0008                 & -0.0159                \\
61                         & 0.2752    & 0.6705    & 1.2540   & 0.3435    & 0.3694    & 0.6794   & 0.4947    & 0.6219    & 0.6653   & 1.8323    & 2.0596    & 2.6435   & 0.5705    & 1.0240    & 1.0245   & 0.0002                 & -0.0125                \\
62                         & 0.5394    & 0.8442    & 0.9058   & 0.2464    & 0.4151    & 0.8748   & 0.4006    & 0.4336    & 0.5201   & 1.4211    & 2.0460    & 2.5739   & 0.6099    & 0.8006    & 0.8025   & 0.0000                 & -0.0124                \\
63                         & 0.3482    & 0.5758    & 1.1654   & 0.2574    & 0.5546    & 0.7363   & 0.4602    & 0.5630    & 0.6205   & 0.8435    & 1.4912    & 2.0275   & 0.6037    & 0.8732    & 0.9055   & 0.0004                 & -0.0138                \\
64                         & 0.5232    & 0.7976    & 1.4313   & 0.2704    & 0.3840    & 0.6304   & 0.4939    & 0.6580    & 0.9009   & 1.3276    & 2.4321    & 2.6656   & 0.2669    & 0.5250    & 0.7893   & 0.0002                 & -0.0146                \\
65                         & 0.5505    & 0.9618    & 1.6547   & 0.2303    & 0.4123    & 0.5301   & 0.4576    & 0.4725    & 0.8002   & 0.9049    & 1.9744    & 2.1879   & 0.4991    & 0.5811    & 0.7742   & 0.0011                 & -0.0151                \\
66                         & 0.3769    & 0.5298    & 0.9082   & 0.3362    & 0.4677    & 0.8314   & 0.5228    & 0.6313    & 0.8156   & 2.3436    & 3.0670    & 3.3131   & 0.4849    & 0.7788    & 0.9847   & 0.0003                 & -0.0127                \\
67                         & 0.4356    & 0.5240    & 0.8872   & 0.1678    & 0.3787    & 0.6959   & 0.3634    & 0.7740    & 0.8174   & 1.6478    & 2.1969    & 2.2979   & 0.4681    & 0.5283    & 0.9946   & 0.0002                 & -0.0123                \\
68                         & 0.3118    & 0.6679    & 1.5616   & 0.3860    & 0.4820    & 0.4979   & 0.2078    & 0.4822    & 0.5792   & 1.2193    & 1.9989    & 2.3429   & 0.3744    & 0.9744    & 1.3631   & 0.0004                 & -0.0157                \\
69                         & 0.2657    & 0.6089    & 1.6889   & 0.1591    & 0.9703    & 0.9885   & 0.4122    & 0.4430    & 0.8520   & 0.8134    & 2.1776    & 2.3155   & 0.2556    & 0.5265    & 1.0022   & 0.0013                 & -0.0209                \\
70                         & 0.4423    & 0.5132    & 0.7706   & 0.1712    & 0.5039    & 1.0065   & 0.5436    & 0.5836    & 0.9676   & 1.4960    & 2.2634    & 2.3728   & 0.3965    & 0.8037    & 1.0035   & 0.0003                 & -0.0138                \\
71                         & 0.3297    & 0.7421    & 1.2518   & 0.1603    & 0.3319    & 0.5611   & 0.1924    & 0.8502    & 0.8750   & 1.7441    & 2.1764    & 2.5123   & 0.5175    & 0.5403    & 0.9272   & 0.0003                 & -0.0149                \\
72                         & 0.2564    & 0.5190    & 1.1865   & 0.2189    & 0.4867    & 0.6087   & 0.5105    & 0.8978    & 0.9660   & 1.1188    & 1.8853    & 2.0543   & 0.6320    & 0.7064    & 1.1830   & 0.0005                 & -0.0146                \\
73                         & 0.3384    & 0.7741    & 1.3467   & 0.2493    & 0.4600    & 0.9359   & 0.1618    & 0.3644    & 0.4621   & 1.1029    & 2.0296    & 2.3918   & 0.3245    & 0.5799    & 1.0733   & 0.0005                 & -0.0177                \\
74                         & 0.3201    & 0.8348    & 1.3631   & 0.2663    & 0.3856    & 0.7575   & 0.3115    & 0.3148    & 0.5241   & 1.6320    & 2.1945    & 2.4871   & 0.5233    & 0.6655    & 0.8155   & 0.0008                 & -0.0137                \\
75                         & 0.3393    & 0.5659    & 0.8701   & 0.4631    & 0.5129    & 0.6990   & 0.2164    & 0.6220    & 0.6942   & 2.0101    & 2.3466    & 2.5101   & 0.4732    & 0.5487    & 0.8021   & 0.0003                 & -0.0105                \\
76                         & 0.2837    & 0.5582    & 0.9444   & 0.3950    & 0.4421    & 0.5431   & 0.1751    & 0.4020    & 0.5619   & 1.6870    & 2.1174    & 2.1781   & 0.5018    & 0.6336    & 0.9021   & 0.0007                 & -0.0106                \\
77                         & 0.4445    & 0.5323    & 0.7503   & 0.2206    & 0.4059    & 0.8056   & 0.3763    & 0.3983    & 0.4701   & 1.8090    & 2.2843    & 2.6355   & 0.2987    & 0.7023    & 0.8220   & 0.0000                 & -0.0112                \\
78                         & 0.2737    & 0.8505    & 1.1569   & 0.1880    & 0.3924    & 0.7670   & 0.7138    & 0.7606    & 1.3360   & 1.4112    & 1.6786    & 2.0875   & 0.5226    & 0.7758    & 0.7810   & 0.0007                 & -0.0147                \\
79                         & 0.3165    & 0.5898    & 1.3197   & 0.1675    & 0.4617    & 0.6047   & 0.3838    & 0.5225    & 0.5382   & 0.9748    & 1.8538    & 2.4536   & 0.2708    & 0.6711    & 0.7817   & 0.0001                 & -0.0152                \\
80                         & 0.3461    & 0.6702    & 1.1763   & 0.3022    & 0.3520    & 0.4520   & 0.6569    & 0.6681    & 0.8353   & 1.2370    & 1.8857    & 2.0325   & 0.5897    & 0.9315    & 1.1326   & 0.0006                 & -0.0115                \\
81                         & 0.2459    & 0.4239    & 0.7045   & 0.4204    & 0.5318    & 0.8716   & 0.2177    & 0.5555    & 0.8846   & 0.4222    & 0.8135    & 0.9481   & 1.6865    & 1.9811    & 2.0368   & 0.0001                 & -0.0113                \\
82                         & 0.3403    & 0.7612    & 1.1092   & 0.4334    & 0.5144    & 0.8808   & 0.3328    & 0.4029    & 0.4705   & 0.2591    & 0.9665    & 1.1519   & 1.6442    & 1.6646    & 2.0056   & 0.0006                 & -0.0127                \\
83                         & 0.3354    & 0.3361    & 0.6449   & 0.3112    & 0.5724    & 0.7814   & 0.4403    & 0.6122    & 0.6972   & 0.4071    & 0.6715    & 0.7985   & 1.7579    & 1.8083    & 2.0356   & 0.0005                 & -0.0095                \\
84                         & 0.2541    & 0.3121    & 0.7328   & 0.2750    & 0.5147    & 0.8158   & 0.3647    & 0.4705    & 0.5364   & 0.3361    & 0.5745    & 0.8569   & 2.2493    & 2.4768    & 2.5976   & 0.0001                 & -0.0100                \\
85                         & 0.2563    & 0.4254    & 1.0037   & 0.4894    & 0.7675    & 0.8830   & 0.2683    & 0.5378    & 0.7885   & 0.2721    & 0.6194    & 0.9705   & 1.2170    & 1.9553    & 2.0052   & 0.0000                 & -0.0133                \\
86                         & 0.1603    & 0.5165    & 1.1611   & 0.4163    & 0.5575    & 0.8069   & 0.3047    & 0.5098    & 0.9244   & 0.2789    & 0.8314    & 0.9753   & 1.7725    & 2.2779    & 2.4726   & 0.0007                 & -0.0148                \\
87                         & 0.3161    & 0.4044    & 0.7211   & 0.3770    & 0.7031    & 0.8956   & 0.4161    & 0.5143    & 0.5263   & 0.3237    & 0.6373    & 1.0797   & 1.9258    & 2.2111    & 2.2455   & 0.0003                 & -0.0111                \\
88                         & 0.4376    & 0.5056    & 0.7913   & 0.2568    & 0.5165    & 0.9247   & 0.2155    & 0.4740    & 0.4888   & 0.3181    & 0.5118    & 0.8131   & 2.0283    & 2.1050    & 2.2869   & 0.0001                 & -0.0106                \\
89                         & 0.2895    & 0.3685    & 0.7559   & 0.2739    & 0.5406    & 0.7596   & 0.3940    & 0.5141    & 0.5443   & 0.2570    & 0.7168    & 0.8646   & 1.8757    & 1.9357    & 2.2471   & 0.0004                 & -0.0105                \\
90                         & 0.2170    & 0.3617    & 0.5975   & 0.2808    & 0.5693    & 0.8554   & 0.1646    & 0.4508    & 0.4667   & 0.2827    & 0.7997    & 0.9626   & 1.6258    & 1.6817    & 2.0345   & 0.0001                 & -0.0116                \\
91                         & 0.2638    & 0.4569    & 0.7386   & 0.2954    & 0.6196    & 0.7953   & 0.3071    & 0.4220    & 0.5941   & 0.3591    & 0.5705    & 1.0592   & 1.8357    & 1.9041    & 2.0103   & 0.0001                 & -0.0109                \\
92                         & 0.3158    & 0.5257    & 1.1027   & 0.2588    & 0.5364    & 0.7614   & 0.4068    & 0.6060    & 1.0271   & 0.3747    & 0.9332    & 1.1276   & 1.9984    & 2.8086    & 2.8937   & 0.0001                 & -0.0143                \\
93                         & 0.4308    & 0.5728    & 0.8886   & 0.2888    & 0.5566    & 0.8214   & 0.2593    & 0.5867    & 0.7546   & 0.6151    & 0.9165    & 1.1822   & 1.6965    & 1.9645    & 2.1680   & 0.0004                 & -0.0126                \\
94                         & 0.1748    & 0.4155    & 0.8630   & 0.2779    & 0.5554    & 0.8499   & 0.1869    & 0.5531    & 1.1592   & 0.3854    & 0.7768    & 0.8426   & 1.7063    & 2.0255    & 2.0379   & 0.0008                 & -0.0141                \\
95                         & 0.2196    & 0.3065    & 0.5833   & 0.3354    & 0.6263    & 1.0339   & 0.3134    & 0.5304    & 0.7728   & 0.5450    & 0.8895    & 1.1553   & 1.6988    & 2.0882    & 2.1519   & 0.0001                 & -0.0122                \\
96                         & 0.1547    & 0.5107    & 0.9194   & 0.3245    & 0.5424    & 0.9488   & 0.1901    & 0.3365    & 0.5089   & 0.3243    & 0.6567    & 1.0052   & 1.7396    & 1.9338    & 2.0386   & 0.0005                 & -0.0131                \\
97                         & 0.3071    & 0.3632    & 0.7293   & 0.2860    & 0.7325    & 0.7813   & 0.2021    & 0.4419    & 0.7421   & 0.5488    & 0.7767    & 0.9969   & 2.0031    & 2.2271    & 2.2947   & 0.0008                 & -0.0119                \\
98                         & 0.2236    & 0.3228    & 0.5170   & 0.5646    & 0.6119    & 0.8154   & 0.5764    & 0.7404    & 0.7591   & 0.4923    & 0.8150    & 0.9543   & 2.1270    & 2.3045    & 2.4905   & 0.0002                 & -0.0084                \\
99                         & 0.2402    & 0.6636    & 1.4234   & 0.3437    & 0.7044    & 0.8298   & 0.4224    & 0.5731    & 1.1834   & 0.3568    & 1.1051    & 1.2835   & 1.4515    & 1.8981    & 2.0670   & 0.0014                 & -0.0180                \\
100                        & 0.1564    & 0.5005    & 1.0582   & 0.2623    & 0.5409    & 0.8707   & 0.2704    & 0.5185    & 0.5988   & 0.7348    & 0.8885    & 1.4047   & 2.1797    & 2.5681    & 2.6065   & 0.0001                 & -0.0133                \\ \hline
\end{tabular}}
\label{tab:Utility_part2}
\end{table}

%
%
%






\end{document}